\documentclass[runningheads]{llncs}
\usepackage[T1]{fontenc}
\usepackage{graphicx,verbatim}
\usepackage{graphicx}
\usepackage{amsmath}
\usepackage{amssymb}
\usepackage{tabularx}
\usepackage{booktabs}
\usepackage[dvipsnames,table,xcdraw]{xcolor}
\usepackage[utf8]{inputenc} %
\usepackage[T1]{fontenc}    %
\usepackage{url}            %
\usepackage{amsfonts}       %
\usepackage{nicefrac}       %
\usepackage{microtype}      %
\usepackage{multirow}
\usepackage{bbm}
\usepackage{rotating}
\usepackage[english]{babel}
\usepackage{pifont}%
\usepackage{comment}
\usepackage{tikz}
\usetikzlibrary{spy,backgrounds}
\usepackage{soul}
\usepackage{algorithm}
\usepackage{algpseudocode}
\usepackage{xspace}
\usepackage{capt-of} %
\usepackage{listings}
\usepackage{tcolorbox}

\definecolor{green}{RGB}{0,150,10}

\newcommand{\mysection}[1]{\noindent\textbf{#1.}}

\newlength\mytmplen

\newcommand{\methodname}{Colon-Bench\xspace}

\usepackage{color}
\definecolor{cvprblue}{rgb}{0.21,0.49,0.74}

\usepackage[pagebackref,breaklinks,colorlinks,allcolors=cvprblue]{hyperref}
\begin{document}
\title{Colon-Bench: An Agentic Workflow for Scalable Dense Lesion Annotation in Full-Procedure Colonoscopy Videos}
\titlerunning{Colon-Bench: An Agentic Workflow for Colonoscopy Videos}
\author{Abdullah Hamdi \and Changchun Yang \and Xin Gao}
\authorrunning{H. Abdullah et al.}
\institute{King Abdullah University of Science and Technology (KAUST) \\
\email{\{abdullah.hamdi@kaust.edu.sa\}}
}
  
\maketitle              %

\begin{abstract}
Early screening via colonoscopy is critical for colon cancer prevention, yet developing robust AI systems for this domain is hindered by the lack of densely annotated, long-sequence video datasets. Existing datasets predominantly focus on single-class polyp detection and lack the rich spatial, temporal, and linguistic annotations required to evaluate modern Multimodal Large Language Models (MLLMs). To address this critical gap, we introduce Colon-Bench, generated via a novel multi-stage agentic workflow. Our pipeline seamlessly integrates temporal proposals, bounding-box tracking, AI-driven visual confirmation, and human-in-the-loop review to scalably annotate full-procedure videos. The resulting verified benchmark is unprecedented in scope, encompassing 528 videos, 14 distinct lesion categories (including polyps, ulcers, and bleeding), over 300,000 bounding boxes, 213,000 segmentation masks, and 133,000 words of clinical descriptions. We utilize Colon-Bench to rigorously evaluate state-of-the-art MLLMs across lesion classification, Open-Vocabulary Video Object Segmentation (OV-VOS), and video Visual Question Answering (VQA). The MLLM results demonstrate surprisingly high localization performance in medical domains compared to SAM-3. Finally, we analyze common VQA errors from MLLMs to introduce a novel "colon-skill" prompting strategy, improving zero-shot MLLM performance by up to 9.7\% across most MLLMs. The dataset and the code are available at ~\url{https://abdullahamdi.com/colon-bench}
\end{abstract}

\section{Introduction} \label{sec:introduction}

Colon cancer is the second leading cause of cancer death worldwide \cite{Sung2021GlobalCancerStats}, and many of these deaths are preventable with early screening and removal of precancerous lesions; however, colonoscopy remains expensive, invasive, and logistically difficult to arrange, requiring specialized clinicians and often long procedures (up to two hours) that increase facility and staffing costs. Automated AI analysis promises to ease these burdens, but lesions are sparse and frequently obscured by motion blur, occlusion, debris, stool, or fluids, and by camera contact with the colon wall, making dense visual analysis challenging. Moreover, manual dense annotation of colonoscopy videos is labor-intensive and inconsistent, which motivates our work on a scalable, affordable pipeline for dense colonoscopy video annotation, facilitating AI research, evaluation, and applications in colonoscopy.

To address the visual challenges of colonoscopy videos, recent methods specialize architectures for pixel-level precision and subtle lesion identification \cite{zeng2025segment,li2025survey}. Because procedures are long, efficient spatiotemporal analysis is also critical; state-space models \cite{tian2025endomamba} and token merging \cite{wang2025improving} have been introduced to capture long-range dependencies while reducing redundancy. Many methods rely on self-supervised pretraining \cite{jong2025gastronet} or use limited annotated examples with text \cite{He2025EndoCLIP}, bounding boxes \cite{choudhuri2025polypsegtrack,li2021colonoscopy}, or segmentation masks \cite{jha2020kvasirseg,choudhuri2025polypsegtrack}. These workarounds highlight the gap our work addresses: the need for a comprehensive, densely annotated, long-sequence dataset to ground spatiotemporal analysis of real colonoscopies. 

REAL-COLON \cite{biffi2024realcolon} introduced 60 long colonoscopy sequences labeled with 351,264 polyp bounding boxes, while CAS-COLON \cite{song2025cascolon} provides anatomical temporal classification over 78 videos, 9 hours, and around 2 million images. Other datasets are either too small for comprehensive training and evaluation \cite{ma2021ldpolypvideo} or large but limited in task scope and lesion diversity \cite{jha2020kvasirseg,misawa2021development,ma2021ldpolypvideo,ali2023multi}. EndoBench \cite{liu2026endobench}, a contemporary related work, evaluates \textit{static-image} VQA in general endoscopy. As Table~\ref{tbl:dataset_comparison} shows, \methodname provides dense lesion annotations with text descriptions and segmentation mask tracklets across diverse video clips, enabling benchmarks for Video Question Answering, lesion clip classification, and Open-Vocabulary Video Segmentation in colonoscopy videos.

\begin{table}[t]
\centering
\caption{\textbf{Comparison of Colonoscopy Datasets.} Relative to prior datasets that emphasize single-class polyp detection or anatomical segmentation, \textit{\methodname} provides a broader lesion taxonomy and richer supervision. It combines long-sequence coverage with dense boxes, masks, and clinical text, enabling multi-task evaluation across detection, video segmentation, and language-based understanding.}
\resizebox{1.0\linewidth}{!}{
\tabcolsep=0.1cm
\begin{tabular}{l|cccccc}
\toprule
\textbf{Attribute} & \textbf{Kvasir-SEG} \cite{jha2020kvasirseg} & \textbf{SUN} \cite{misawa2021development} & \textbf{PolypGen} \cite{ali2023multi} & \textbf{REAL-Col.} \cite{biffi2024realcolon} & \textbf{CAS-Col.} \cite{song2025cascolon} & \textbf{\methodname (Ours)} \\
\midrule
Year & 2020 & 2021 & 2023 & 2024 & 2025 & 2026 \\
\rowcolor[HTML]{EFEFEF} 
Focus & Segmentation & Detection & Segmentation & Detection & Anatomy & \textbf{Lesion, Video Seg., \& Text} \\
Number of Videos & N/A & N/A & N/A & 60 (Full-Procedure) & 78 (Clip) & \textbf{528 (Clip)} \\
\rowcolor[HTML]{EFEFEF} 
Total Frames/Images & 1,000 & 158,690 & 6,282 & \textbf{2,757,723} & 1,961,100 & 464,035 \\
Lesion Classes & 1 (Polyp) & 1 (Polyp) & 1 (Polyp) & 1 (Polyp) & N/A & \textbf{14 (Bleeding, Polyps, ...)} \\
\rowcolor[HTML]{EFEFEF} 
Bounding Boxes & N/A & 158,690 & 6,282 & \textbf{351,264} & N/A & 300,132 \\
Segmentation Masks & 1,000 & N/A & 6,282 & N/A & Yes & \textbf{213,067} \\
\rowcolor[HTML]{EFEFEF} 
Language Descriptions & N/A & N/A & N/A & N/A & N/A & \textbf{Yes (133k Words)} \\
\bottomrule
\end{tabular}}
    \label{tbl:dataset_comparison}
\end{table}

\begin{figure}[t]
  \centering
  \includegraphics[width=0.98\textwidth,trim=0 0 0 0,clip]{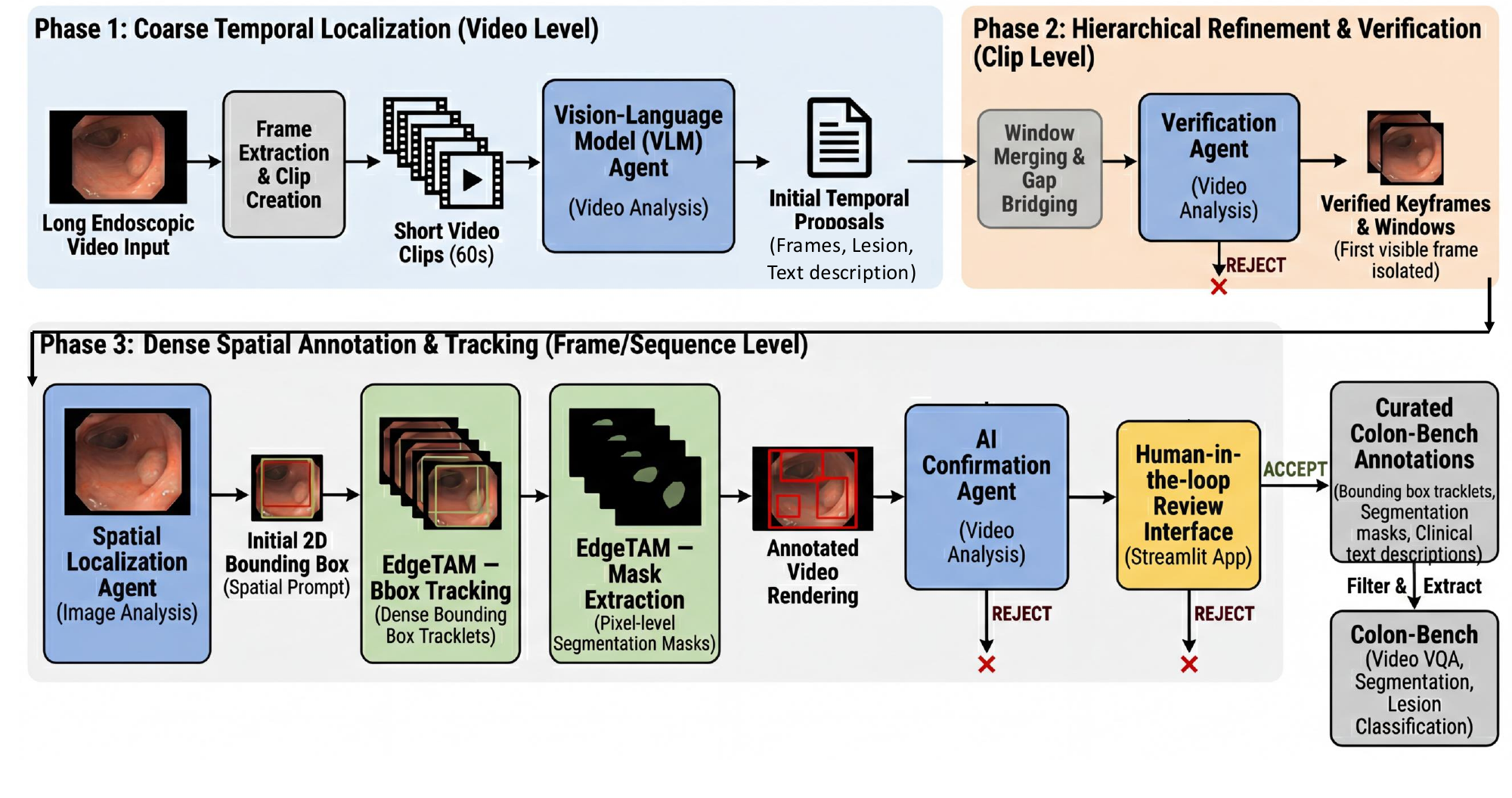}
\caption{\textbf{\methodname Annotation Pipeline.} Overview of the multi-stage agentic workflow used to build the dataset, from VLM proposal generation through verification, spatial tracking, AI confirmation, and clinician review. The accepted annotations support multiple video tasks: Visual Quesion Answering (VQA), binary lesion classification, and Open-Vocabulary Video Object Segmentation (OV-VOS).}
  \label{fig:pullfigure}
\end{figure}

Large Language Models (LLMs) \cite{gpt4,gemini2024gemini15} and foundational vision-language models \cite{CLIP} have accelerated Multimodal Large Language Models (MLLMs) for complex visual reasoning \cite{bai2025qwen3,deitke2025molmo,ge2023planting,glm4_2024,gemini2024gemini15}. Despite strong zero-shot performance in general domains, their ability to interpret long, occluded, noisy medical sequences remains largely untested, motivating \textit{\methodname} as a systematic benchmark for colonoscopy video understanding. Recent work on agentic MLLM pipelines for long-form videos \cite{fan2024videoagent,wang2024internvid} and hybrid automatic labeling with human verification for large medical image datasets \cite{chambon2024chexpert} further motivates our pipeline in Fig. \ref{fig:pullfigure}. 

\noindent\textbf{Contributions:} 
\textbf{(i)} We introduce a novel, multi-stage agentic workflow for dense annotation of full-procedure colonoscopy videos. By combining temporal proposals, bounding-box tracking, AI-driven confirmation with visual cues, and an efficient human-in-the-loop review interface, our pipeline reduces manual labor while yielding high-quality spatial, temporal, and textual annotations.
\textbf{(ii)} We present a comprehensive, multi-task video benchmark (\textit{\methodname}) for colonoscopy understanding. Spanning 14 distinct lesion categories (including polyps, bleeding, and ulcers), it includes over 464k frames, 300k bounding boxes, 213k segmentation masks, and 133k words of clinical descriptions, all verified by humans and an experienced surgeon. It supports four rigorous evaluation tasks: binary lesion classification, Open-Vocabulary Video Object Segmentation (OV-VOS), and two difficulty tiers of Video Visual Question Answering (VQA).
\textbf{(iii)} We conduct a large-scale evaluation of lesion-focused colonoscopy visual reasoning and localization in frontier MLLMs. By analyzing cross-model error patterns, we propose a novel ``colon-skill'' that defeines a textual guidance prompt, improving most MLLM performance by up to 9.7\% without additional training.

\section{Methodology} \label{sec:methodology}
\subsection{Agentic Workflow for Colonoscopy Annotation } \label{sec:pipeline}
\begin{table}[t]
\centering
\caption{\textbf{Annotation Pipeline Stages.} We show \methodname annotations after each major processing stage. As automated filters and human review progressively discard false-positive windows (reducing total frames), the temporal precision, F1 score, and specificity of the automatically identified lesions demonstrably increase on the surrogate polyp labels of REAL-COLON \cite{biffi2024realcolon} (polyps are one type of lesion in \methodname).}
\label{tab:pipeline_master}
\resizebox{0.93\linewidth}{!}{
\begin{tabular}{l | c | c | c | c}
\toprule
& \textbf{Initial VLM} & \textbf{+ verification} & \textbf{+ Cued AI} & \textbf{Final Human-} \\
\textbf{Metric} & \textbf{Proposals} & \textbf{Agent Filtering} & \textbf{Confirmation Agent} & \textbf{Curated Annots.} \\
\midrule
Total Windows          & 1,325   & 903     & 597     & \textbf{528}     \\
Total Frames           & 826,763 & 648,440 & 492,606 & \textbf{464,035} \\
Duration (Hours)       & 22.97   & 18.01   & 13.68   & \textbf{12.89}   \\
\midrule
Total Bounding Boxes   & -       & -       & 314,408 & \textbf{300,132} \\
Total Seg. Masks       & -       & -       & 227,343 & \textbf{213,067} \\
Total Clinical Words   & -       & -       & 145,515 & \textbf{133,289} \\
\midrule
Precision              & 30.9    & 36.2    & 53.1    & \textbf{55.4}    \\
Recall                 & 67.5    & 68.7    & 48.8    & \textbf{48.6}    \\
F1 Score               & 42.4    & 47.4    & 50.8    & \textbf{51.8}    \\
Specificity            & 78.6    & 82.9    & 93.9    & \textbf{94.4}    \\
\midrule
\end{tabular}
}
\end{table}

\mysection{Automatic annotations with quality control}
As illustrated in Fig.~\ref{fig:pullfigure}, we built dense colonoscopy annotations by applying a multi-stage agentic pipeline to 60 REAL-COLON videos \cite{biffi2024realcolon}. An initial vision-language model proposed 1,325 candidate lesion windows (22.97 hours), which were progressively refined by verification filtering, EdgeTAM \cite{zhou2025edgetam} bounding-box tracking, cued AI confirmation using lesion-box overlays, and final human review (Table~\ref{tab:pipeline_master}).
Tracking and AI confirmation produced the spatial annotations (over 314k initial bounding boxes) while pruning weak temporal boundaries. Stage-wise evaluation on REAL-COLON surrogate polyp labels shows that the automated AI stages provide the main precision, F1, and specificity gains; because many \methodname categories (ulcers, bleeding, angiectasia, etc.) are absent from REAL-COLON, these labels serve as quality control rather than exhaustive recall targets. Motivated by Gemini-3's recent medical-domain performance \cite{bose2026multi}, we used Gemini-2.5-flash-lite for proposals, Gemini-3-pro for video verification, and Gemini-3-flash for bounding boxes, balancing performance and cost (Table~\ref{tab:combined_results}).

\mysection{Human review}
Human review served as the final quality gate. To enable efficient review at scale, the pipeline pre-rendered short clips with spatial overlays into an interactive web interface. A reviewing physician rejected only 69 of the 597 presented windows (11.6\%), demonstrating strong agreement with the AI filters. Ultimately, the pipeline retained 528 curated windows (39.8\% retention) spanning 464,035 frames. The final dataset yields a dense, multi-modal benchmark comprising 300,132 bounding boxes, 213,067 segmentation masks, and over 133k words of verified clinical text (averaging 252.4 words per window).

\begin{figure}[t]
  \centering
  \includegraphics[width=0.8\linewidth]{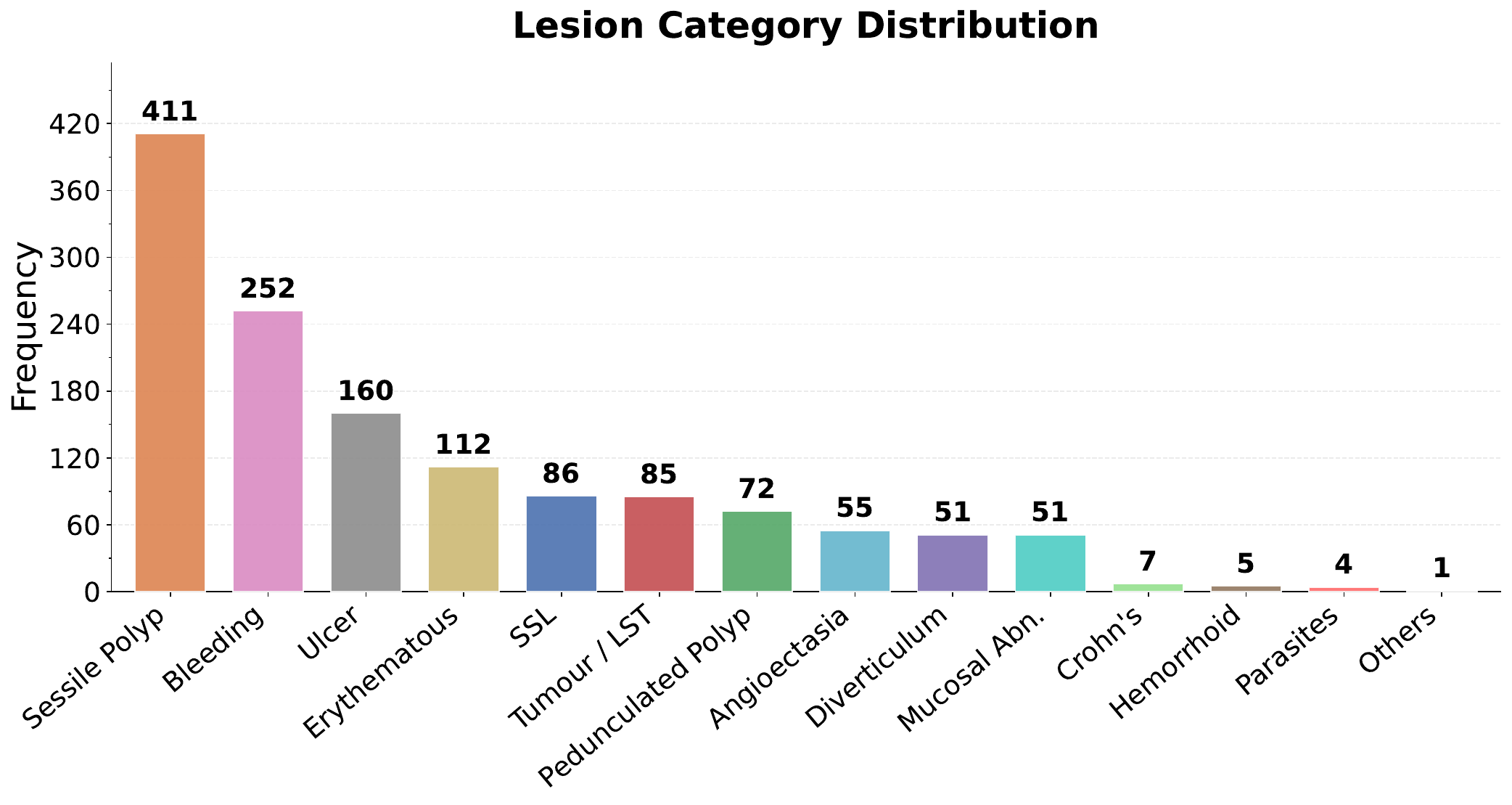}
  \caption{%
    \textbf{Lesion Category Distribution.} Long-tailed lesion category distribution in \methodname, highlighting the diversity of lesions.%
  }
  \label{fig:lesion_distribution}
\end{figure}

\mysection{Lesion types}
The dataset spans 14 lesion categories identified through multi-label keyword matching over clinician-verified text fields (Fig.~\ref{fig:lesion_distribution}). The distribution is heavily long-tailed: sessile polyps dominate (411 windows), followed by bleeding (252), ulcers (160), and erythematous lesions (112). Because Colon-becnh descriptions are free-form text (one window can have multiple lesions), a single review window may contribute to more than one category.

\begin{table*}[t]
\centering
\caption{\textbf{Colon-Bench Combined Results.} Summary of model performance across the four benchmark tasks: visual prompted and unprompted video Visual Question Answering (VQA) accuracy (with visual box prompts and without them), binary lesion classification precision/recall/F1, and open-vocabulary video segmentation IoU/Dice. The video segemtnation is based on 3 box detections for each MLLM prompting the same EdgeTAM tracker \cite{zhou2025edgetam}. Best scores per metric are shown in \textbf{bold}.}
\label{tab:combined_results}
\resizebox{0.99\textwidth}{!}{%
\tabcolsep=0.18cm
\begin{tabular}{l cc ccc cc}
\toprule
& \multicolumn{2}{c}{\textbf{VQA Accuracy}} & \multicolumn{3}{c}{\textbf{Lesion Cls.}} & \multicolumn{2}{c}{\textbf{Segmentation}} \\
\cmidrule(lr){2-3} \cmidrule(lr){4-6} \cmidrule(lr){7-8}
Model & Prompted & Unprompted & Precision & Recall & F1 & IoU & Dice \\
\midrule
CLIP~\cite{CLIP}                 & -            & -            & 34.5          & \textbf{100} & 51.3          & -            & - \\
ViCLIP~\cite{wang2024internvid}                & -            & -            & 34.8          & 98.5           & 51.4          & -            & - \\
Endo-CLIP~\cite{He2025EndoCLIP}             & -            & -            & 41.9          & 95.2           & 58.2          & -            & - \\
Colon-ViCLIP~\cite{wang2024internvid}         & -            & -            & 49.0          & 84.2           & 62.0          & -            & - \\ \midrule
SAM 3~\cite{carion2025sam}        & -            & -            & -            & -             & -            & 2.5           & 2.9 \\
GPT-4o                & -            & -            & -            & -             & -            & 0.5           & 0.8 \\
Claude Opus 4.6       & -            & -            & -            & -             & -            & 16.1          & 20.5 \\
\midrule
Qwen3-VL 8B           & 32.9          & 38.3          & 34.4          & \textbf{100} & 51.2          & 10.4          & 13.1 \\
Seed 1.6 Flash        & 38.1          & 45.4          & \textbf{94.2} & 24.3           & 38.6          & 2.6           & 3.5 \\
Qwen-VL Max           & 39.1          & 45.4          & 0.0           & 0.0            & 0.0           & 25.6          & 29.6 \\
Qwen3-VL 32B          & 39.3          & 44.4          & 0.0           & 0.0            & 0.0           & 12.7          & 15.9 \\
Qwen3.5 Plus          & 44.3          & 60.5          & 36.2          & 24.6           & 29.3          & 16.7          & 21.0 \\
Molmo 2-8B            & 46.1          & 53.4          & 52.9          & 46.7           & 49.6          & 2.6           & 3.8 \\
Qwen3-VL 235B         & 46.7          & 56.6          & 22.2          & 0.7            & 1.4           & 13.6          & 16.9 \\
Gemini 2.5 F. Li. & 47.8          & 46.8          & 42.3          & 95.9           & 58.7          & 19.9          & 24.3 \\
Qwen3.5 397B          & 49.0          & 60.1          & 10.0          & 0.4            & 0.7           & 16.6          & 21.0 \\
GLM-4.6V              & 55.7          & 53.8          & 46.5          & 94.1           & 62.2          & 12.5          & 16.1 \\
Seed 1.6              & 62.9          & 72.0          & 85.0          & 58.7           & 69.4          & 12.6          & 16.1 \\
GPT-5.2               & 62.0	           & 71.4            & 74.2	            & 79.4	             & 76.7            & 30.7          & 36.5 \\
GPT-5.4               & 59.4	           & 70.8            & 77.0	            & 78.7	            & 77.8            & 34.5          & 41.1 \\
Gemini 3.1 F. Li.     & 69.2          & 67.7          & 72.6          & 90.8           & \textbf{80.7}          & 37.4          & 43.4 \\

Gemini 3 Flash        & 76.6          & 76.0          & 55.3          & 97.1           & 70.5          & \textbf{48.3} & \textbf{54.7} \\
Gemini 3 Pro          & \textbf{78.6} & \textbf{82.5} & 66.1          & 93.0           & 77.3 & 45.0          & 51.3 \\
\bottomrule
\end{tabular}
}
\end{table*}
\subsection{\methodname Evaluation Benchmark} \label{sec:becnh}
\mysection{Filtering}
\methodname is a multi-task video benchmark for colonoscopy understanding comprising 1{,}597 clips from 60 patients (955{,}126 frames), spanning five tasks: binary classification (790 clips), detection (272 clips, 61{,}538 per-frame bounding boxes), instance segmentation (264 clips, 57{,}550 per-frame masks), and VQA at two difficulty tiers. The binary lesion classes and segmentation masks follow directly from Sec.~\ref{sec:pipeline} to establish the first colonoscopy Open-Vocabulary Video Object Segmentation (OV-VOS) benchmark.

\mysection{VQA (prompted \& unprompted) }
The \emph{prompted} VQA split contains 1{,}485 five-choice questions over 499 clips whose videos include bounding-box overlays on confirmed lesion windows; the \emph{unprompted} split contains 2{,}740 questions over 918 clips rendered from raw frames, fully encompassing all detection and segmentation videos. Across both splits the questions are dominated by lesion type (39\%/43\%, prompted/unprompted), color \& surface appearance (38\%/44\%), and anatomical location (30\%/35\%), followed by morphology (20\%/20\%), with explicit temporal references largely confined to the prompted split (59\% vs.\ 2\%). For instance, an unprompted question for the clip in Fig.~\ref{fig:qualitative-examples} first row asks:
\textit{``How is the morphology and anatomical location of the identified lesion described?''}
with distractors spanning pedunculated, depressed, and flat lesions at various anatomical sites;
the correct answer is \textit{(D)~a sessile polyp on a haustral fold}.

\mysection{Debiasing \methodname}
For each clip, three five-way MCQs are generated with Gemini-3, covering lesion identification, clinical characteristics, and temporal reasoning.
To mitigate text-only shortcuts, we apply two-stage debiasing: (i)~adversarial distractors are regenerated in a separate LLM call receiving only the stem and correct answer; (ii)~a blind text-only stress-test identifies and reverts any newly introduced bias.
Post-debiasing blind accuracy is $44.6\%$ (prompted) and $37.1\%$ (unprompted) vs.\ the $20\%$ random baseline, with residual margins reflecting dataset priors rather than surface cues. Up to 10\% of the VQA pairs were manually checked by the authors after debiasing with 98\% correctness.

\mysection{Baselines}
Methods reported in this work include a set of Multimodal Large Language Models (MLLMs) such as different variants of GPTs, Gemini, Qwen, Seed, GLM, and Molmo \cite{gpt4,bai2025qwen3,deitke2025molmo,ge2023planting,glm4_2024,gemini2024gemini15}. Also we report (for some corresponding benchmarks) SAM-3 \cite{carion2025sam}, CLIP \cite{CLIP}, Video CLIP (ViCLIP) \cite{wang2024internvid} and its fine-tuned version on Colon-Bench text (Colon-CLIP), and a recent model Endo-CLIP \cite{He2025EndoCLIP}. Modles that do not have video API support (\textit{E.g.} GPT-series) are only evluated on the segemntation task. All models in segemtnation (Except SAM-3) are based on 3-frame box detections prompting EdgeTAM tracker \cite{zhou2025edgetam}.

\subsection{\methodname Results}
\begin{figure*}[t]
  \centering
  \includegraphics[width=0.96\linewidth]{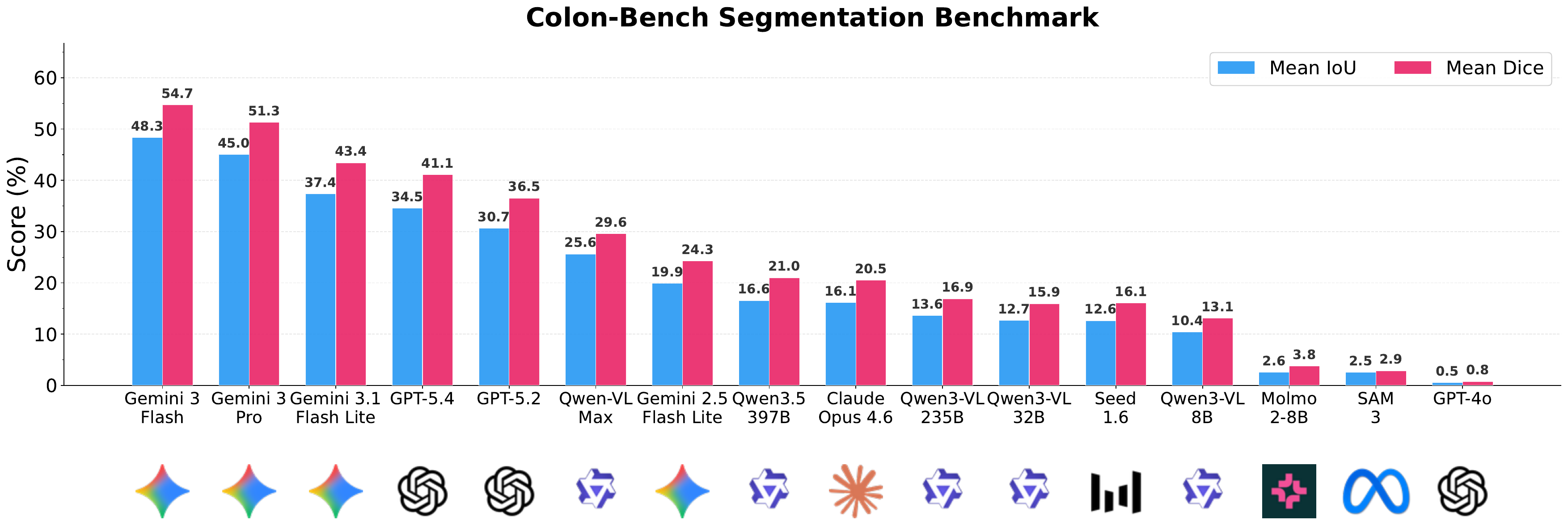}
  \caption{%
    \textbf{Multi-Modal Large Language Models for Lesion Segmentation.} Video object segmentation results on the 272-video benchmark, where MLLMs \cite{gpt4,bai2025qwen3,deitke2025molmo,ge2023planting,glm4_2024,gemini2024gemini15} first localize the target lesion with 3 bounding boxes and a EdgeTAM tracker \cite{zhou2025edgetam} propagates masks.
  }
  \label{fig:segmentation_metrics}
\end{figure*}

Table~\ref{tab:combined_results} reveals a clear performance hierarchy across all four Colon-Bench tasks.
Gemini~3 Pro and Gemini~3 Flash consistently dominate, achieving the highest VQA accuracy on both prompted and unprompted splits and the best segmentation quality (IoU/Dice of 45-48\%/51-55\%), while Gemini~3.1 Flash Lite leads in classification F1 (80.7\%).
Among open-weight models, Seed~1.6 is the strongest overall, ranking third in VQA and delivering competitive classification F1 (69.4\%) with the highest precision among full-benchmark models (85.0\%).
The Qwen family shows mixed behavior: larger variants (Qwen3.5~397B, Qwen3-VL~235B) perform well on VQA yet nearly collapse on classification (R\,$<$\,1\%, F1\,$<$\,2\%), suggesting they default to the majority (negative) benign class rather than detecting lesion presence.
For segmentation, performance drops sharply beyond the top three models.
Overall, our \methodname demonstrates how these strong MLLM models can be used for lesion detection in short clips, beating specialized models like Endo-CLIP \cite{He2025EndoCLIP} by 30\%. For the task of open-vocabulary lesion segmentation in video clips, an MLLM like GPT-5.4 (when combined with box-promptable EdgeTAM tracker \cite{zhou2025edgetam}) beats SAM-3 \cite{carion2025sam} by 32.0\% mIoU (see Fig.~\ref{fig:segmentation_metrics}).

\newcommand{\snap}[1]{\includegraphics[width=\linewidth,keepaspectratio]{#1}}
\newcommand{\colhead}[1]{%
  \colorbox{white}{%
    \parbox[c][1.6\baselineskip][c]{\linewidth}{\centering\small\bfseries #1}%
  }%
}
\newcommand{\snapcell}[1]{\includegraphics[width=\linewidth,keepaspectratio]{#1}}
\begin{figure*}[!t]
  \centering
  \begin{minipage}{\textwidth}
  \setlength{\tabcolsep}{1pt}
  \renewcommand{\arraystretch}{0.6}
  {\setlength{\tabcolsep}{1pt}%
  \renewcommand{\arraystretch}{1.0}%
    \resizebox{\textwidth}{!}{%
  \begin{tabular}{*{9}{p{\dimexpr\textwidth/9\relax}}}
    \colhead{Input} &
    \colhead{GT} &
    \colhead{GPT-5.2} &
    \colhead{Opus 4.6} &
    \colhead{Gemini 3 Flash} &
    \colhead{Qwen 3.5-Plus} &
    \colhead{Qwen VL-Max} &
    \colhead{Qwen 3 VL-235B} &
    \colhead{SAM\,3} \\
  \end{tabular}}\par}
  \vspace{2pt}
  \resizebox{\textwidth}{!}{%
  \begin{tabular}{ccccccccc}
    \snap{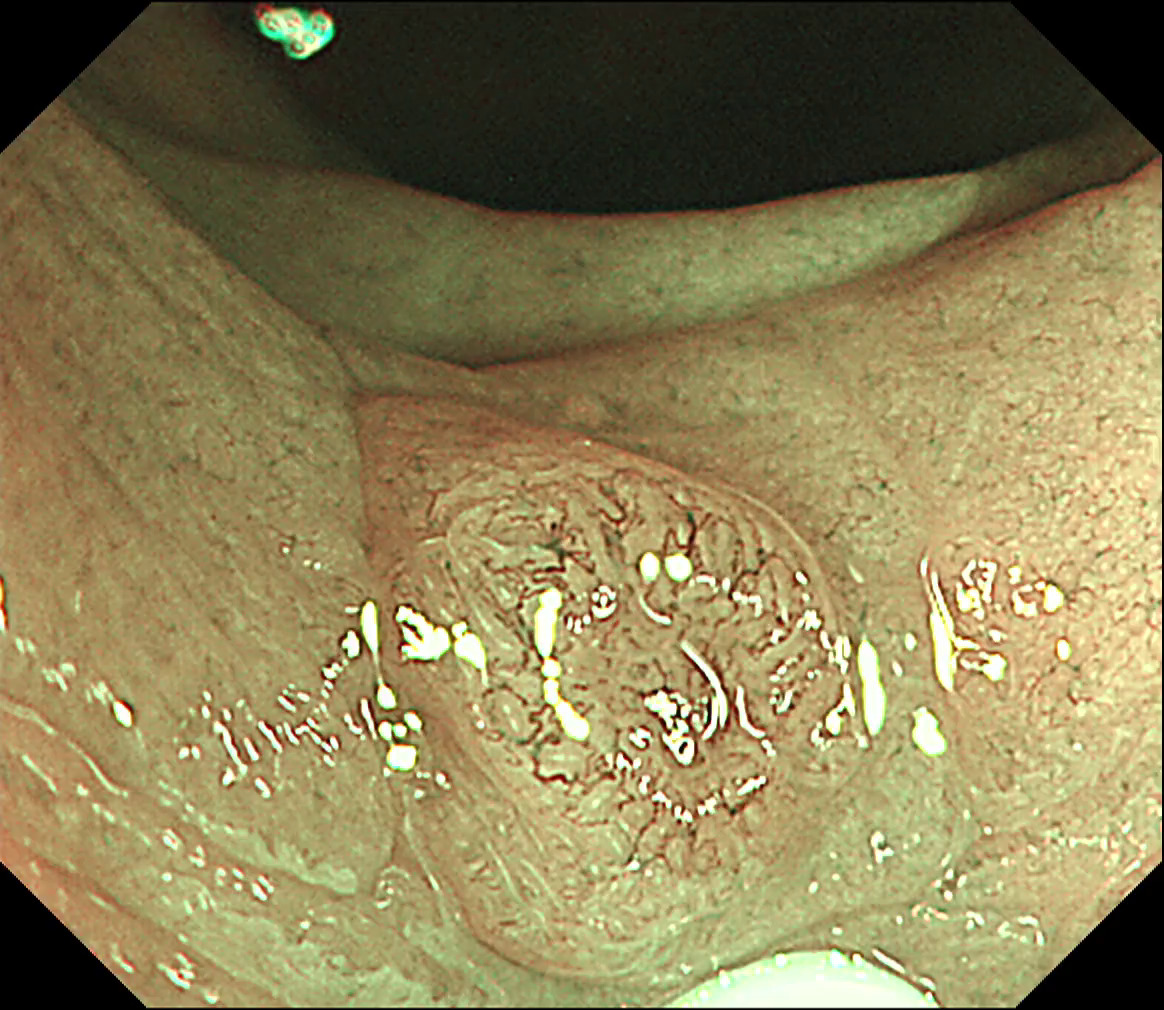} &
    \snap{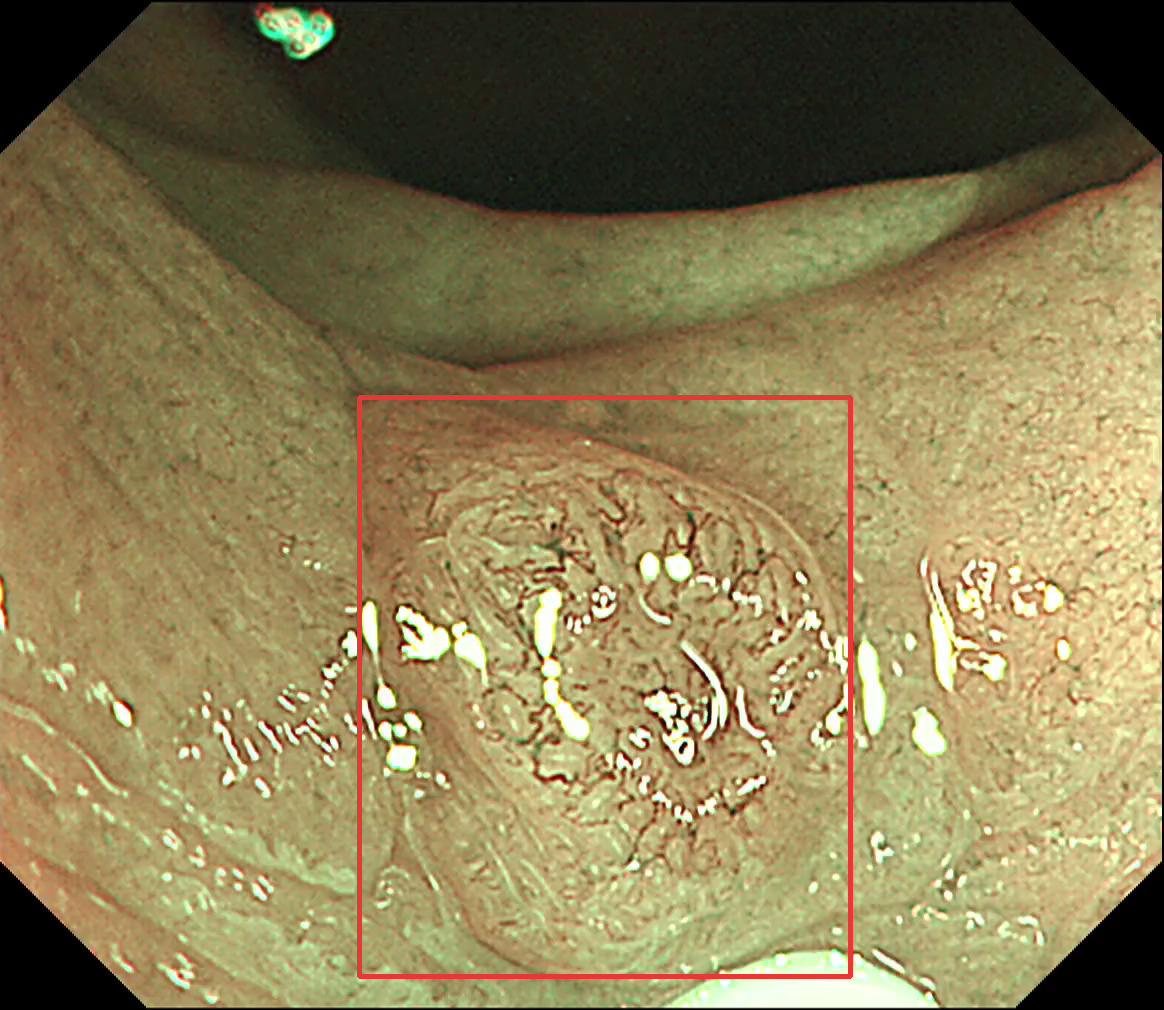} &
    \snap{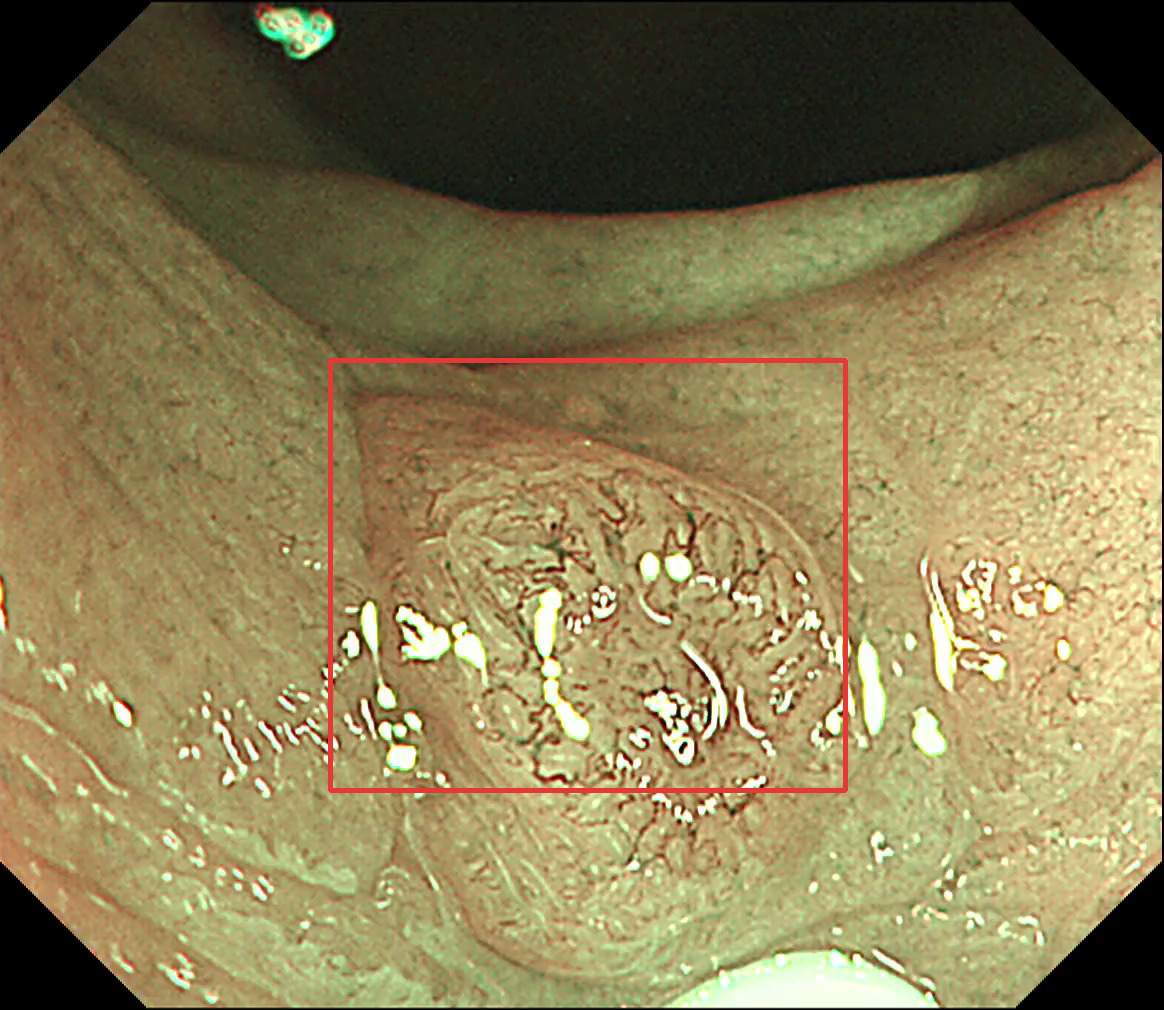} &
    \snap{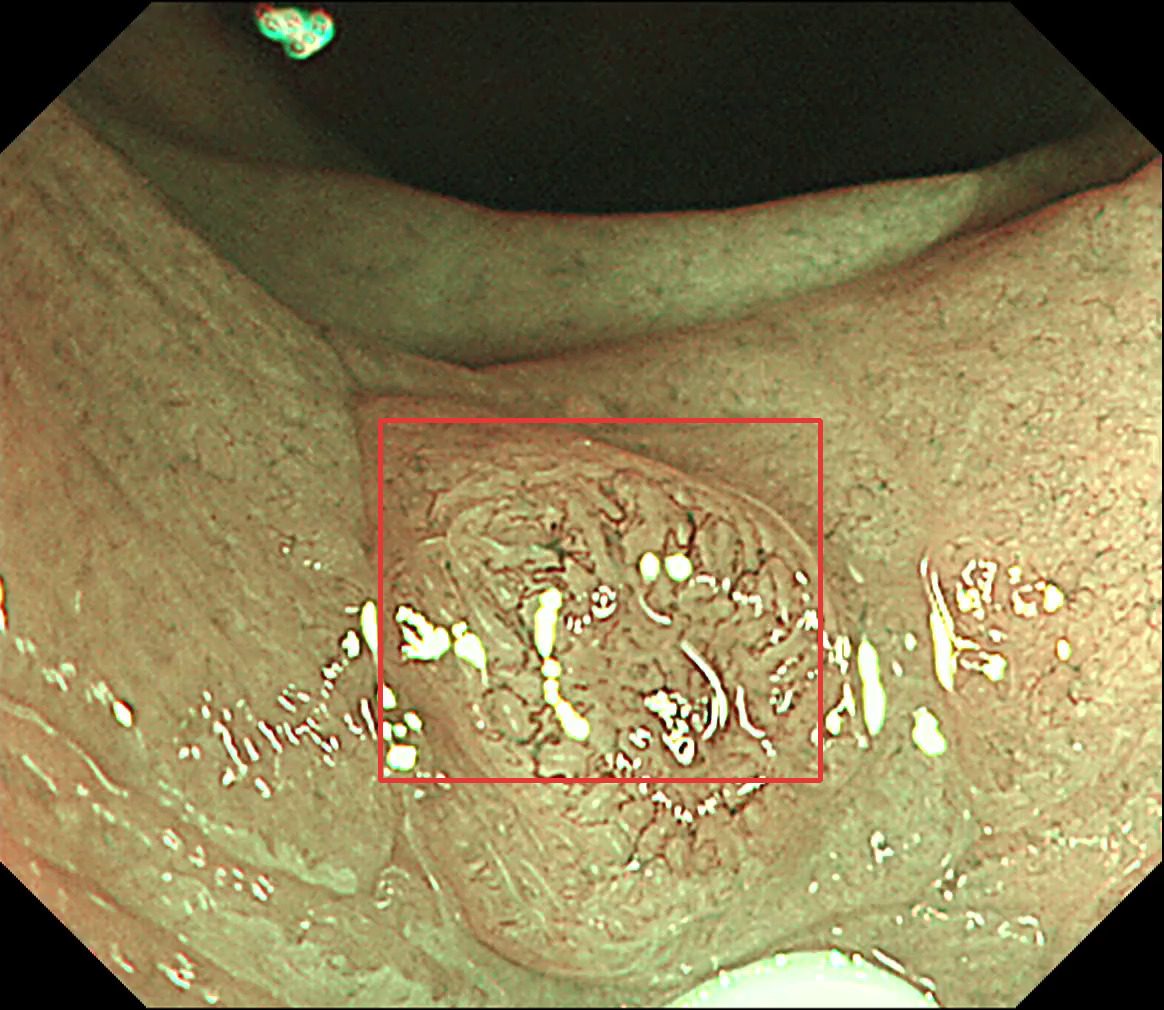} &
    \snap{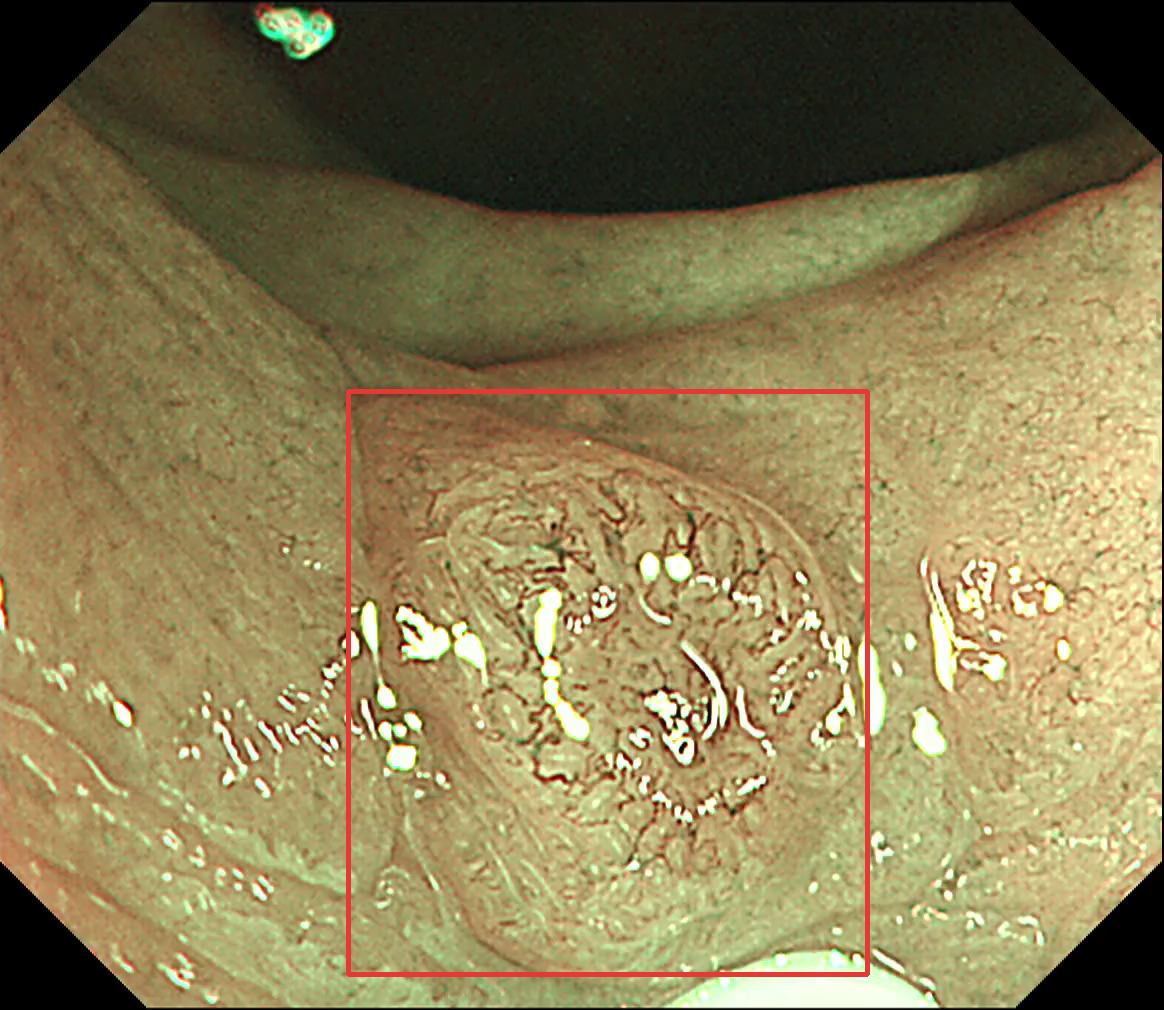} &
    \snap{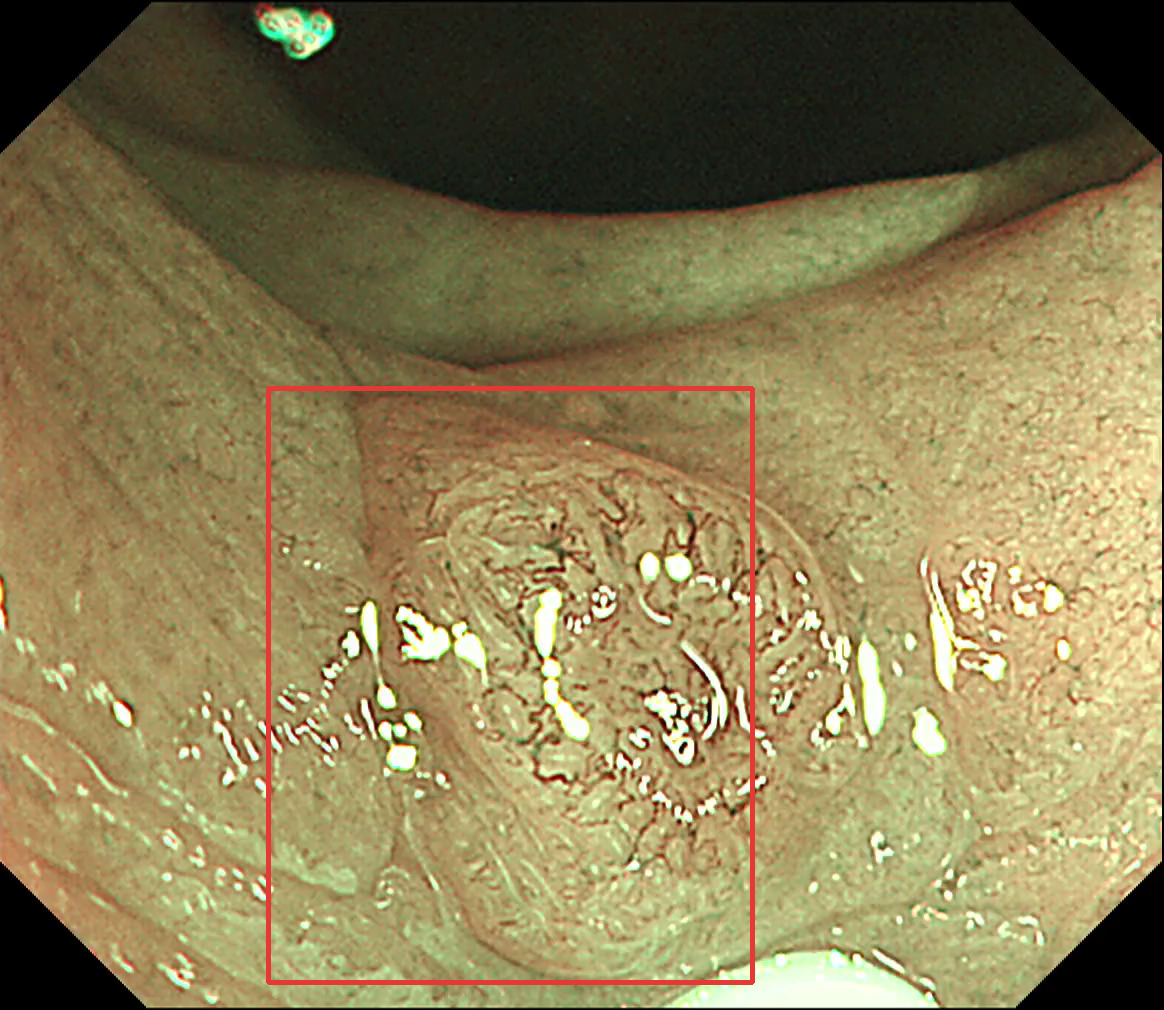} &
    \snap{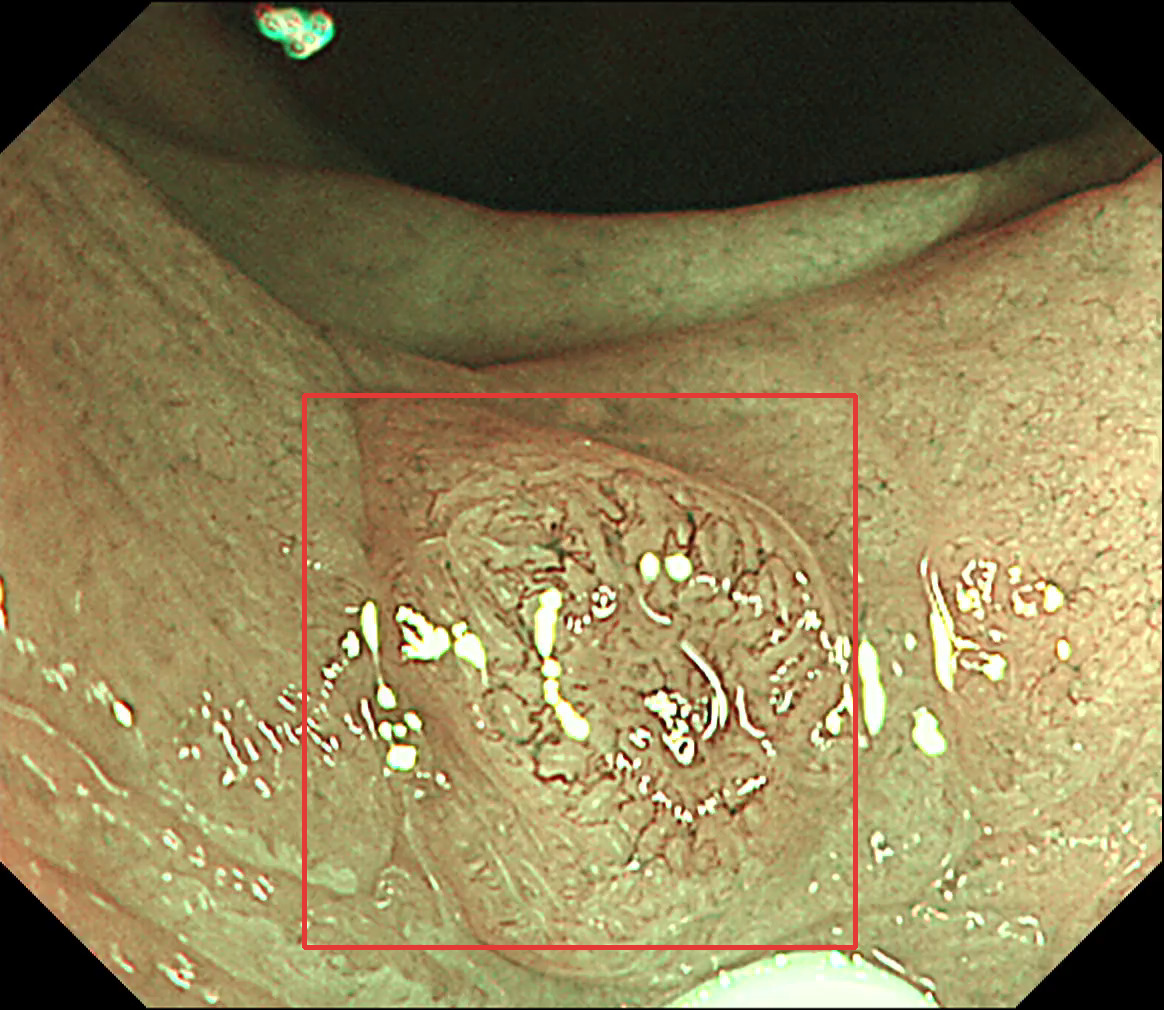} &
    \snap{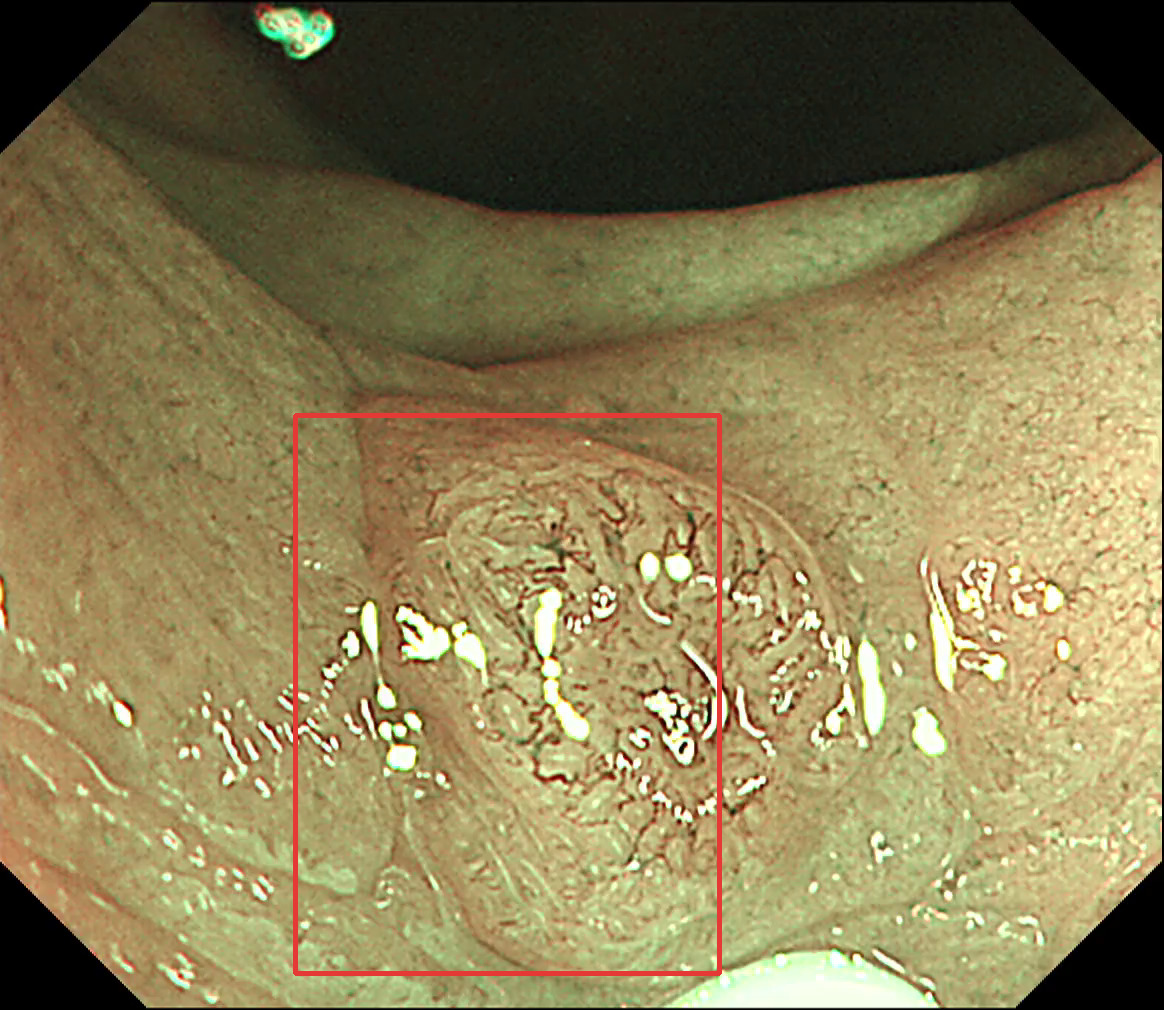} &
    \snap{images/snapshots/detection/144f8236c7f34e94badd9496dfc36886/frame_000000/input.jpg} \\[1pt]
\snap{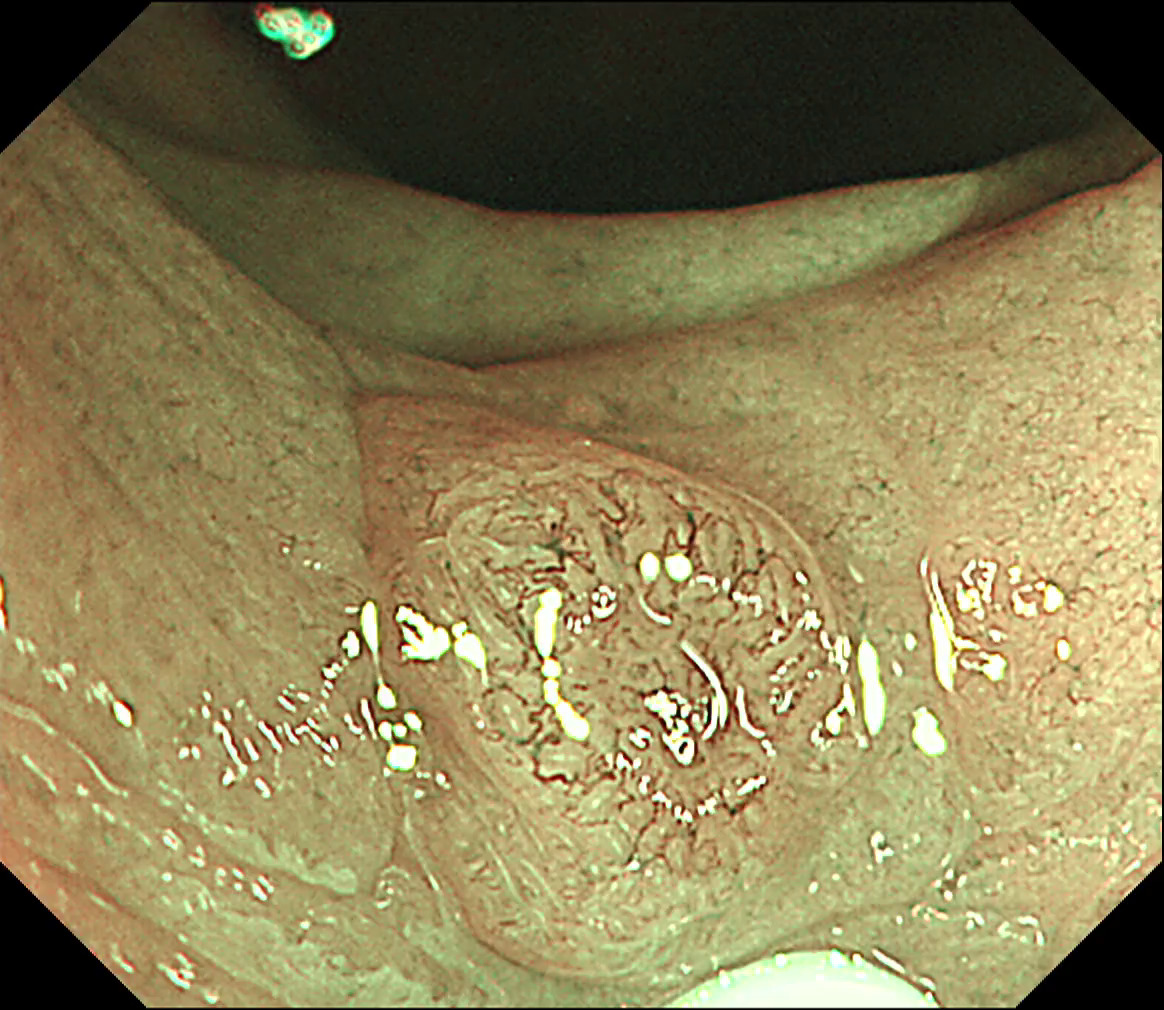} &
\snap{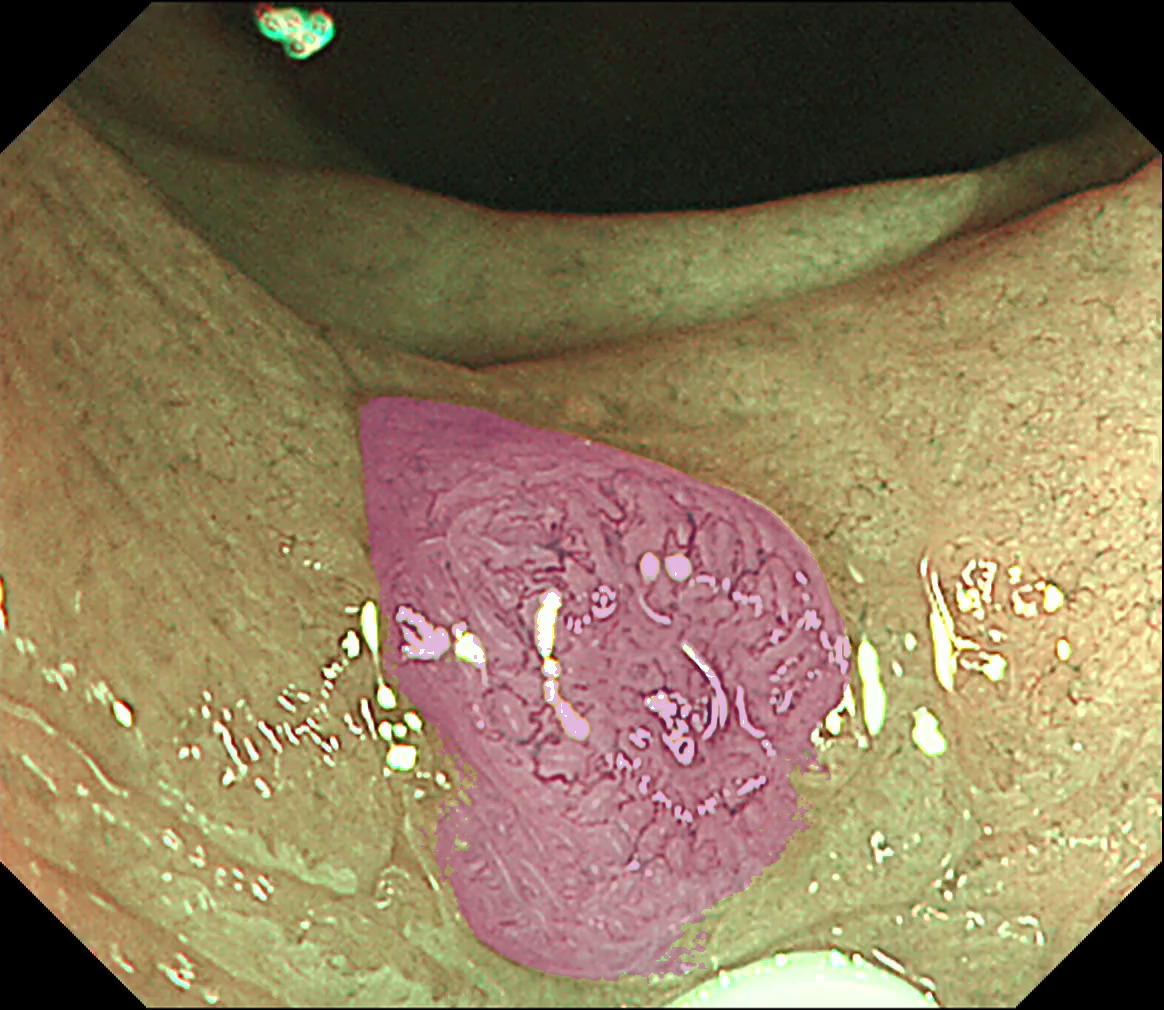} &
\snap{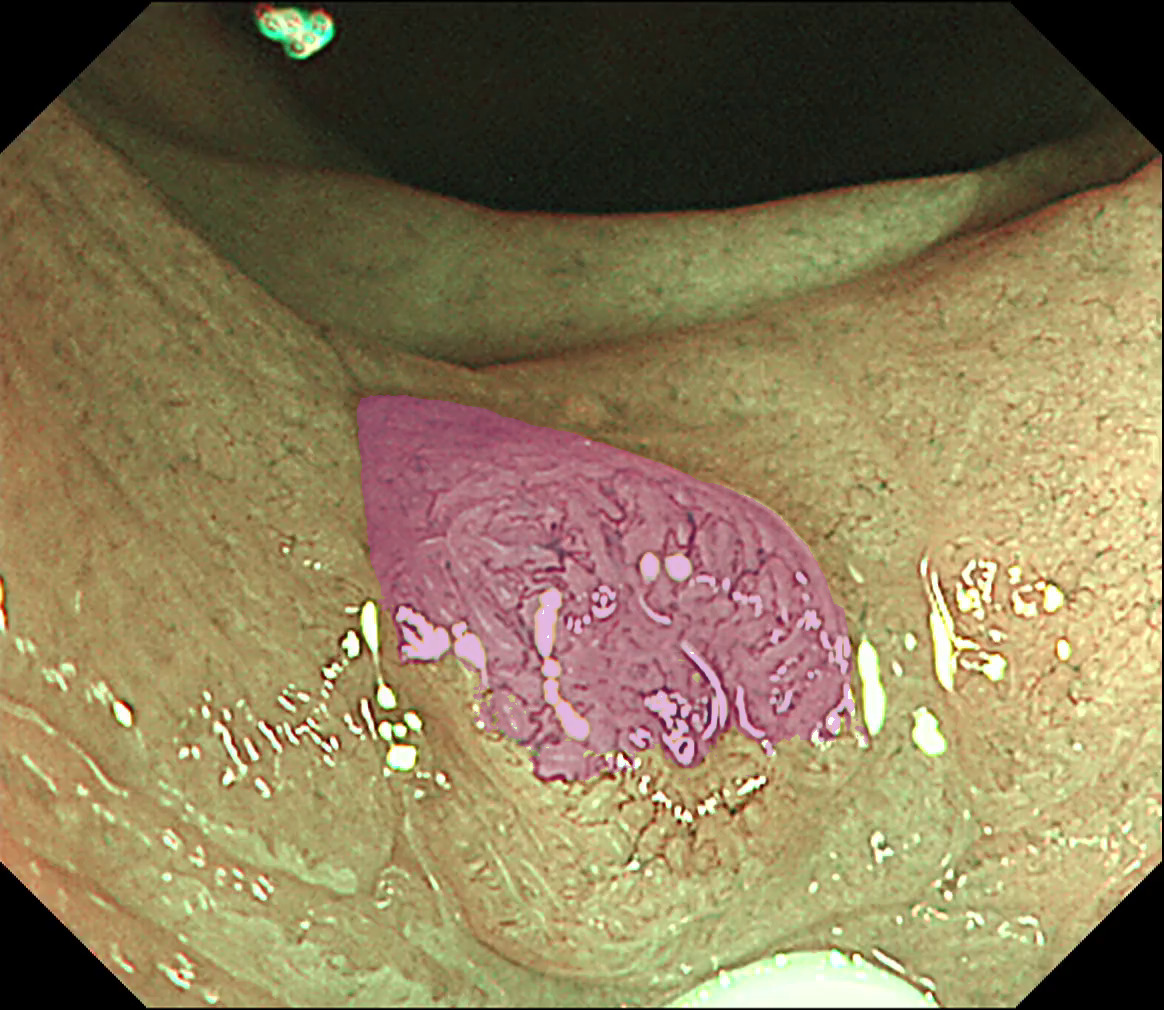} &
\snap{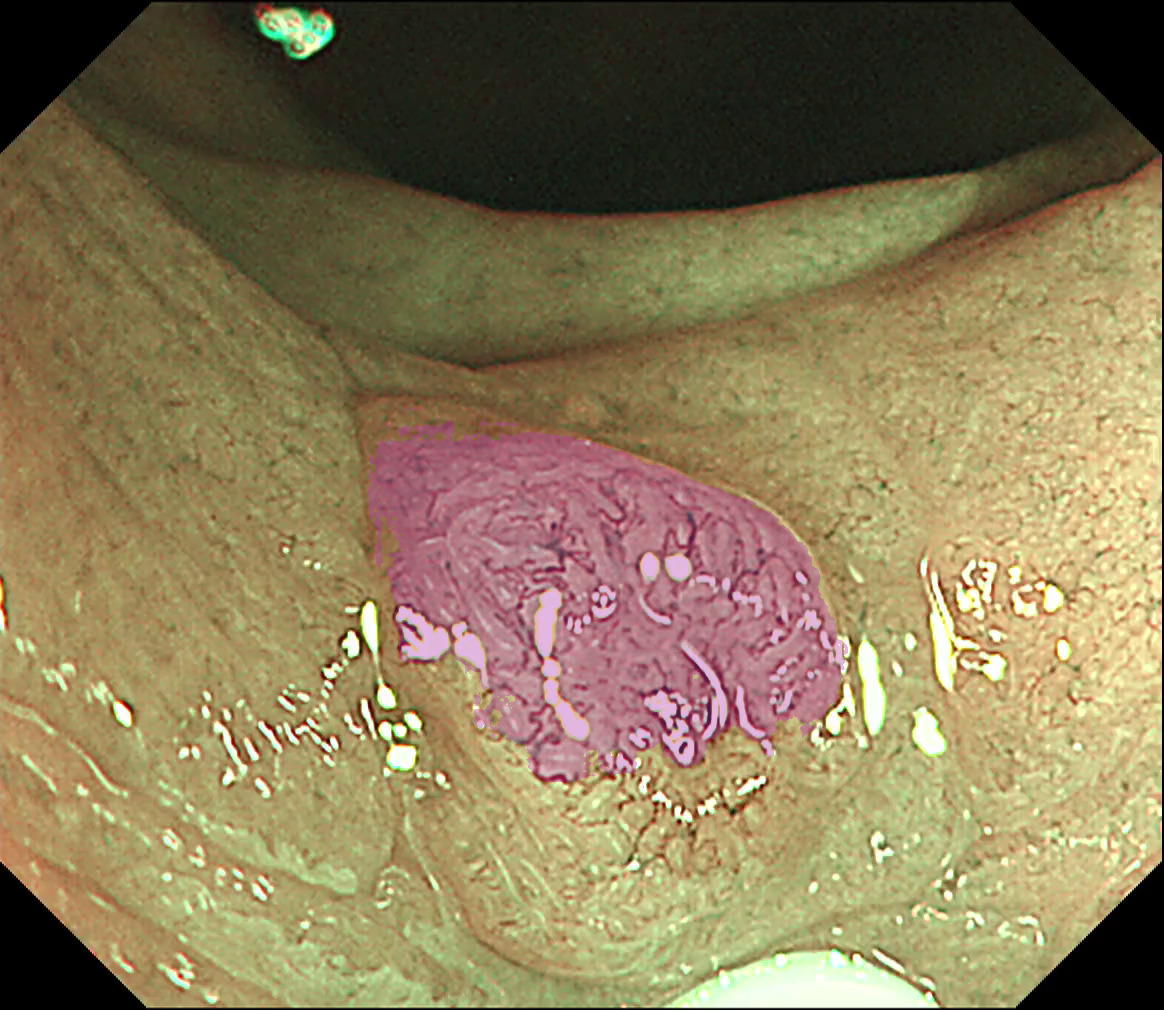} &
\snap{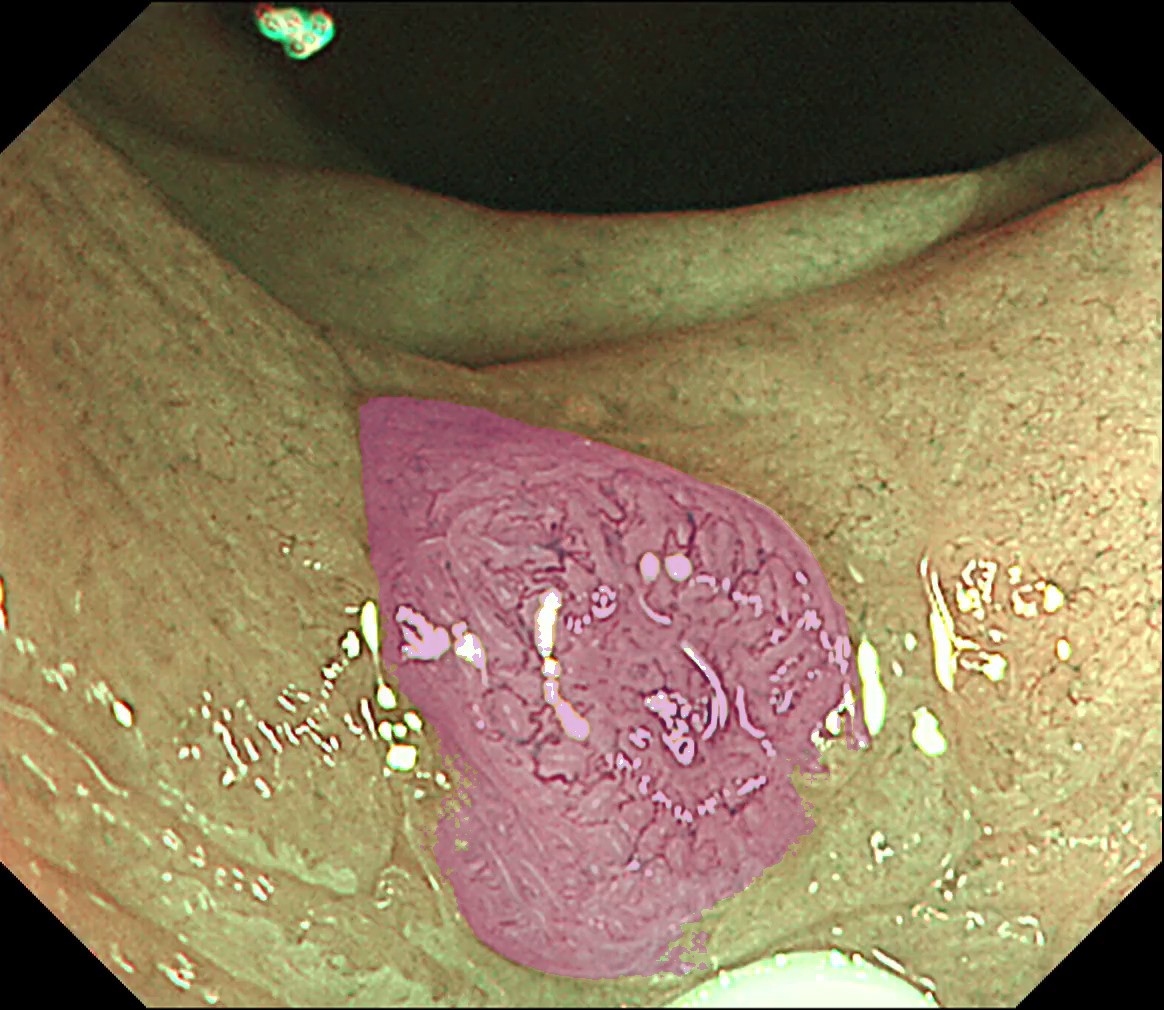} &
\snap{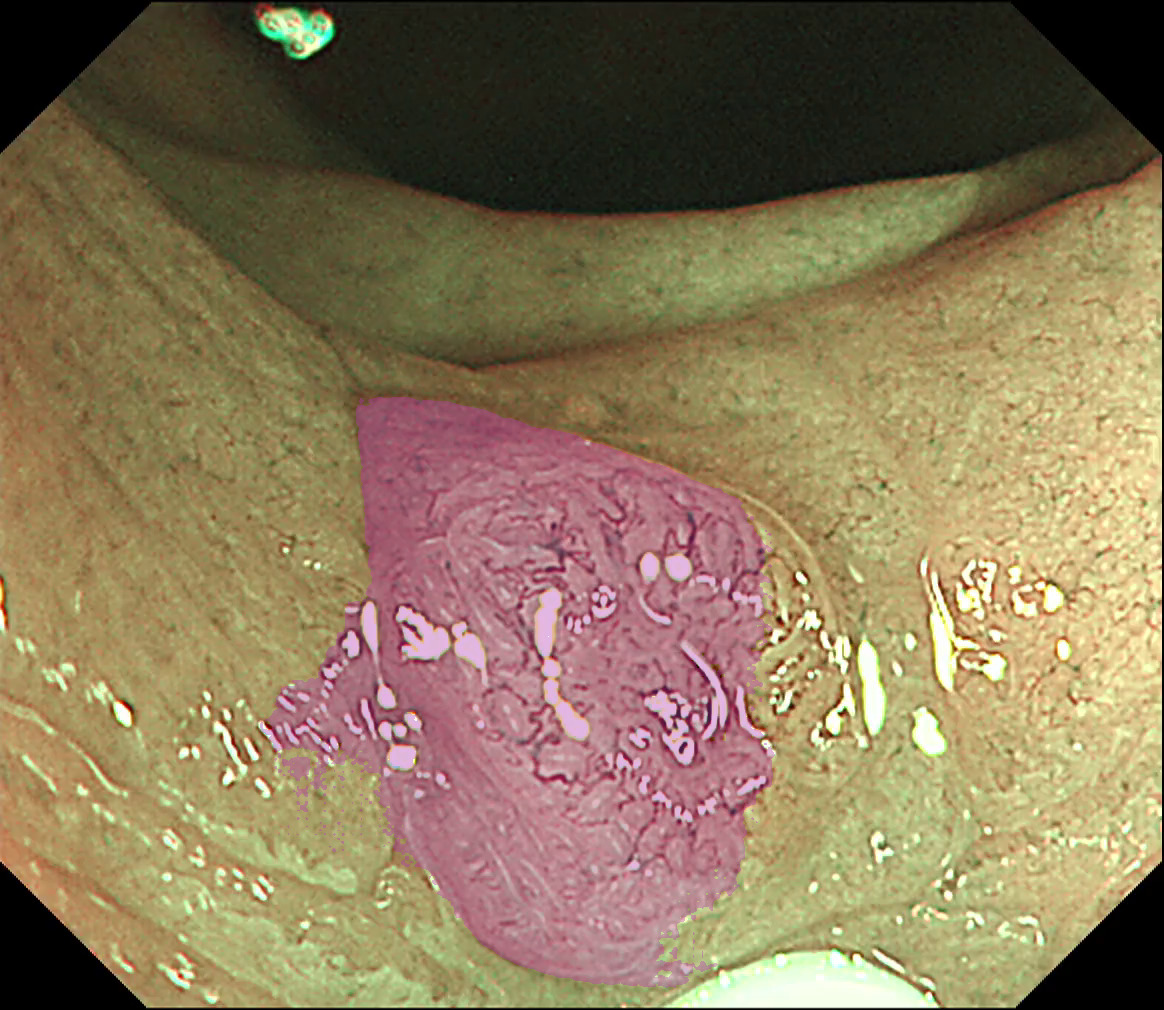} &
\snap{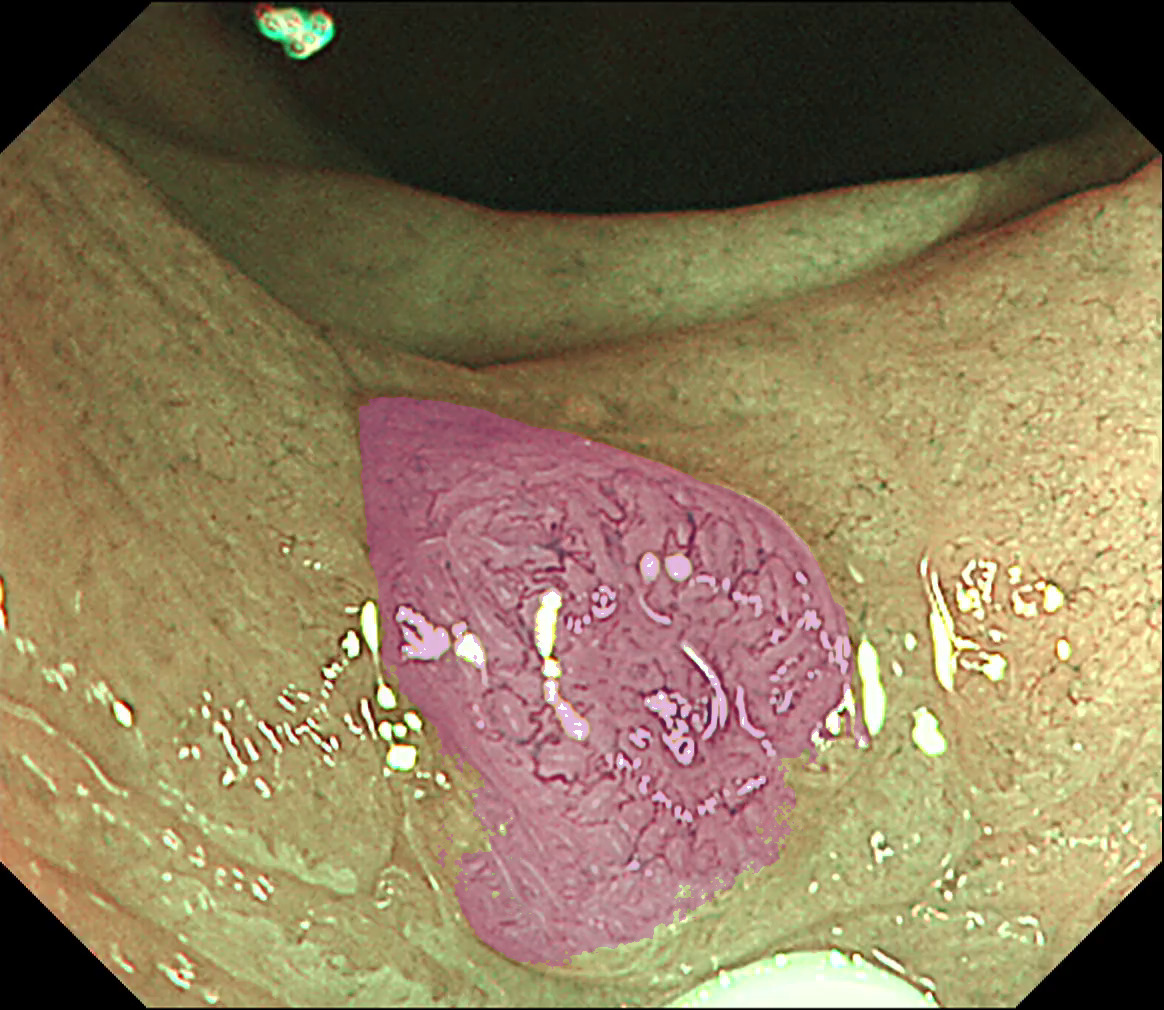} &
\snap{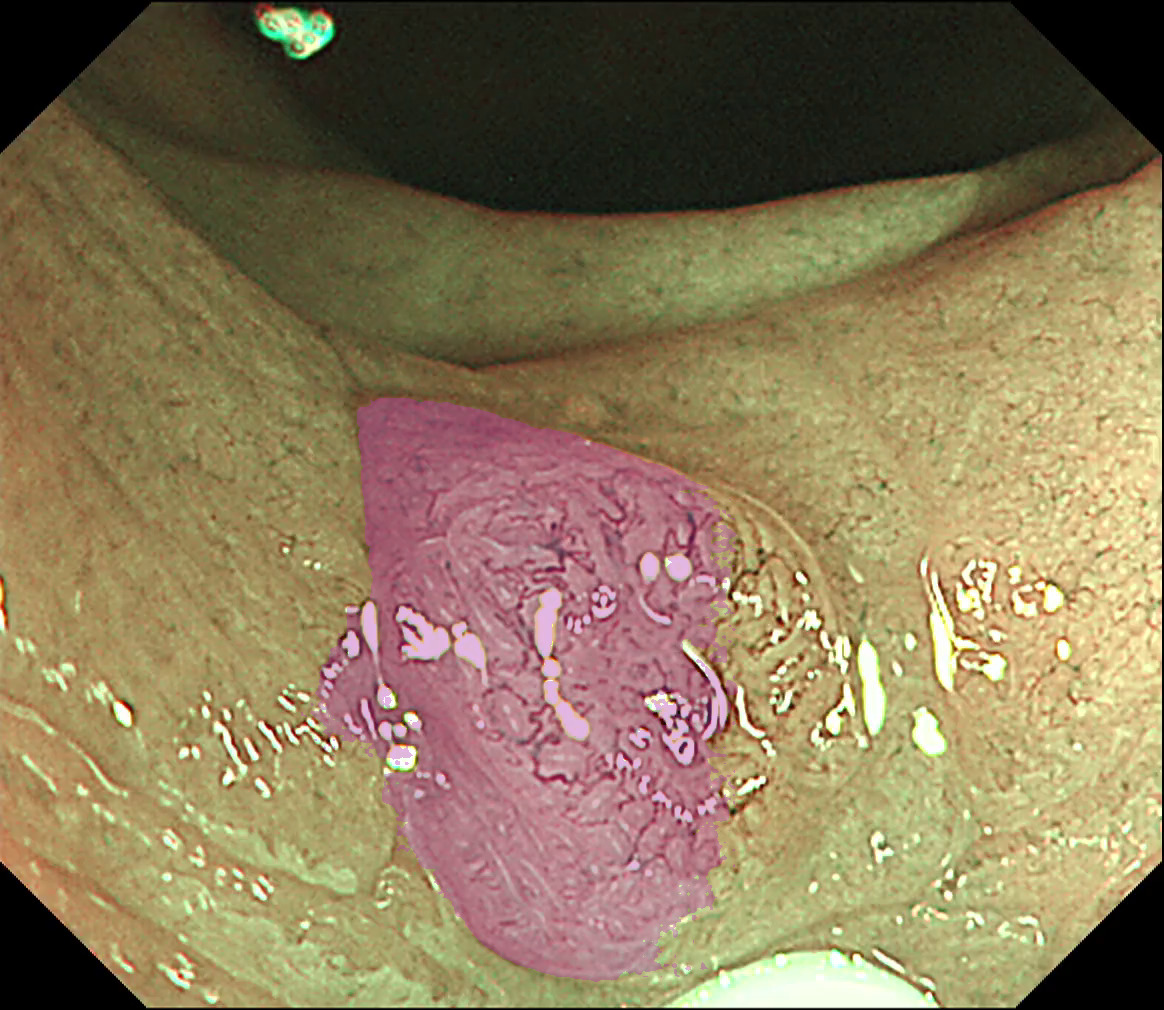} &
\snap{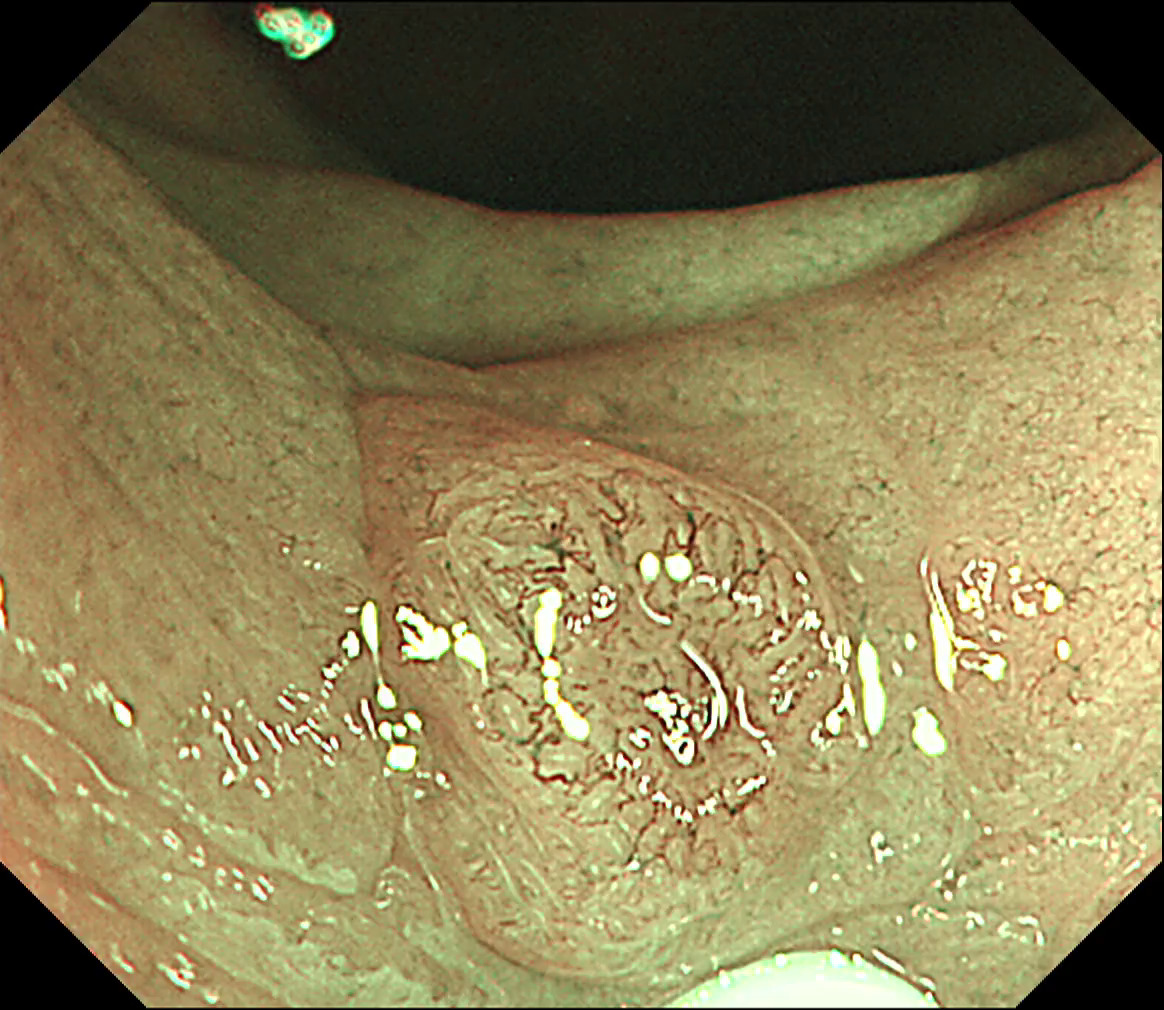} \\[1pt]
  \end{tabular}}%

  \vspace{1pt}

  \resizebox{\textwidth}{!}{%
  \begin{tabular}{ccccccccc}
    \snap{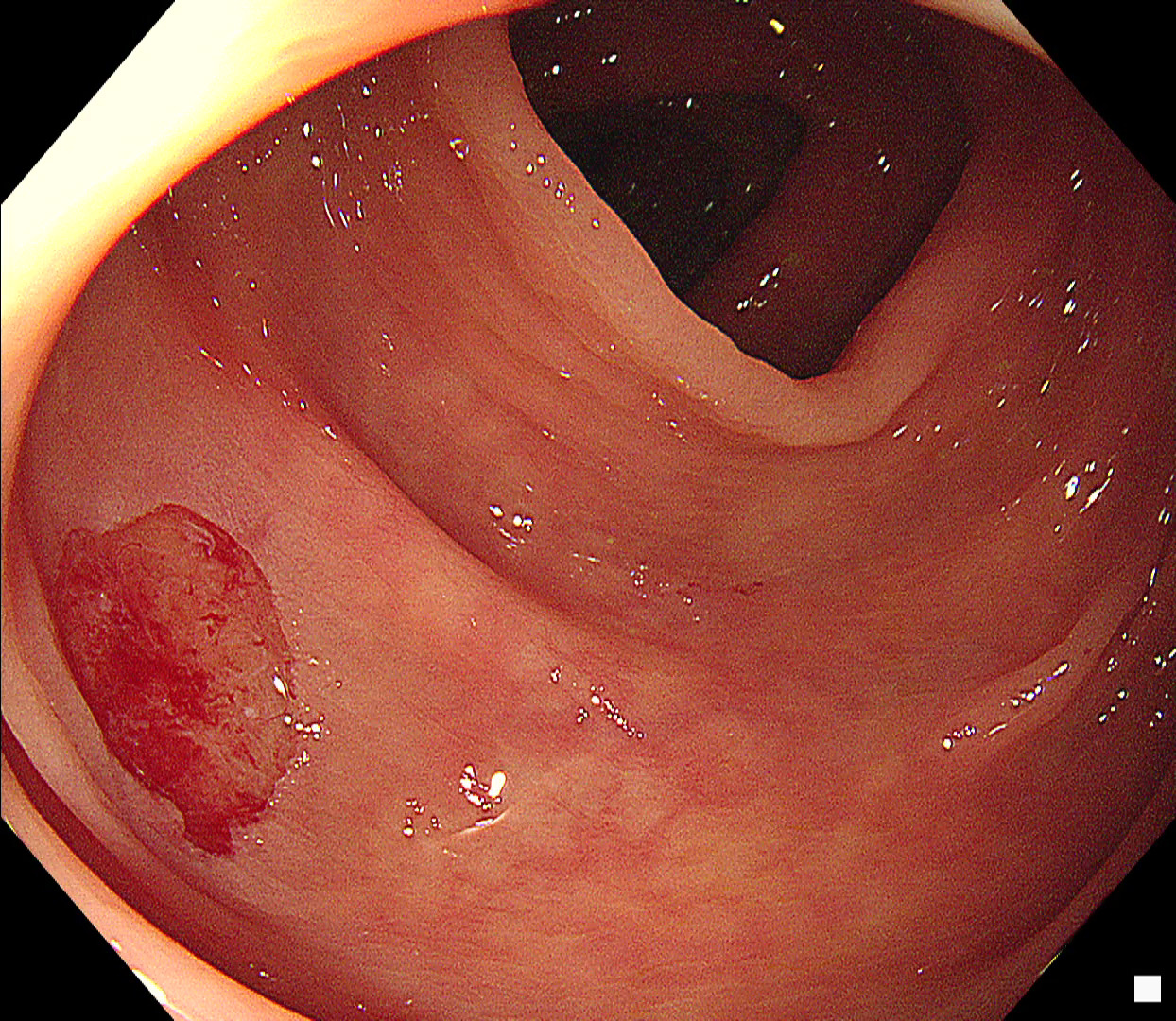} &
    \snap{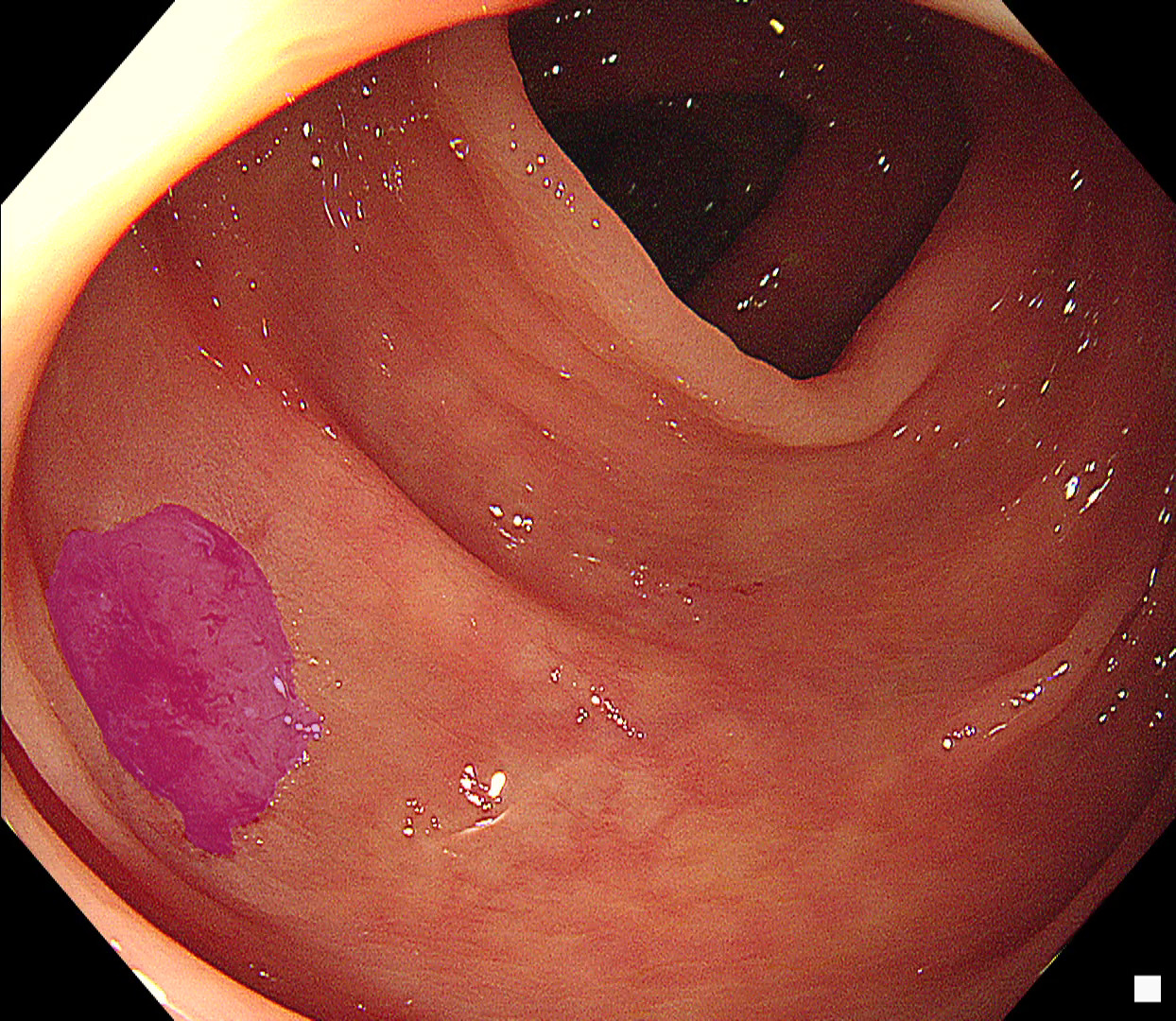} &
    \snap{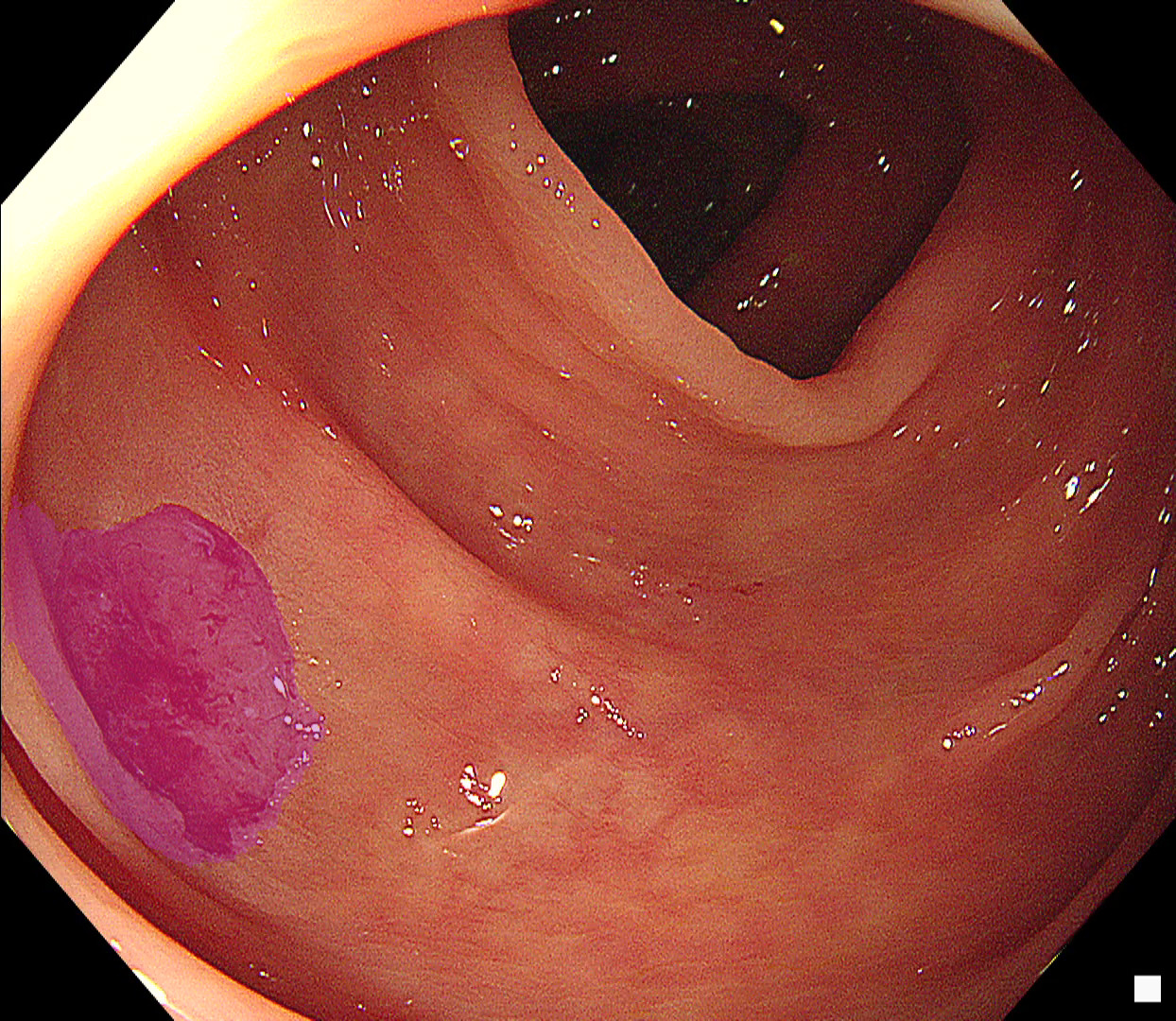} &
    \snap{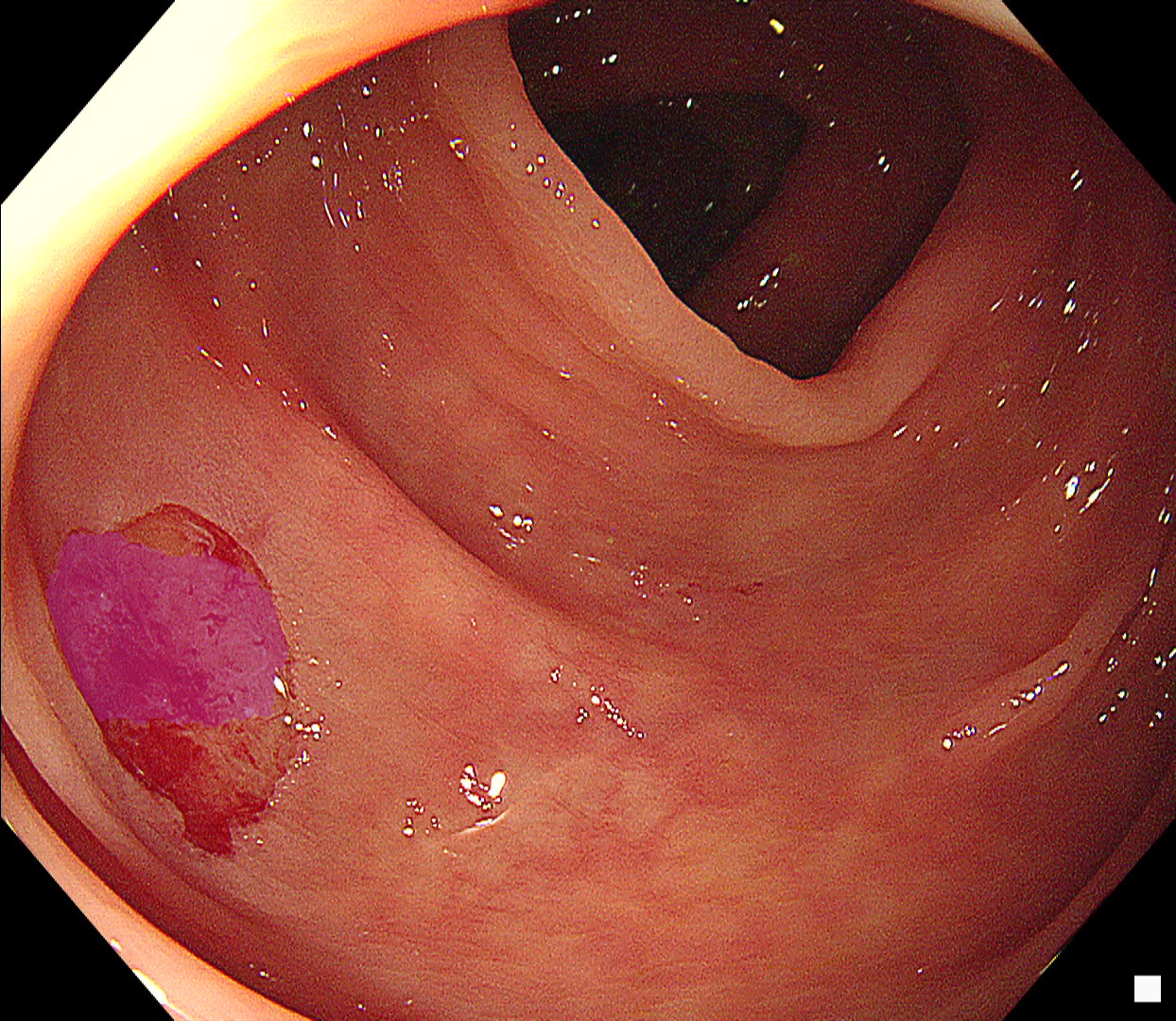} &
    \snap{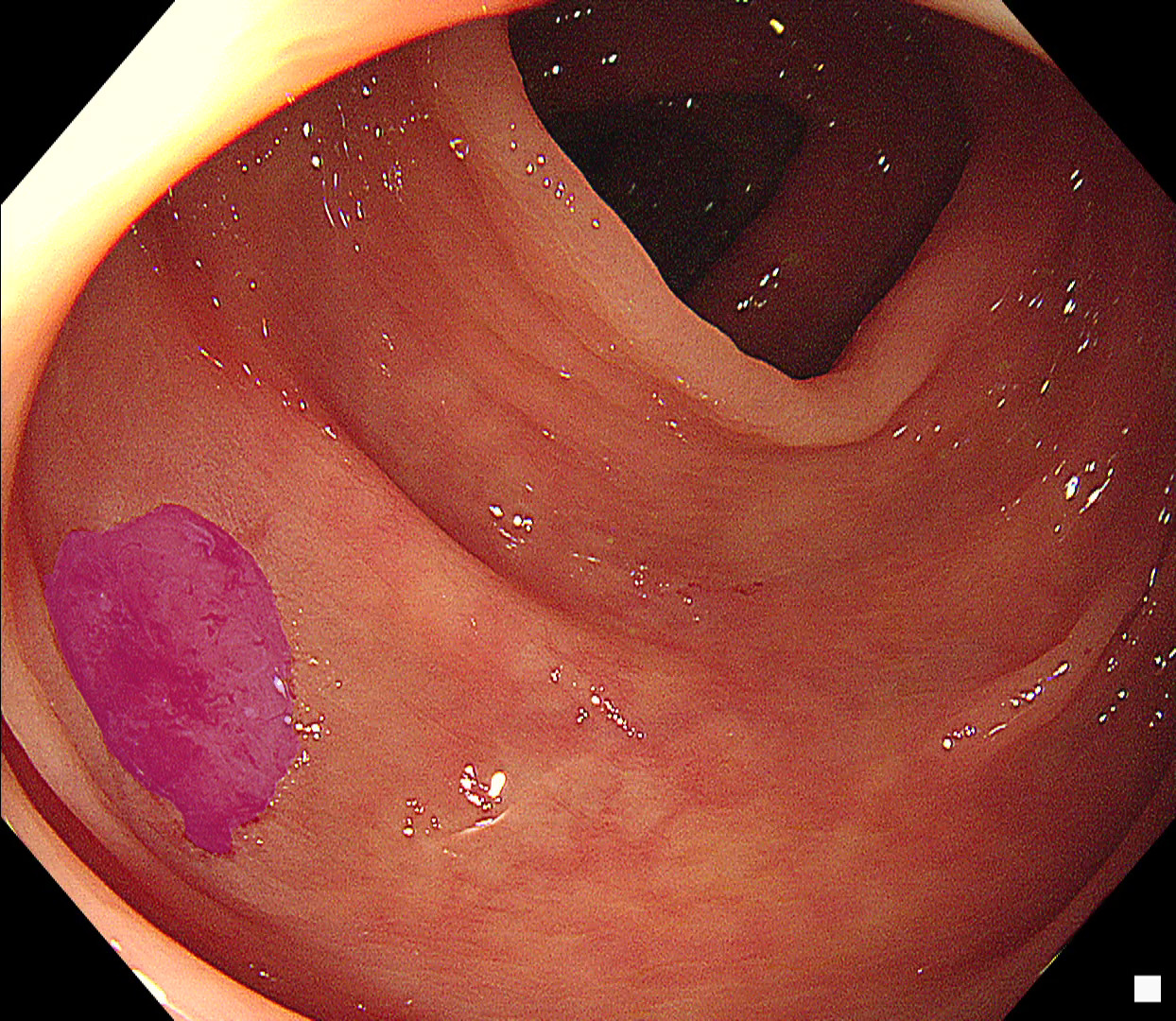} &
    \snap{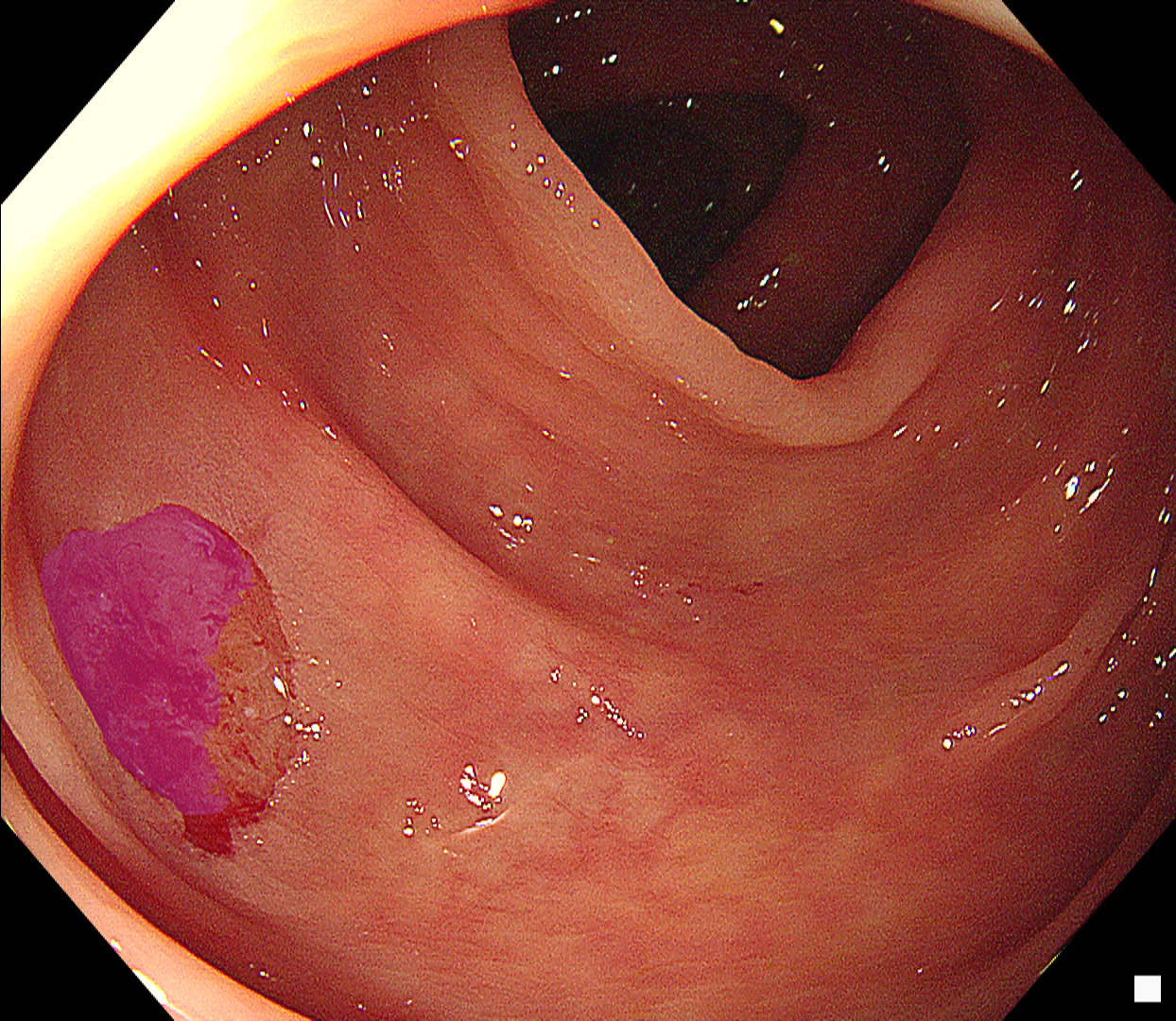} &
    \snap{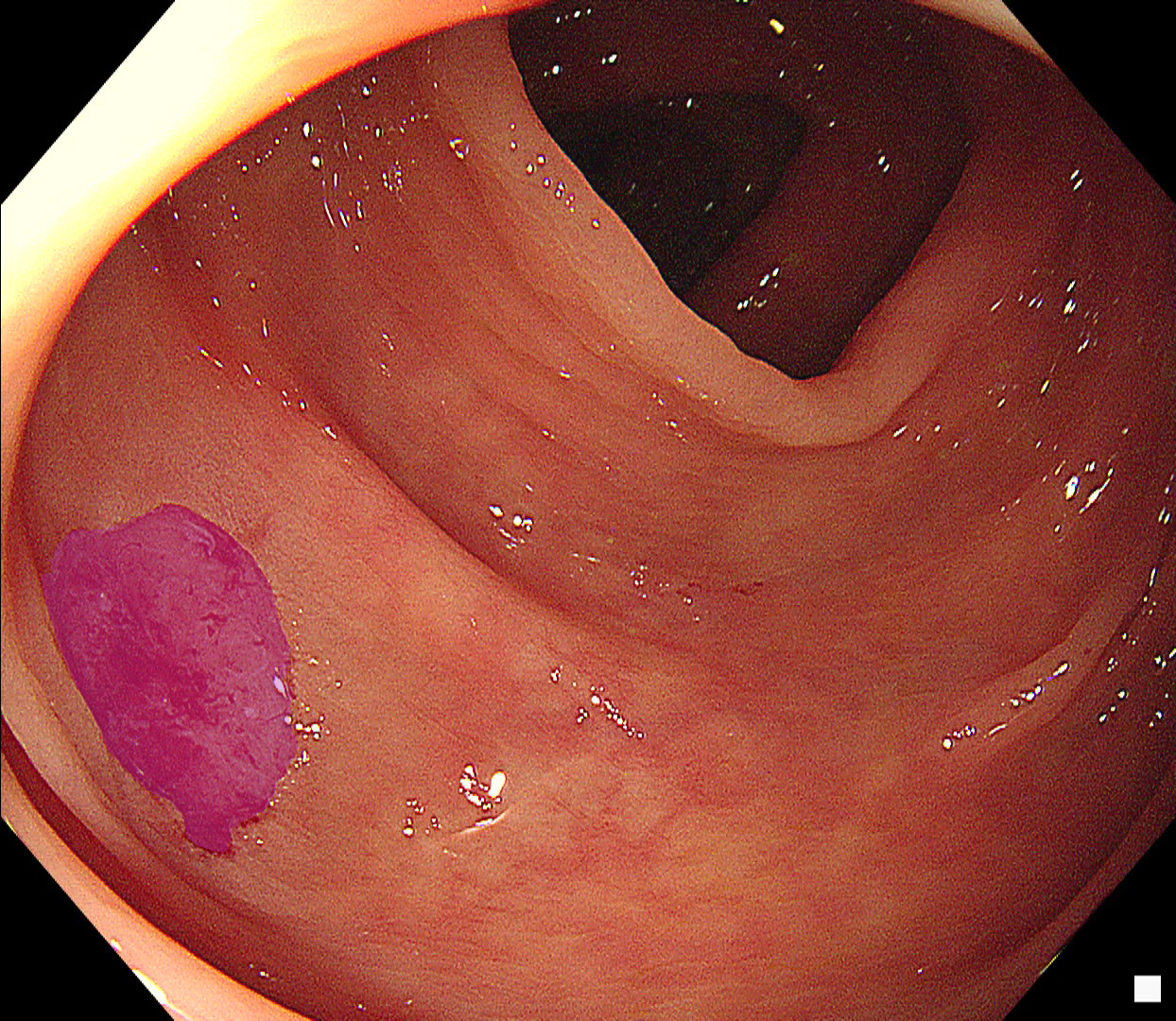} &
    \snap{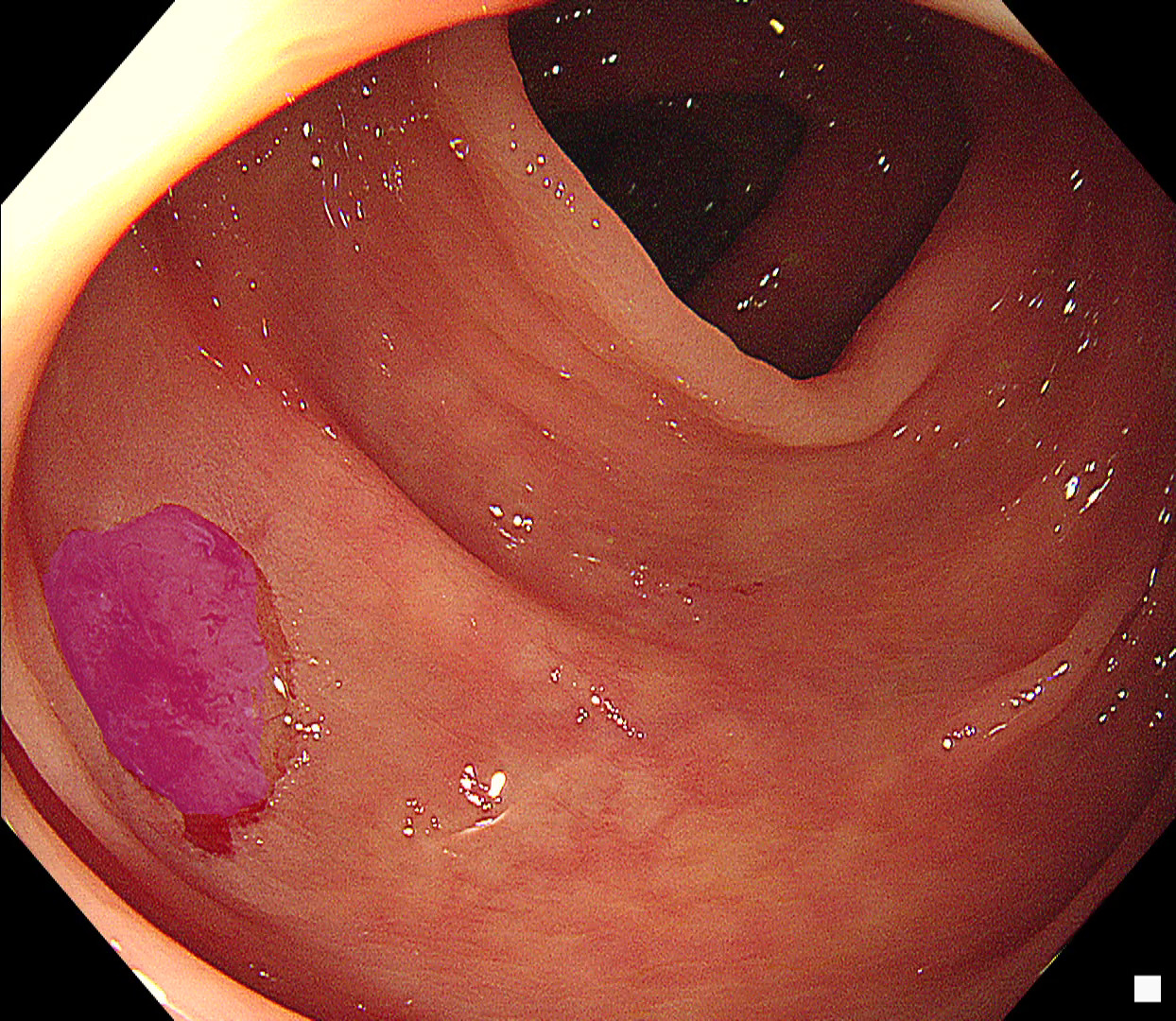} &
    \snap{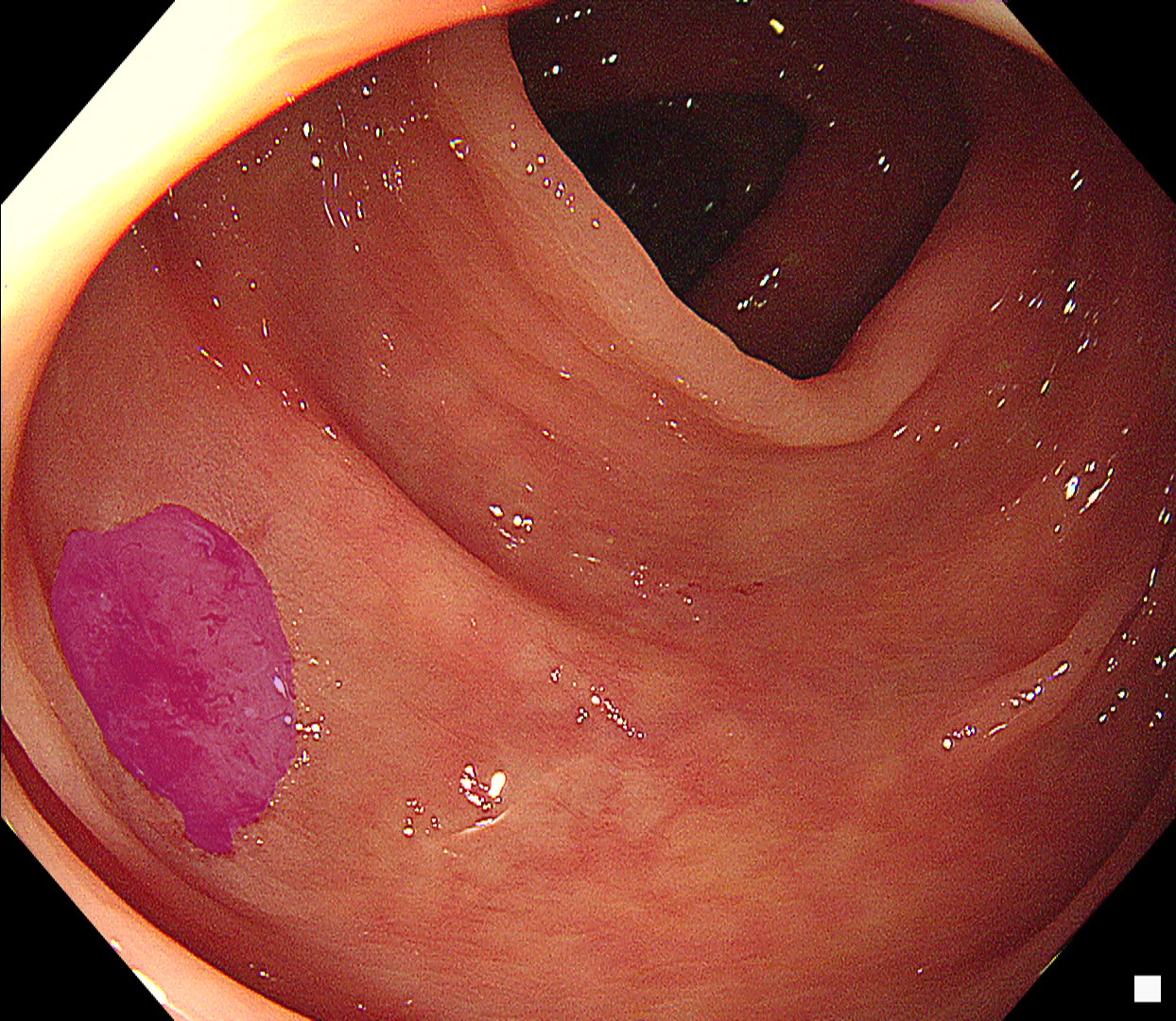} \\[1pt]
  \end{tabular}}%
  \caption{%
    \textbf{Qualitative Comparisons.} The first row shows detection (\textit{Sessile Polyp}); the remaining rows show segmentation (\textit{Sessile Polyp \& Erythematous Region}). Columns show the input, ground truth, and model outputs. SAM-3 underperforms compared to modern MLLMs when paired with a prompted tracker such as EdgeTAM \cite{zhou2025edgetam}.
  }
  \label{fig:qualitative-examples}
  \end{minipage}
\end{figure*}

\section{Analysis and Insights} \label{sec:analysis}
\subsection{Colon-Skill for MLLMs}
\begin{figure}[t]
  \centering
  \includegraphics[width=0.6\linewidth]{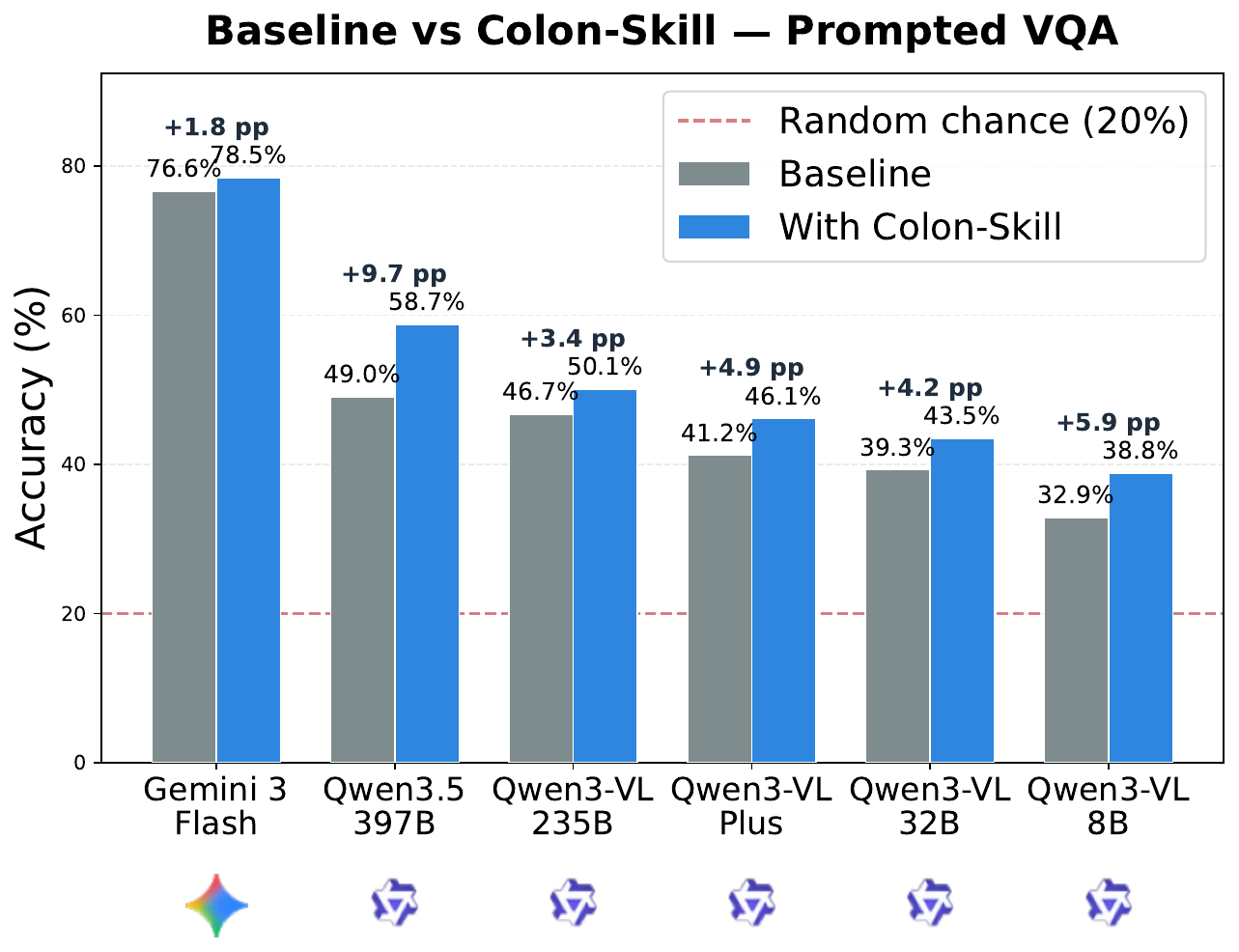}
  \caption{%
    \textbf{Colon-Skill.} Skill-augmented prompting on the prompted VQA split, where a distilled error-based skill context yields consistent accuracy improvements.%
  }
  \label{fig:exp6_skill_context_easy}
\end{figure}

To test whether Multimodal Large Language Models (MLLMs) benefit from structured domain knowledge at inference time, we use a two-stage skill-extraction and prompt-augmentation pipeline. First, we collect per-model VQA predictions on the full benchmark and stratify errors by lesion category (Fig.~\ref{fig:lesion_distribution}), retaining only questions that a majority of models answer incorrectly. These shared failure cases, along with their question/answer context and category metadata, are fed to a frontier large language model which synthesises a concise, natural-language \emph{Colon-Skill}: morphological cues, common confusion traps, and a decision checklist tailored to colonoscopic VQA. Second, we prepend this Colon-Skill to every VQA prompt and re-evaluate all models under identical conditions, yielding consistent improvements up to $+9.7\%$ (Fig.~\ref{fig:exp6_skill_context_easy}). This suggests that distilling cross-model error patterns into structured textual guidance is an effective, training-free way to improve MLLM performance on specialised medical VQA tasks, particularly when models can integrate the added context.

\subsection{Ablation Study} \label{sec:ablation}
\mysection{Effect of frame count on segmentation} Increasing the number of input frames for detection yields steady gains in downstream segmentation quality. For instance, expanding the context from 1 to 7 frames boosts the mean IoU of Gemini~3 Flash from 43.1\% to 54.4\% (mDice: 48.8\% to 61.5\%), and similarly improves Qwen-VL Max from 19.7\% to 33.7\% mIoU. However, this comes with a significant cost as these are per-window detections. We pick 3 equally-spaced frames per window as a good trade-off in all of our segmentation results. 

\mysection{Temporal context in VQA} Evaluating temporal context on the prompted VQA split showed that full video generally outperforms single-frame inputs, though the degree of impact is model-dependent. Restricting the input to a single frame reduced accuracy for Qwen3.5 397B ($-7.0$\%), Qwen-VL Max ($-3.4$\%), and Gemini~3 Flash ($-2.5$\%). This highlights the importance of using videos instead of images used in previous polyp detection benchmarks. 

\section{Conclusions} \label{sec:conclusion}
We presented \methodname, a novel agentic workflow and a comprehensive multi-task benchmark for full-procedure colonoscopy video understanding. By scalably and densely annotating 14 distinct lesion categories with masks, bounding boxes, and clinical text, we demonstrated that state-of-the-art MLLMs excel at high-level colonoscopy reasoning and spatial grounding with engineered context and agentic workflows.

\mysection{Acknowledgments}
We thank Dr. Jamal Hamdi, a surgeon with over 30 years of experience, who reviewed the accuracy of the \methodname annotations together with the authors. This work was supported by the King Abdullah University of Science and Technology (KAUST) Center of Excellence for Smart Health. Additional support came from the KAUST Ibn Rushd Postdoctoral Fellowship.

\mysection{License}
 Video copyrights remain with their owners \cite{biffi2024realcolon}. The \methodname dataset is licensed under Creative Commons Attribution (CC BY), consistent with REAL-COLON \cite{biffi2024realcolon}.

\mysection{Disclosure of Interests}
 The authors have no competing interests to declare that are relevant to the content of this article.

{\small
\bibliographystyle{splncs04}
\bibliography{egbib}
}
\clearpage \clearpage
\appendix

\section{Additional Dataset Details} \label{supsec:dataset}
\subsection{Annotation Pipeline}

The benchmark was constructed through a multi-stage filtering pipeline applied to 60 video sequences from the REAL-colon dataset \cite{biffi2024realcolon}. A vision-language model identified 1{,}325 candidate lesion windows spanning 22.97 hours of video. Successive verification agent filtering, AI confirmation agent with cued video from the tracking, and human review retained 528 windows (39.8\%) covering 464{,}035 frames (12.89 hours) across 59 sequences (Table~\ref{tab:pipeline-funnel}). The largest reduction occurred at verification filtering (422 windows removed), followed by AI confirmation (306 windows), with human review removing only 69 windows (11.6\%), indicating strong agreement with automated filtering.

\begin{figure}[t]
  \centering
  \includegraphics[width=0.99\linewidth]{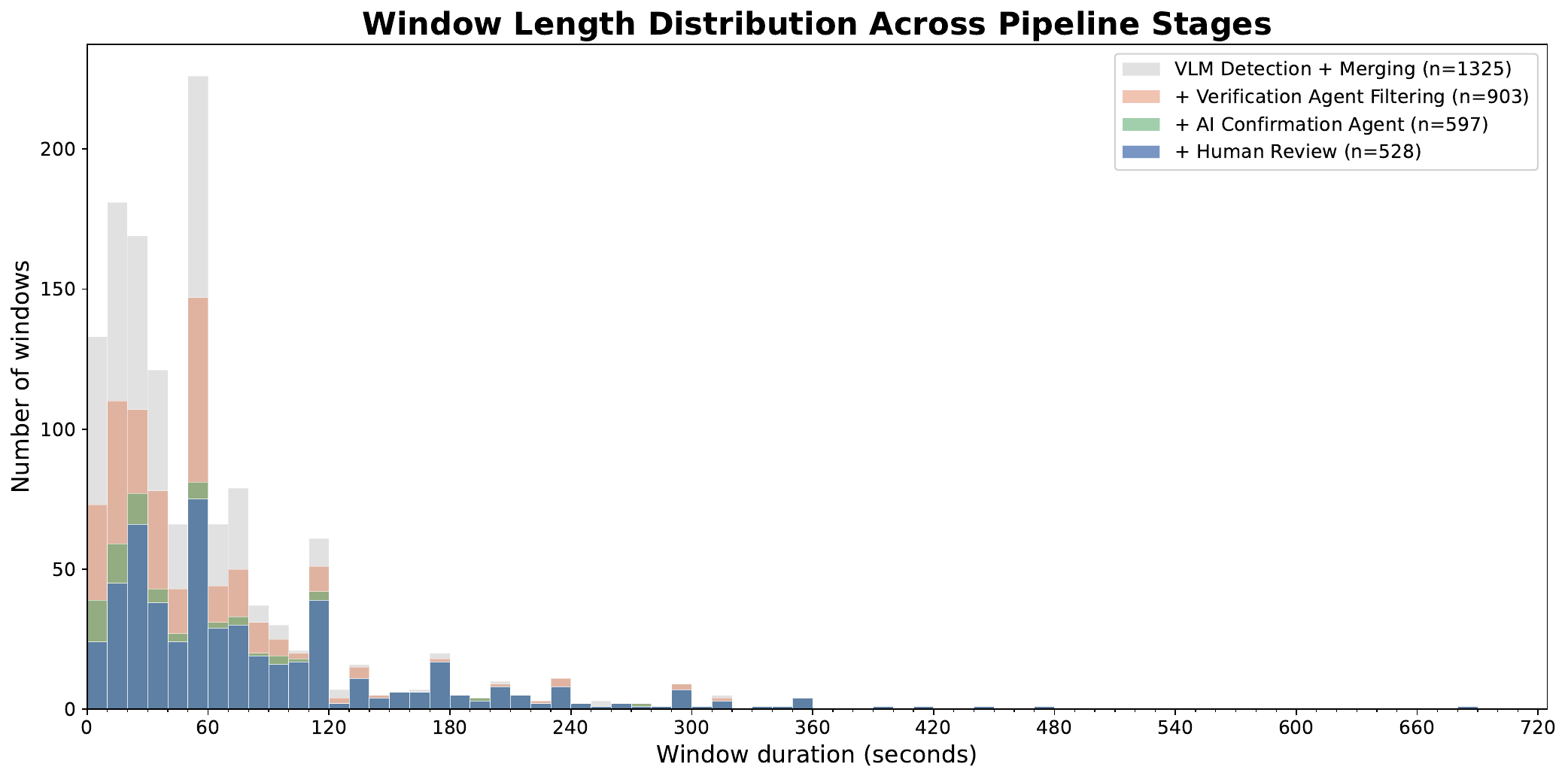}
  \caption{\textbf{Window Duration Distributions Across Stages.} Overlaid distributions of lesion window durations (seconds at 10\,FPS) are shown at successive pipeline filtering stages: VLM-detected, after verification, after AI confirmation with visual box overlays, and the final human-curated set. Each stage is a strict subset of the previous
    one.%
  }
  \label{fig:pipeline_stage_distributions}
\end{figure}

Curated windows average 878.9 frames compared to 414-509 for rejected ones, and carry richer annotations averaging 252.4 words per window versus 145-190 for rejected windows (Table~\ref{tab:per-stage-stats}). The curated set comprises 314{,}408 bounding boxes, 227{,}343 segmentation masks, and 145{,}515 words of clinical text (Table~\ref{tab:confirmed-vs-rejected}). Window duration distributions across pipeline stages are shown in Fig.~\ref{fig:pipeline_stage_distributions}.

\subsection{Validating the Annotations}

The agentic pipeline produced 597 lesion videos, each reviewed by a clinician. Of these, 528 were accepted (88.4\% success rate), totalling 464{,}035 frames.

\begin{figure}[t]
  \centering
  \includegraphics[width=0.99\linewidth]{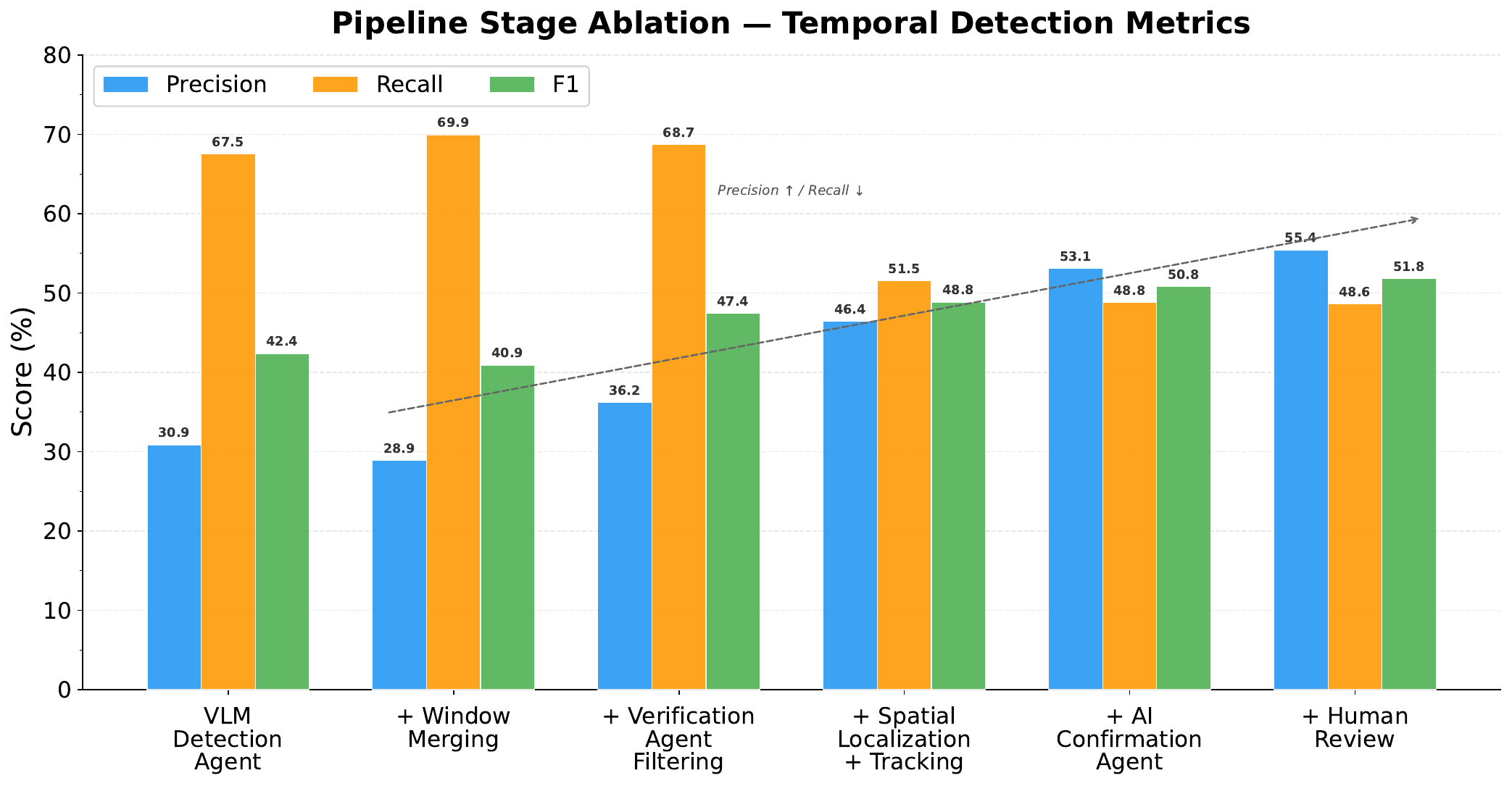}
  \caption{\textbf{Polyp-Based \methodname Pipeline Ablation.} Temporal frame-level Precision, Recall, and F1 are reported at
    each processing stage, from raw VLM detection through human review. Annotated values are percentages from the polyp annotations of REAL-COLON\cite{biffi2024realcolon}.%
  }
  \label{fig:pipeline_temporal}
\end{figure}

We ablate each pipeline stage on the full REAL-colon dataset box annotations for polyps (Table~\ref{tab:pipeline_ablation}, Fig.~\ref{fig:pipeline_temporal}). Verification filtering, in which a a video agent reviews the candidate window and confirms whether a lesion is present, yields the largest single-step precision gain (+7.3\%) with minimal recall loss. Tracking restricts positive predictions to frames with actual bounding boxes, sharply increasing precision (+10.2\%) at the cost of recall ($-$17.2\%). AI confirmation, where a second video agent re-examines the tracked window with bounding-box overlays to validate the localised lesion, removes 34\% of remaining windows and improves precision by +6.7\% and F1 by +2.0pp. Human review provides a further marginal but consistent improvement (+2.4\% precision, +1.0\% F1), while spatial metrics remain nearly unchanged, indicating that the rejected windows had minimal impact on localisation quality.

\begin{table}[t]
\centering
\caption{\textbf{Curated \methodname Dataset Statistics.} Windows that passed all filtering stages (Confirmed, $n{=}528$) are compared against those rejected during human review (Rejected, $n{=}69$).}
\label{tab:confirmed-vs-rejected}
\small
\begin{tabular}{lrrr}
\toprule
\textbf{Metric} & \textbf{Confirmed (528)} & \textbf{Rejected (69)} & \textbf{Total (597)} \\
\midrule
Sequences covered          & 59          & 37          & 60         \\
Total frames               & 464,035     & 28,571      & 492,606    \\
Avg.\ frames / window      & 878.9       & 414.1       & -        \\
Duration (hours, 10\,fps)  & 12.89       & 0.79        & 13.68      \\
\midrule
Total bounding boxes       & 300,132     & 14,276      & 314,408    \\
Avg.\ bboxes / window      & 568.4       & 206.9       & -        \\
Bbox coverage (\%)         & 64.7        & 50.0        & -        \\
\midrule
Total segmentation masks   & 213,067     & 14,276      & 227,343    \\
Windows with masks         & 465 / 528   & 69 / 69     & 534 / 597  \\
Mask coverage (\%)         & 45.9        & 50.0        & -        \\
\midrule
Avg.\ words - initial desc.       & 162.8 & 88.6  & -   \\
Avg.\ words - verified desc.      & 55.5  & 53.0  & -   \\
Avg.\ words - confirmation note   & 34.2  & 35.7  & -   \\
Avg.\ words - all text combined   & 252.4 & 177.2 & -   \\
Total words (all text)               & 133,289 & 12,226 & 145,515 \\
\bottomrule
\end{tabular}
\end{table}

\begin{table}[t]
\centering
\caption{\textbf{\methodname Filtering Funnel Overview.} Each row shows the number of windows, total frames, and video duration at successive pipeline stages. Dropped rows indicate windows removed by the corresponding filter.}
\label{tab:pipeline-funnel}
\small
\begin{tabular}{lcccc}
\toprule
\textbf{Pipeline Stage} & \textbf{Windows} & \textbf{Frames} & \textbf{Hours} & \textbf{Retention (\%)} \\
\midrule
VLM Detection Proposals + Merging         & 1,325 & 826,763 & 22.97 & 100.0 \\
\quad$-$ Rejected by Verification Agent       &  $-$422 & $-$178,323 & $-$4.95 & -     \\
+ Verification Agent Filtering                   &   903 & 648,440 & 18.01 &  68.2 \\
\quad$-$rejected by AI confirmation agent    &  $-$306 & $-$155,834 & $-$4.33 & -     \\
+ Cued AI Confirmation Agent                &   597 & 492,606 & 13.68 &  45.1 \\
\quad$-$ Human Review Rejected       &   $-$69 &  $-$28,571 & $-$0.79 & -     \\
\textbf{Final Curated Set}           &  \textbf{528} & \textbf{464,035} & \textbf{12.89} & \textbf{39.8} \\
\bottomrule
\end{tabular}
\end{table}

\begin{table}[t]
\centering
\caption{\textbf{\methodname Annotations Stage-Wise Rejection Characteristics.} Statistics include window duration (in frames at 10\,fps) and text description lengths (in words). At earlier stages, only a subset of text fields is available.}
\label{tab:per-stage-stats}
\small
\resizebox{1.0\linewidth}{!}{
\begin{tabular}{lcccc}
\toprule
& \multicolumn{1}{c}{\textbf{Verification}} & \multicolumn{1}{c}{\textbf{AI Confirmation}} & \multicolumn{1}{c}{\textbf{Human Review}} & \multicolumn{1}{c}{\textbf{Final Curated}} \\
& \multicolumn{1}{c}{\textbf{Rejected (422)}} & \multicolumn{1}{c}{\textbf{Rejected (306)}} & \multicolumn{1}{c}{\textbf{Rejected (69)}} & \multicolumn{1}{c}{\textbf{Confirmed (528)}} \\
\midrule
Sequences           & 60        & 56        & 37        & 59        \\
Total frames        & 178,323   & 155,834   & 28,571    & 464,035   \\
Avg.\ frames/window  & 422.6     & 509.3     & 414.1     & 878.9     \\
Median frames/window & 339.5     & 417.0     & 259.0     & 599.0     \\
Duration (hours)    & 4.95      & 4.33      & 0.79      & 12.89     \\
\midrule
\multicolumn{5}{l}{\textit{Text descriptions (avg.\ words per window)}} \\
\quad Initial desc.        & 84.9  & 101.3 & 88.6  & 162.8 \\
\quad Verified desc.       & 60.4  & 53.8  & 53.0  & 55.5  \\
\quad Confirmation note    & -   & 34.4  & 35.7  & 34.2  \\
\quad All text combined    & 145.3 & 189.5 & 177.2 & 252.4 \\
\midrule
Total words          & 61,326 & 57,976 & 12,226 & 133,289 \\
\bottomrule
\end{tabular} }
\end{table}

\subsection{\methodname Suite of Evaluation Benchmarks}
Colon-Bench is a multi-task video benchmark for colonoscopy understanding spanning five tasks: binary lesion classification, lesion detection, instance segmentation, and visual question answering at two difficulty levels (prompted and unprompted). The benchmark comprises 1{,}597 unique video clips from 60 patient sequences totalling 955{,}126 frames (Table~\ref{tab:colon_bench_stats}). Classification covers 790 clips (518 lesion-free, 272 lesion-positive), while detection and segmentation use the 272 and 264 lesion-positive clips respectively, providing 61{,}538 per-frame bounding boxes and 57{,}550 per-frame masks. The prompted VQA split contains 1{,}485 five-choice questions over 499 clips, while the unprompted split contains 2{,}740 questions over 918 clips and fully encompasses all detection and segmentation videos, enabling direct comparison of spatial localisation against open-ended clinical reasoning (see Table~\ref{tab:vqa_examples} for examples).

\subsection{Benchmark Formation and Blind Tests}
For each clip, three five-way multiple-choice questions are generated using Gemini 3 Flash, covering lesion identification, clinical characteristics, and temporal reasoning. The \emph{prompted} set uses videos with bounding-box and mask overlays on confirmed lesion windows, while the \emph{unprompted} set uses raw frames and additionally includes non-lesion windows. Both sets undergo structural validation and are randomised with a fixed seed for reproducibility. To mitigate text-only exploitability, we apply two-stage debiasing: (i) for every question, four adversarial distractors are regenerated in a separate LLM call receiving only the question stem and correct answer, producing length- and style-matched alternatives with re-randomised option positions; (ii) a blind text-only stress-test identifies cases where debiasing introduced new textual shortcuts, and these questions revert to their original formulation. After this procedure, blind-only accuracy is 44.6\% (prompted) and 37.1\% (unprompted) on Gemini-3 Flash versus the 20\% random baseline, with residual margins attributable to skewed lesion-type distributions rather than surface-level cues (see Fig.~\ref{fig:lesion_distribution}).

\begin{table*}[t]
\centering
\caption{\textbf{\methodname Pipeline Stage Ablation on REAL-colon \cite{biffi2024realcolon}.} Temporal metrics are frame-level detection rates. Spatial metrics (available only after tracking) report bounding-box F1 at IoU thresholds 0.25, 0.50, 0.75 and the mean IoU of matched boxes.}
\label{tab:pipeline_ablation}
\resizebox{0.99\textwidth}{!}{%
\begin{tabular}{l  c c c c  c c c c}
\toprule
 & \multicolumn{4}{c}{\textbf{Temporal (Frame-level)}} & \multicolumn{4}{c}{\textbf{Spatial (Bounding Box)}} \\
\cmidrule(lr){2-5} \cmidrule(lr){6-9}
Pipeline Stage & Prec. & Rec. & F1 & Spec. & F1@.25 & F1@.50 & F1@.75 & mIoU \\
\midrule
VLM Detection Agent & 30.9 & 67.5 & 42.4 & 78.6 & - & - & - & - \\
+ Window Merging & 28.9 & \textbf{69.9} & 40.9 & 75.6 & - & - & - & - \\
+ Verification Agent Filtering & 36.2 & 68.7 & 47.4 & 82.9 & - & - & - & - \\
+ Spatial Localization + Tracking & 46.4 & 51.5 & 48.8 & 91.6 & \textbf{34.2} & 27.7 & 15.1 & 34.4 \\
+ AI Confirmation Agent & 53.1 & 48.8 & 50.8 & 93.9 & 34.2 & 27.8 & 15.3 & 35.2 \\
+ Human Review & \textbf{55.4} & 48.6 & \textbf{51.8} & \textbf{94.4} & 34.2 & \textbf{27.8} & \textbf{15.3} & \textbf{35.2} \\
\bottomrule
\end{tabular}}
\end{table*}

\begin{table}[t]
\centering
\caption{\textbf{\methodname Example VQA Items.} Correct answers are \textbf{bolded}.}
\label{tab:vqa_examples}
\footnotesize
\setlength{\tabcolsep}{0pt}
\renewcommand{\arraystretch}{1.05}
\begin{tabular}{@{}p{0.97\linewidth}@{}}
\toprule
\textbf{Unprompted} \textit{(raw video; same clip as Fig.~\ref{fig:qualitative-detection}\,/\,Fig.~\ref{fig:qualitative-segmentation})} \\
\midrule
\textit{Q: How is the morphology and anatomical location of the identified lesion described?} \\[1pt]
(A)~Pedunculated polyp on a rectal fold \;\;
(B)~Depressed lesion on the cecal base \;\;
(C)~Semi-pedunculated polyp in the rectum \;\;
\textbf{(D)~Sessile polyp on a haustral fold} \;\;
(E)~Flat lesion on the splenic flexure \\[4pt]
\textbf{Prompted} \textit{(video with bbox\,/\,mask overlays)} \\
\midrule
\textit{Q: At the 1.0\,s mark, which clinical finding is first clearly visible?} \\[1pt]
(A)~Flat mucosal lesion (Paris IIb) \;\;
(B)~Sessile polyp (Paris Is) \;\;
(C)~Subepithelial bulge \;\;
(D)~No abnormality detected \;\;
\textbf{(E)~Pedunculated polyp} \\
\bottomrule
\end{tabular}
\end{table}

\begin{table}[t]
\centering
\caption{\textbf{Colon-Bench Task Statistics.} Frame counts are derived from the name mapping using per-clip start/end frame indices. Bounding boxes and masks are per-frame annotations. VQA questions are 5-choice MCQ with approximately 3 questions per video.}
\label{tab:colon_bench_stats}
\setlength{\tabcolsep}{6pt}
\resizebox{1.0\linewidth}{!}{
\begin{tabular}{lccccc}
\toprule
\textbf{Statistic} & \textbf{Classification} & \textbf{Detection} & \textbf{Segmentation} & \textbf{VQA Prompted} & \textbf{VQA Unprompted} \\
\midrule
Total clips / questions      & 790       & 272       & 264       & 1,485     & 2,740     \\
Unique videos                & 790       & 272       & 264       & 499       & 918       \\
Total frames                 & 241,996   & 92,311    & 87,631    & 367,572   & 538,115   \\
Avg.\ clip length (frames)   & 306       & 339       & 332       & 737       & 586       \\
Total bounding boxes         & -       & 61,538    & -       & -       & -       \\
Total mask annotations       & -       & -       & 57,550    & -       & -       \\
MCQ choices per question     & -       & -       & -       & 5         & 5         \\
Avg.\ questions per video    & -       & -       & -       & 3.0       & 3.0       \\
Unique patient sequences     & 60        & 57        & 57        & 59        & 60        \\
\midrule
\textit{Lesion / No-lesion split} & 272 / 518 & - & - & - & - \\
\bottomrule
\end{tabular} }
\end{table}

\section{Additional Results} \label{supsec:results}
\begin{figure*}[t]
  \centering
  \includegraphics[width=0.99\linewidth]{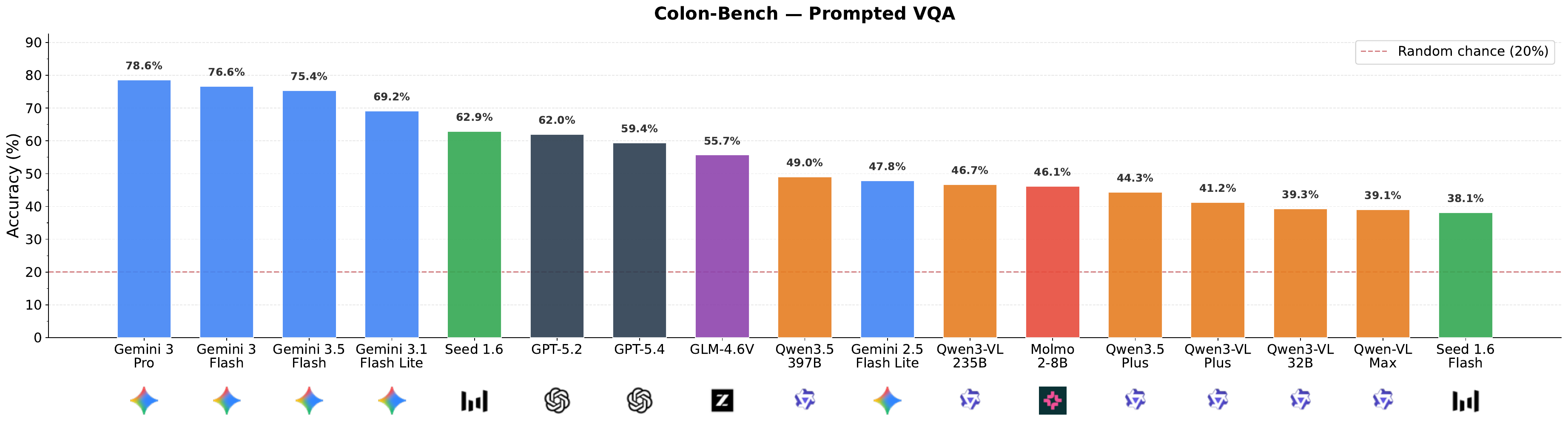}
  \caption{\textbf{\methodname Prompted VQA Accuracy.} VQA accuracy (\%) on the Colon-Bench \textbf{prompted} benchmark (1{,}485 questions) is computed as
    correct answers divided by the total number of questions so that
    unanswered questions count as incorrect.
    The dashed red line indicates the 20\% random-chance baseline.
    Models are ordered by accuracy (highest first), and colours denote model family.%
  }
  \label{fig:vqa_accuracy_easy}
\end{figure*}

\begin{figure*}[t]
  \centering
  \includegraphics[width=0.99\linewidth]{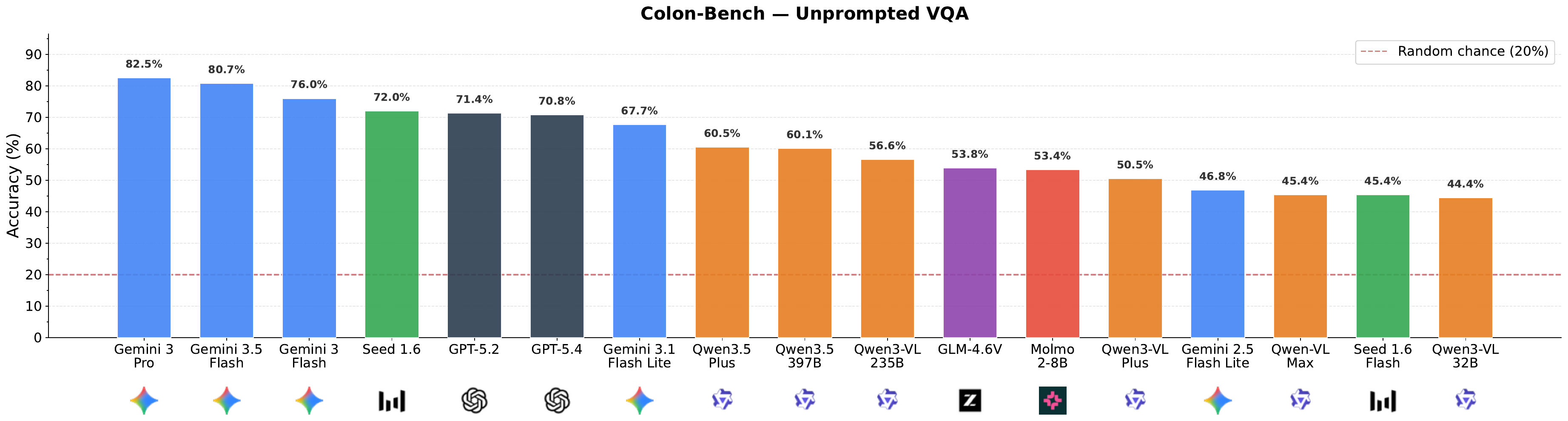}
  \caption{\textbf{\methodname Unprompted VQA Accuracy.} VQA accuracy (\%) on the Colon-Bench \textbf{unprompted} benchmark (2{,}740 questions) follows the same protocol as the prompted benchmark: all unanswered
    questions are treated as incorrect.
    The dashed red line indicates the 20\% random-chance baseline.
    Models are ordered by accuracy (highest first), and colours denote model family.%
  }
  \label{fig:vqa_accuracy_hard}
\end{figure*}

\begin{figure*}[t]
  \centering
  \includegraphics[width=0.99\linewidth]{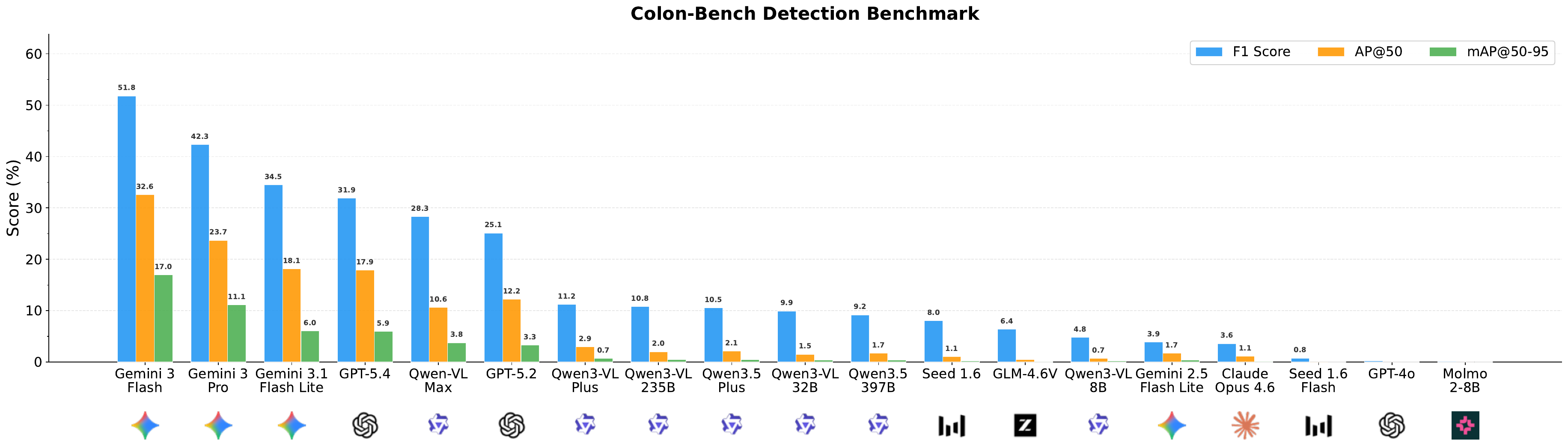}
  \caption{\textbf{\methodname Detection Metric Breakdown.} Object detection results on the Colon-Bench (272 videos) evaluate each model's ability to localise lesions via bounding-box prediction. These detections are used in the main segmentation benchmark. 
    Three metrics are reported for each model:
    \textbf{F1 Score} (primary sort key),
    \textbf{AP@50} (average precision at IoU$\,{\geq}\,0.50$), and
    \textbf{mAP@50-95} (mean average precision averaged over
    IoU thresholds 0.50-0.95).
    Models are ordered by F1 (highest first).%
  }
  \label{fig:detection_metrics}
\end{figure*}

\begin{figure*}[t]
  \centering
  \includegraphics[width=0.99\linewidth]{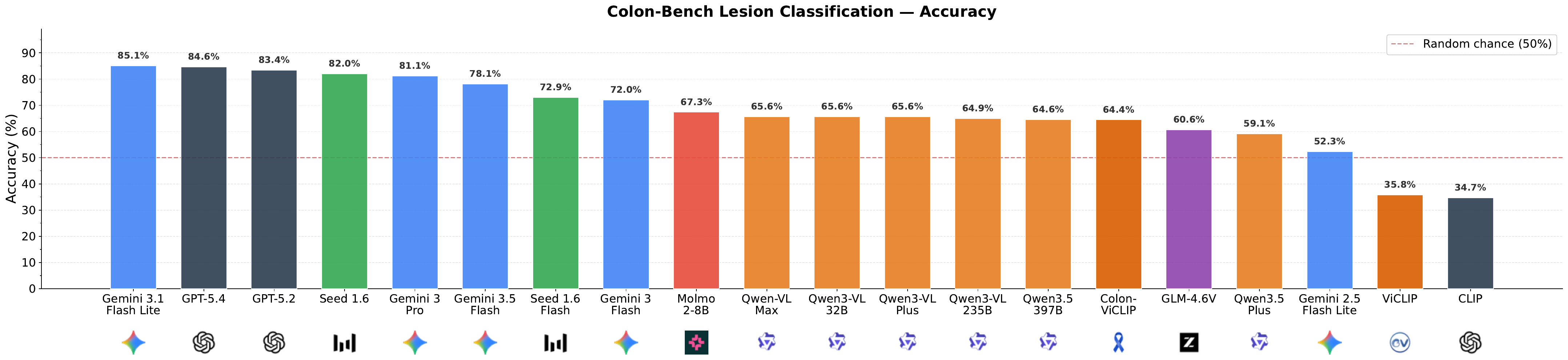}
  \caption{\textbf{\methodname Classification Accuracy Summary.} Binary lesion classification accuracy (\%) on the Colon-Bench classification benchmark (790 records) is computed after each model predicts whether a lesion is present
    (positive) or absent (negative).
    Accuracy is computed over all records; unevaluated records count as
    incorrect.
    The dashed red line indicates the 50\% random-chance baseline.
    Models are ordered by accuracy (highest first), and colours denote model family.%
  }
  \label{fig:cls_accuracy}
\end{figure*}

\begin{figure*}[t]
  \centering
  \includegraphics[width=0.99\linewidth]{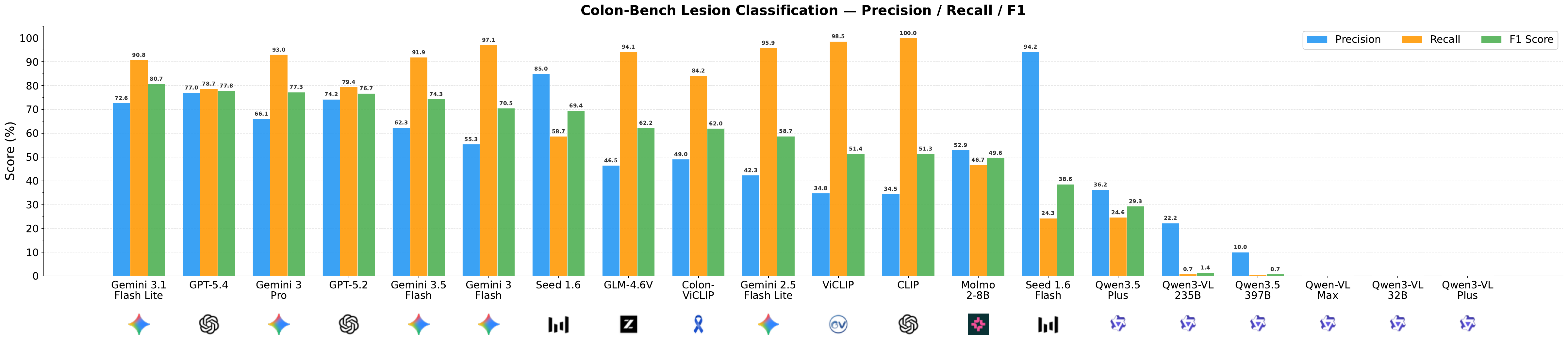}
  \caption{\textbf{\methodname Classification Precision,Recall,F1.} Binary lesion classification \textbf{Precision}, \textbf{Recall}, and \textbf{F1 Score} (\%) are reported on the Colon-Bench classification benchmark (790 records) for the positive (polyp-present) class.
    Models are ordered by F1 (highest first).%
  }
  \label{fig:cls_prf1}
\end{figure*}

Per-task results are reported in Tables~\ref{tab:vqa_accuracy} (VQA), \ref{tab:classification_results} (classification), \ref{tab:detection_results} (detection), and~\ref{tab:segmentation_results} (segmentation), with corresponding visualisations in Figs.~\ref{fig:vqa_accuracy_easy},~\ref{fig:vqa_accuracy_hard} (VQA), Fig.~\ref{fig:cls_accuracy} and~\ref{fig:cls_prf1} (classification), and Fig.~\ref{fig:detection_metrics} (detection). Qualitative detection and segmentation examples are shown in Figs.~\ref{fig:qualitative-detection} and~\ref{fig:qualitative-segmentation}.

\begin{table*}[t]
\centering
\caption{\textbf{\methodname VQA Benchmark Accuracy.} Accuracy (\%) is reported on Colon-Bench Prompted (1485 questions) and Unprompted (2740 questions), computed as correct answers divided by total questions; unanswered questions count as incorrect. Best results per benchmark are shown in \textbf{bold}.}
\label{tab:vqa_accuracy}
\begin{tabular}{l c c}
\toprule
Model & Prompted Accuracy (\%) & Unprompted Accuracy (\%) \\
\midrule
Qwen3-VL 8B & 32.9 & 38.3 \\
Seed 1.6 Flash & 38.1 & 45.4 \\
Qwen-VL Max & 39.1 & 45.4 \\
Qwen3-VL 32B & 39.3 & 44.4 \\
Qwen3-VL Plus & 41.2 & 50.5 \\
Qwen3.5 Plus & 44.3 & 60.5 \\
Molmo 2-8B & 46.1 & 53.4 \\
Qwen3-VL 235B & 46.7 & 56.6 \\
Gemini 2.5 Flash Lite & 47.8 & 46.8 \\
Qwen3.5 397B & 49.0 & 60.1 \\
GLM-4.6V & 55.7 & 53.8 \\
Seed 1.6 & 62.9 & 72.0 \\
GPT-5.2               & 62.0	           & 71.4 \\
GPT-5.4               & 59.4	           & 70.8 \\
Gemini 3.1 Flash Lite & 69.2 & 67.7 \\
Gemini 3 Flash & 76.6 & 76.0 \\
Gemini 3 Pro & \textbf{78.6} & \textbf{82.5} \\
\bottomrule
\end{tabular}
\end{table*}

\begin{table*}[t]
\centering
\caption{\textbf{\methodname Segmentation Benchmark Results.} Video object segmentation results (\%) are reported on the Colon-Bench segmentation benchmark (272 videos). Each model first detects the target lesion via bounding-box prompting, then a SAM-based tracker propagates the mask across the tracking window. Mean IoU and Mean Dice are computed over all evaluated frames. Best results per metric are shown in \textbf{bold}.}
\label{tab:segmentation_results}
  \setlength{\tabcolsep}{3pt}
\begin{tabular}{l c c}
\toprule
Model & mIoU & mDice \\
\midrule
GPT-4o & 0.5 & 0.8 \\
SAM 3 & 2.5 & 2.9 \\
Seed 1.6 Flash & 2.6 & 3.5 \\
Molmo 2-8B & 2.6 & 3.8 \\
Qwen3-VL 8B & 10.4 & 13.1 \\
GLM-4.6V & 12.5 & 16.1 \\
Seed 1.6 & 12.6 & 16.1 \\
Qwen3-VL 32B & 12.7 & 15.9 \\
Qwen3-VL 235B & 13.6 & 16.9 \\
Claude Opus 4.6 & 16.1 & 20.5 \\
Qwen3.5 397B & 16.6 & 21.0 \\
Qwen3.5 Plus & 16.7 & 21.0 \\
Gemini 2.5 Flash Lite & 19.9 & 24.3 \\
Qwen3-VL Plus & 20.4 & 24.6 \\
Qwen-VL Max & 25.6 & 29.6 \\
GPT-5.2 & 30.7 & 36.5 \\
GPT-5.4 & 34.5 & 41.1 \\
Gemini 3.1 Flash Lite & 37.4 & 43.4 \\
Gemini 3 Pro & 45.0 & 51.3 \\
\textbf{Gemini 3 Flash} & \textbf{48.3} & \textbf{54.7} \\
\bottomrule
\end{tabular}
\end{table*}

\begin{table*}[t]
\centering
\caption{\textbf{\methodname Classification Benchmark Results.} Binary lesion classification results (\%) are reported on the Colon-Bench classification benchmark (790 records). Each model receives a video clip and predicts whether a lesion is present (positive) or absent (negative). Accuracy is computed over all records; unevaluated records count as incorrect. Precision, Recall, and F1 are reported for the positive (lesion-present) class. Best results per metric are shown in \textbf{bold}.}
\label{tab:classification_results}
  \setlength{\tabcolsep}{3pt}
\begin{tabular}{l c c c c}
\toprule
Model & Acc. & Prec. & Rec. & F1 \\
\midrule
Qwen3-VL 8B & 34.4 & 34.4 & \textbf{100.0} & 51.2 \\
CLIP & 34.7 & 34.5 & \textbf{100.0} & 51.3 \\
ViCLIP & 35.8 & 34.8 & 98.5 & 51.4 \\
Gemini 2.5 Flash Lite & 52.3 & 42.3 & 95.9 & 58.7 \\
Endo-CLIP & 52.9 & 41.9 & 95.2 & 58.2 \\
Qwen3.5 Plus & 59.1 & 36.2 & 24.6 & 29.3 \\
GLM-4.6V & 60.6 & 46.5 & 94.1 & 62.2 \\
Colon- ViCLIP & 64.4 & 49.0 & 84.2 & 62.0 \\
Qwen3.5 397B & 64.6 & 10.0 & 0.4 & 0.7 \\
Qwen3-VL 235B & 64.9 & 22.2 & 0.7 & 1.4 \\
Qwen-VL Max & 65.6 & 0.0 & 0.0 & 0.0 \\
Qwen3-VL 32B & 65.6 & 0.0 & 0.0 & 0.0 \\
Qwen3-VL Plus & 65.6 & 0.0 & 0.0 & 0.0 \\
Molmo 2-8B & 67.3 & 52.9 & 46.7 & 49.6 \\
Seed 1.6 Flash & 72.9 & \textbf{94.2} & 24.3 & 38.6 \\
Seed 1.6 & 82.0 & 85.0 & 58.7 & 69.4 \\
GPT-5.2               & 83.4  &  74.2	            & 79.4	             & 76.7          \\
GPT-5.4               & 84.6 &  77.0	            & 78.7	            & 77.8            \\
Gemini 3 Flash & 72.0 & 55.3 & 97.1 & 70.5 \\
Gemini 3 Pro & 81.1 & 66.1 & 93.0 & 77.3 \\
\textbf{Gemini 3.1 Flash Lite} & \textbf{85.1} & 72.6 & 90.8 & \textbf{80.7} \\
\bottomrule
\end{tabular}
\end{table*}

\begin{table*}[t]
\centering
\caption{\textbf{\methodname Detection Benchmark Results.} Object detection results (\%) are reported on the Colon-Bench detection benchmark (272 videos). Precision, Recall, and F1 are computed at IoU$\geq$0.50. AP@50 is the average precision at IoU$\geq$0.50; mAP@50-95 averages AP across IoU thresholds 0.50-0.95. Mean IoU is the average intersection-over-union of matched predicted and ground-truth boxes. Best results per metric are shown in \textbf{bold}.}
\label{tab:detection_results}
  \setlength{\tabcolsep}{3pt}
\begin{tabular}{l c c c c c c}
\toprule
Model & Prec. & Rec. & F1 & AP@50 & mAP@50-95 & mIoU \\
\midrule
Molmo 2-8B & 0.2 & 0.1 & 0.1 & 0.0 & 0.0 & 53.2 \\
GPT-4o & 0.6 & 0.1 & 0.2 & 0.0 & 0.0 & 55.4 \\
Seed 1.6 Flash & 1.6 & 0.5 & 0.8 & 0.0 & 0.0 & 55.7 \\
Claude Opus 4.6 & 3.5 & 3.7 & 3.6 & 1.1 & 0.1 & 59.5 \\
Gemini 2.5 Flash Lite & 2.5 & 9.5 & 3.9 & 1.7 & 0.4 & 61.8 \\
Qwen3-VL 8B & 6.6 & 3.8 & 4.8 & 0.7 & 0.2 & 62.6 \\
GLM-4.6V & 6.6 & 6.2 & 6.4 & 0.5 & 0.1 & 61.6 \\
Seed 1.6 & 10.4 & 6.6 & 8.0 & 1.1 & 0.2 & 59.7 \\
Qwen3.5 397B & 9.4 & 8.9 & 9.2 & 1.7 & 0.4 & 62.9 \\
Qwen3-VL 32B & 14.4 & 7.6 & 9.9 & 1.5 & 0.4 & 62.1 \\
Qwen3.5 Plus & 10.7 & 10.3 & 10.5 & 2.1 & 0.4 & 61.5 \\
Qwen3-VL 235B & 15.7 & 8.2 & 10.8 & 2.0 & 0.5 & 62.9 \\
Qwen3-VL Plus & 10.0 & 12.7 & 11.2 & 2.9 & 0.7 & 64.0 \\
GPT-5.2 & 26.1 & 24.2 & 25.1 & 12.2 & 3.3 & 66.1 \\
Qwen-VL Max & 39.0 & 22.2 & 28.3 & 10.6 & 3.8 & 71.4 \\
GPT-5.4 & 33.2 & 30.6 & 31.9 & 17.9 & 5.9 & 68.4 \\
Gemini 3.1 Flash Lite & 36.8 & 32.5 & 34.5 & 18.1 & 6.0 & 69.9 \\
Gemini 3 Pro & 42.2 & 42.5 & 42.3 & 23.7 & 11.1 & 76.3 \\
\textbf{Gemini 3 Flash} & \textbf{50.9} & \textbf{52.7} & \textbf{51.8} & \textbf{32.6} & \textbf{17.0} & \textbf{79.8} \\
\bottomrule
\end{tabular}
\end{table*}

\begin{figure*}[!t]
  \centering
  \begin{minipage}{\textwidth}
  \setlength{\tabcolsep}{1pt}
  \renewcommand{\arraystretch}{0.6}
  {\setlength{\tabcolsep}{1pt}%
  \renewcommand{\arraystretch}{1.0}%
  \resizebox{0.99\textwidth}{!}{%
  \begin{tabular}{*{8}{p{\dimexpr\textwidth/8\relax}}}
    \colhead{Input} &
    \colhead{GT} &
    \colhead{GPT-5.2} &
    \colhead{Opus 4.6} &
    \colhead{Gemini 3 Flash} &
    \colhead{Qwen 3.5-Plus} &
    \colhead{Qwen VL-Max} &
    \colhead{Qwen 3 VL-235B} \\
  \end{tabular}}\par}
  \vspace{2pt}
  \resizebox{0.99\textwidth}{!}{%
  \begin{tabular}{cccccccc}
    \snap{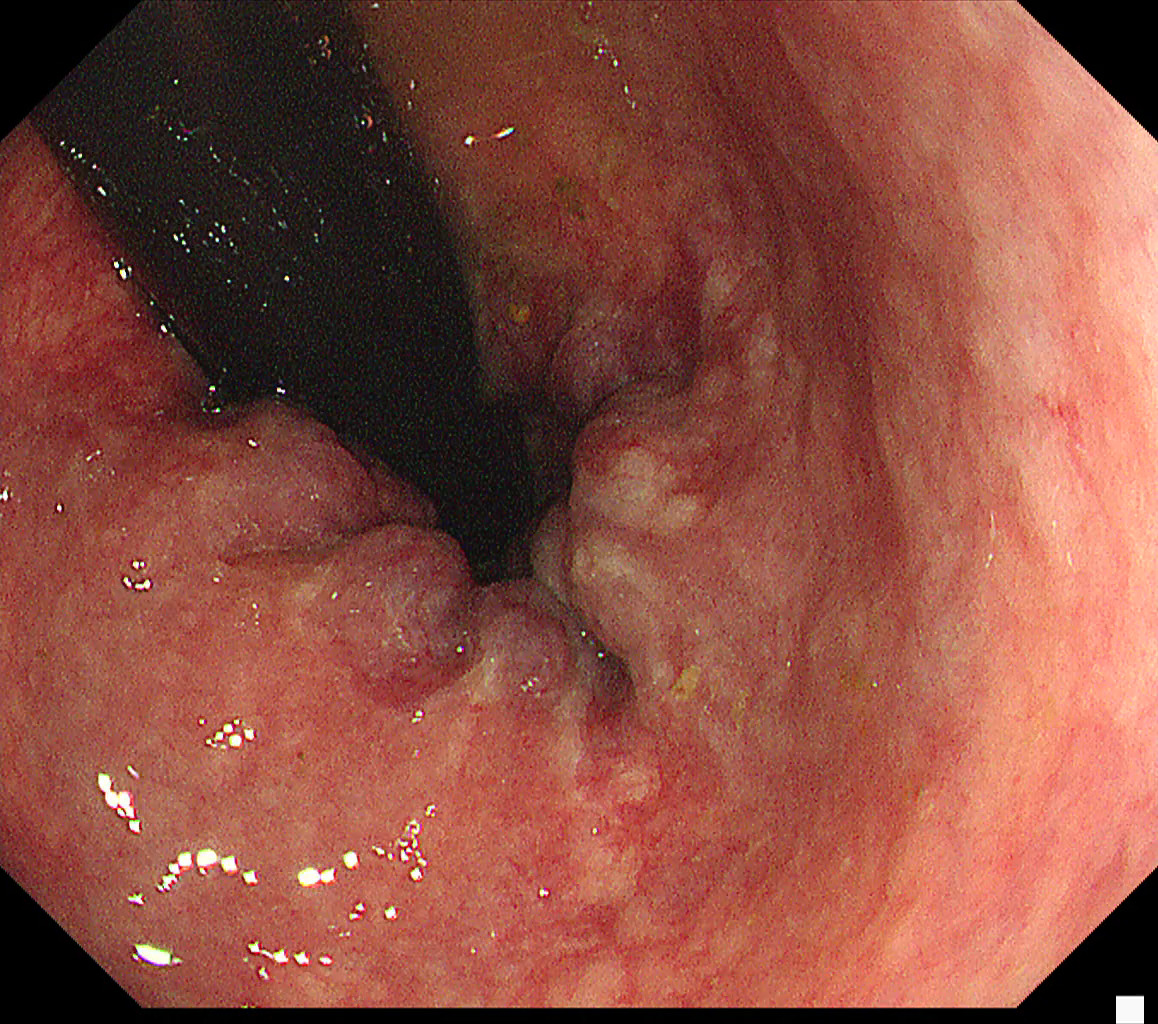} &
    \snap{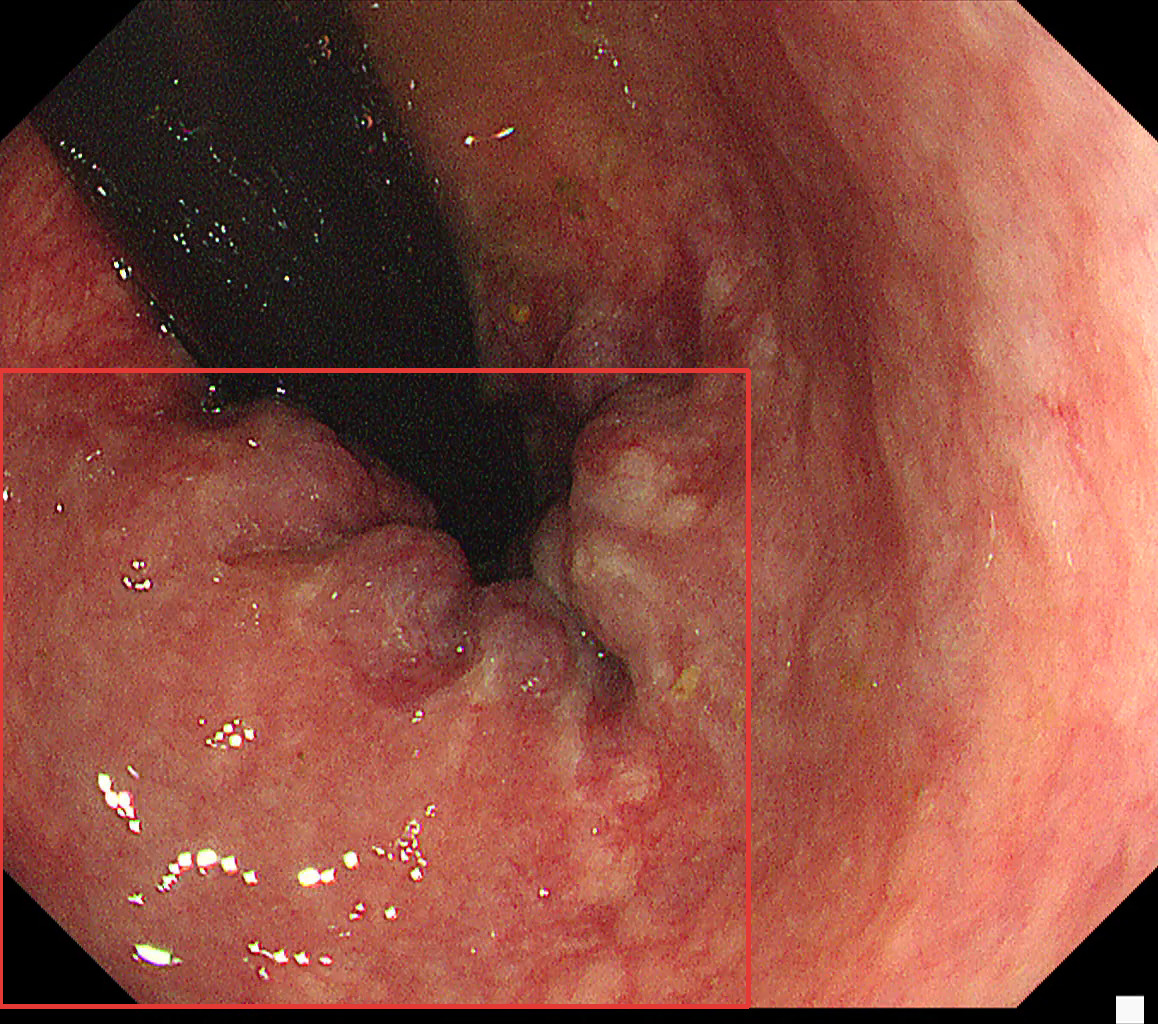} &
    \snap{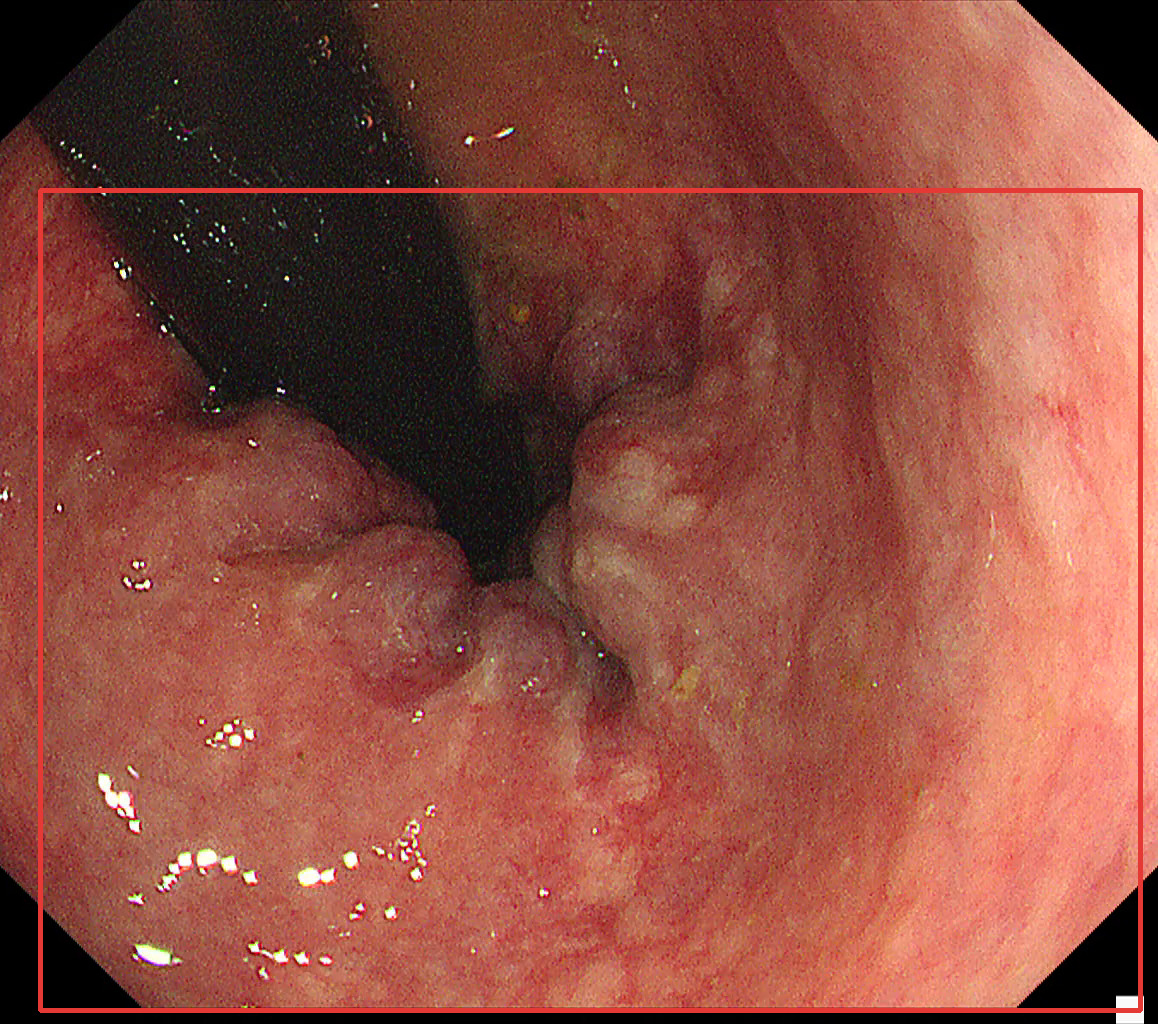} &
    \snap{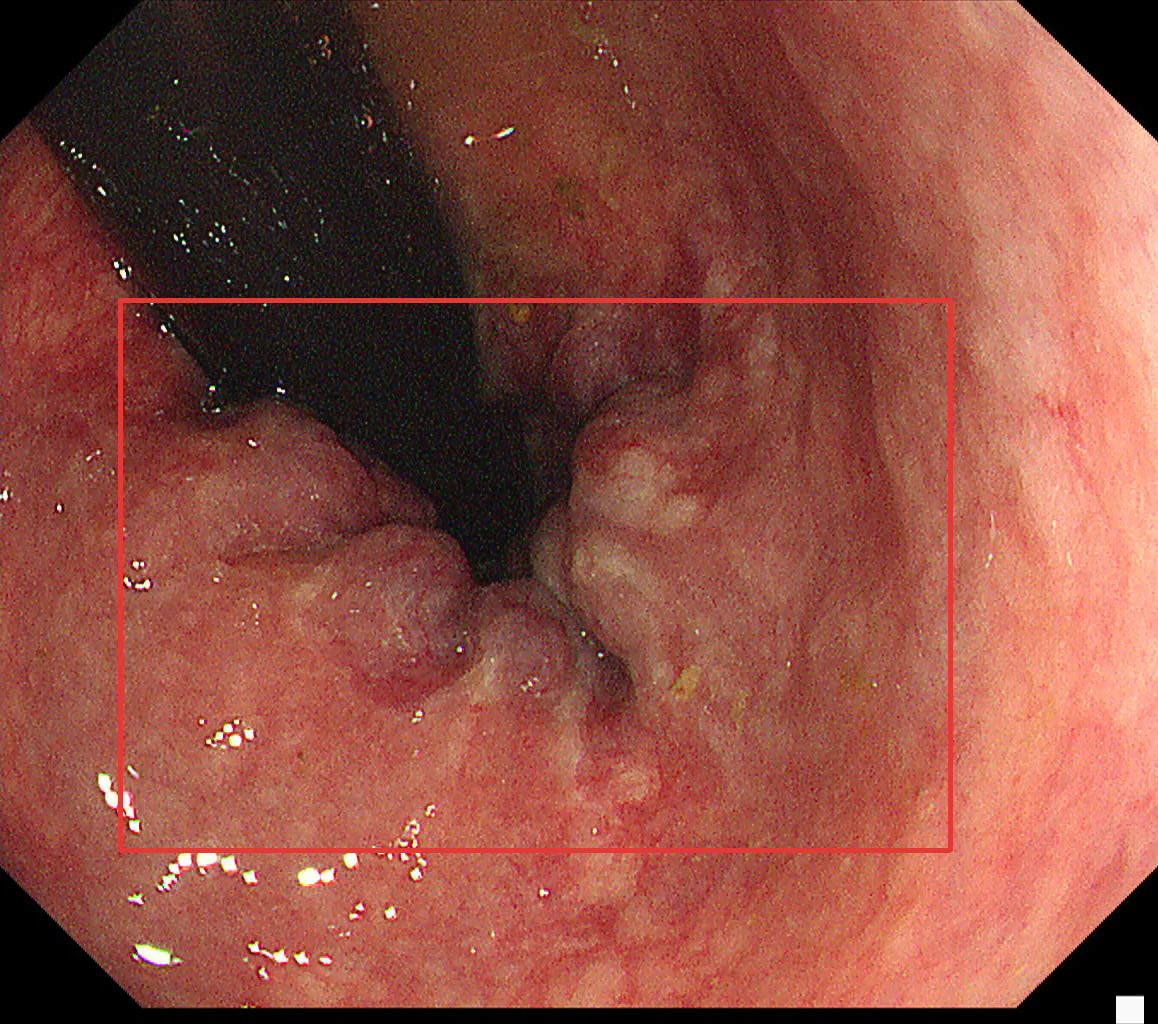} &
    \snap{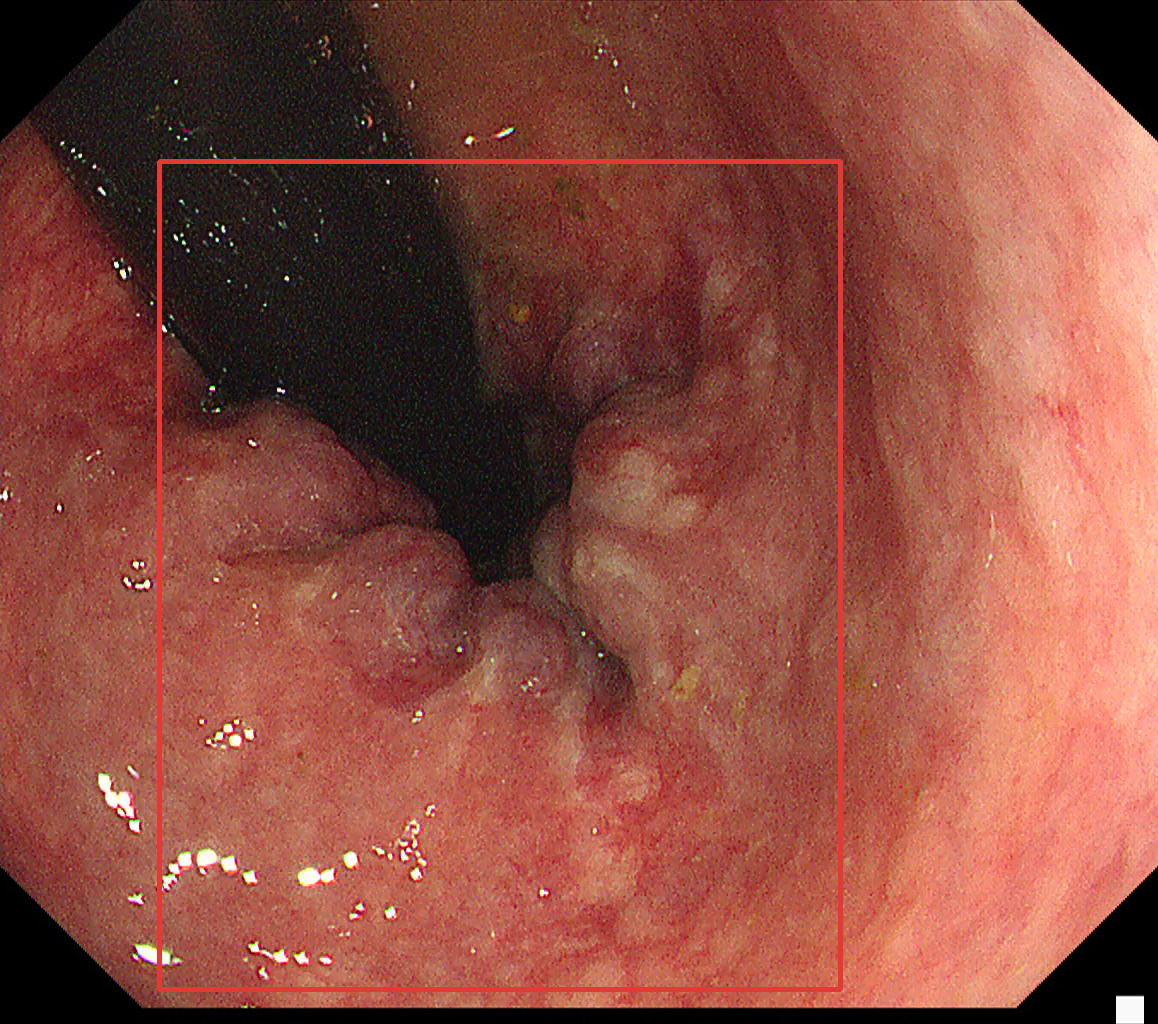} &
    \snap{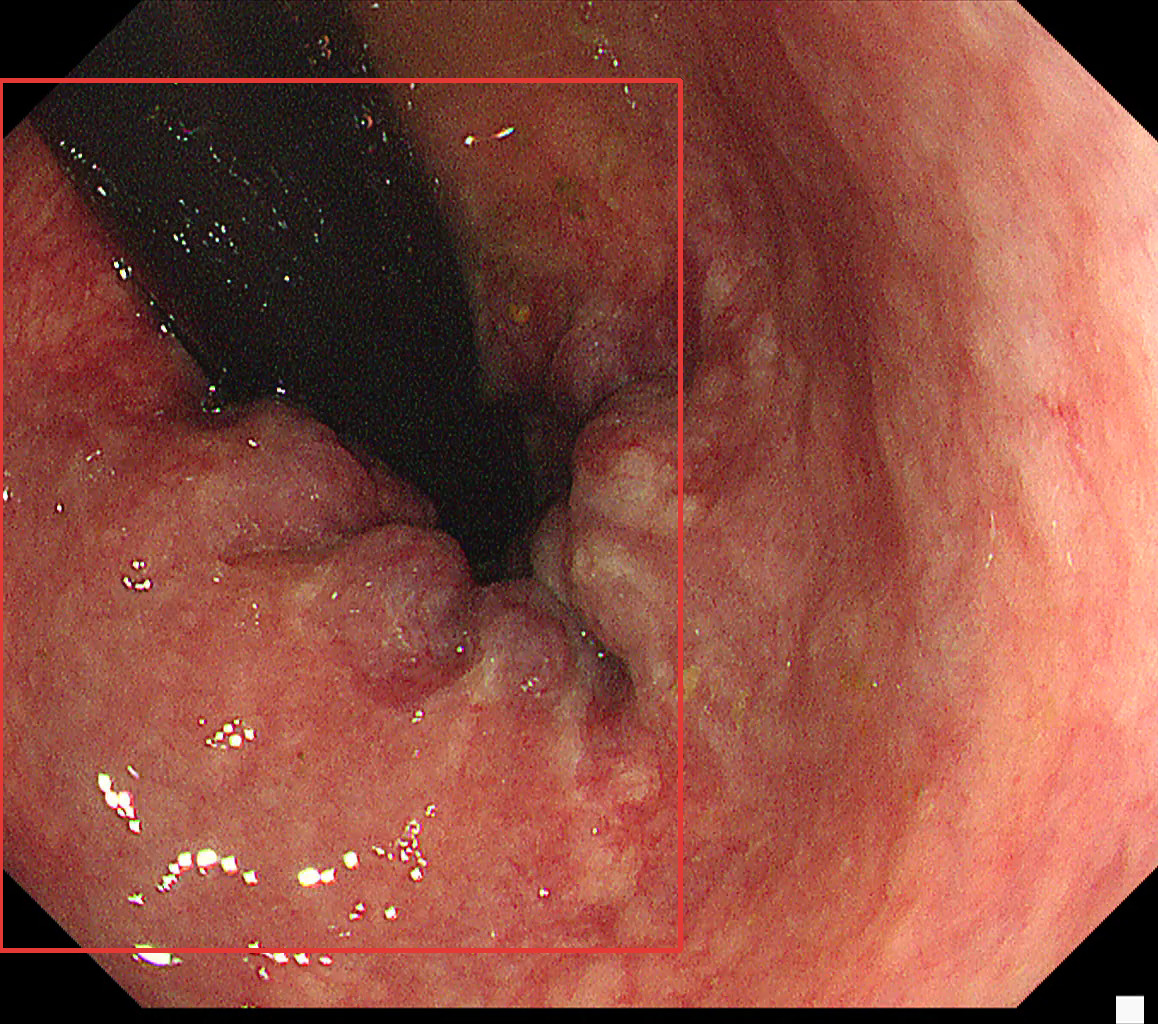} &
    \snap{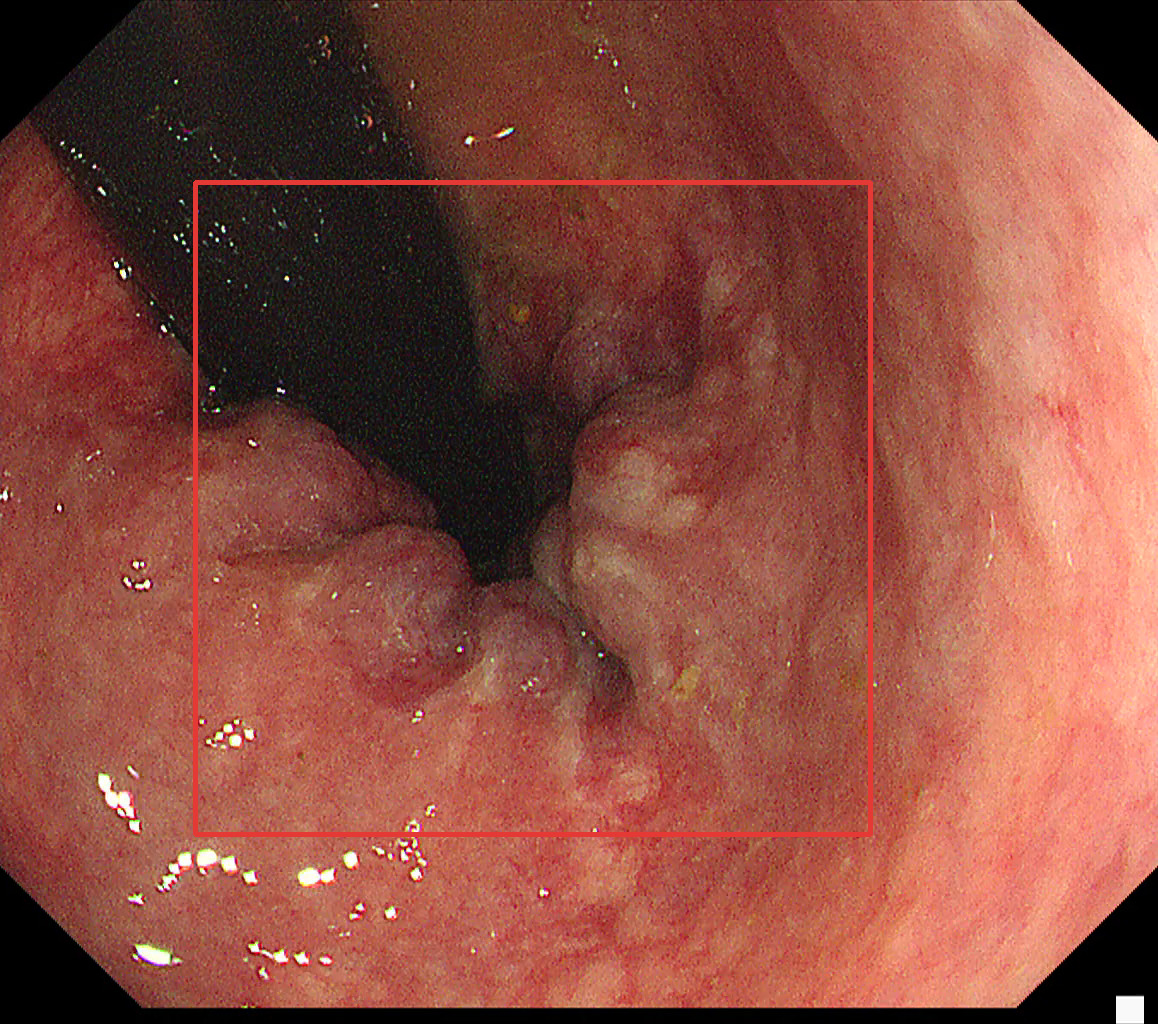} &
    \snap{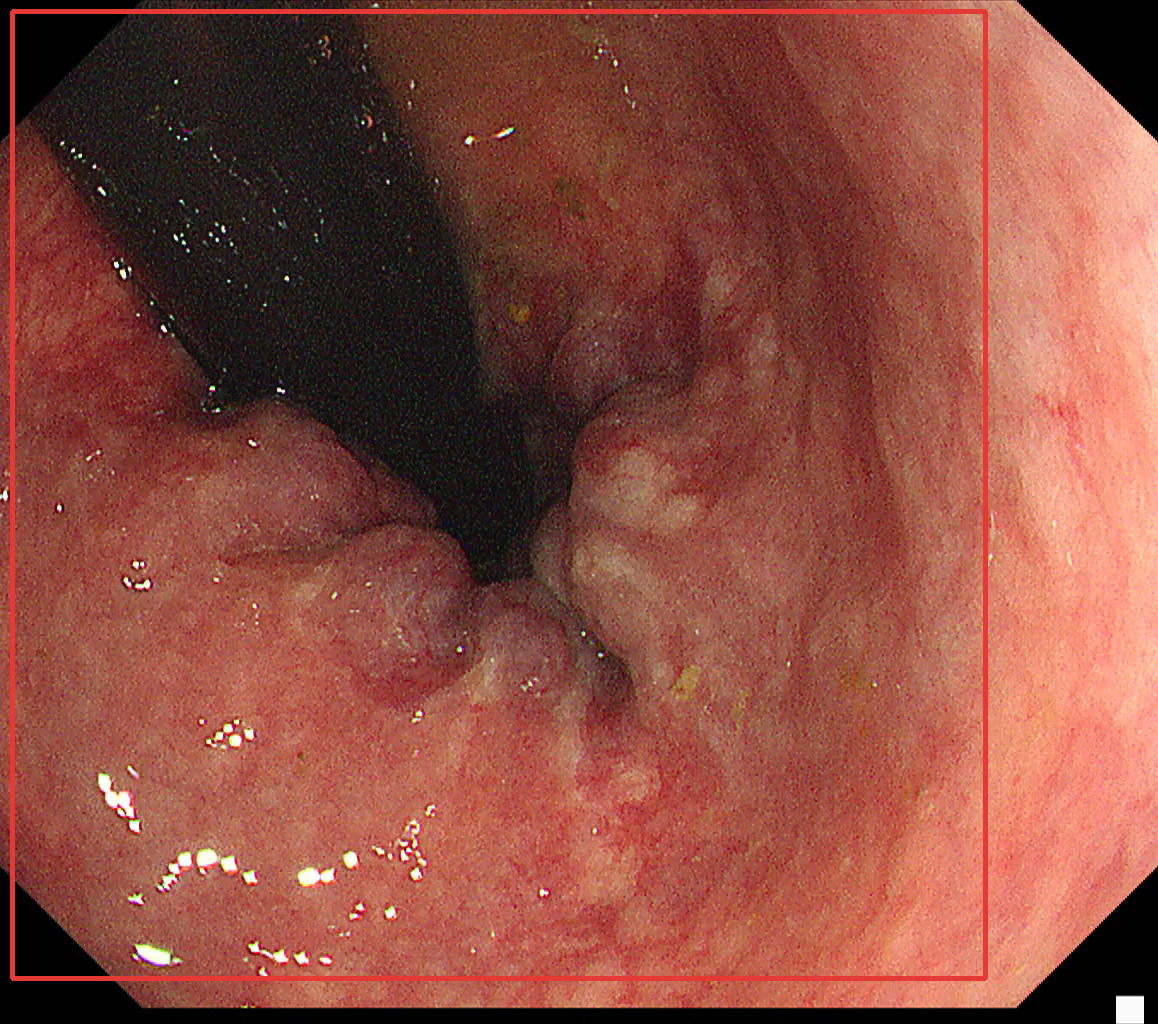} \\[1pt]
    \snap{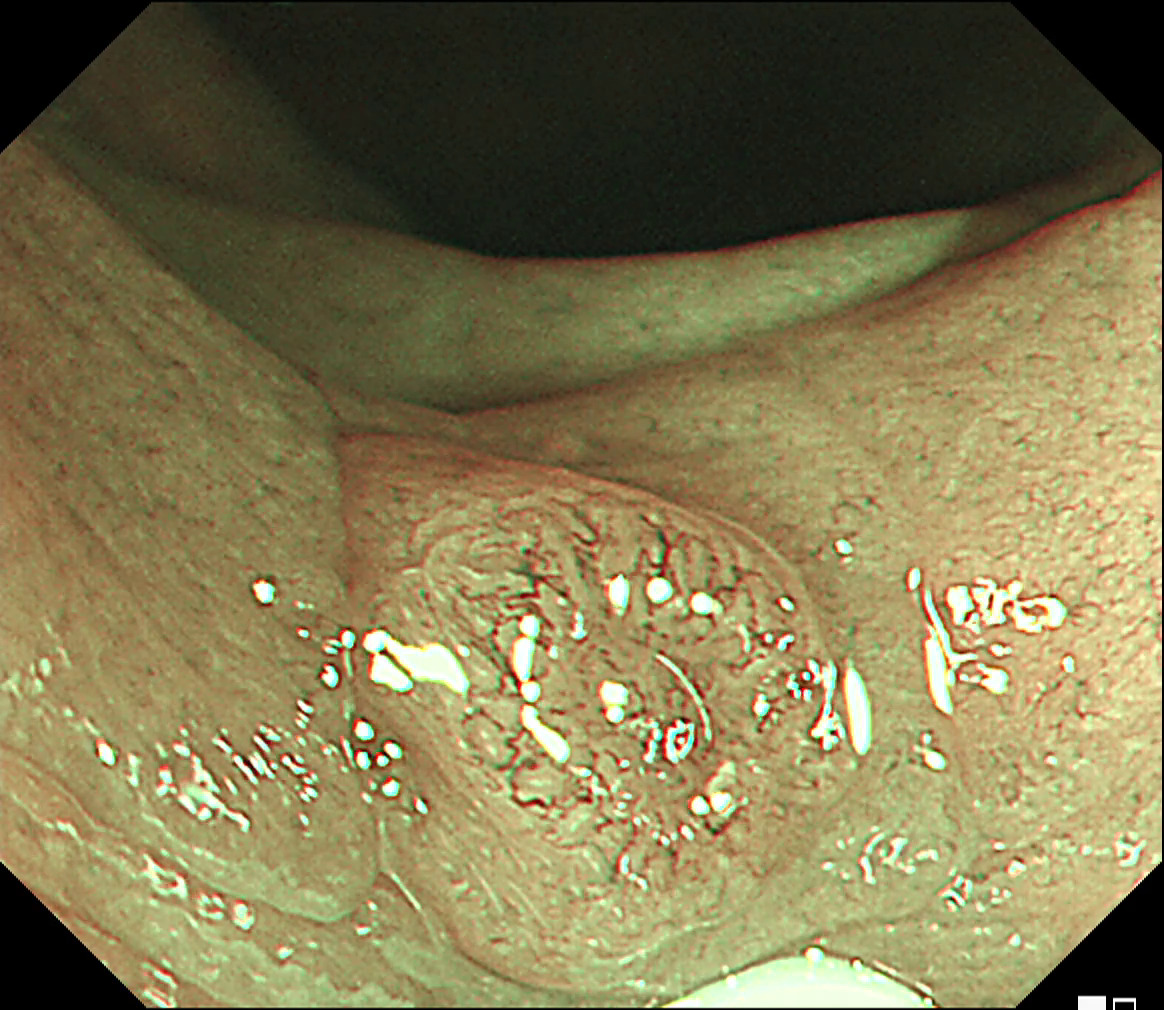} &
    \snap{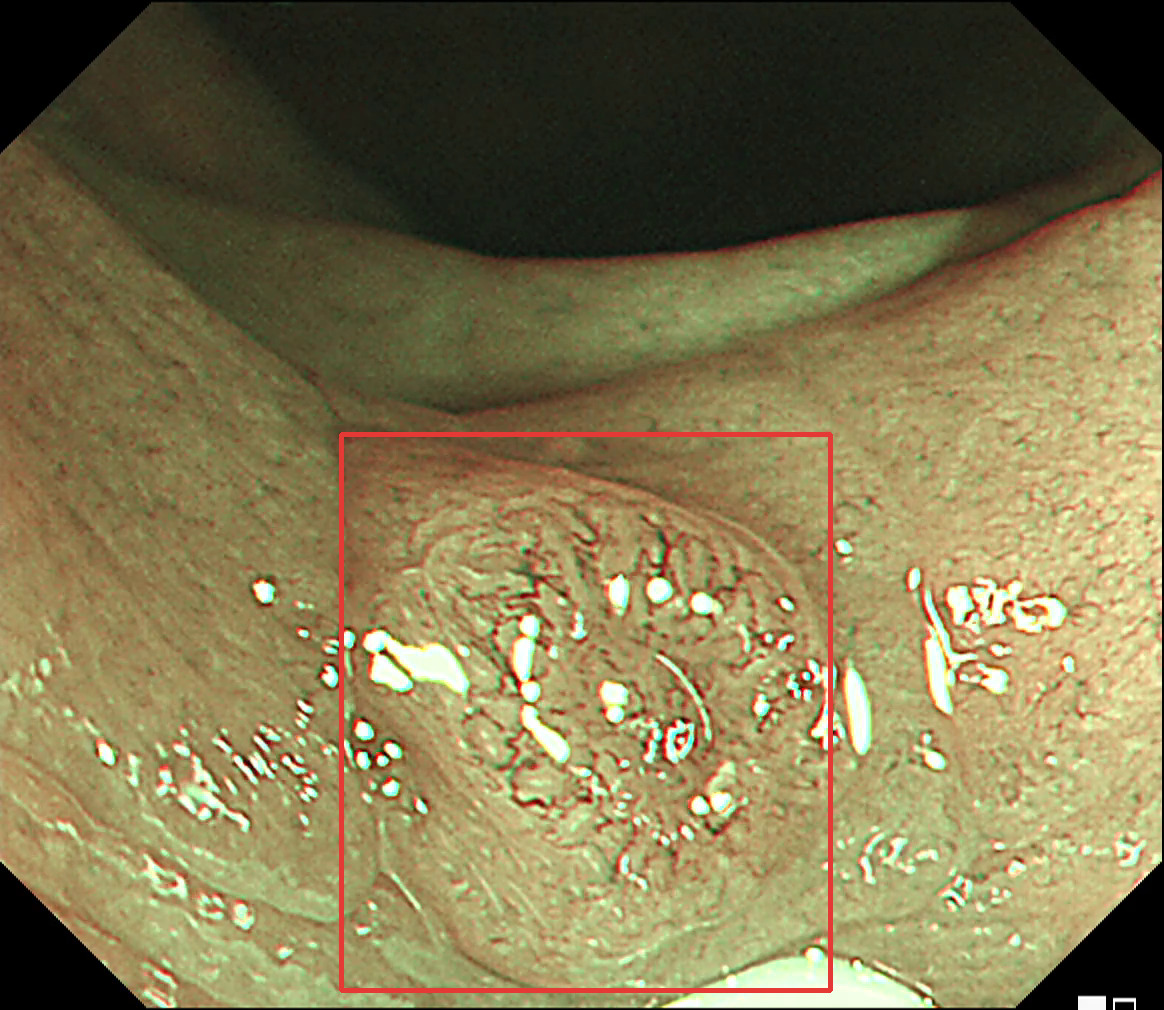} &
    \snap{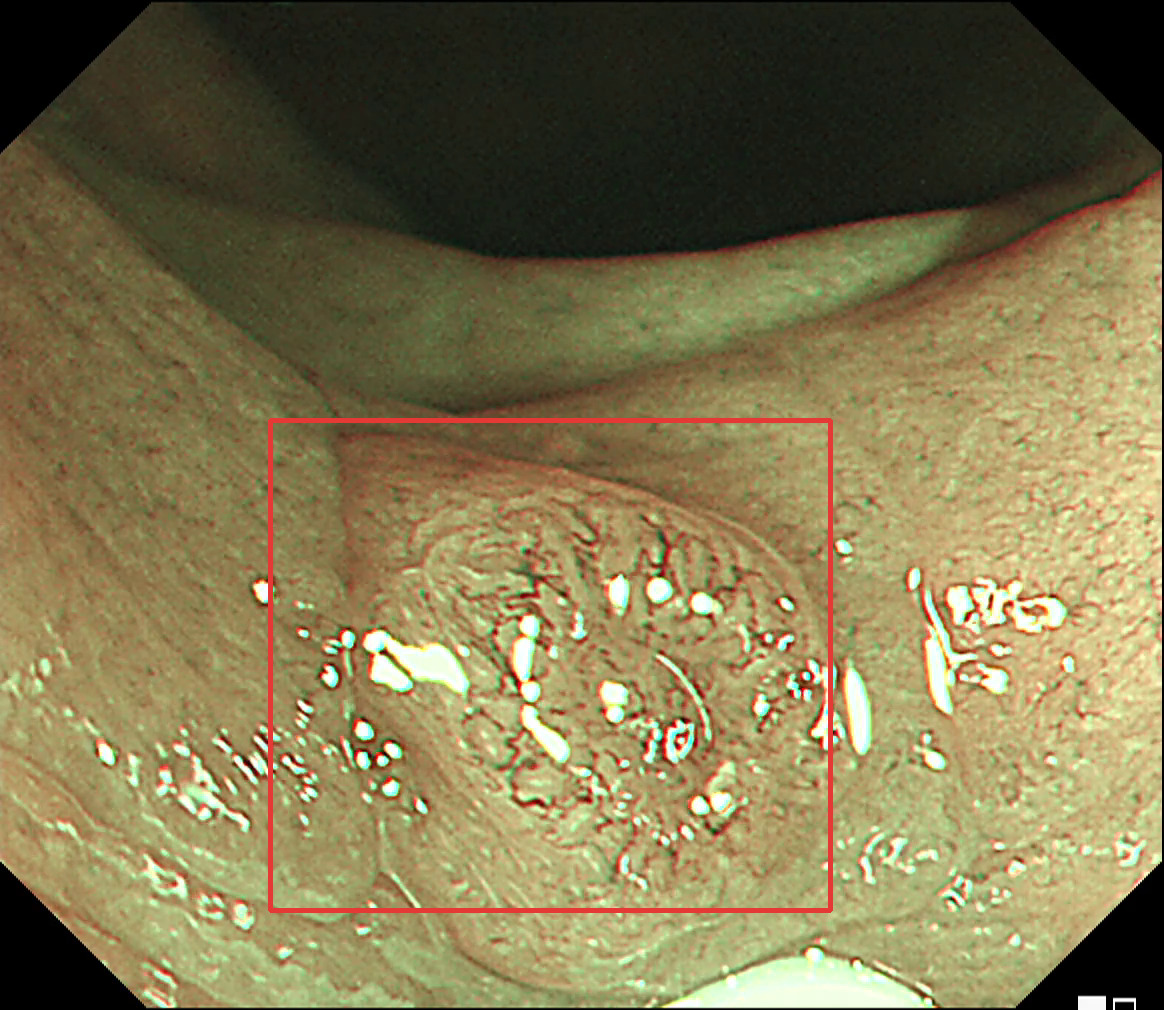} &
    \snap{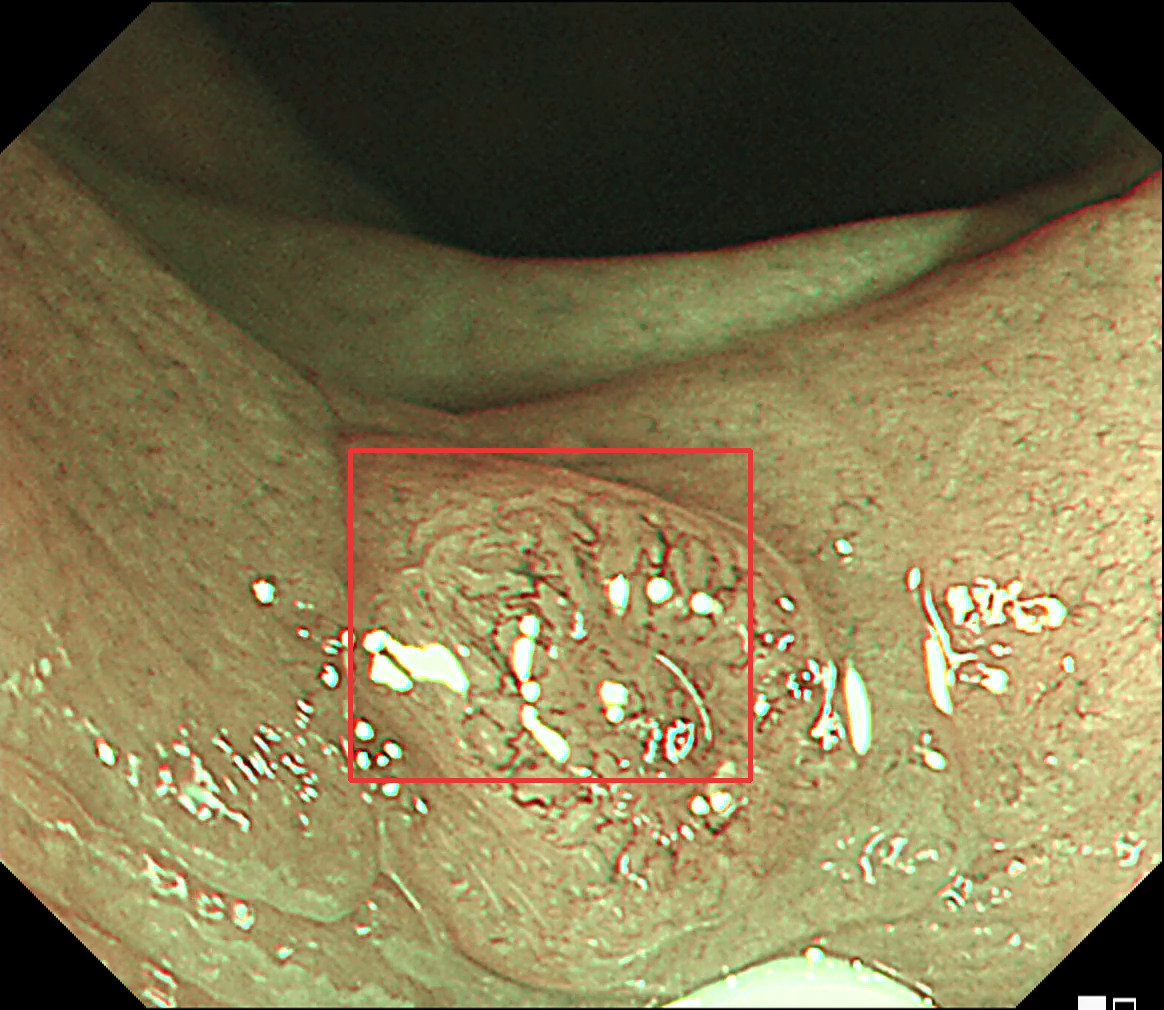} &
    \snap{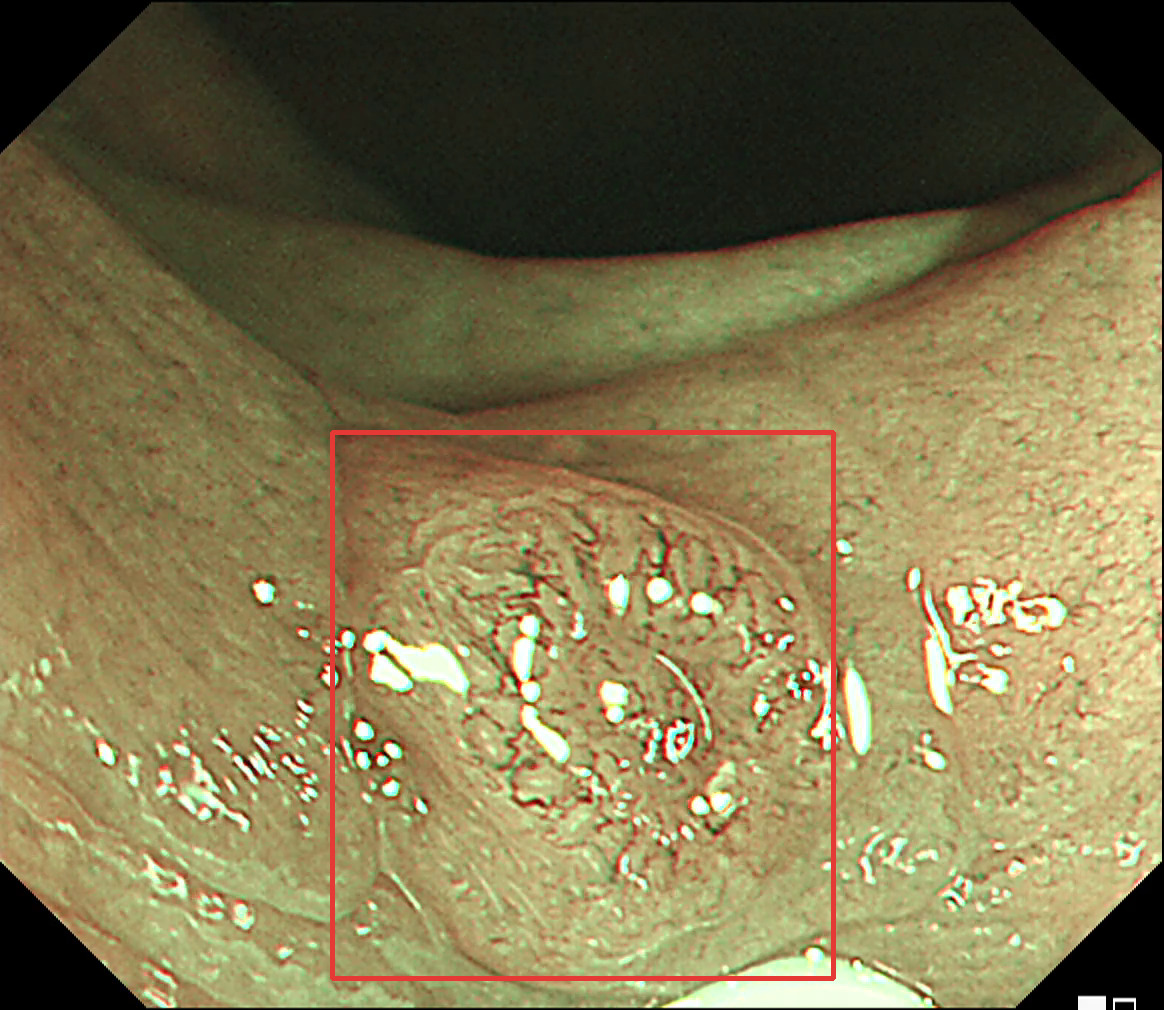} &
    \snap{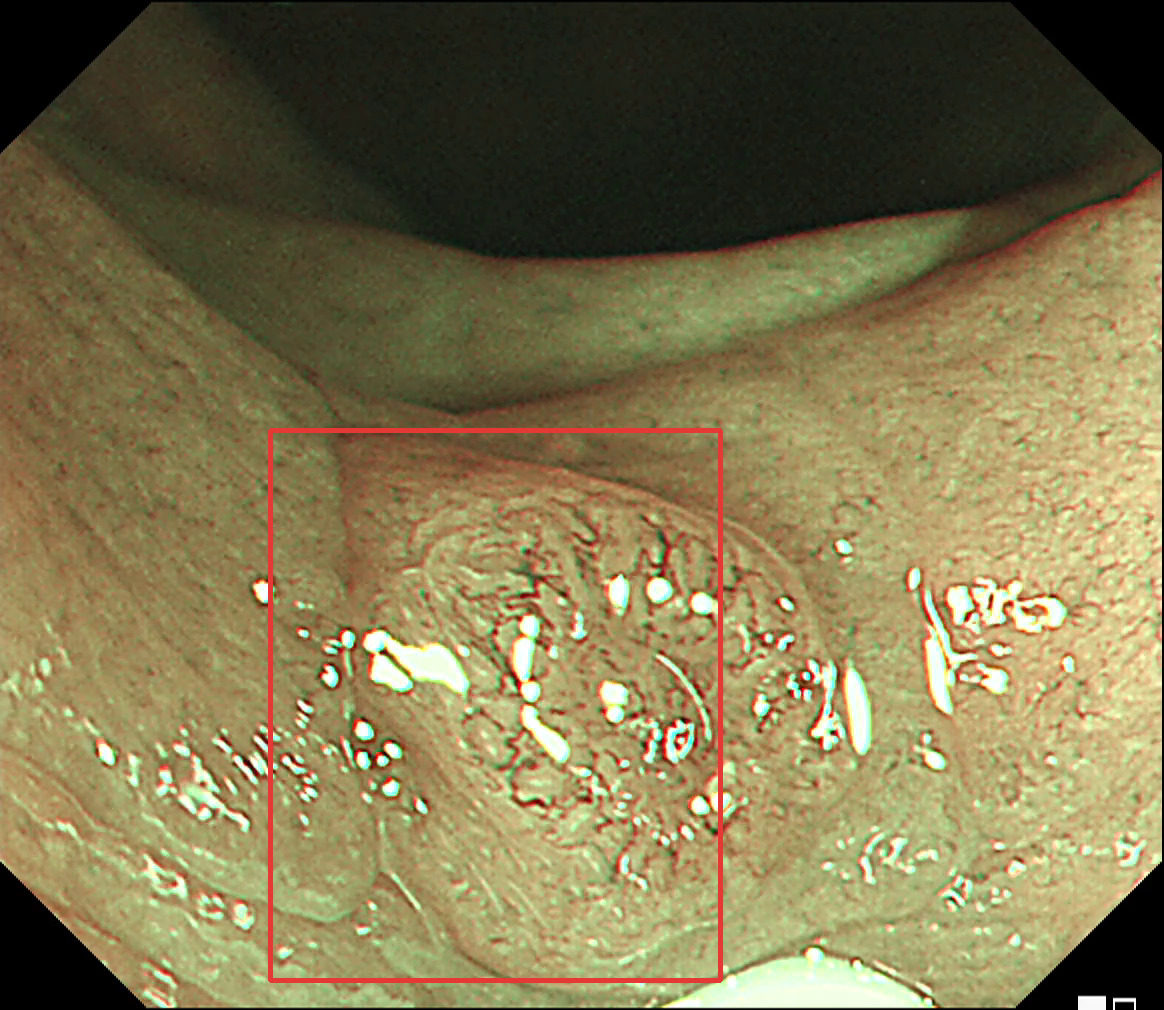} &
    \snap{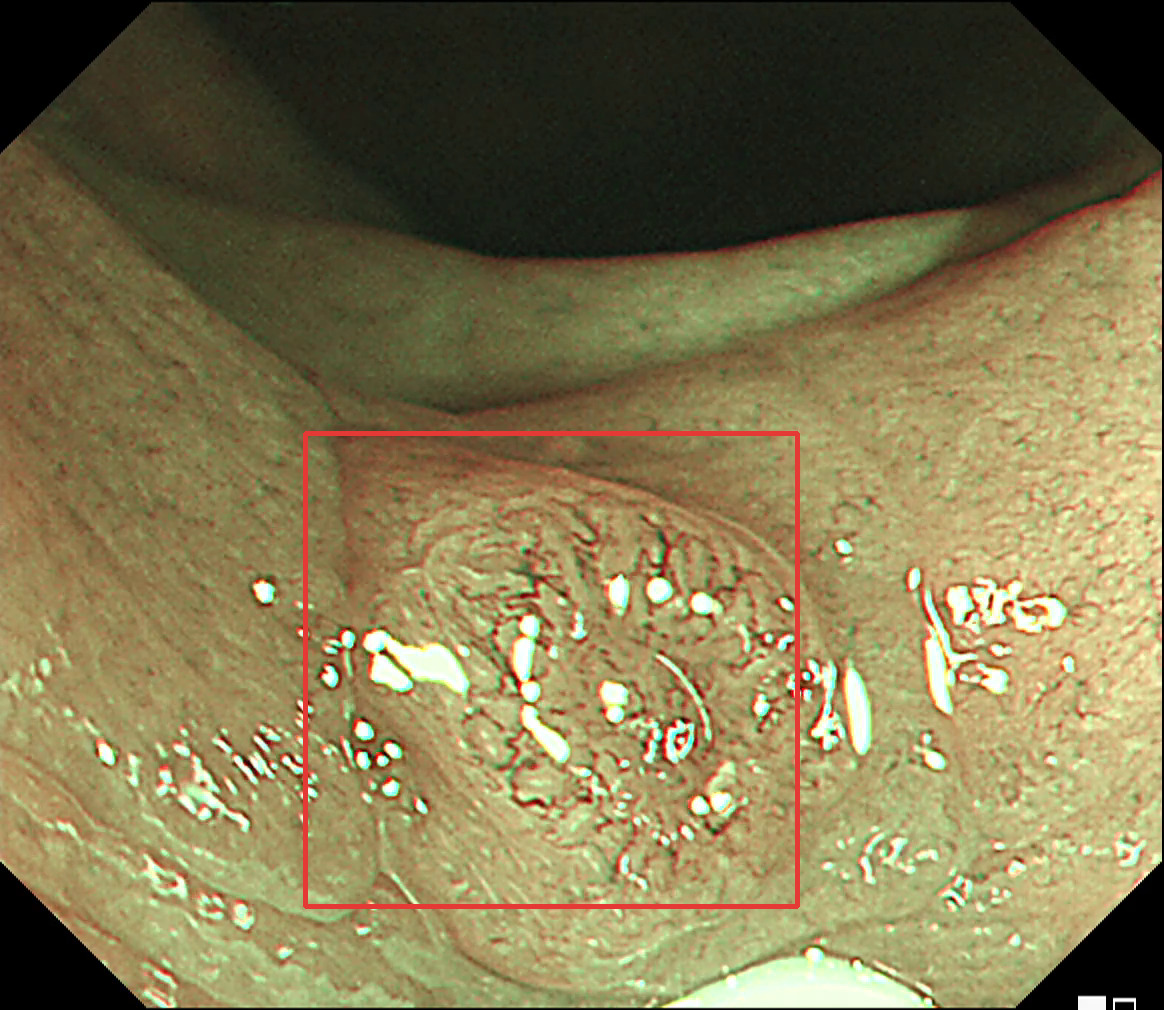} &
    \snap{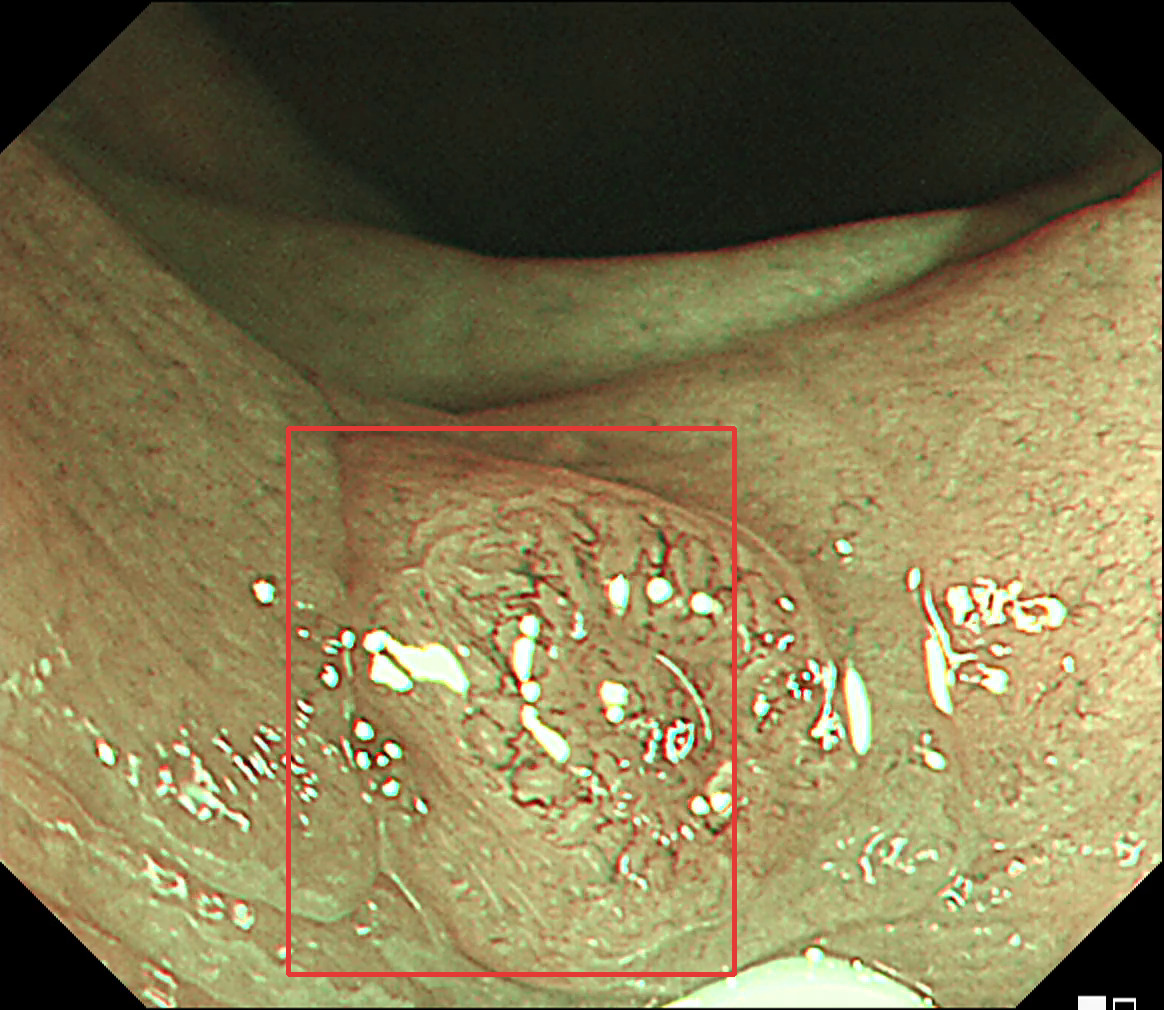} \\[1pt]
    \snap{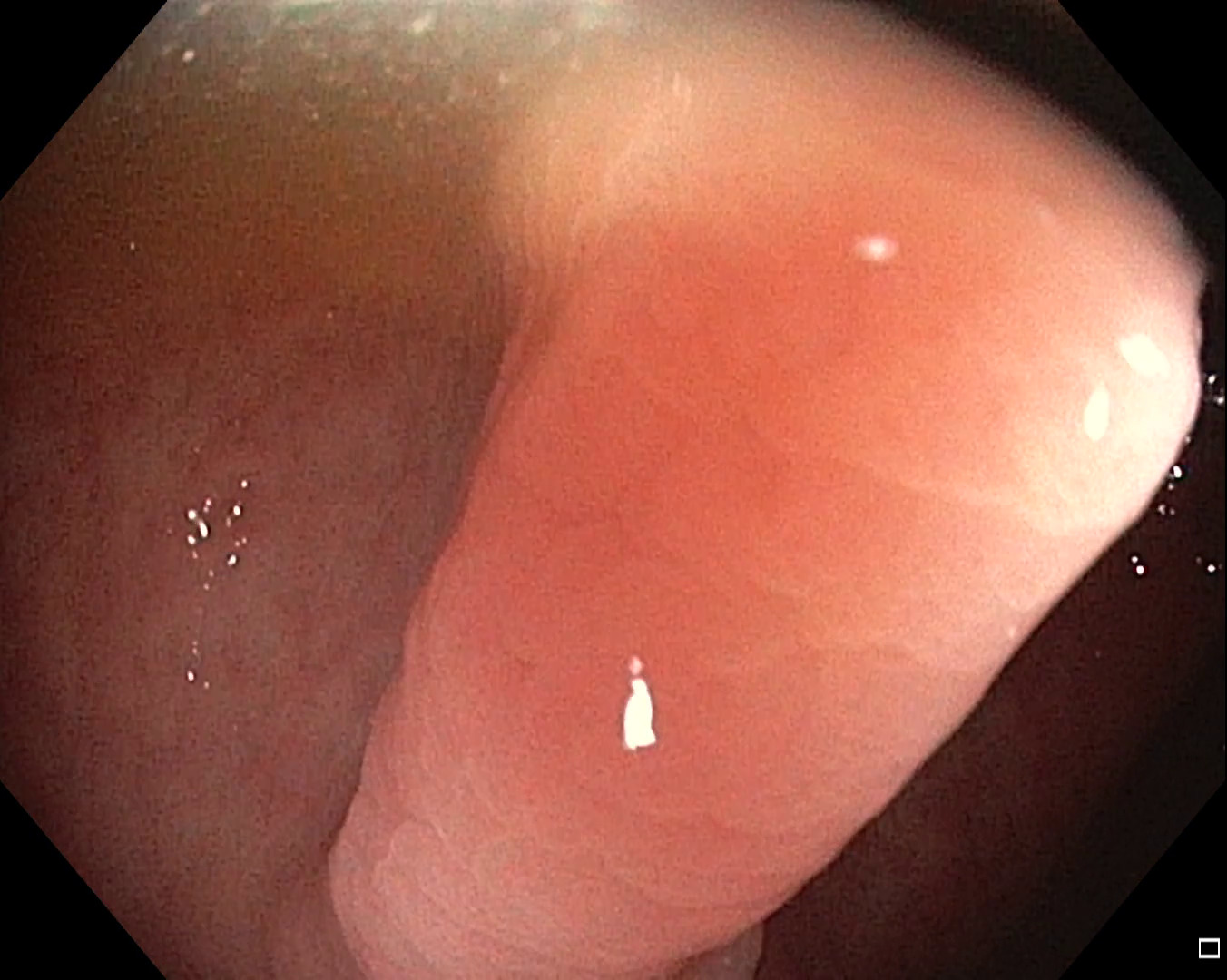} &
    \snap{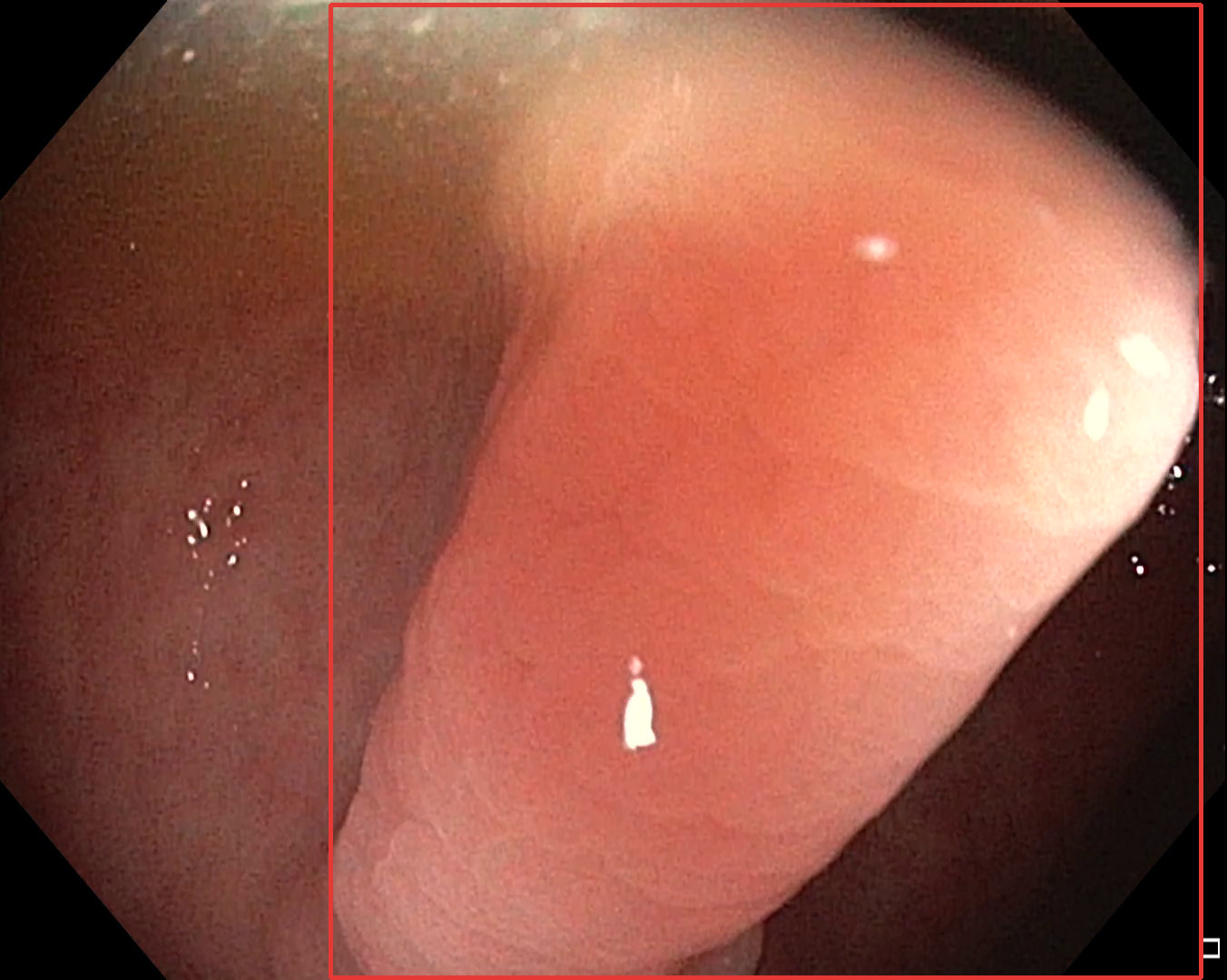} &
    \snap{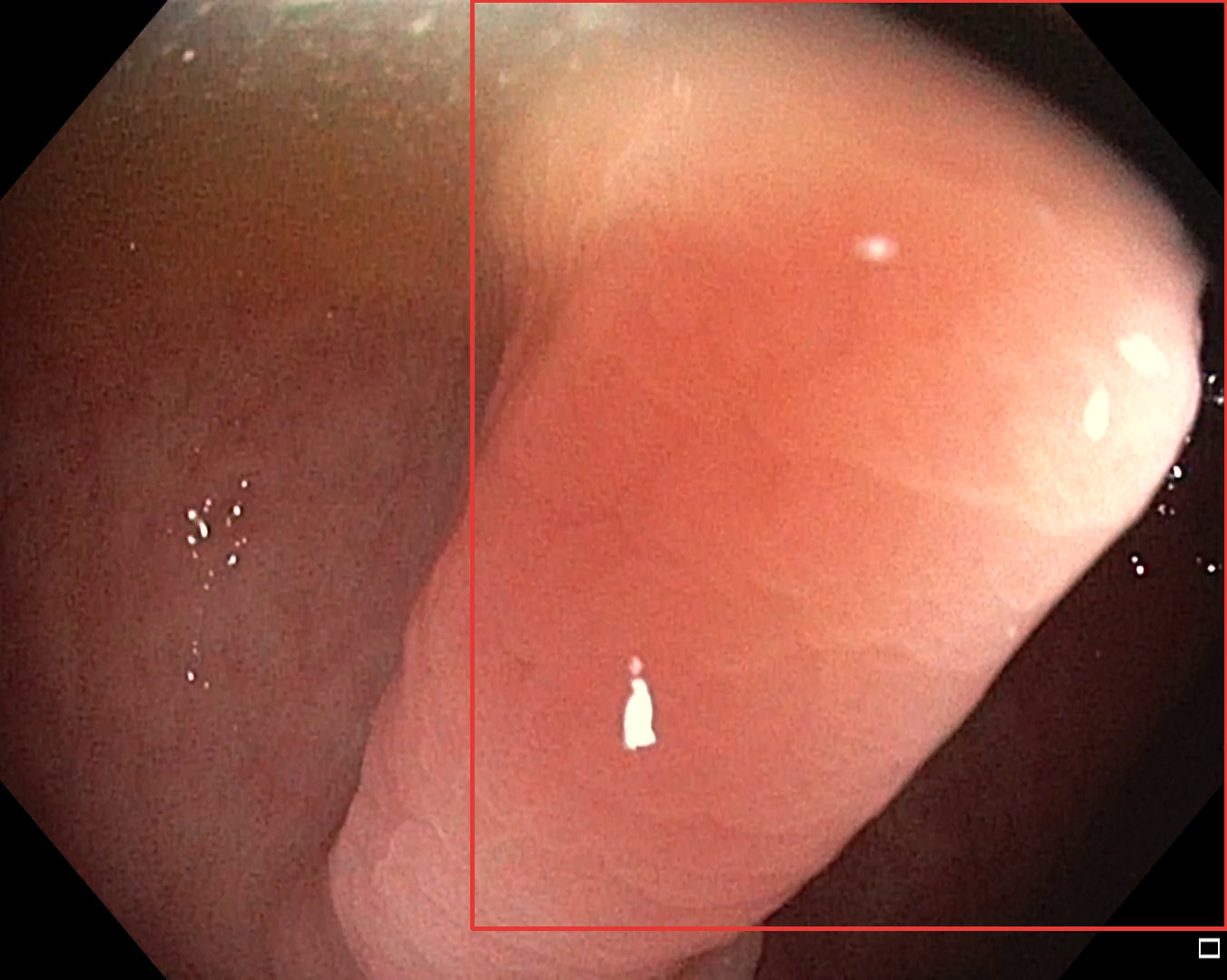} &
    \snap{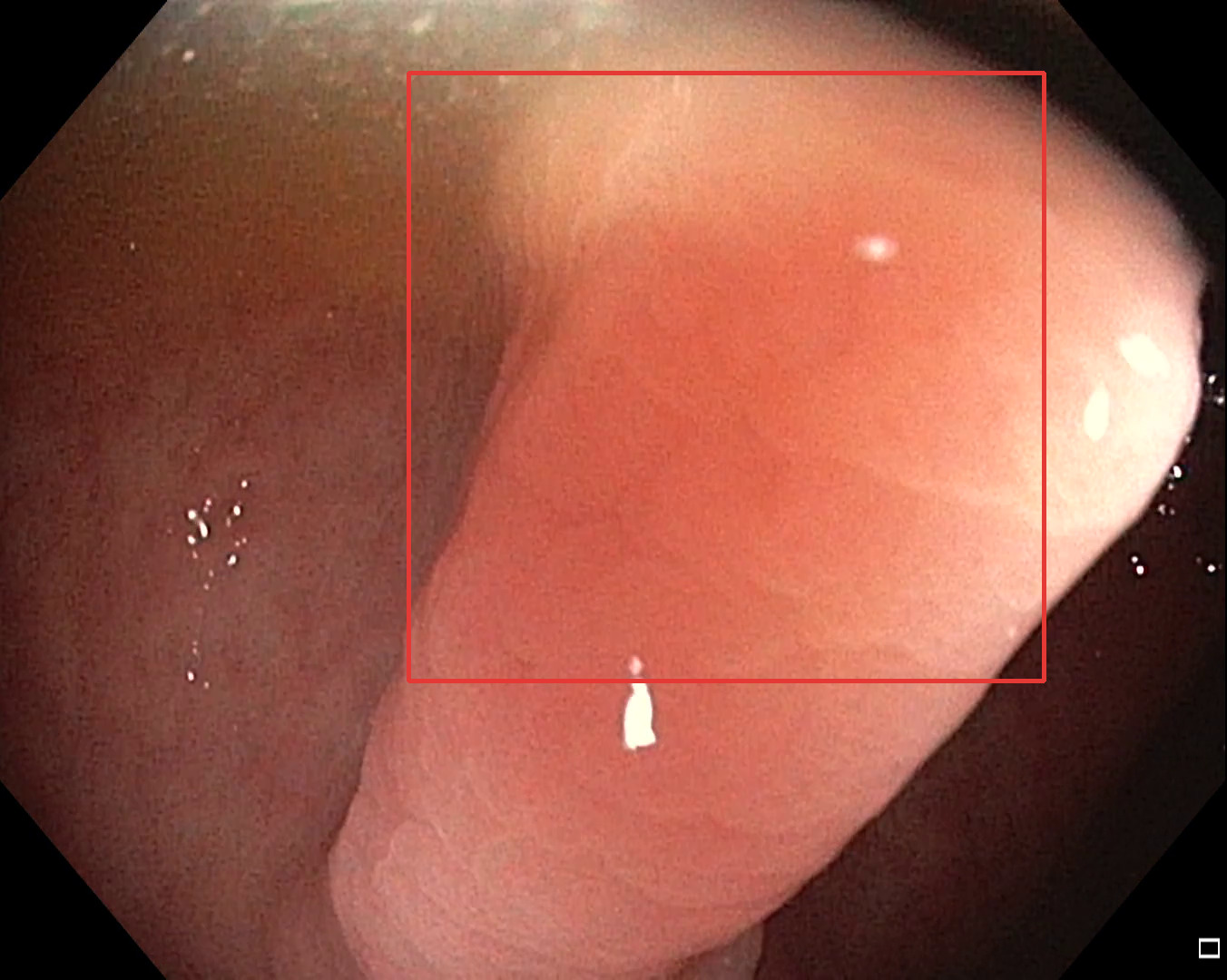} &
    \snap{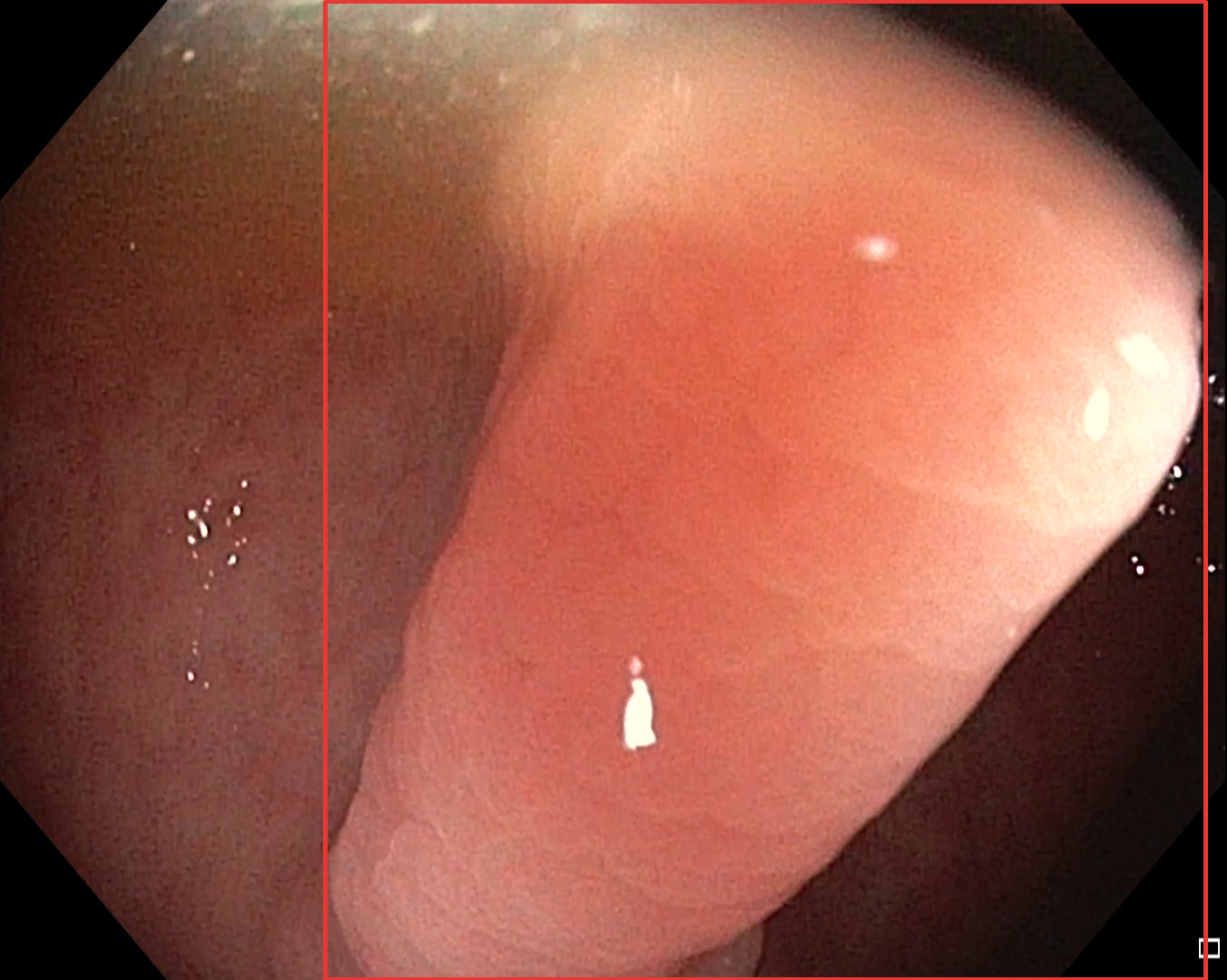} &
    \snap{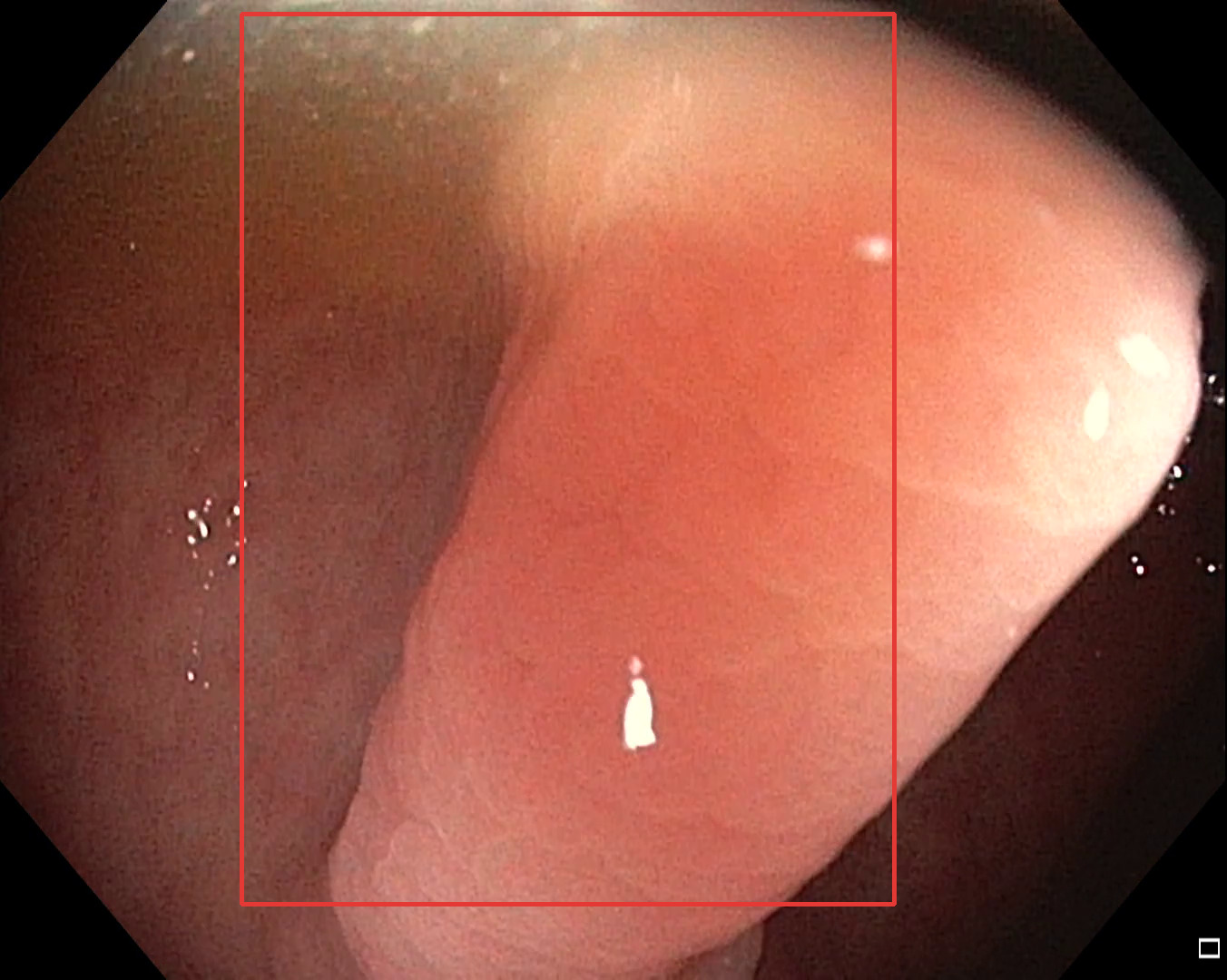} &
    \snap{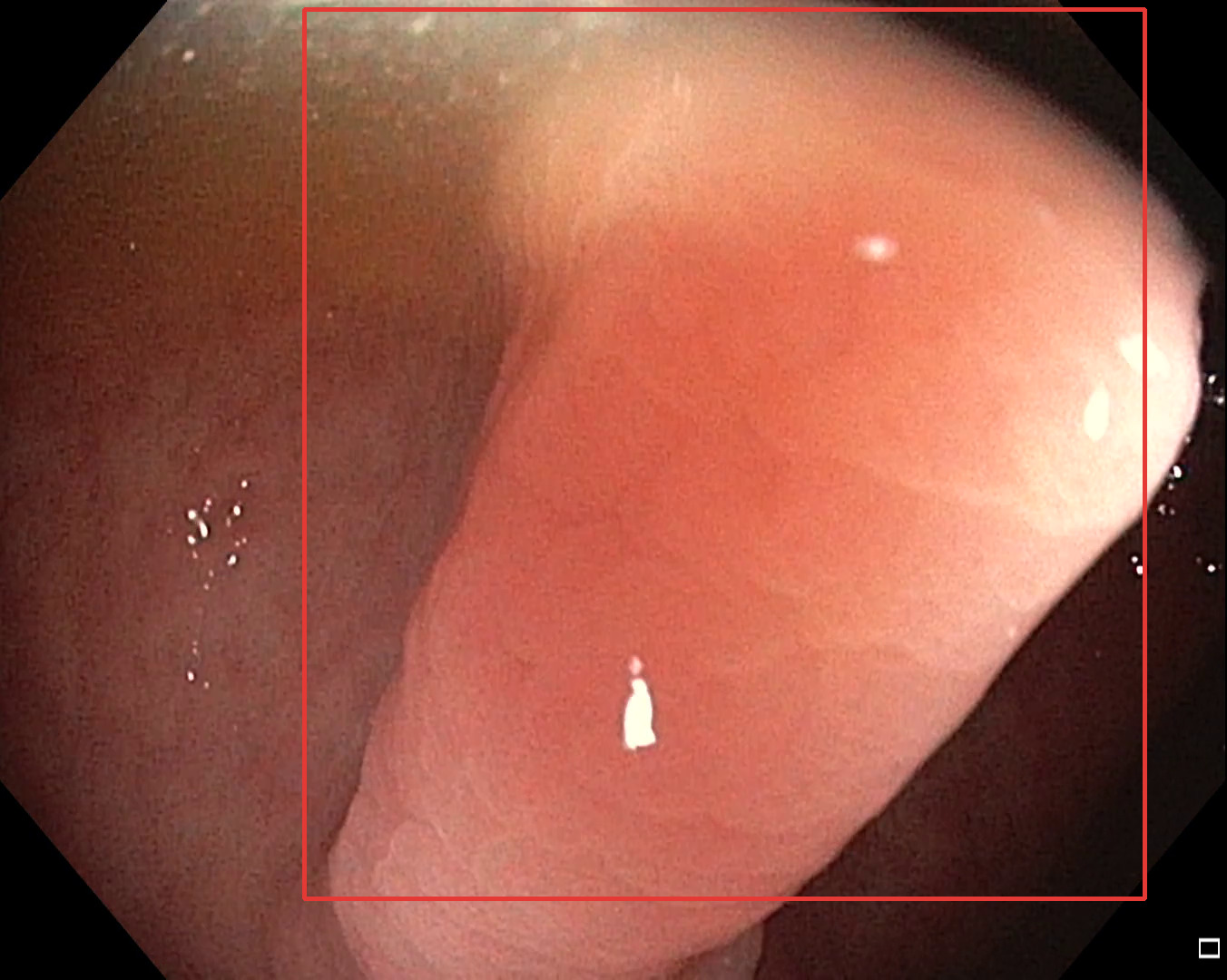} &
    \snap{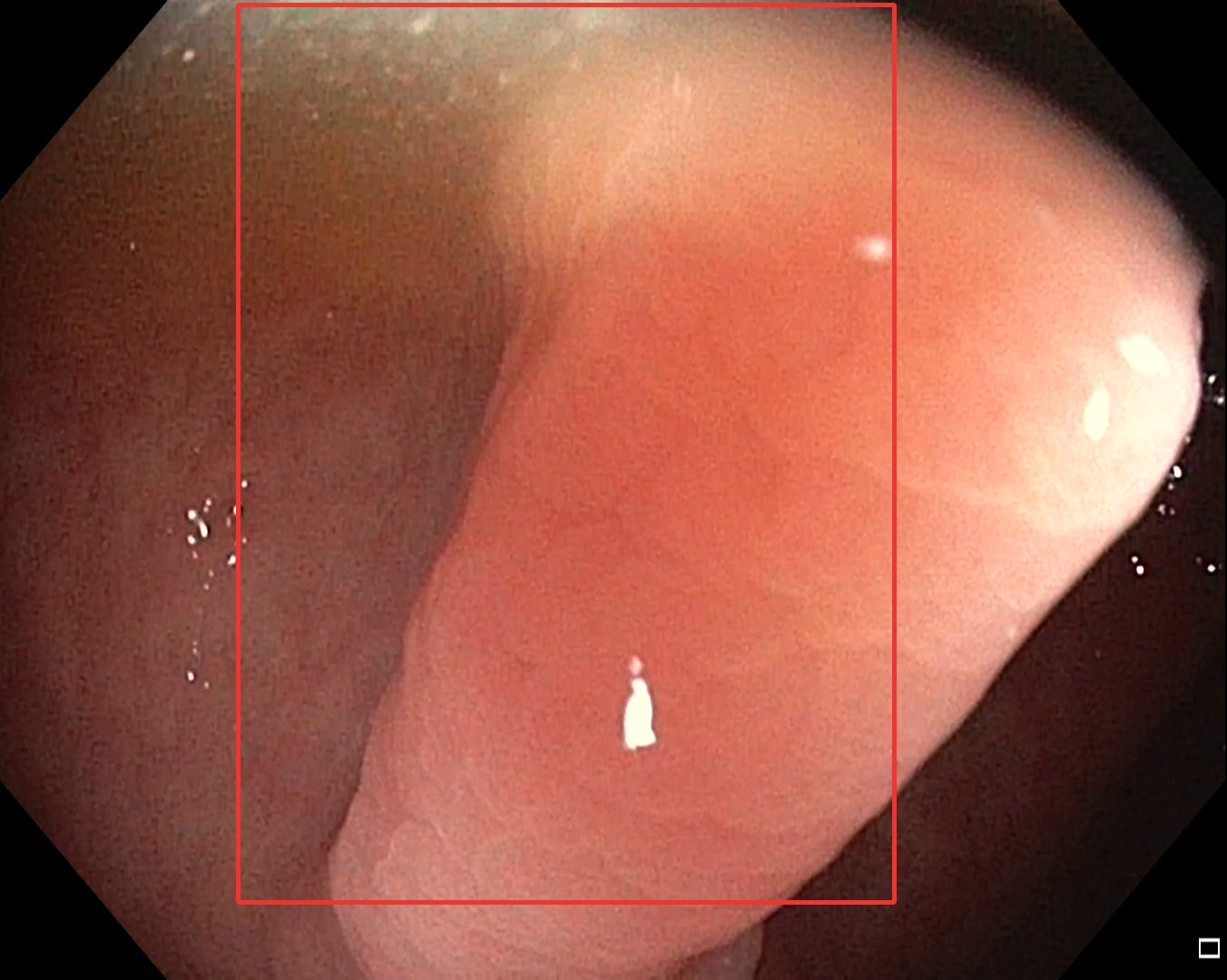} \\ 
    \snap{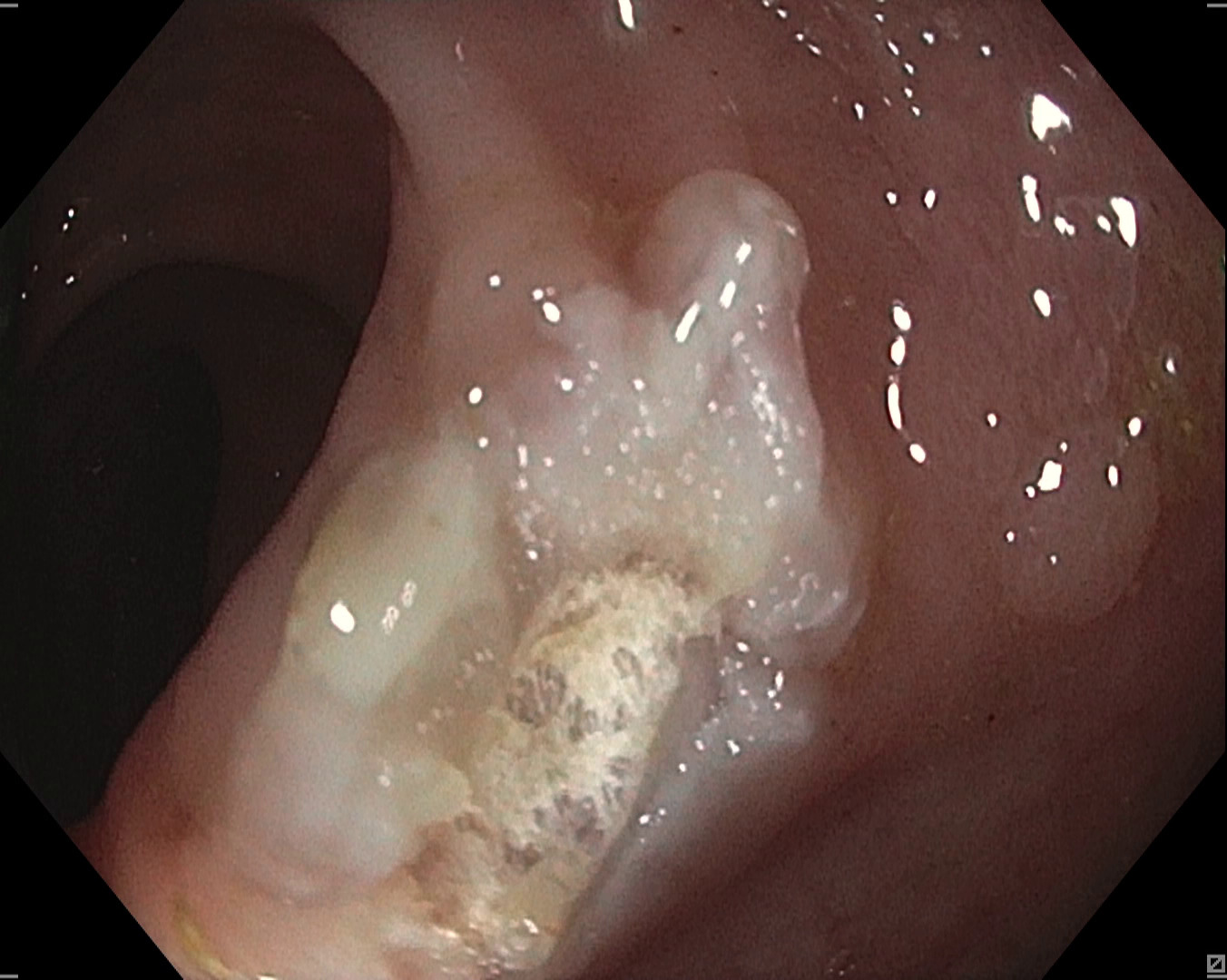} &
    \snap{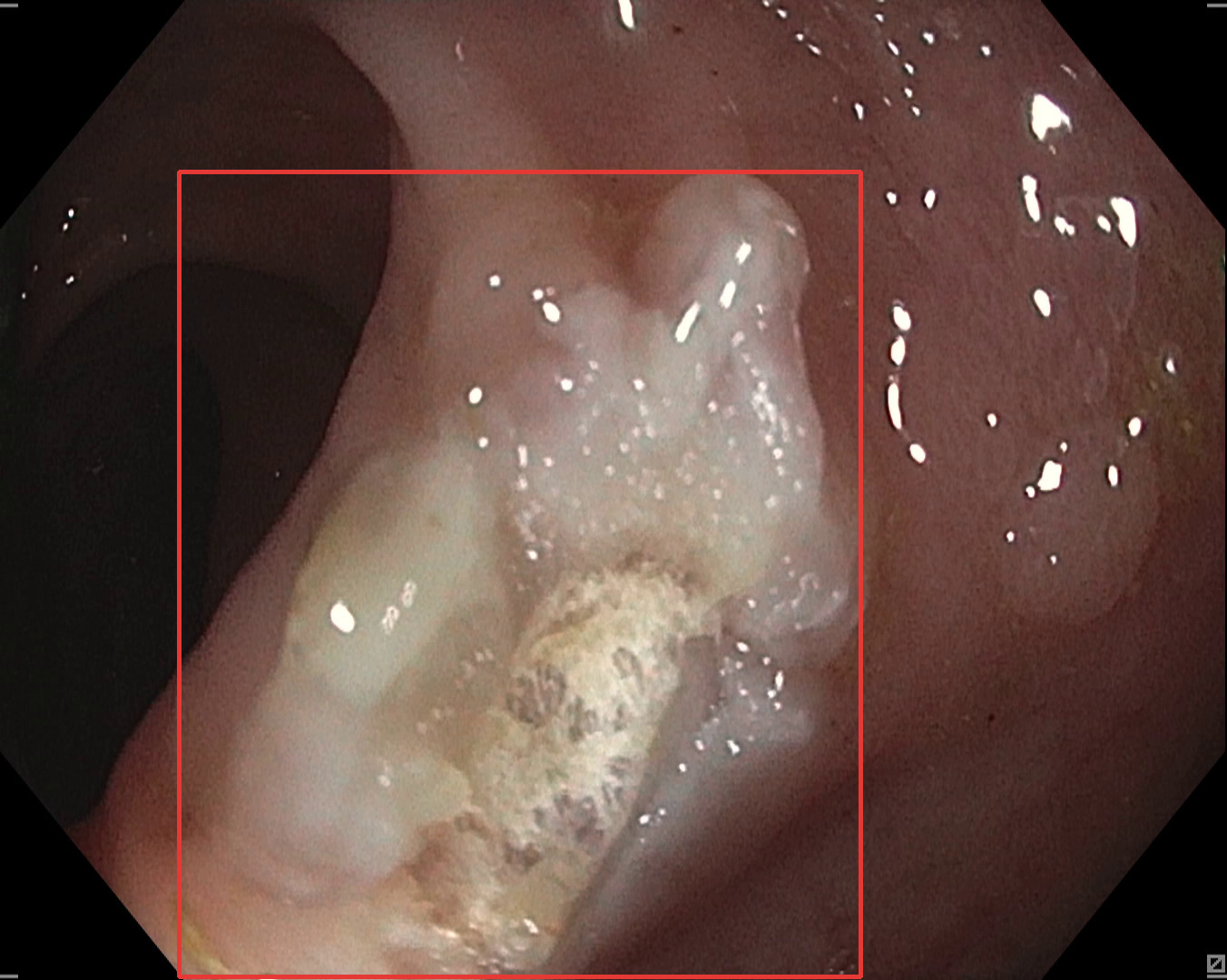} &
    \snap{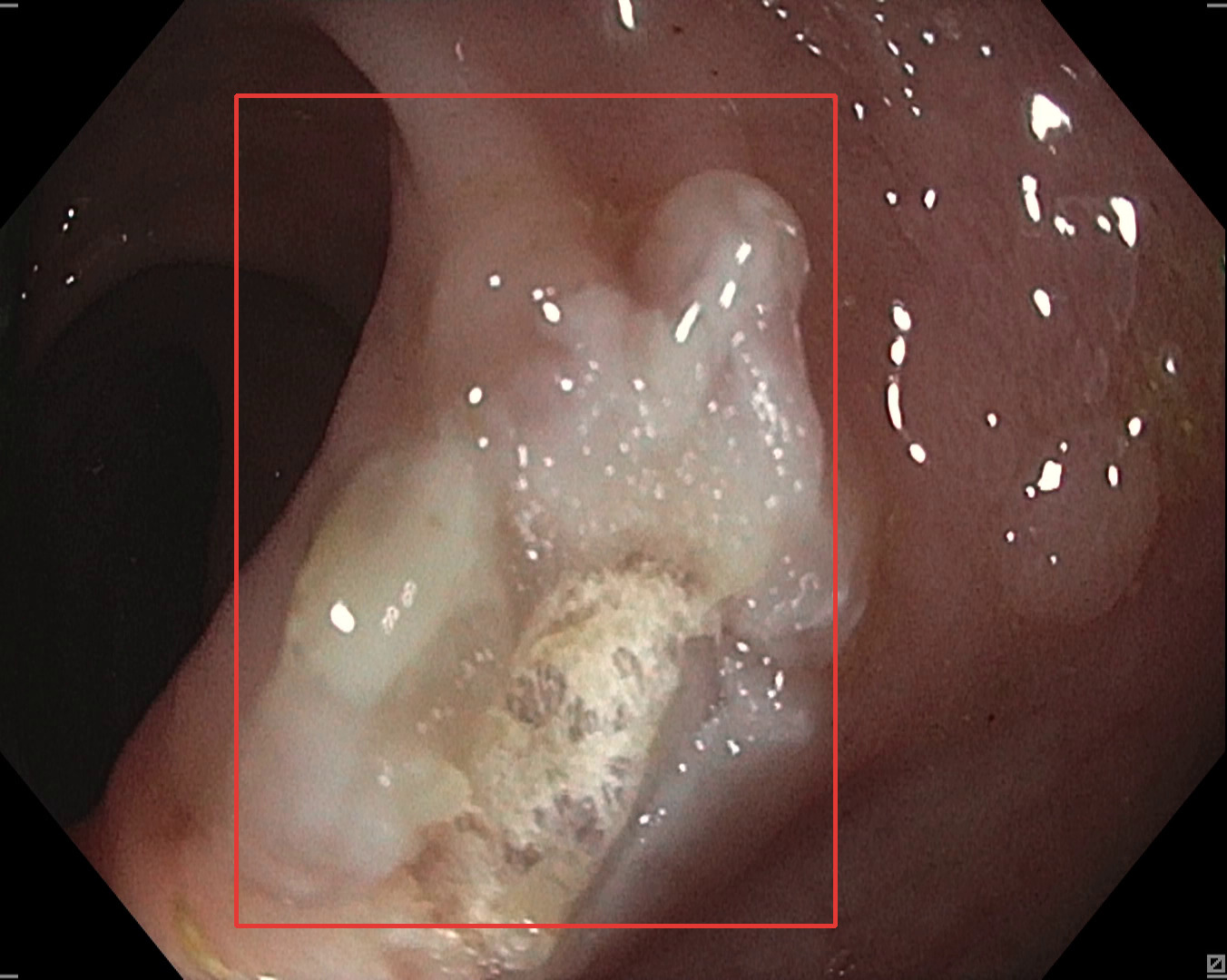} &
    \snap{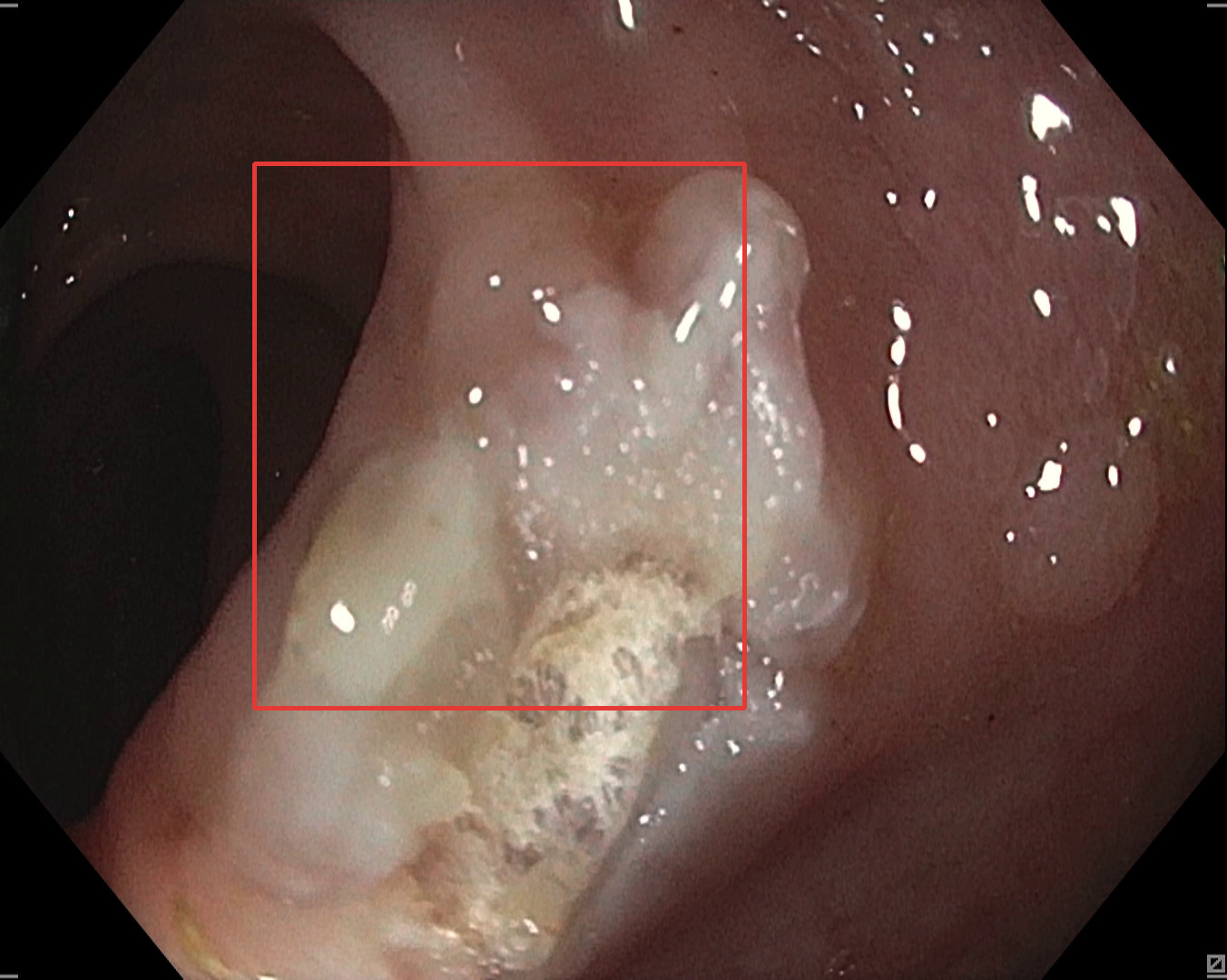} &
    \snap{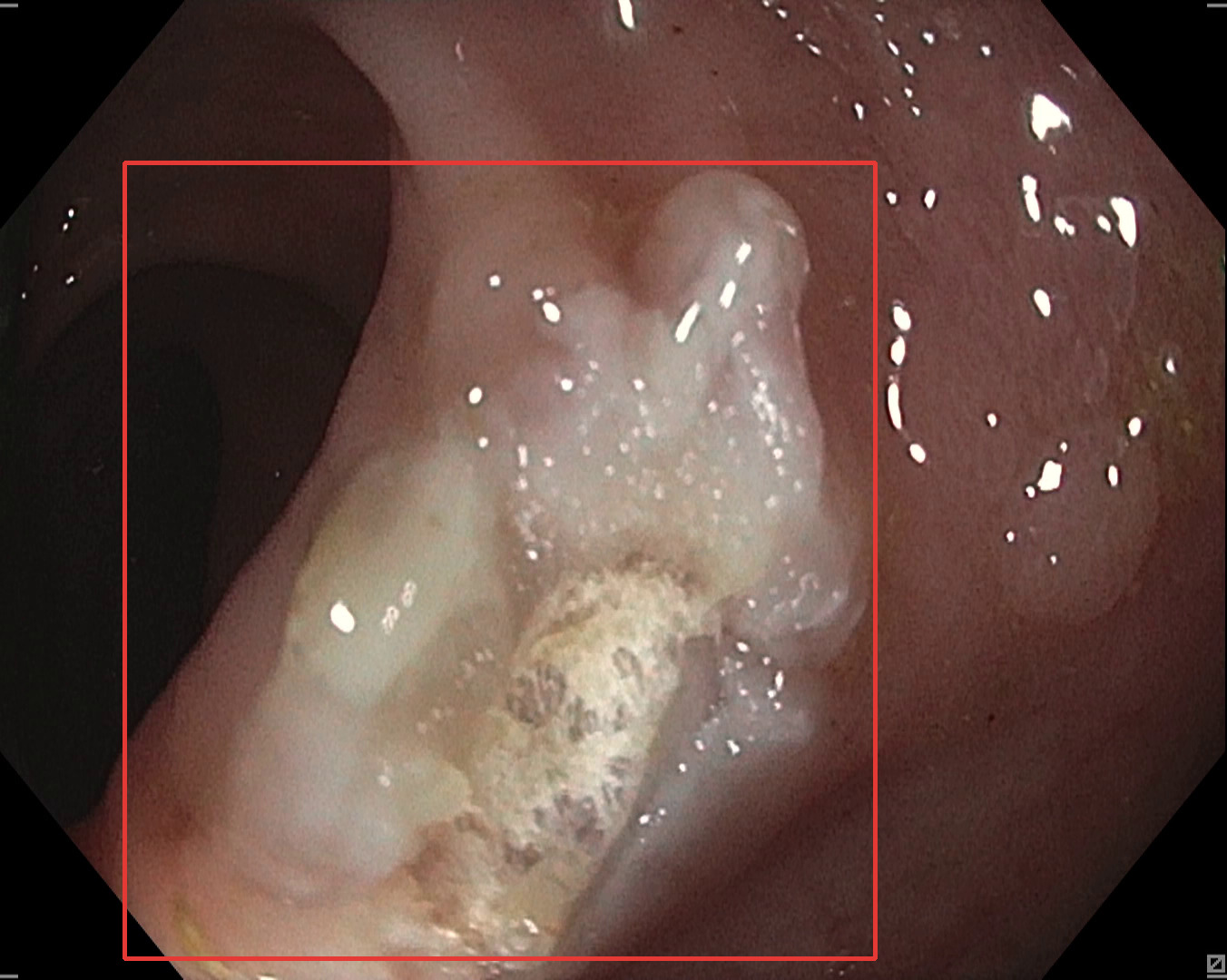} &
    \snap{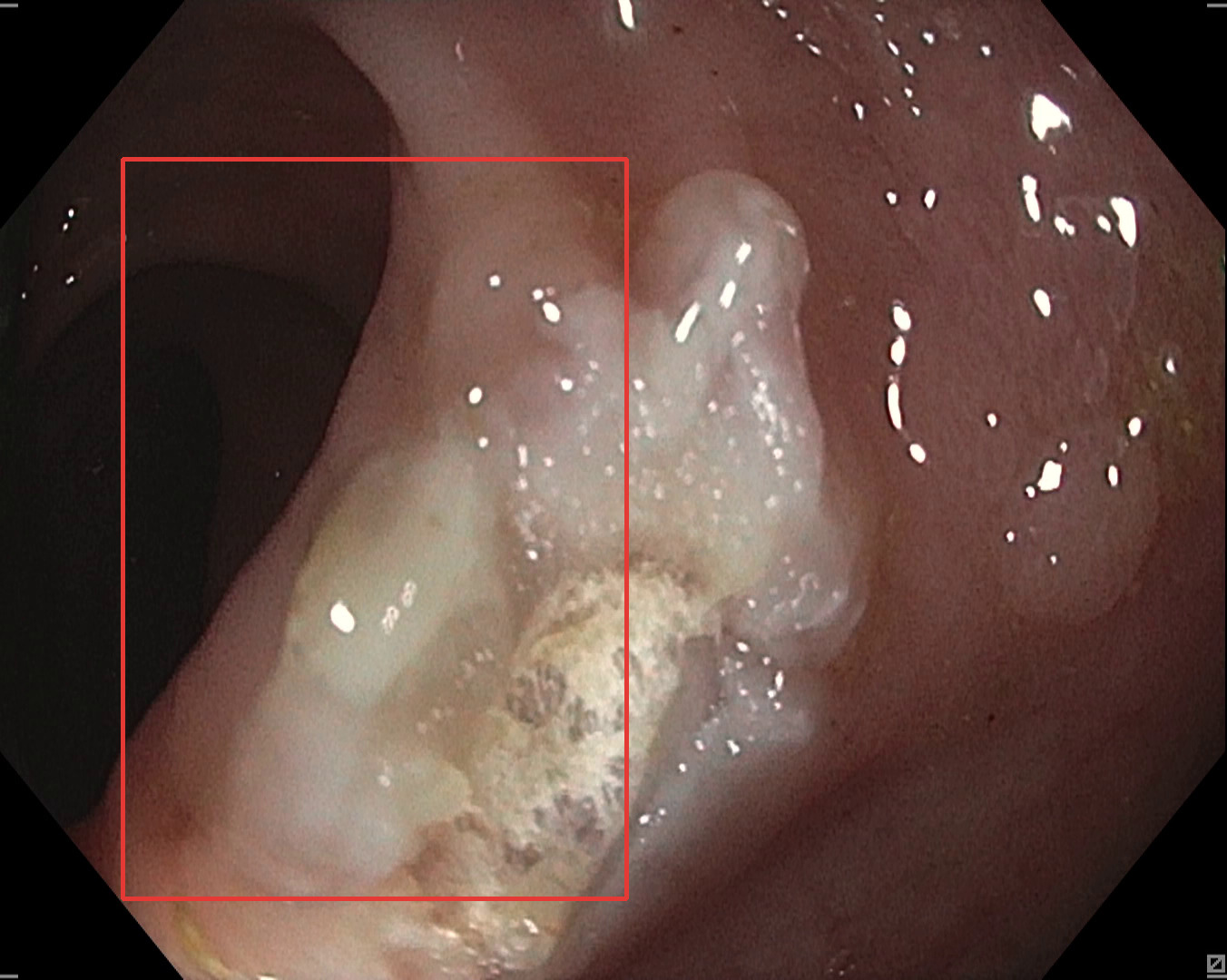} &
    \snap{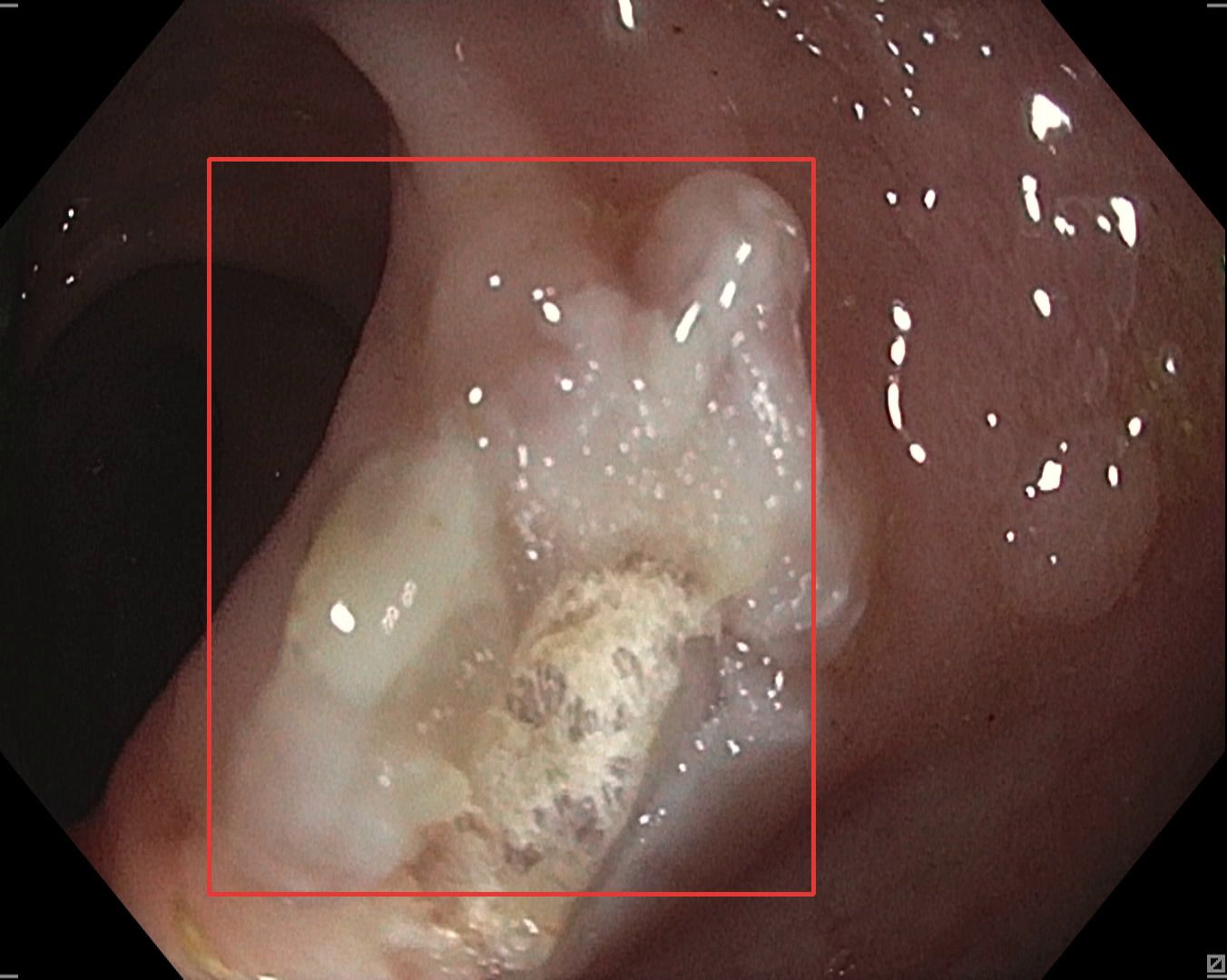} &
    \snap{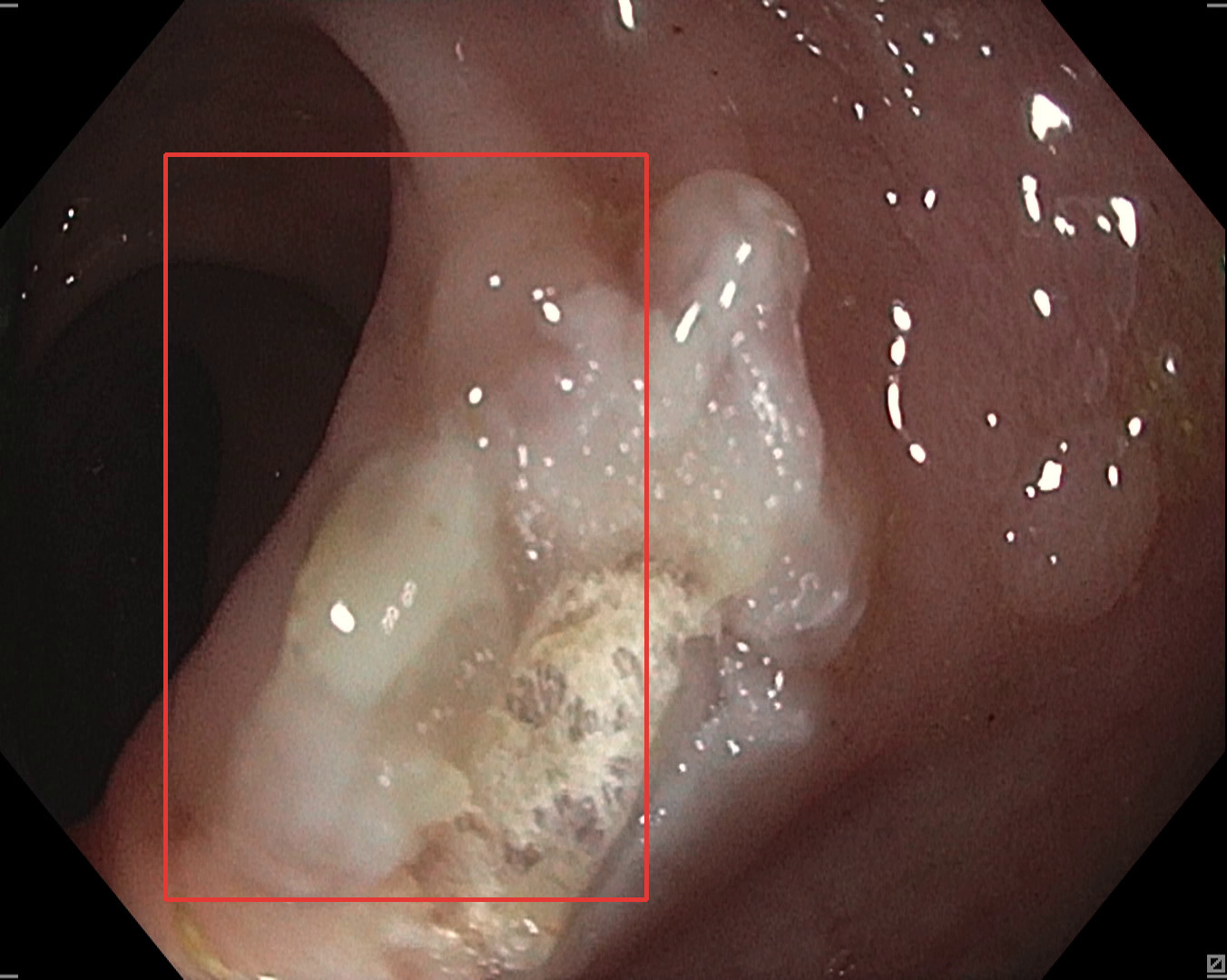} \\[1pt]
\snap{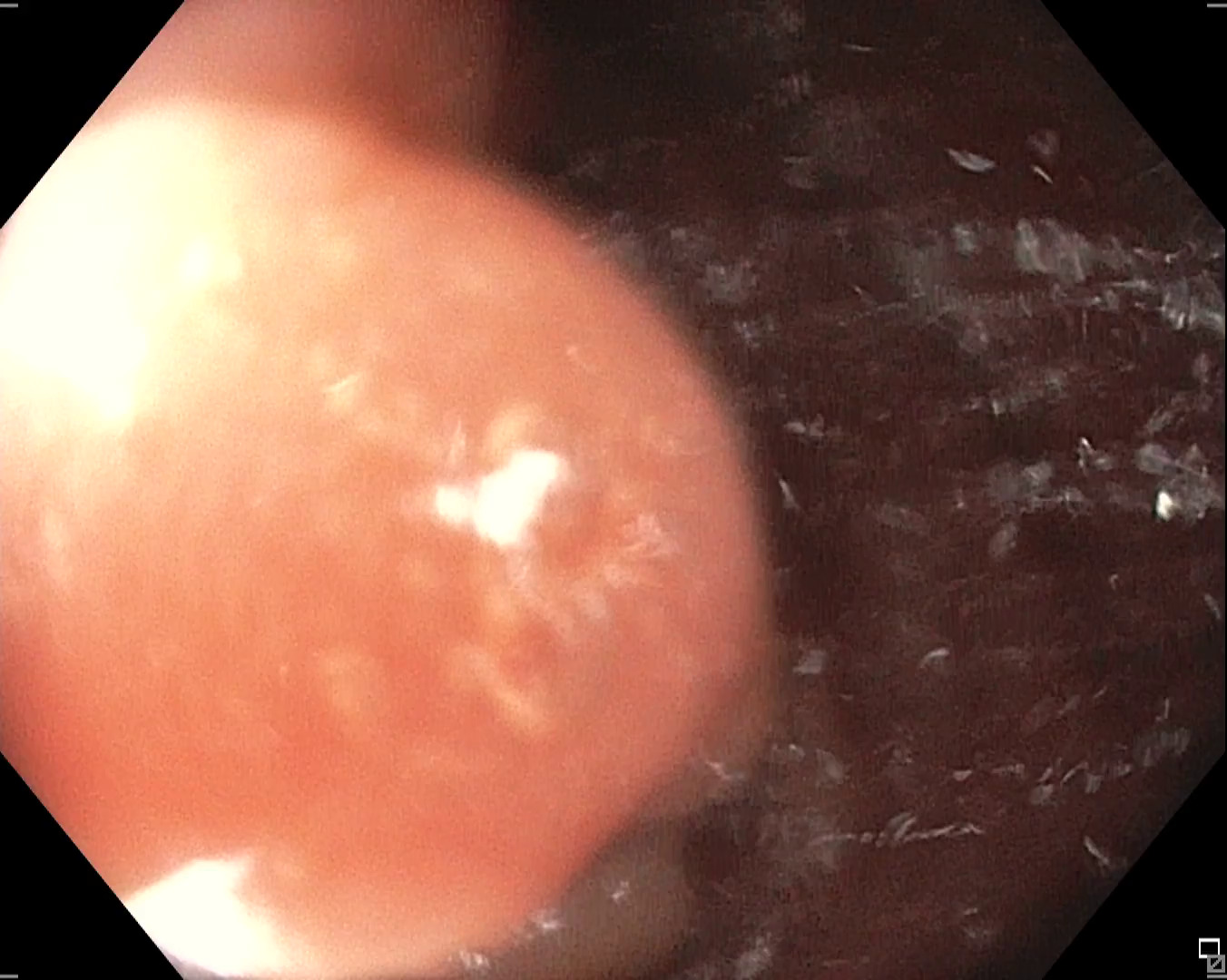} &
\snap{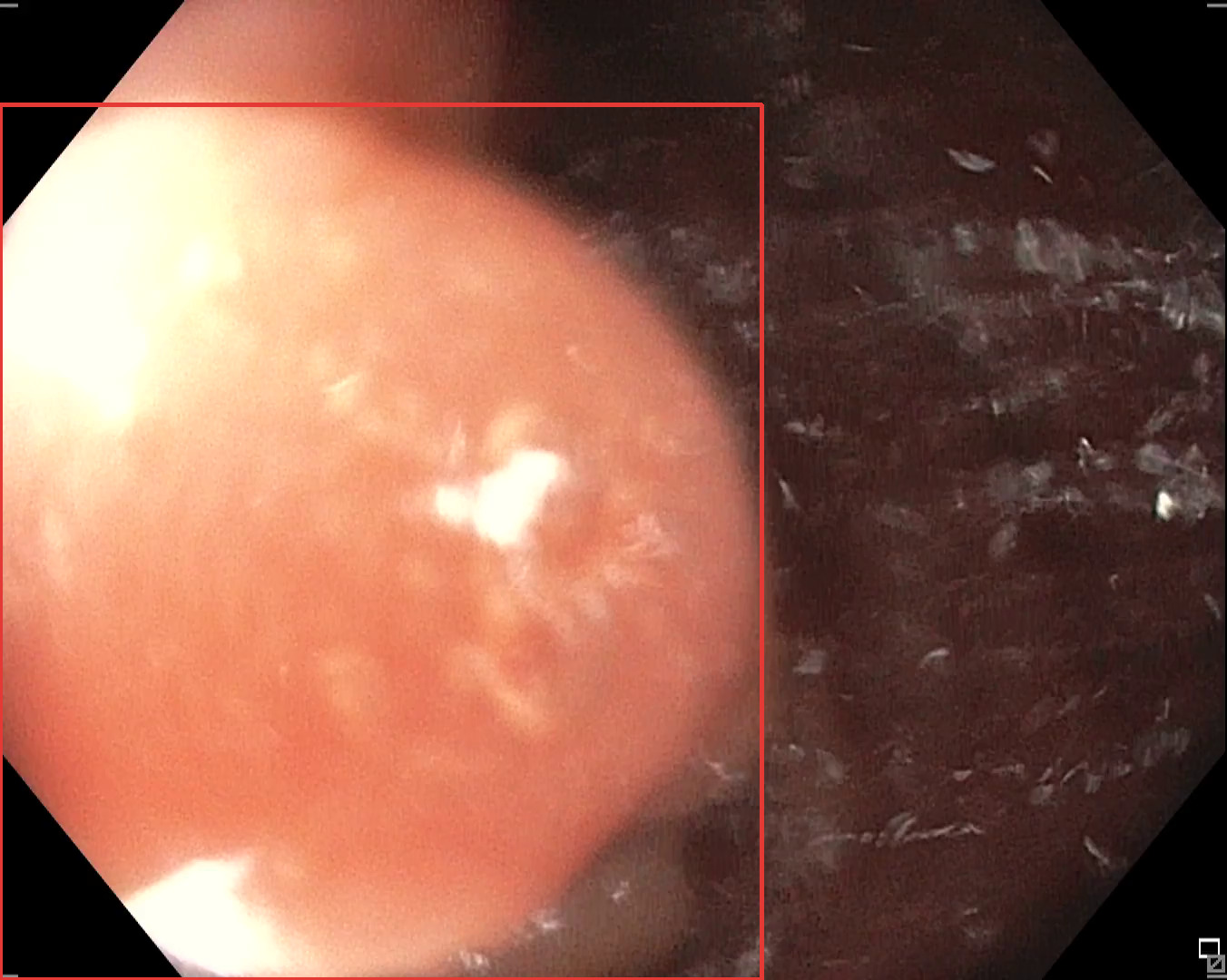} &
\snap{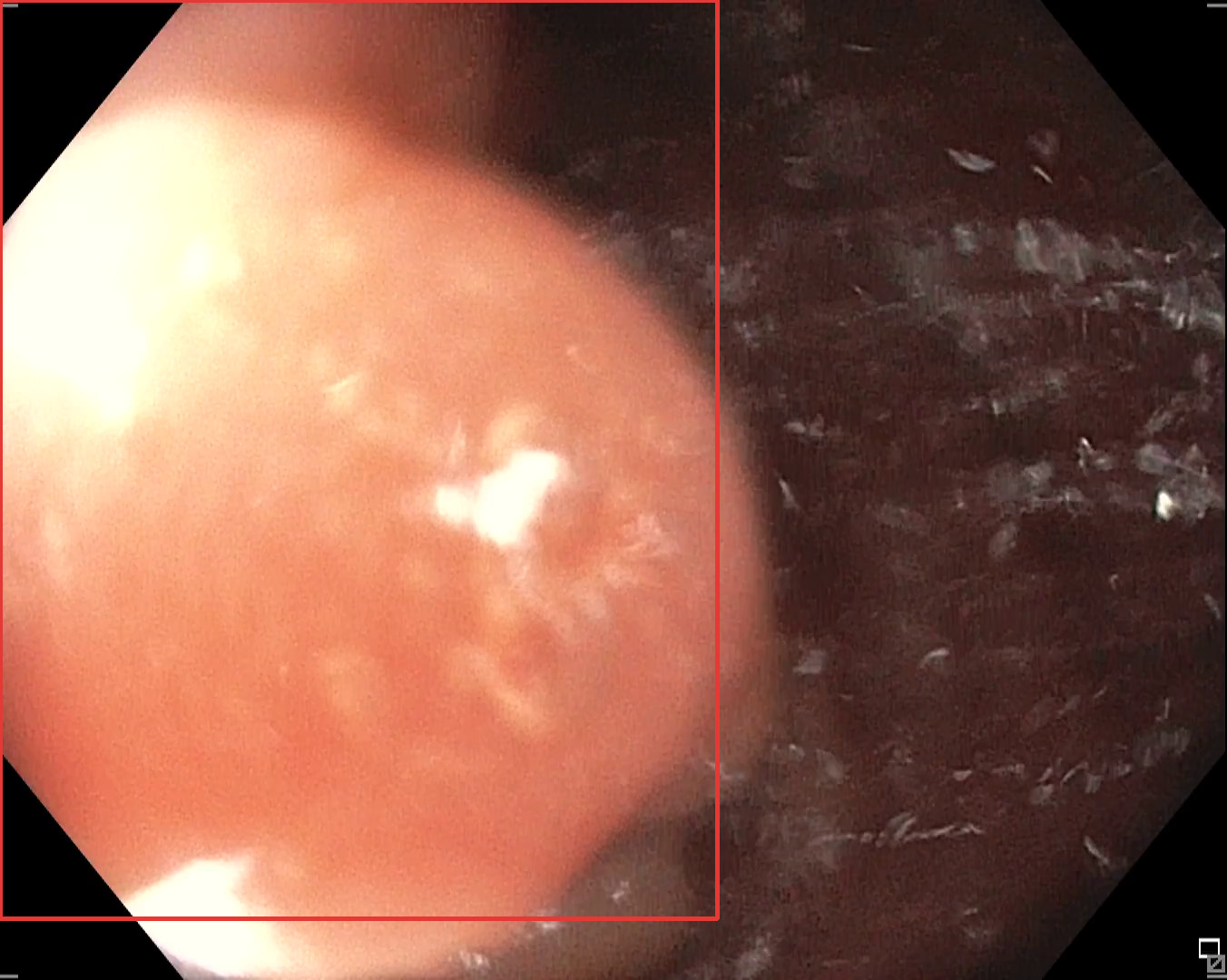} &
\snap{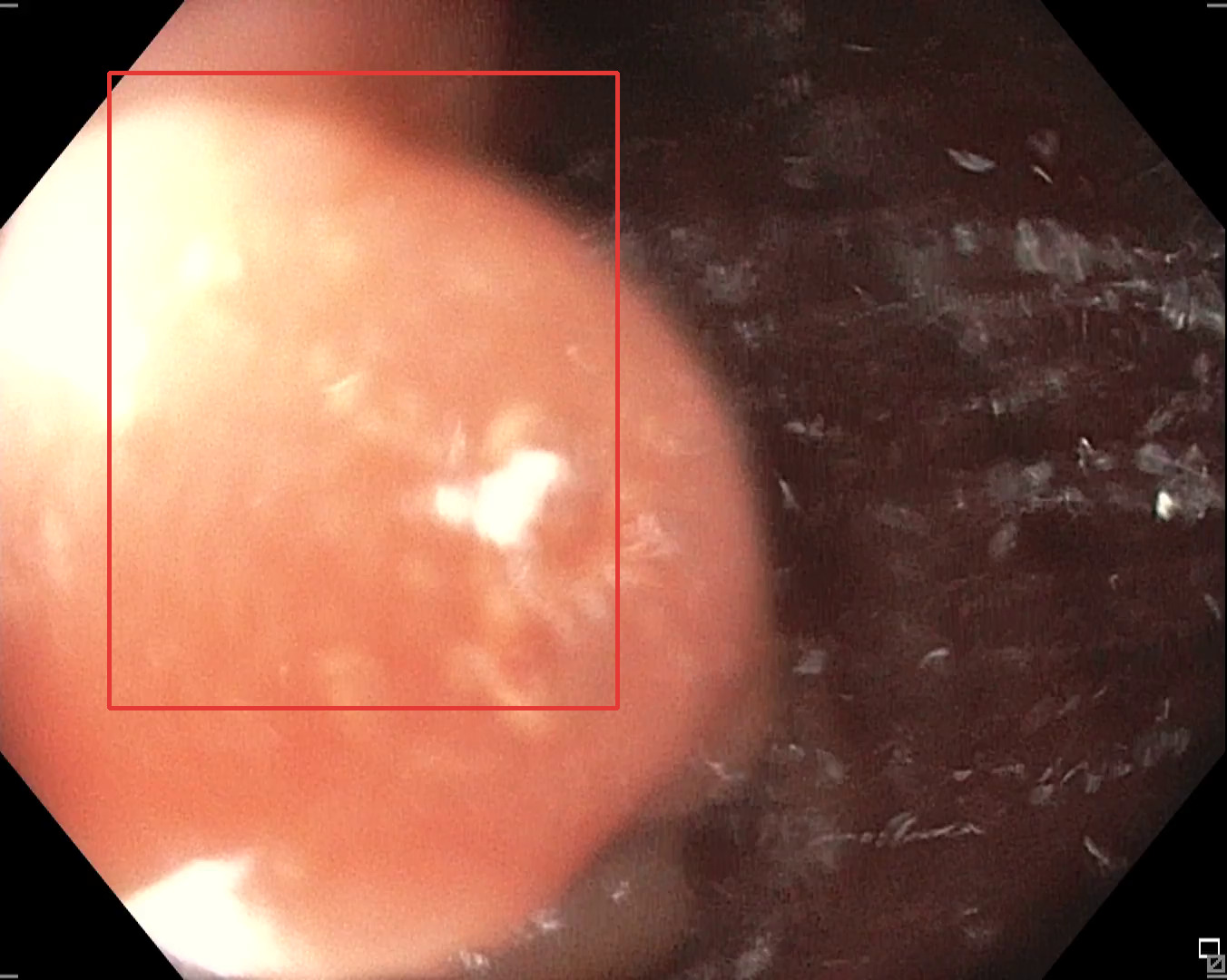} &
\snap{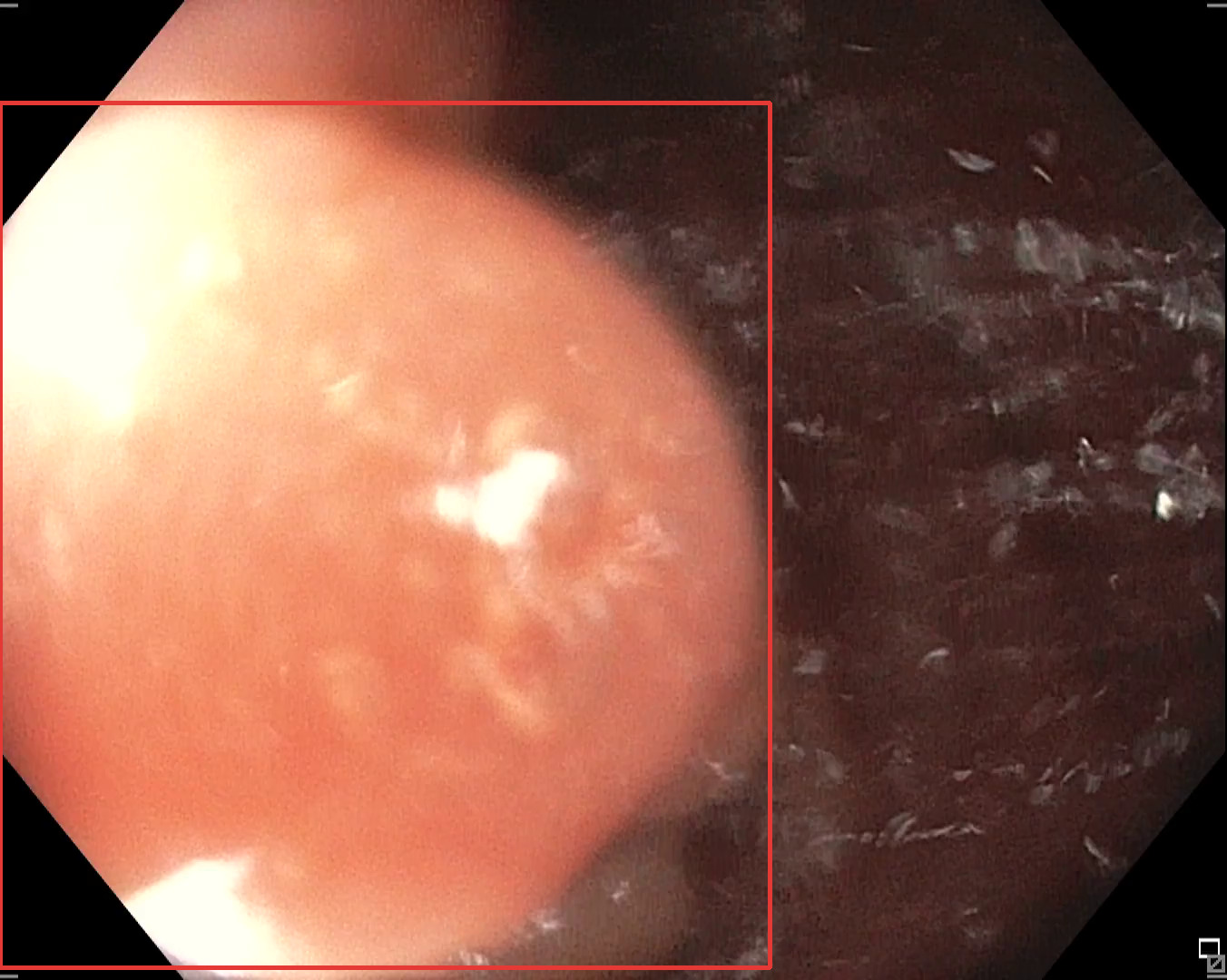} &
\snap{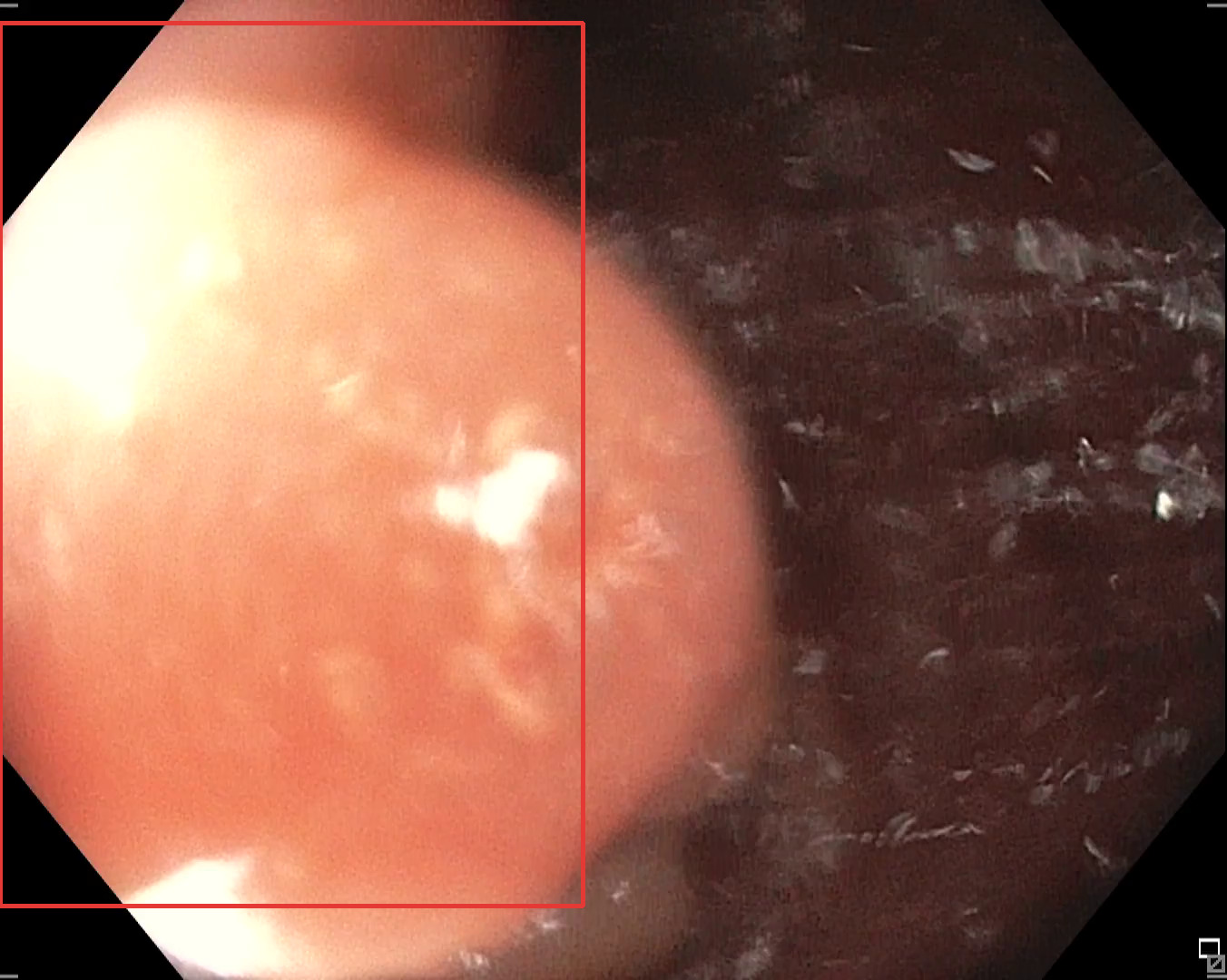} &
\snap{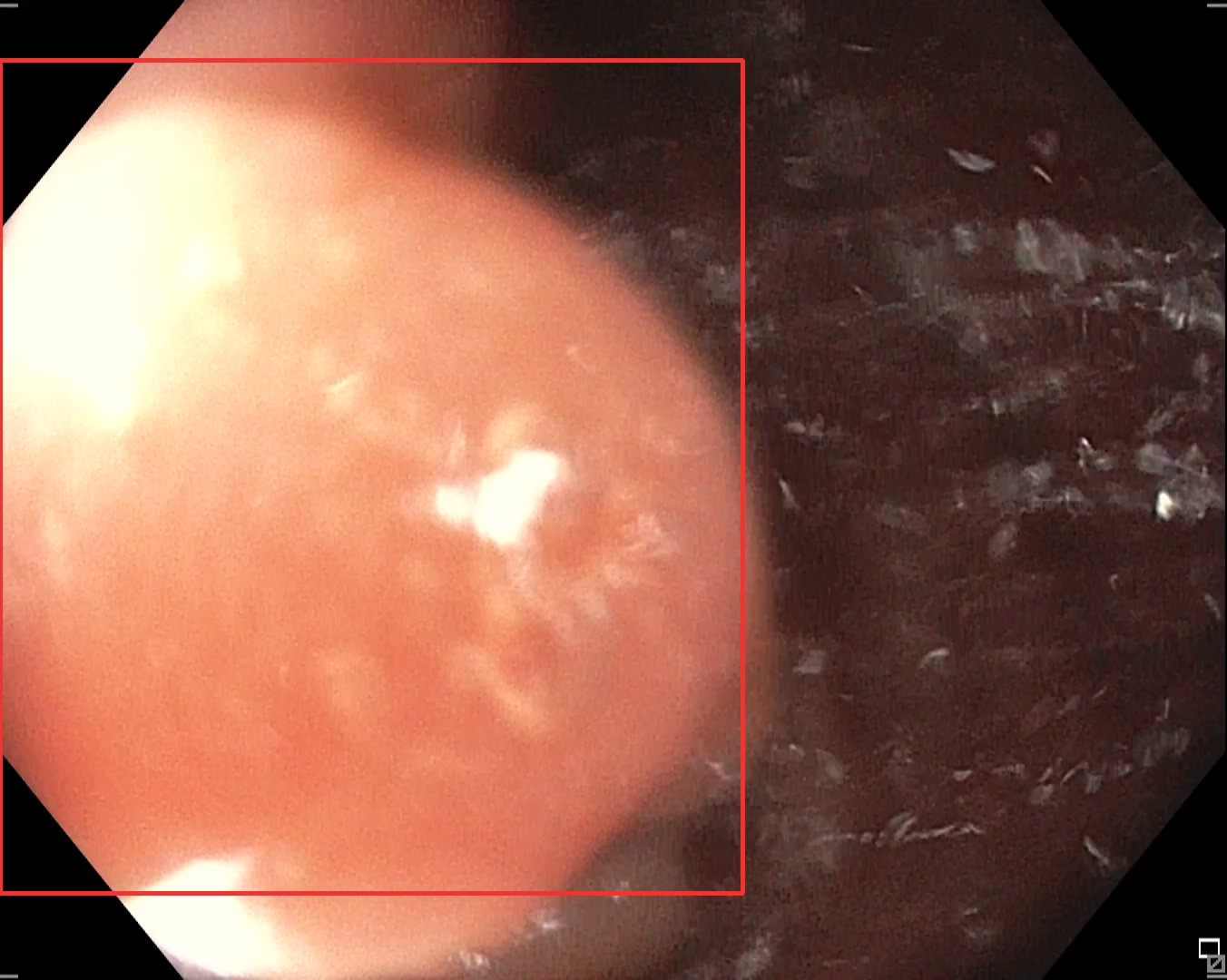} &
\snap{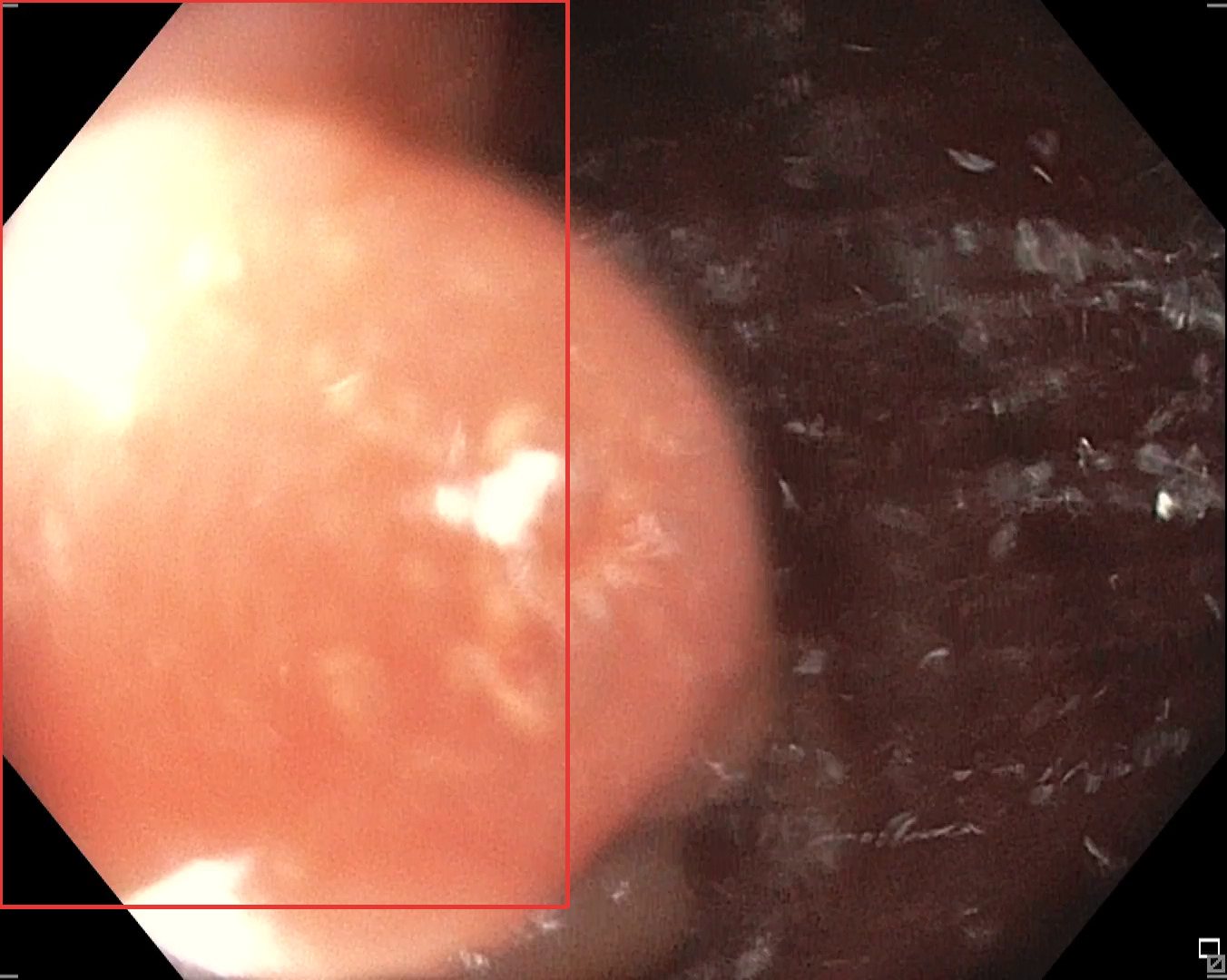} \\[1pt]
\snap{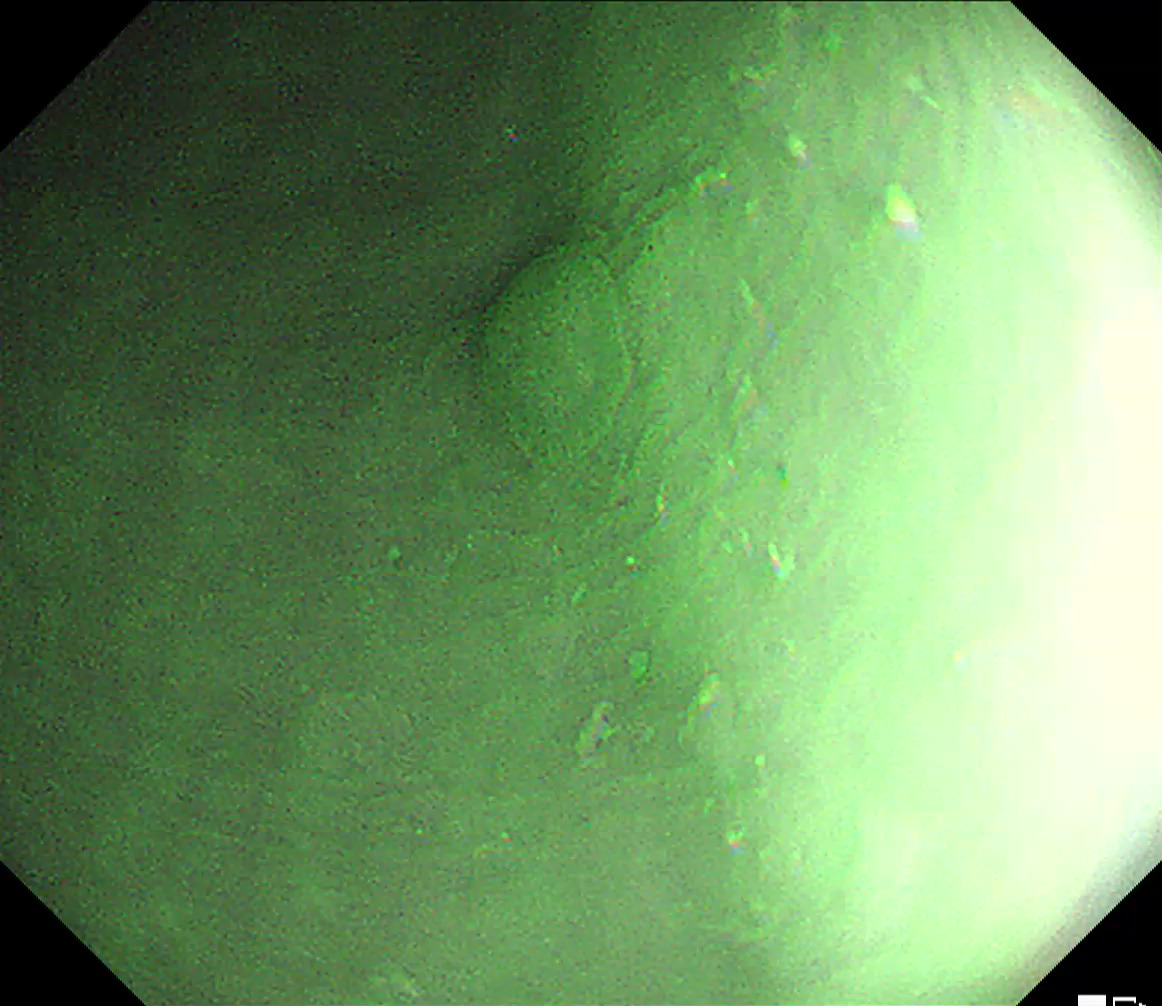} &
\snap{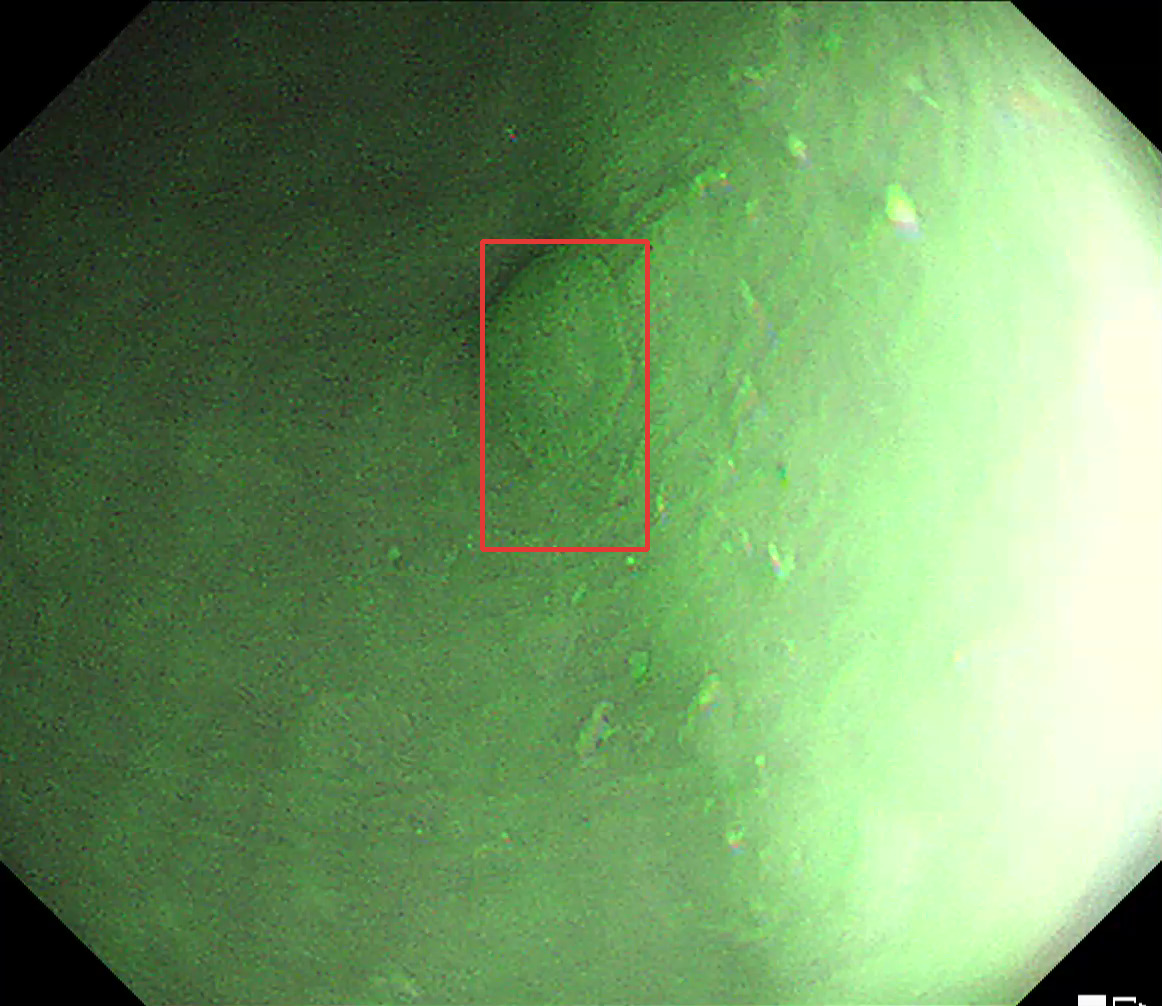} &
\snap{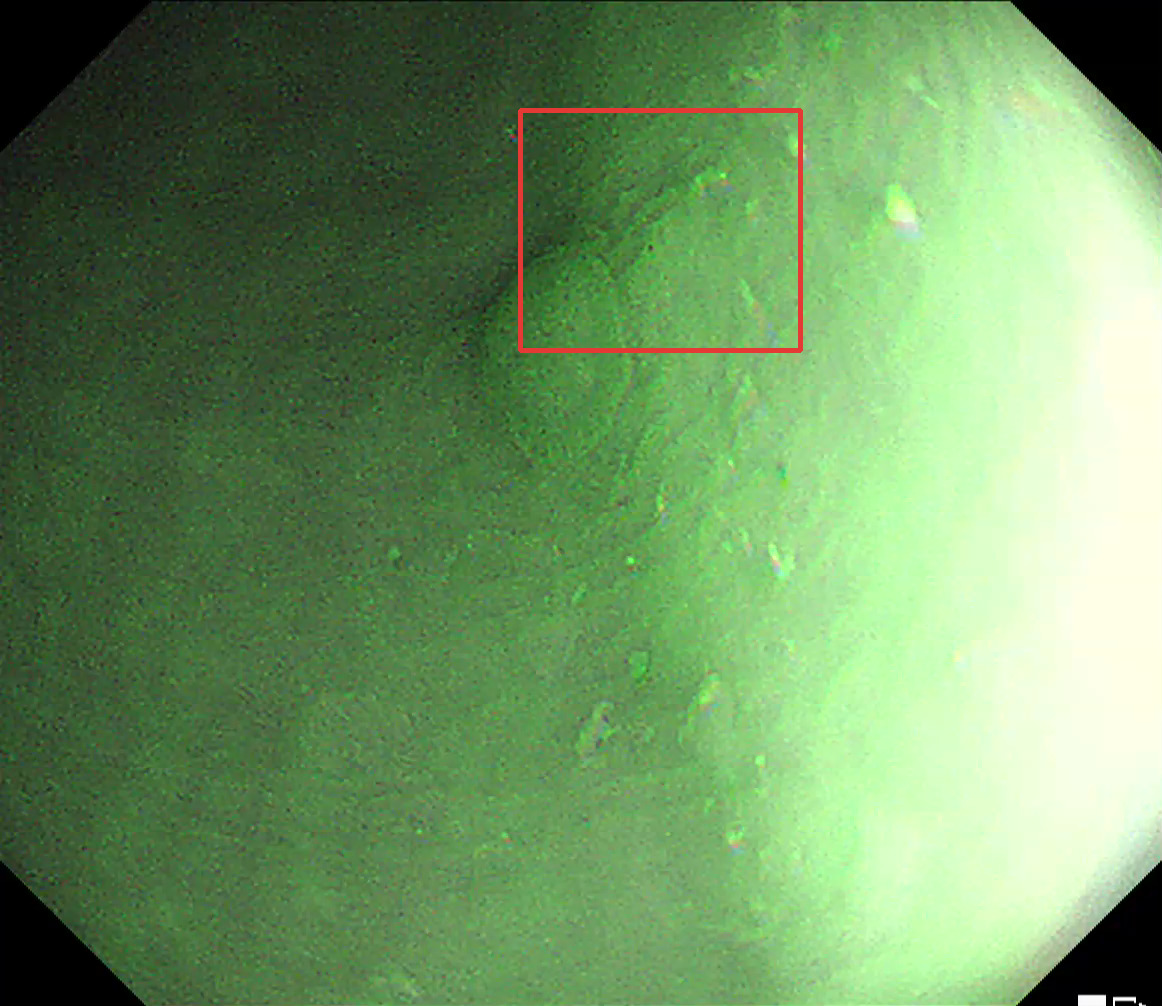} &
\snap{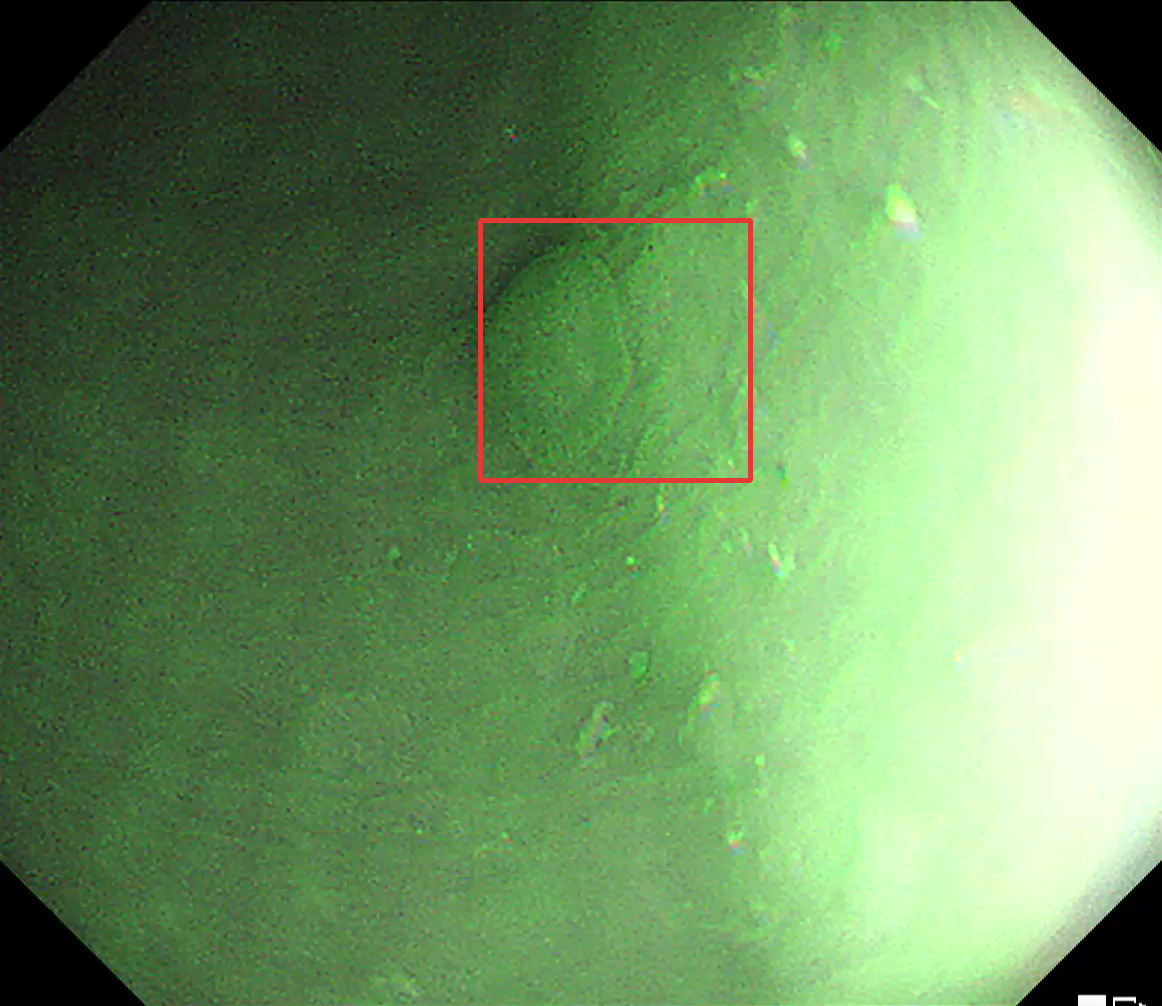} &
\snap{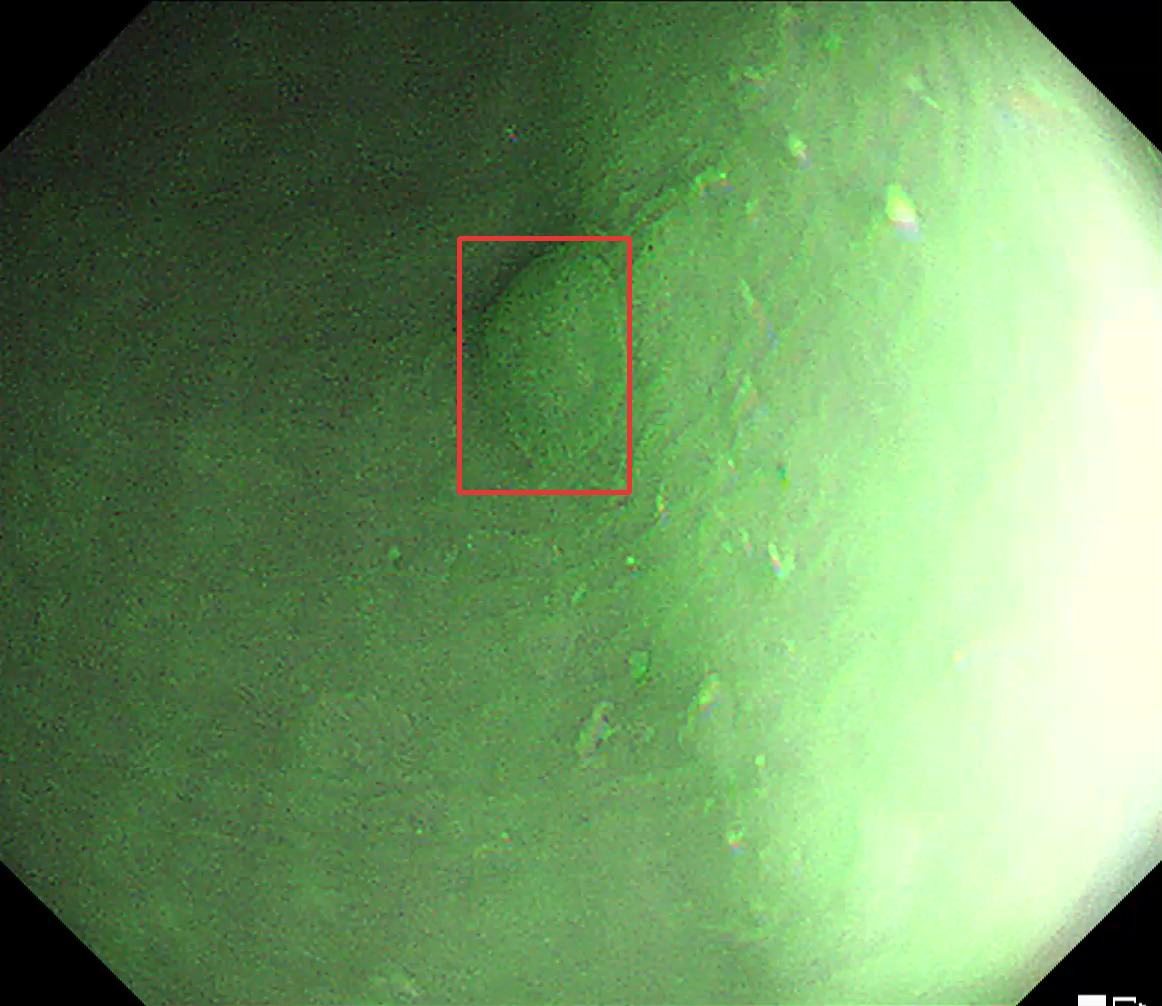} &
\snap{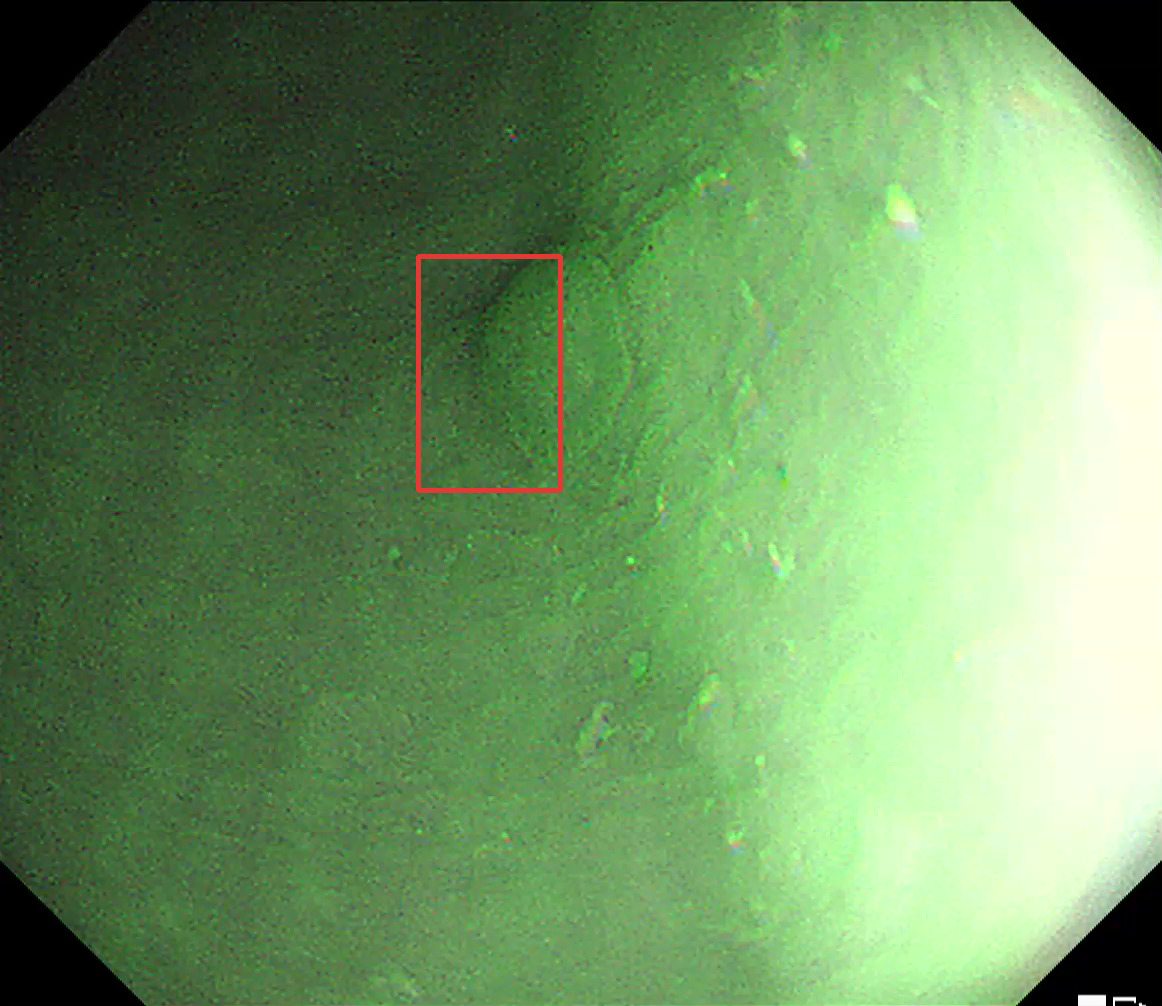} &
\snap{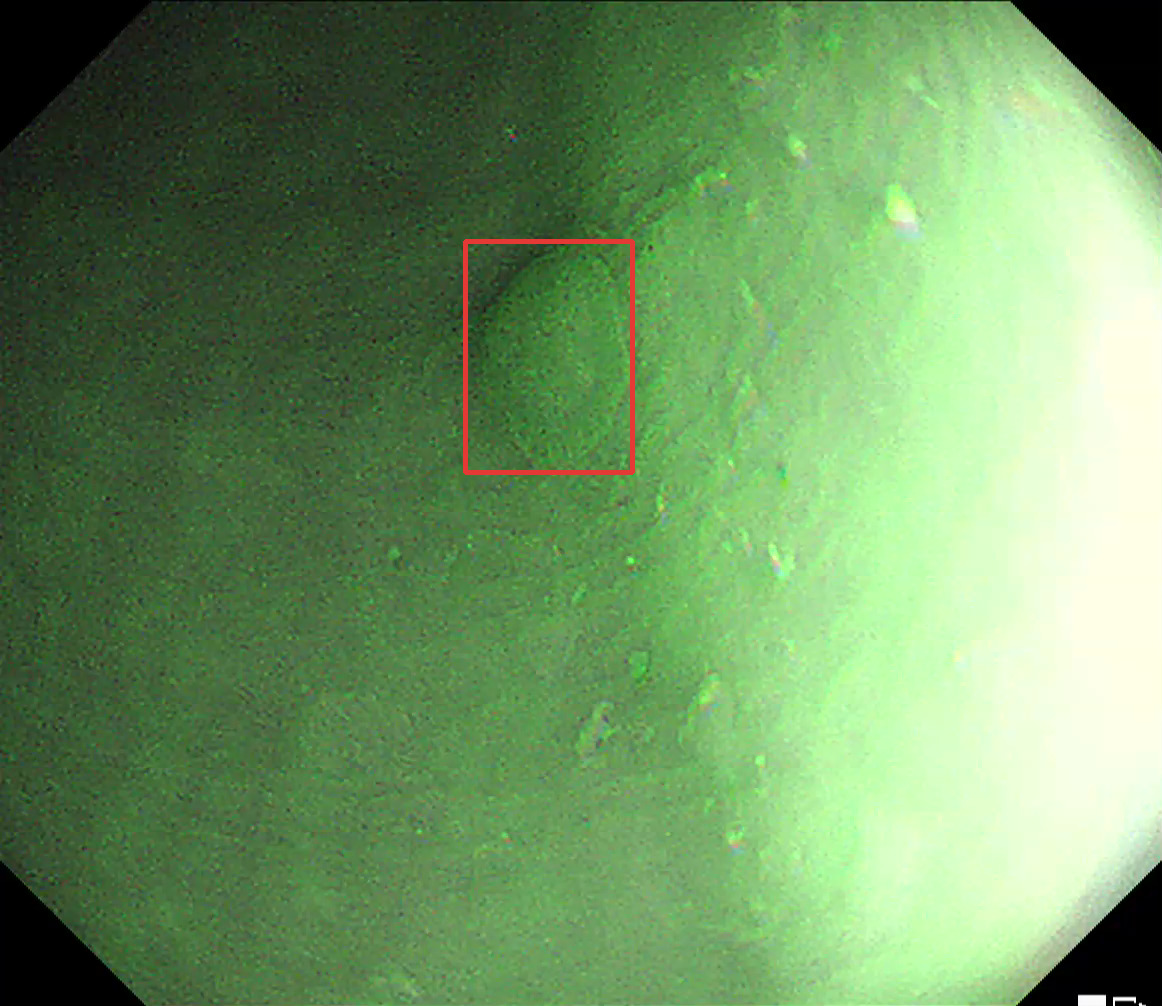} &
\snap{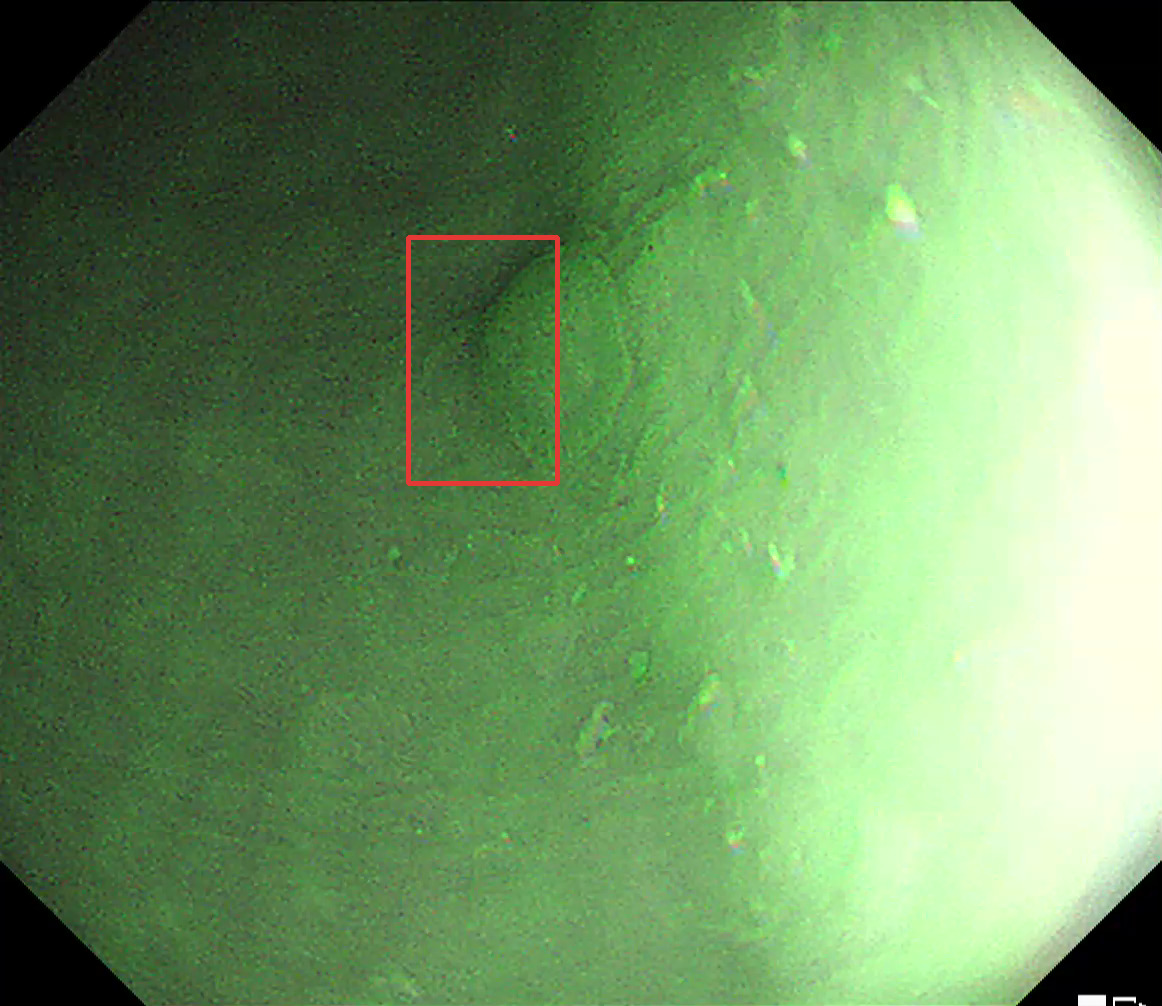} \\
  \end{tabular}}%
  \caption{\textbf{\methodname Qualitative Detection Comparison.} Each row shows a different endoscopy video frame:
    (1)~\textit{mass},
    (2)~\textit{sessile polyp},
    (3)~\textit{pedunculated polyp},
    (4)~\textit{ulcer},
    (5)~\textit{pedunculated polyp},
    (6)~\textit{sessile polyp}.
    Columns show the raw input, ground-truth bounding box, and predicted bounding boxes from each model.
    Bounding boxes are drawn in red. These detections are used in the segmentation benchmark using 3 box detections as prompts for the EdgeTAM tracker~\cite{zhou2025edgetam}.%
  }
  \label{fig:qualitative-detection}
  \end{minipage}
\end{figure*}

\begin{figure*}[!t]
  \centering
  \begin{minipage}{\textwidth}
  \setlength{\tabcolsep}{1pt}
  \renewcommand{\arraystretch}{0.6}
  {\setlength{\tabcolsep}{1pt}%
  \renewcommand{\arraystretch}{1.0}%
  \resizebox{0.99\textwidth}{!}{%
  \begin{tabular}{*{9}{p{\dimexpr\textwidth/9\relax}}}
    \colhead{Input} &
    \colhead{GT} &
    \colhead{GPT-5.2} &
    \colhead{Opus 4.6} &
    \colhead{Gemini 3 Flash} &
    \colhead{Qwen3.5-Plus} &
    \colhead{QwenVL-Max} &
    \colhead{Qwen3-VL-235B} &
    \colhead{SAM\,3} \\
  \end{tabular}}\par}
  \vspace{2pt}
  \resizebox{0.99\textwidth}{!}{%
  \begin{tabular}{ccccccccc}
    \snap{images/snapshots/segmentation/1154f3aa0833463280d8a855bce3425b/frame_000041/input.jpg} &
    \snap{images/snapshots/segmentation/1154f3aa0833463280d8a855bce3425b/frame_000041/gt.jpg} &
    \snap{images/snapshots/segmentation/1154f3aa0833463280d8a855bce3425b/frame_000041/gpt-5.2.jpg} &
    \snap{images/snapshots/segmentation/1154f3aa0833463280d8a855bce3425b/frame_000041/claude-opus-4.6.jpg} &
    \snap{images/snapshots/segmentation/1154f3aa0833463280d8a855bce3425b/frame_000041/gemini-3-flash-preview.jpg} &
    \snap{images/snapshots/segmentation/1154f3aa0833463280d8a855bce3425b/frame_000041/qwen3.5-plus.jpg} &
    \snap{images/snapshots/segmentation/1154f3aa0833463280d8a855bce3425b/frame_000041/qwen-vl-max.jpg} &
    \snap{images/snapshots/segmentation/1154f3aa0833463280d8a855bce3425b/frame_000041/qwen3-vl-235b.jpg} &
    \snap{images/snapshots/segmentation/1154f3aa0833463280d8a855bce3425b/frame_000041/facebook_sam3.jpg} \\[1pt]
    \snap{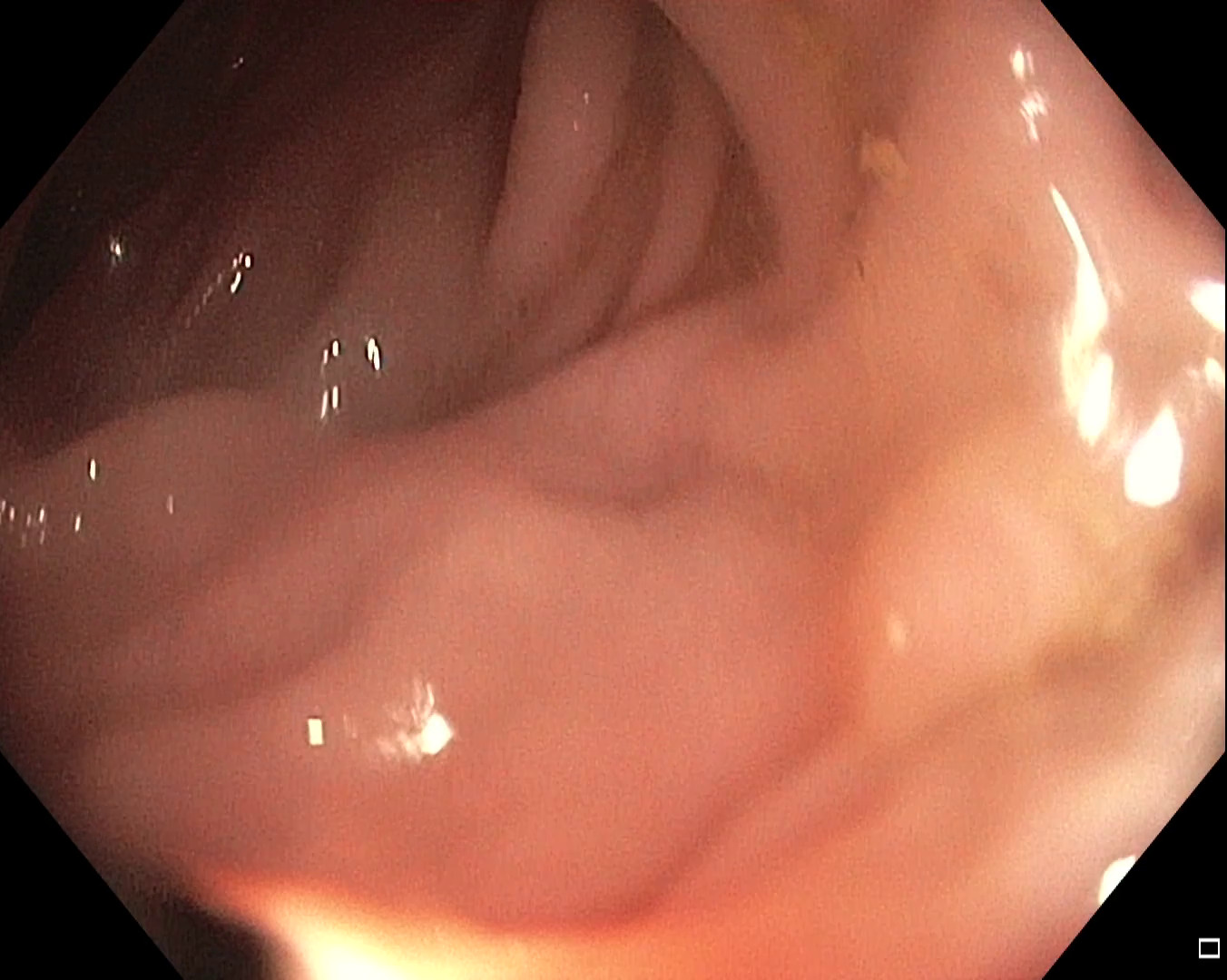} &
    \snap{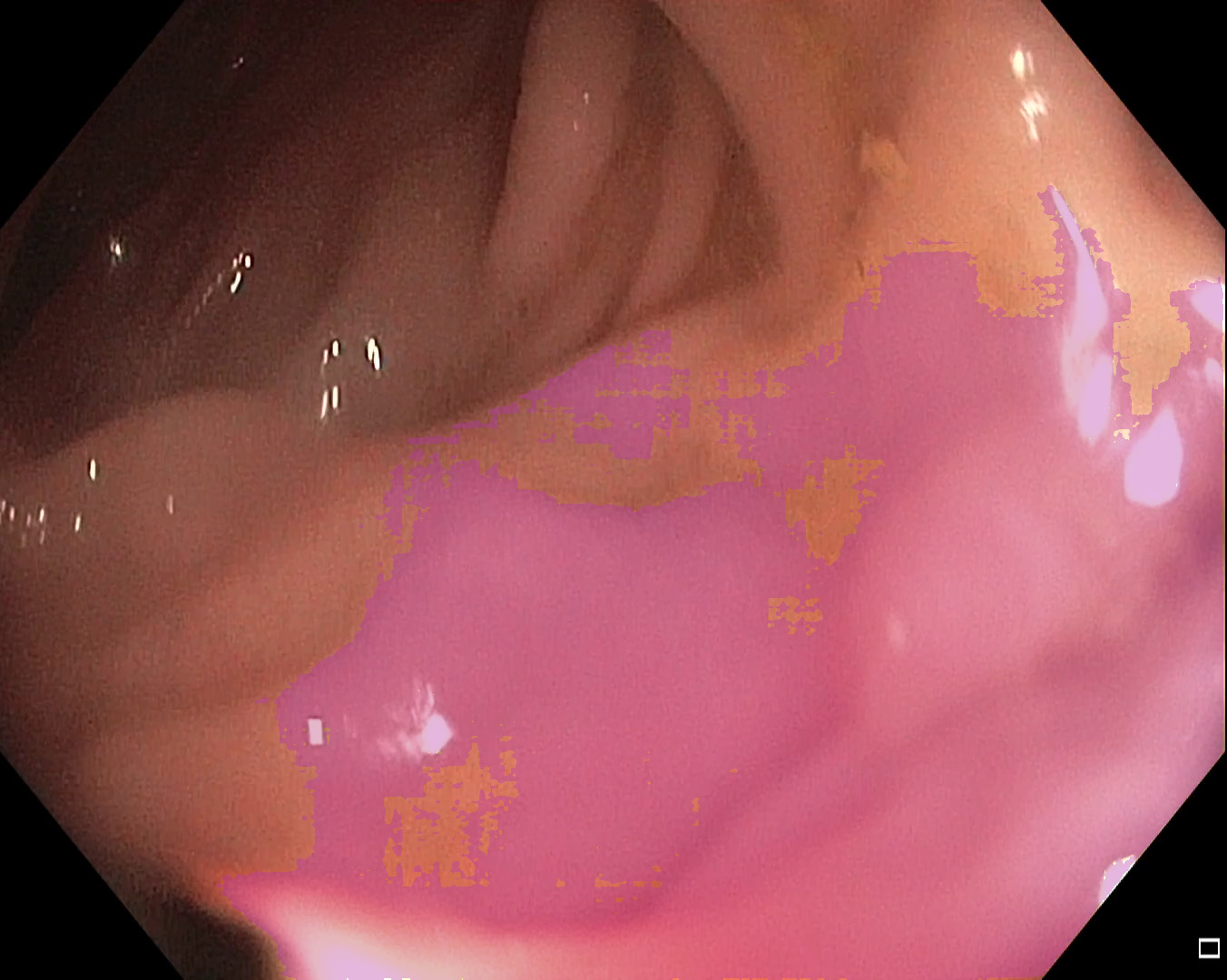} &
    \snap{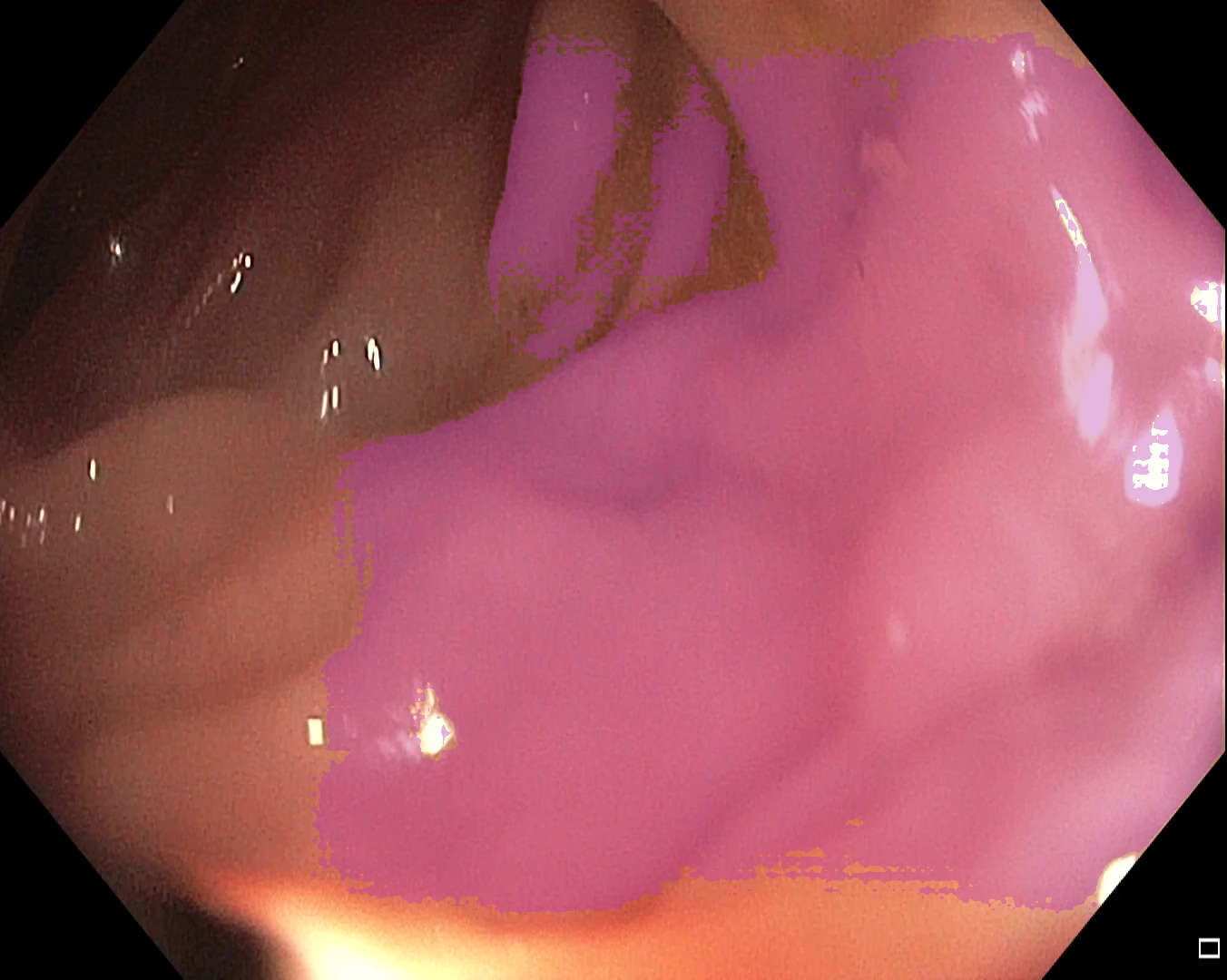} &
    \snap{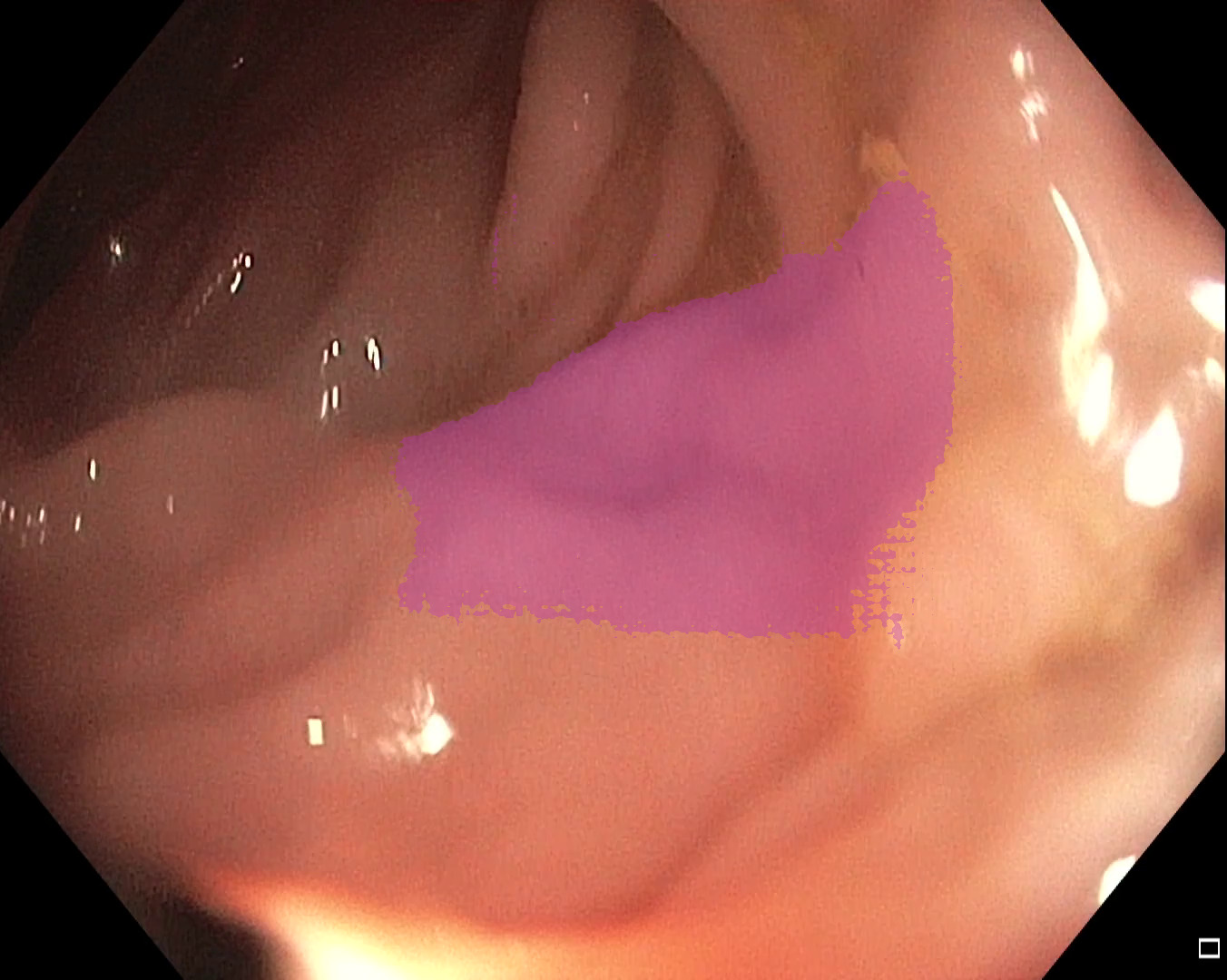} &
    \snap{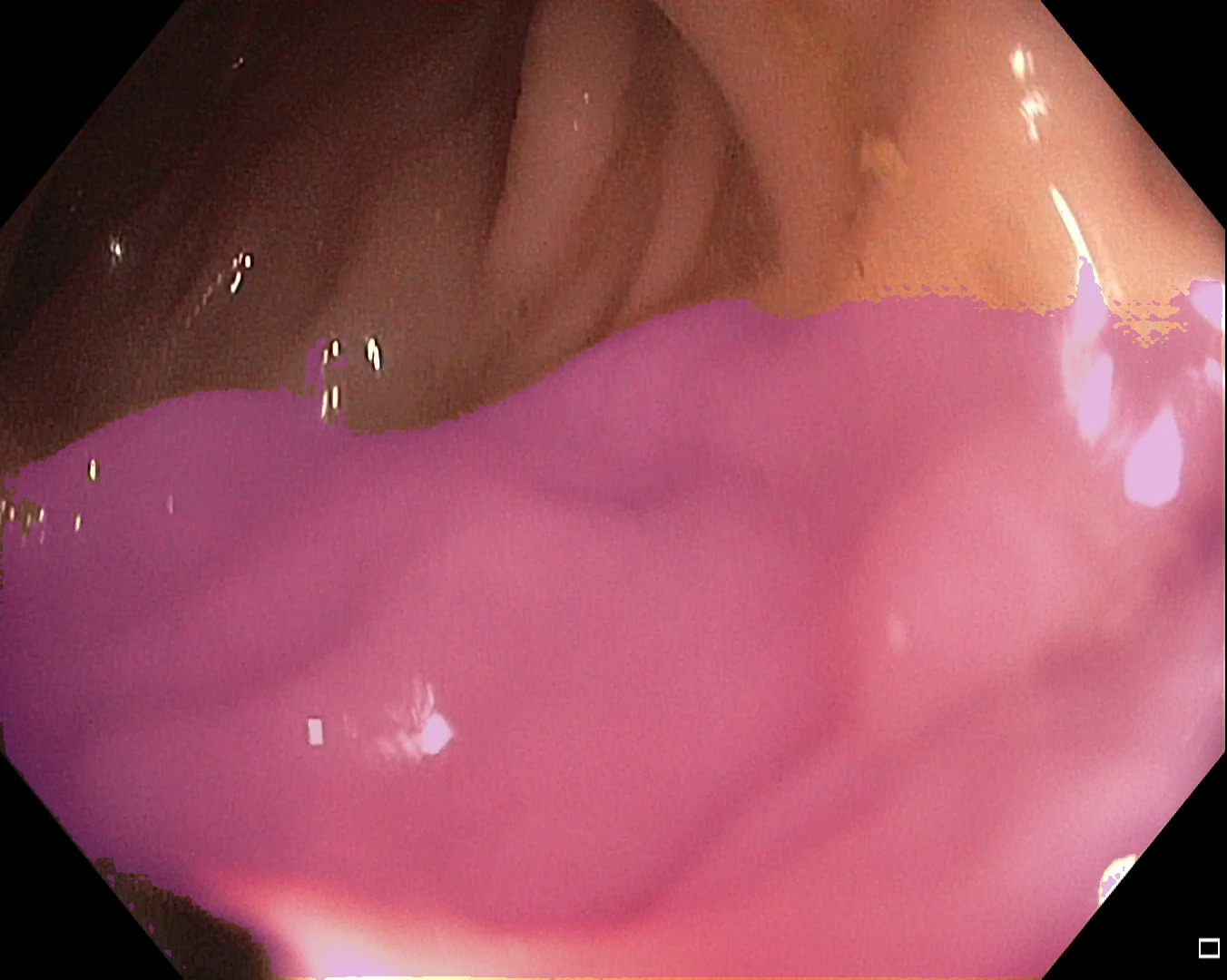} &
    \snap{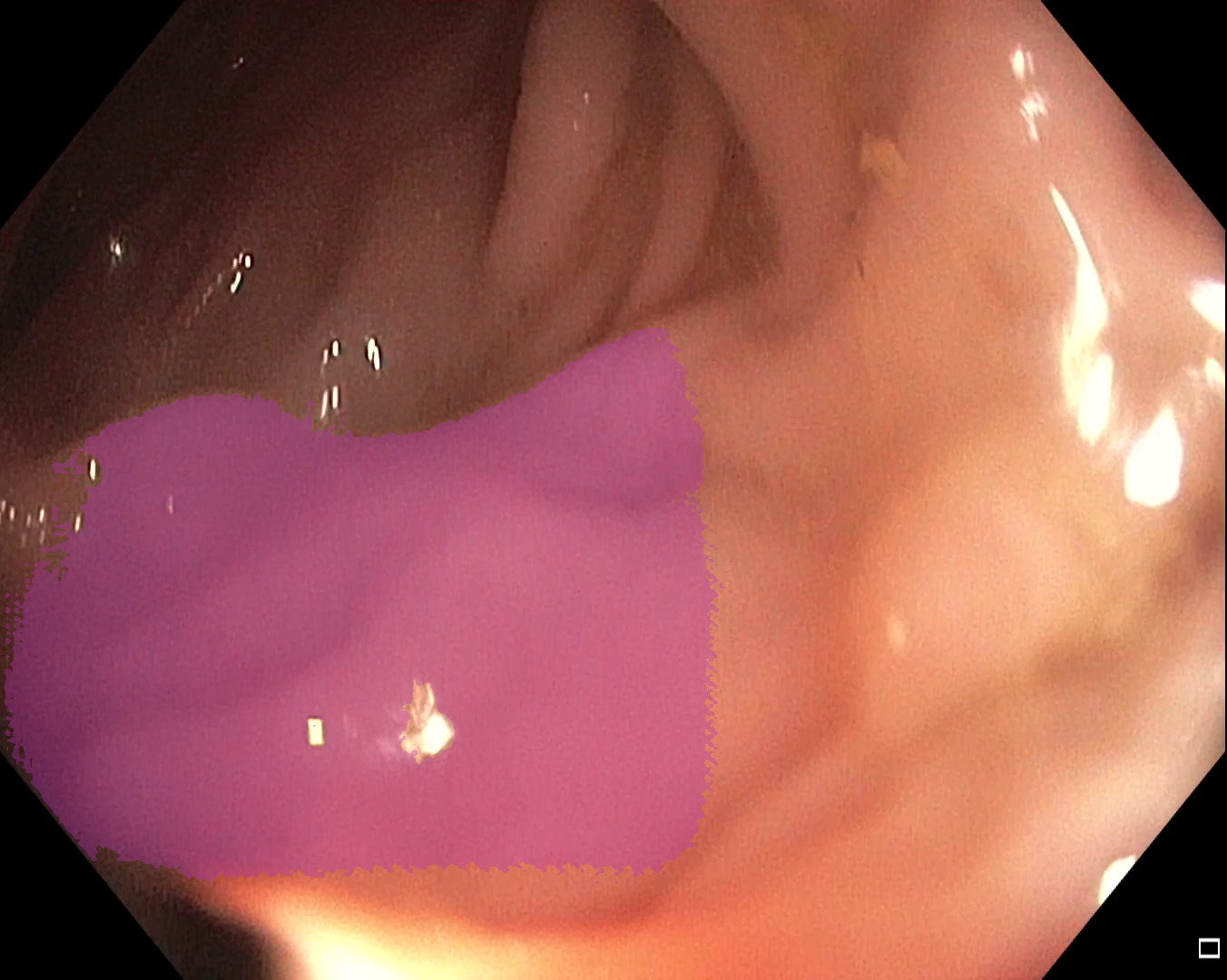} &
    \snap{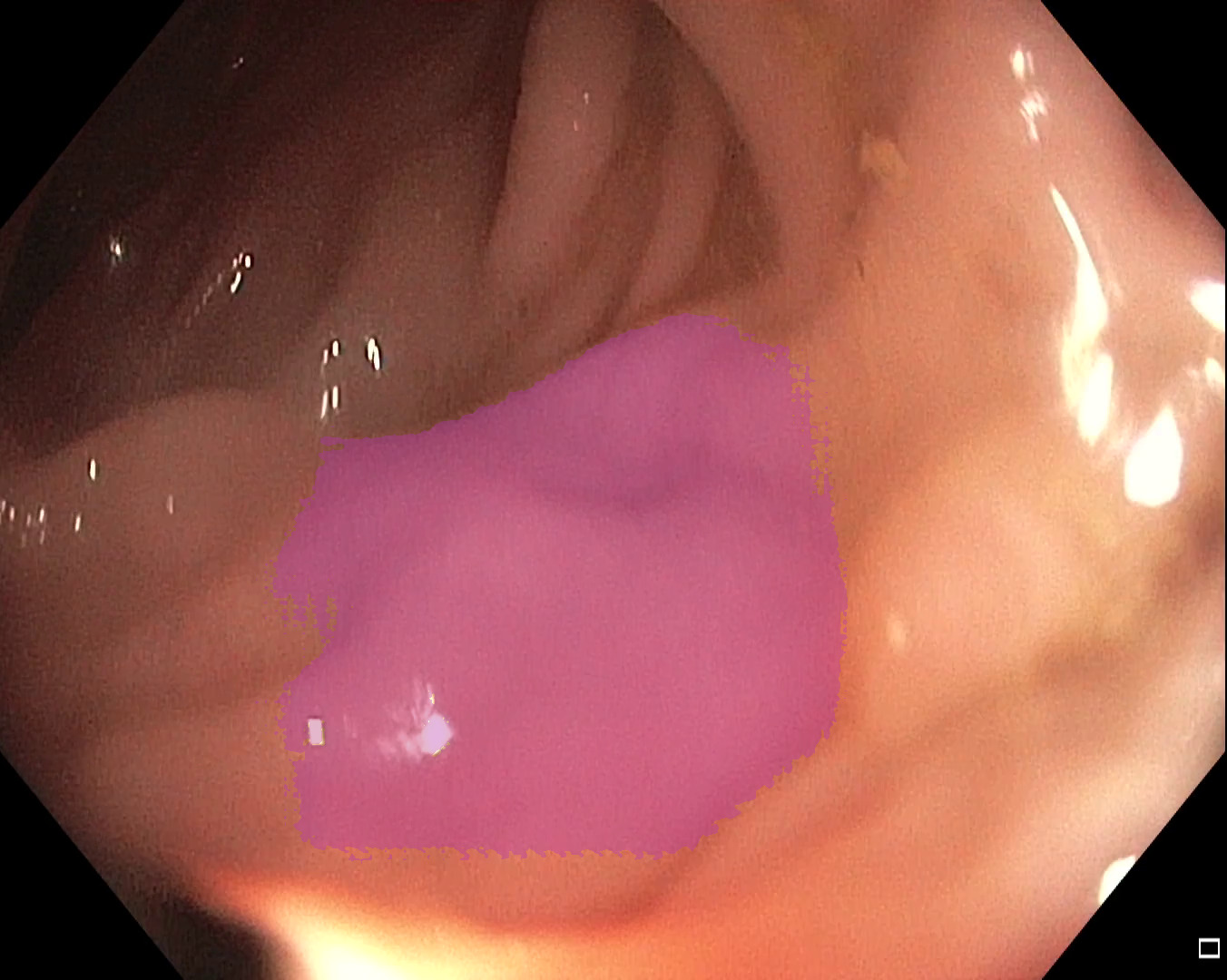} &
    \snap{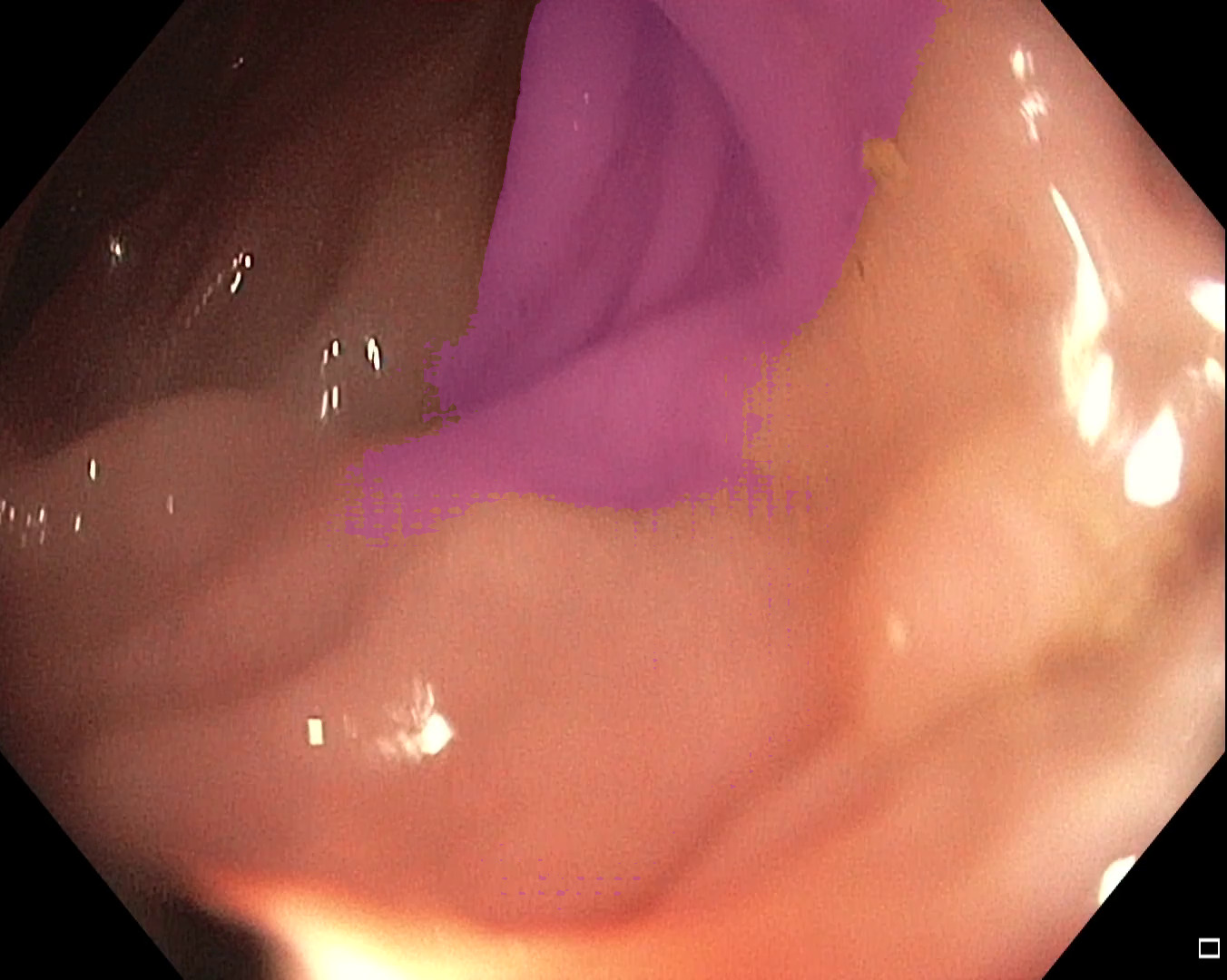} &
    \snap{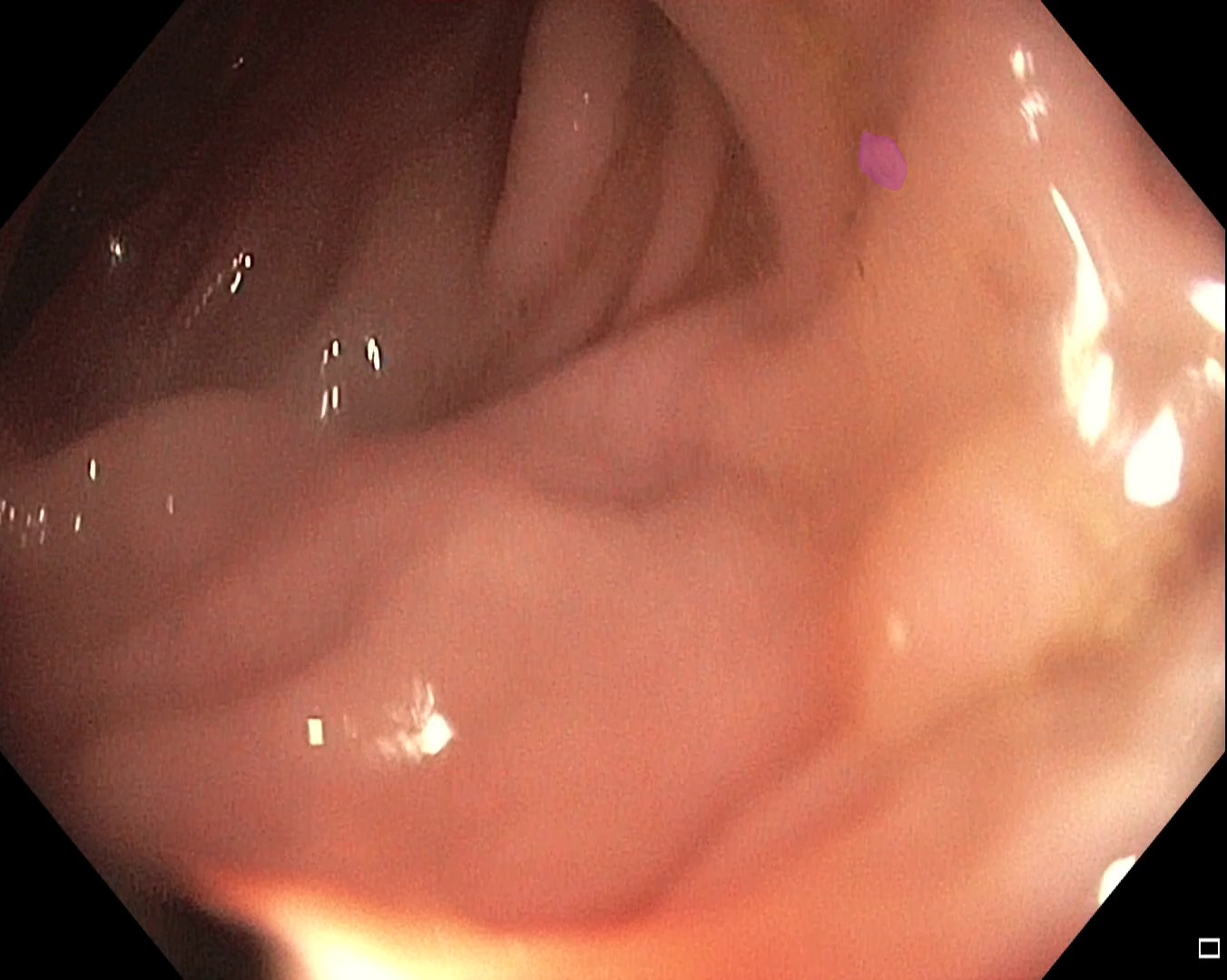} \\
\snap{images/snapshots/segmentation/144f8236c7f34e94badd9496dfc36886/frame_000000/input.jpg} &
\snap{images/snapshots/segmentation/144f8236c7f34e94badd9496dfc36886/frame_000000/gt.jpg} &
\snap{images/snapshots/segmentation/144f8236c7f34e94badd9496dfc36886/frame_000000/gpt-5.2.jpg} &
\snap{images/snapshots/segmentation/144f8236c7f34e94badd9496dfc36886/frame_000000/claude-opus-4.6.jpg} &
\snap{images/snapshots/segmentation/144f8236c7f34e94badd9496dfc36886/frame_000000/gemini-3-flash-preview.jpg} &
\snap{images/snapshots/segmentation/144f8236c7f34e94badd9496dfc36886/frame_000000/qwen3.5-plus.jpg} &
\snap{images/snapshots/segmentation/144f8236c7f34e94badd9496dfc36886/frame_000000/qwen-vl-max.jpg} &
\snap{images/snapshots/segmentation/144f8236c7f34e94badd9496dfc36886/frame_000000/qwen3-vl-235b.jpg} &
\snap{images/snapshots/segmentation/144f8236c7f34e94badd9496dfc36886/frame_000000/facebook_sam3.jpg} \\[1pt]
\snap{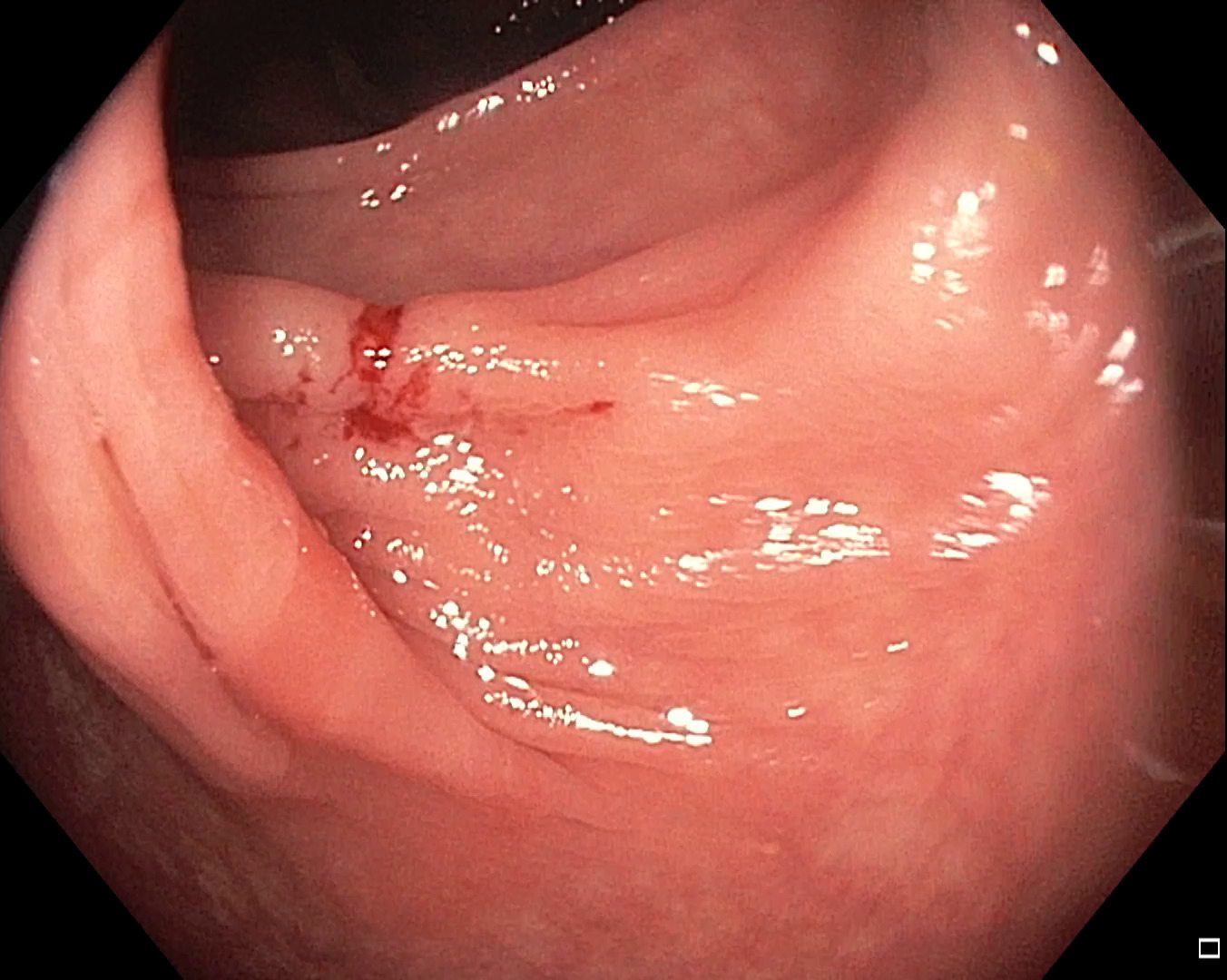} &
\snap{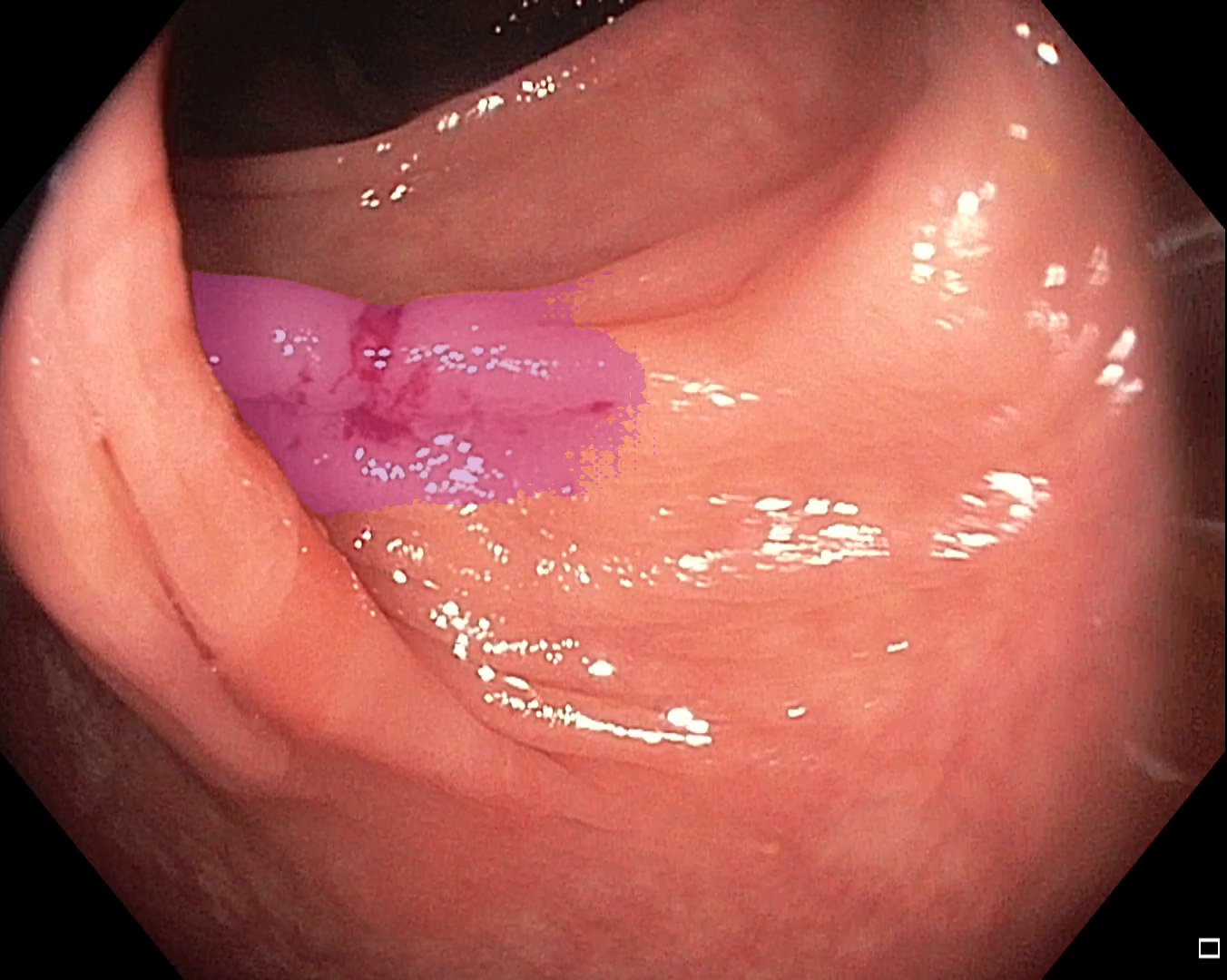} &
\snap{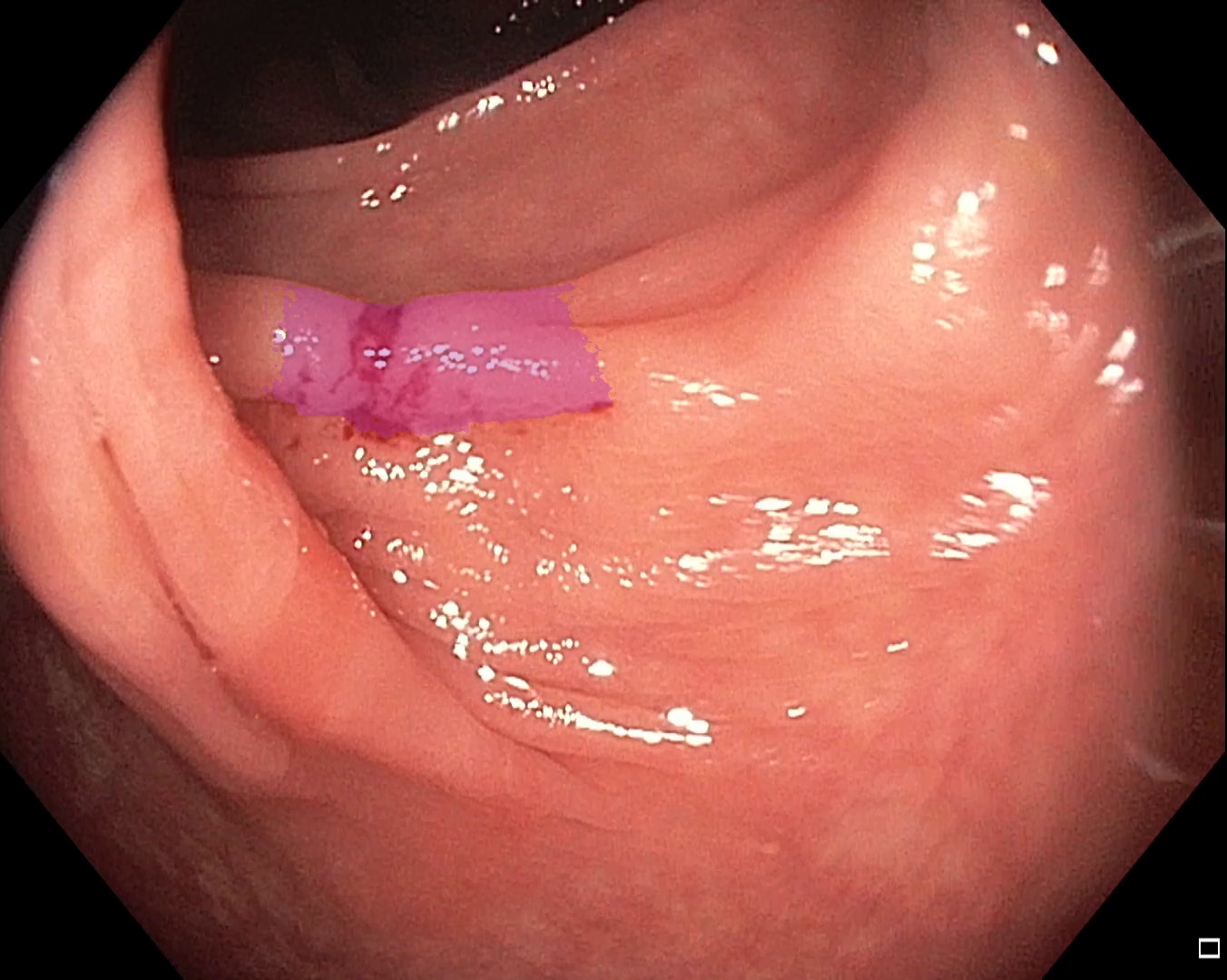} &
\snap{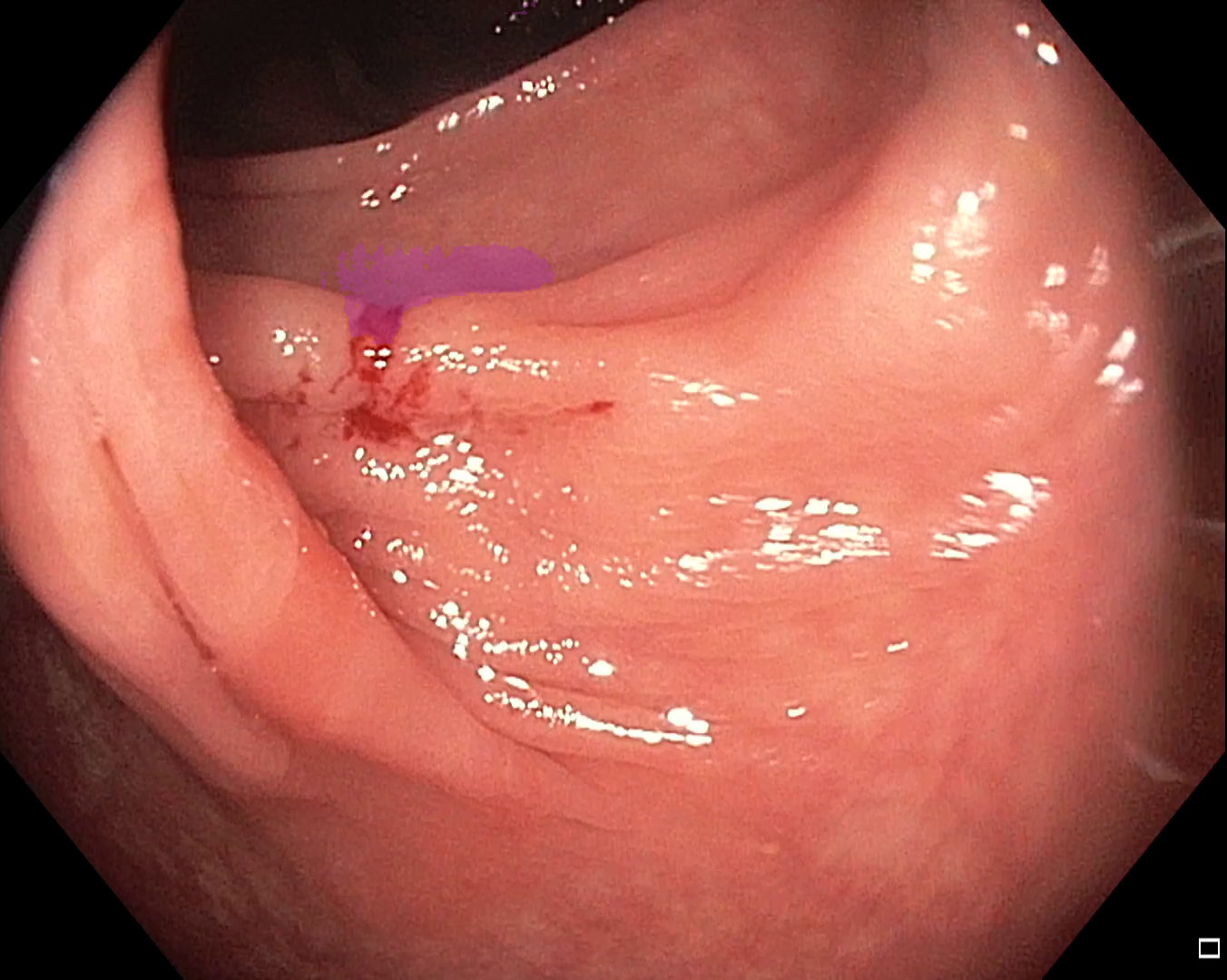} &
\snap{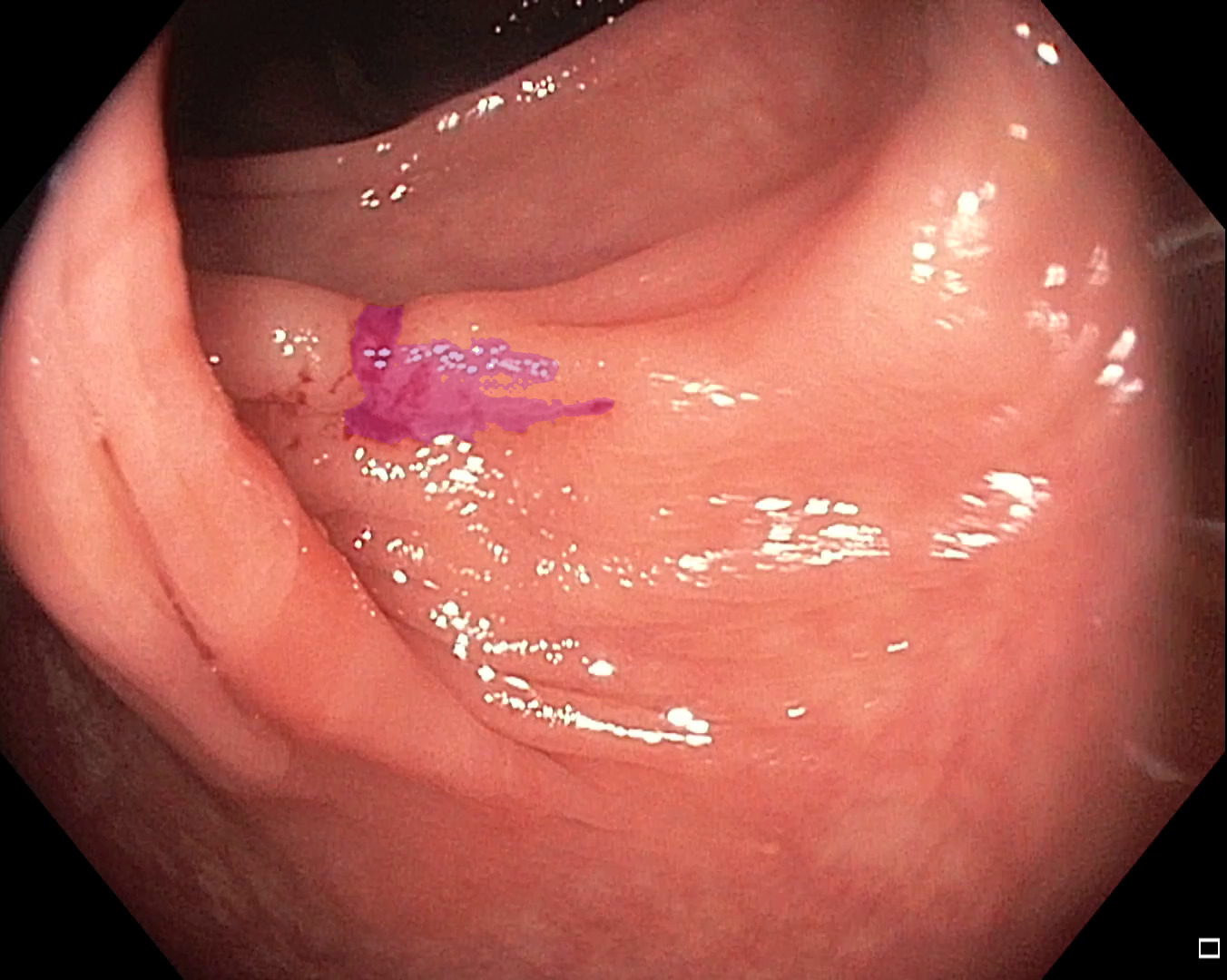} &
\snap{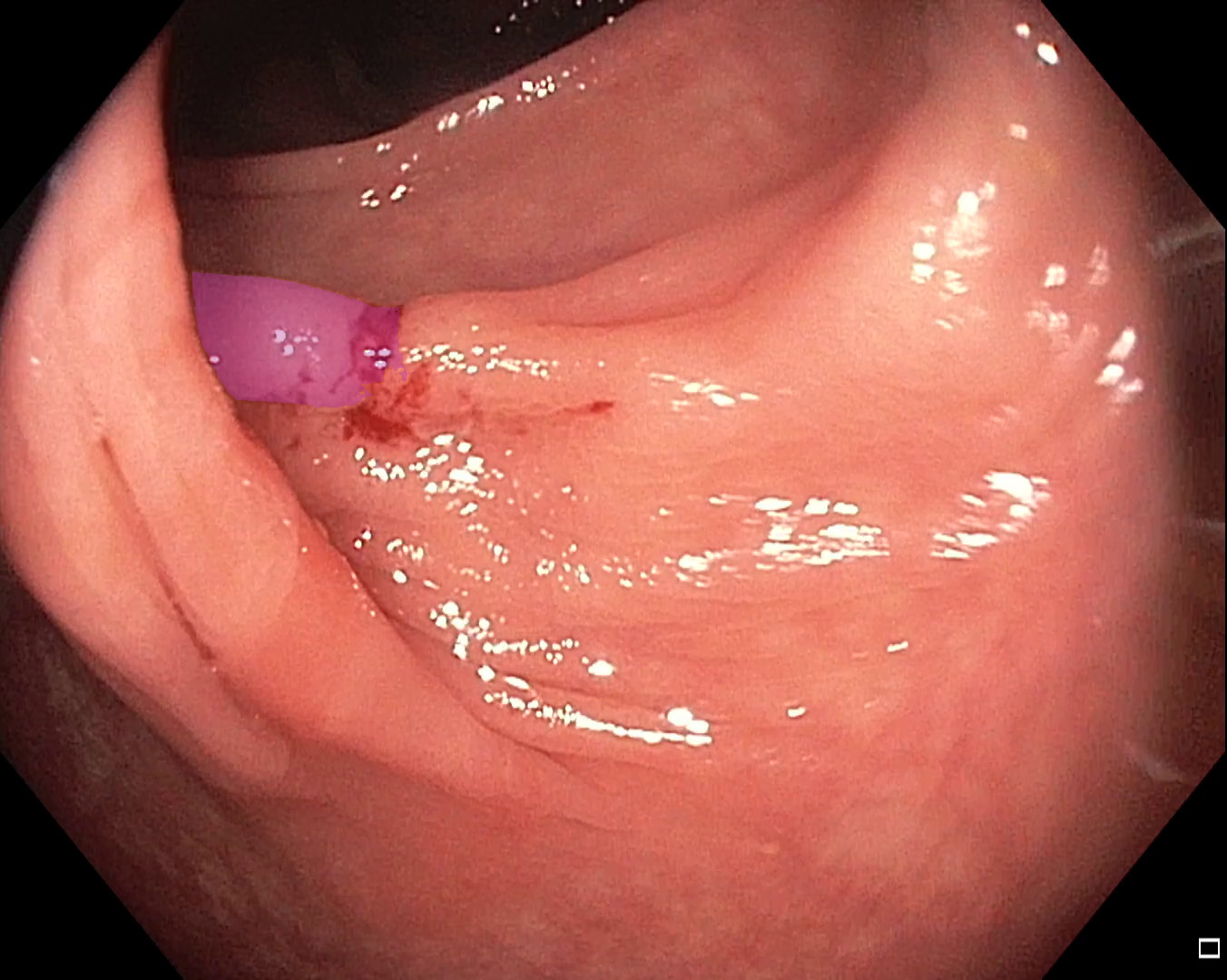} &
\snap{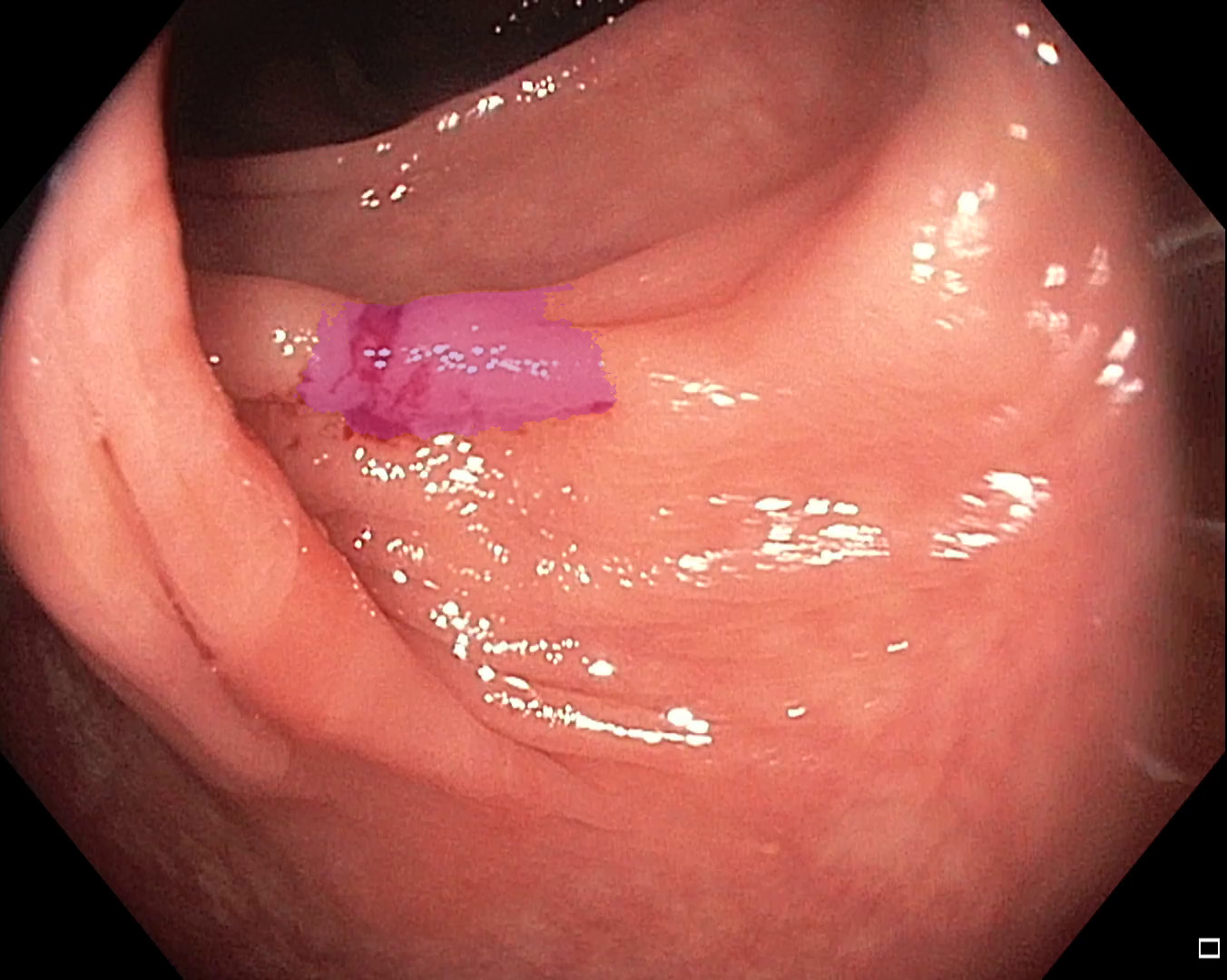} &
\snap{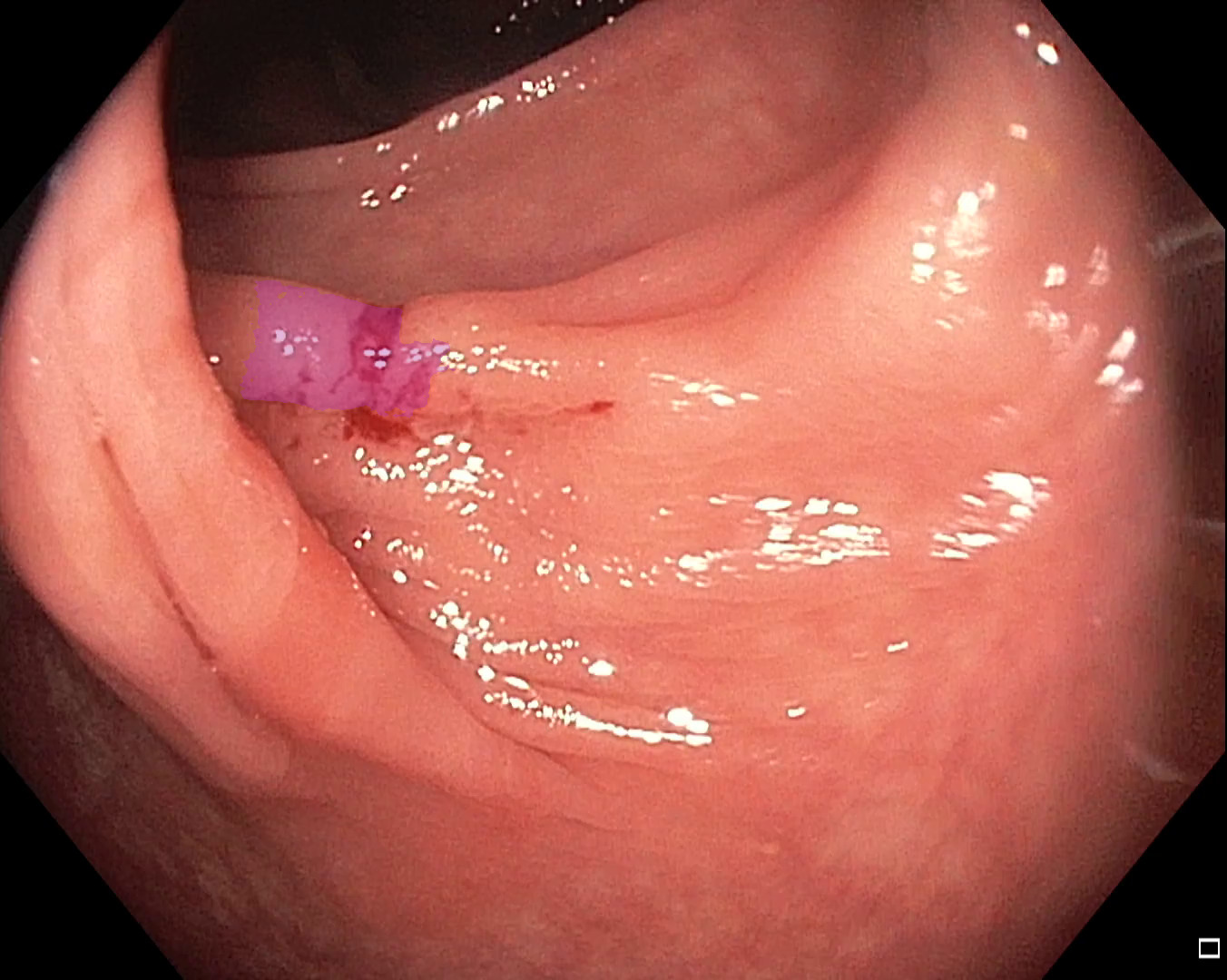} &
\snap{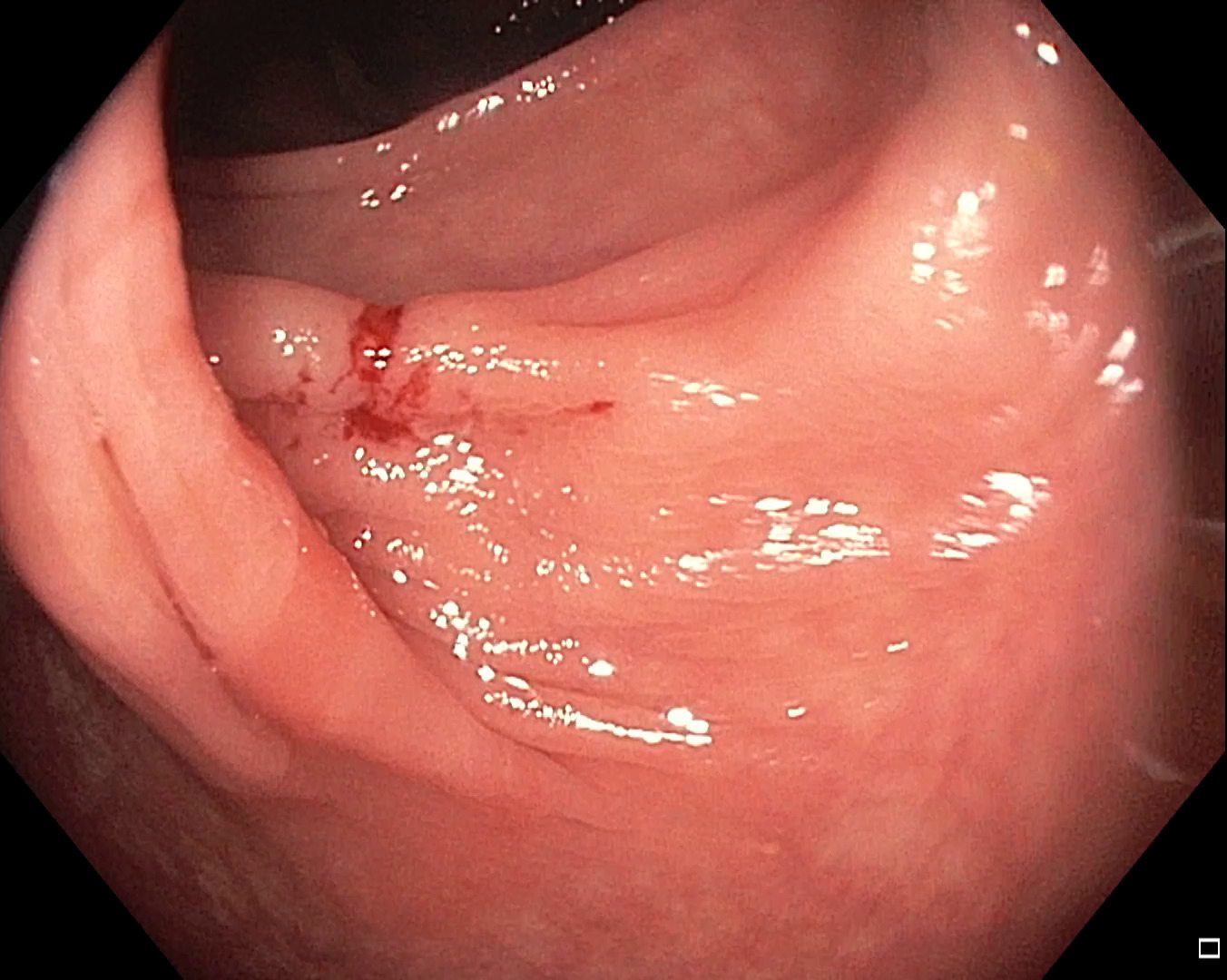} \\[1pt]
    \snap{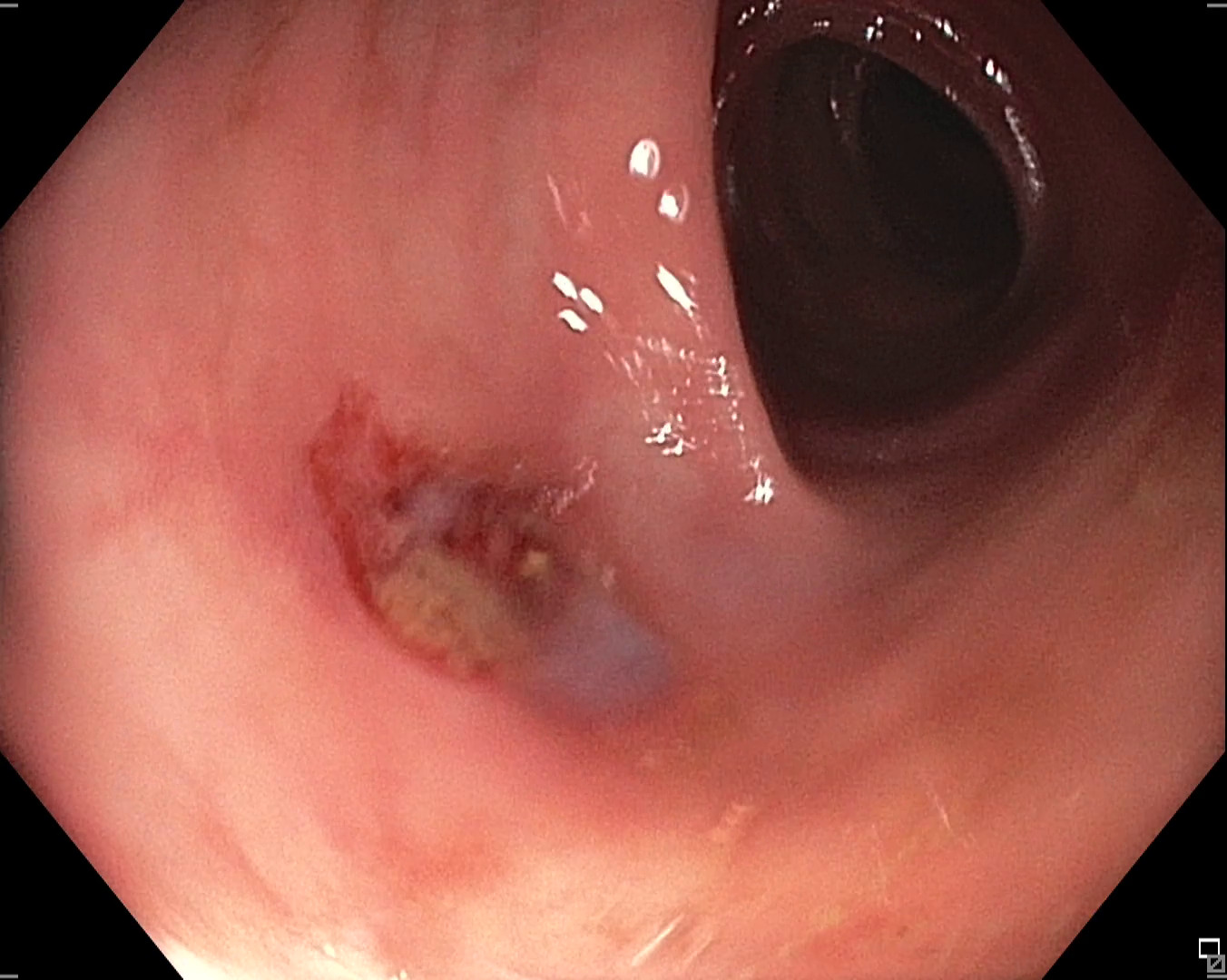} &
    \snap{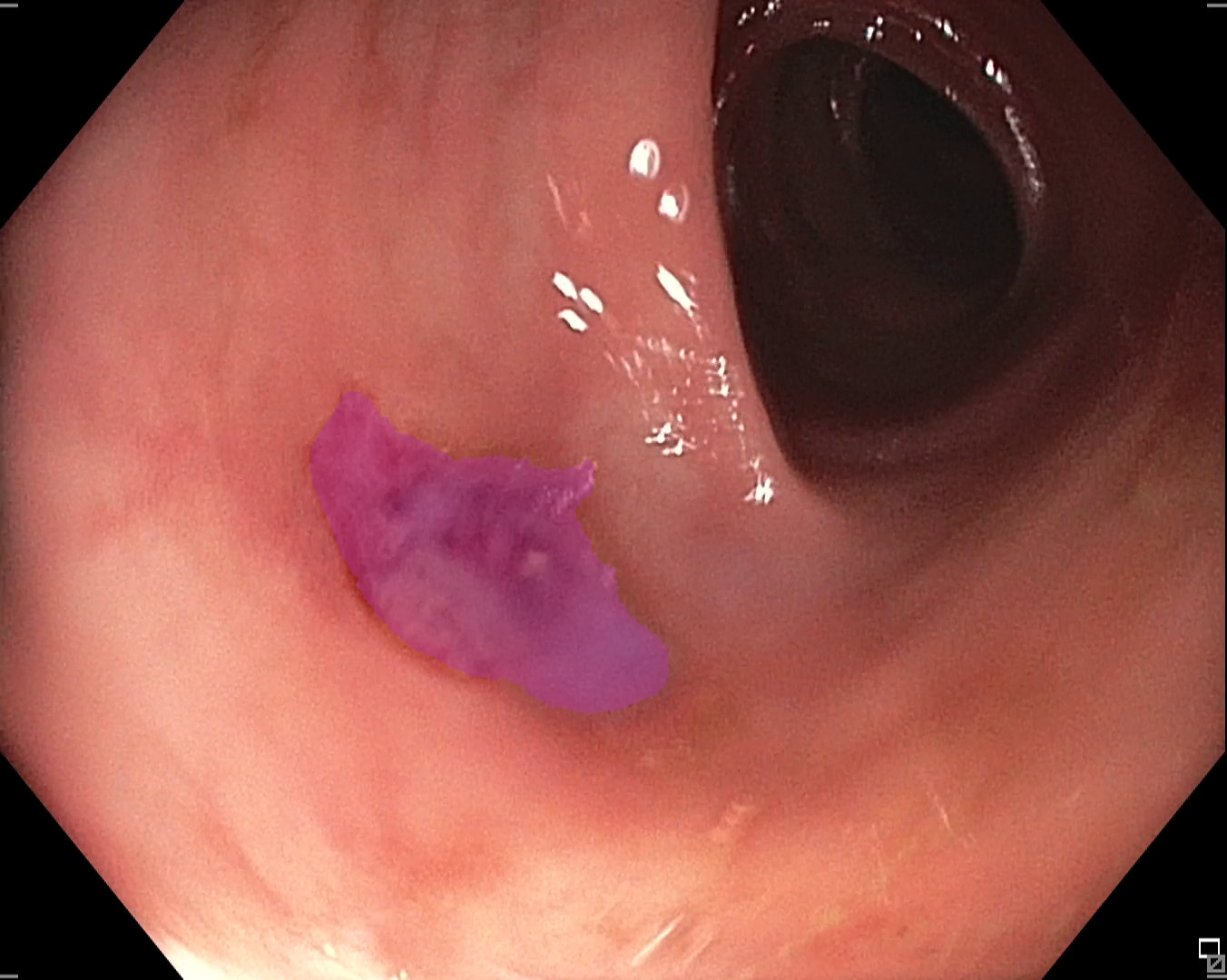} &
    \snap{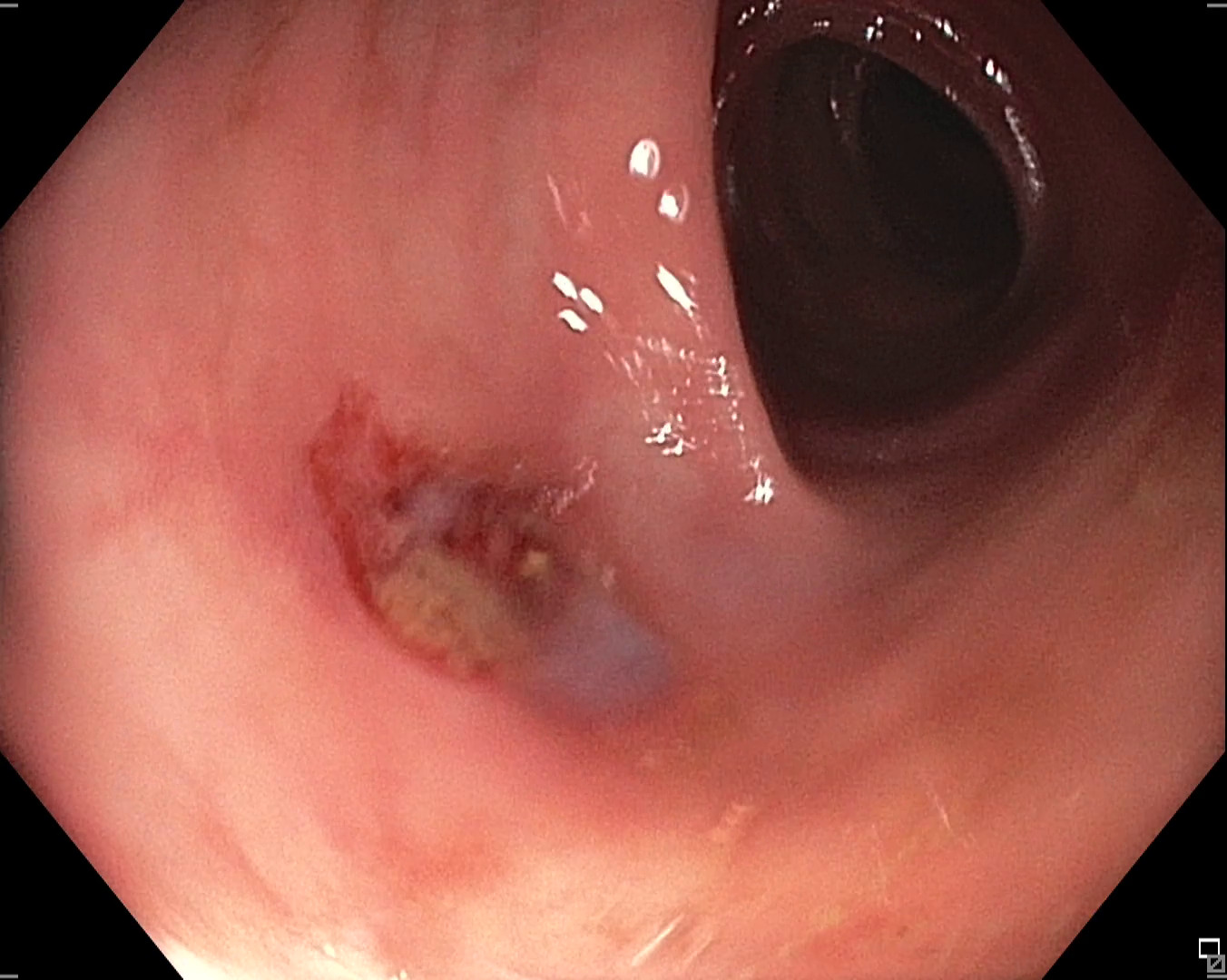} &
    \snap{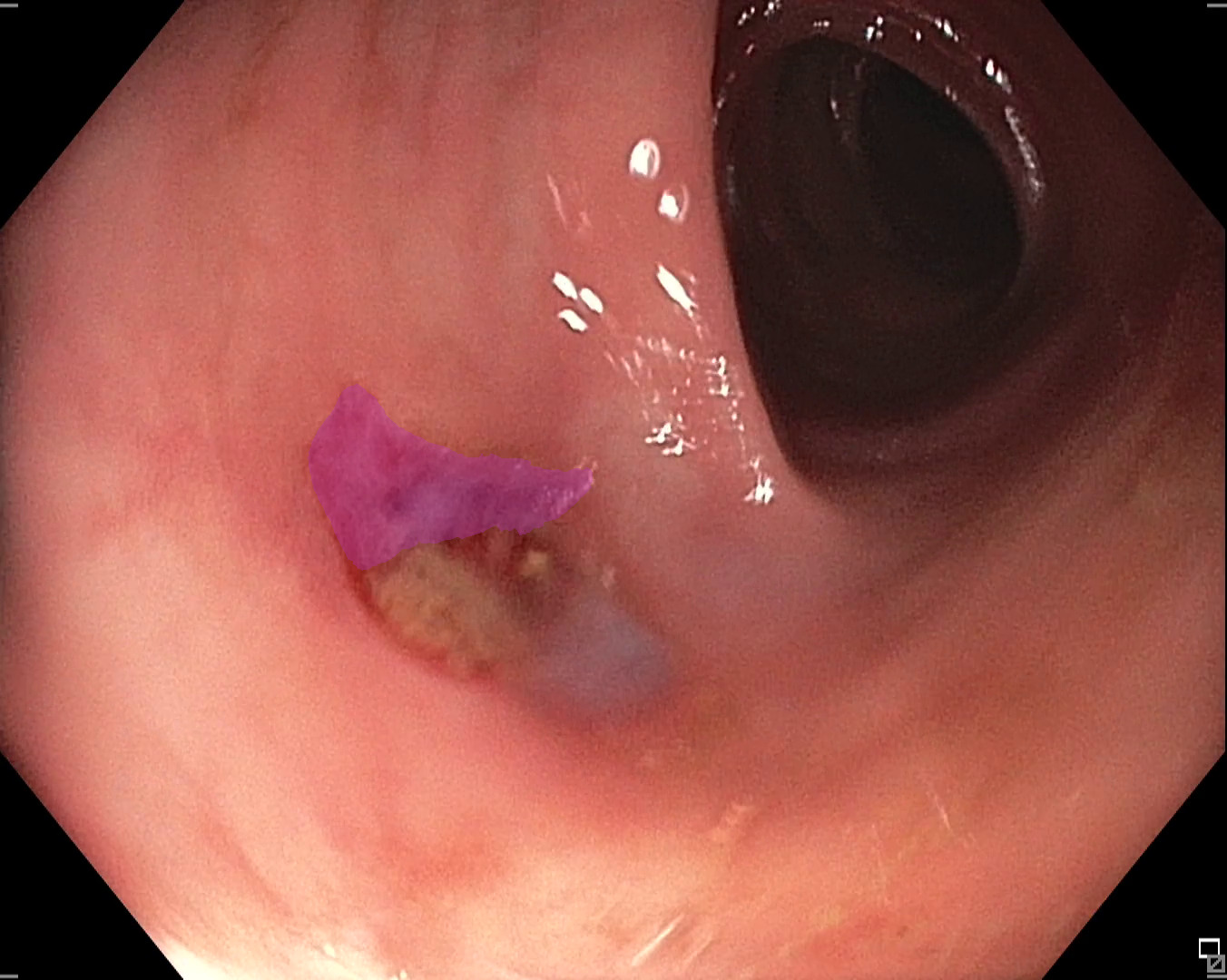} &
    \snap{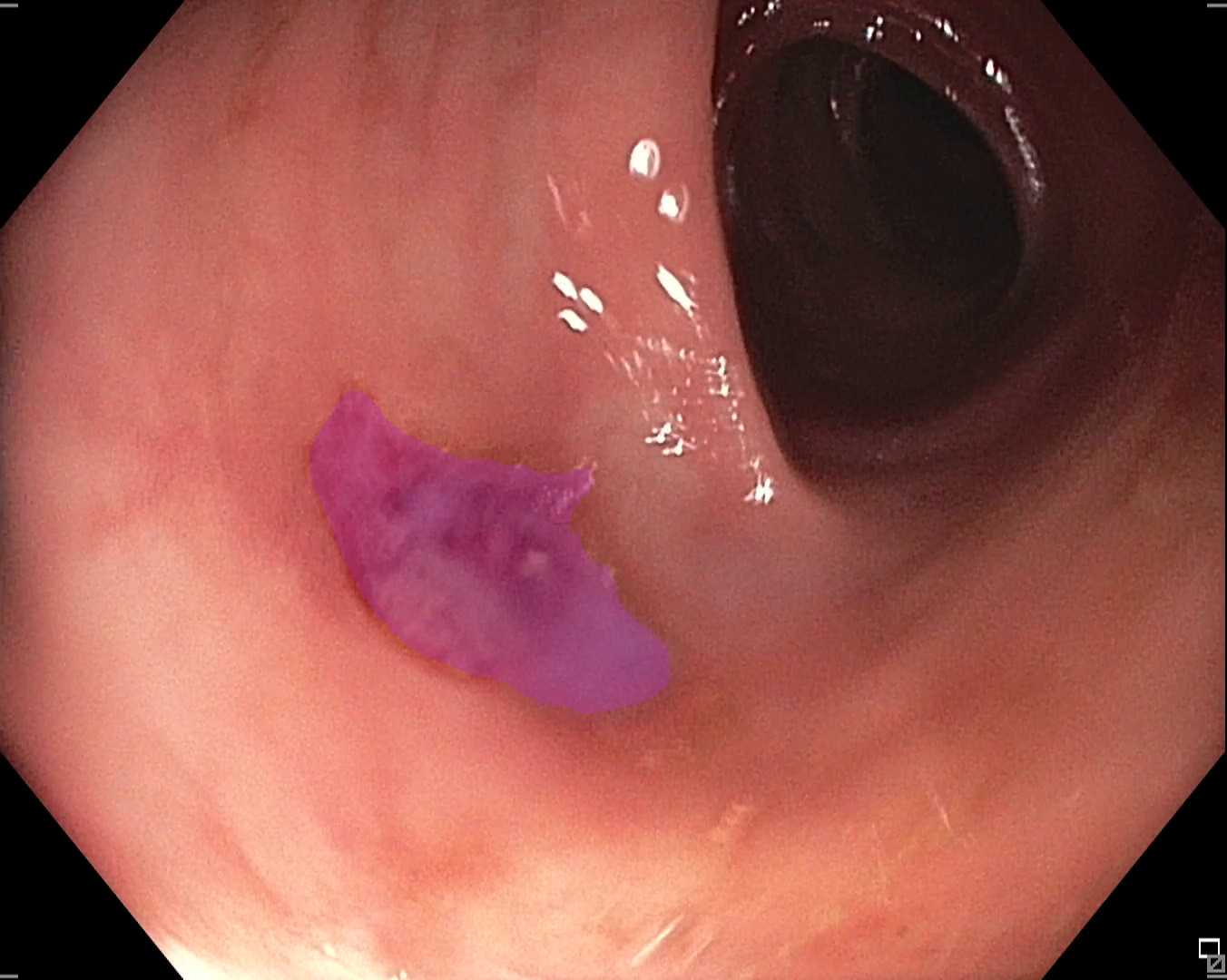} &
    \snap{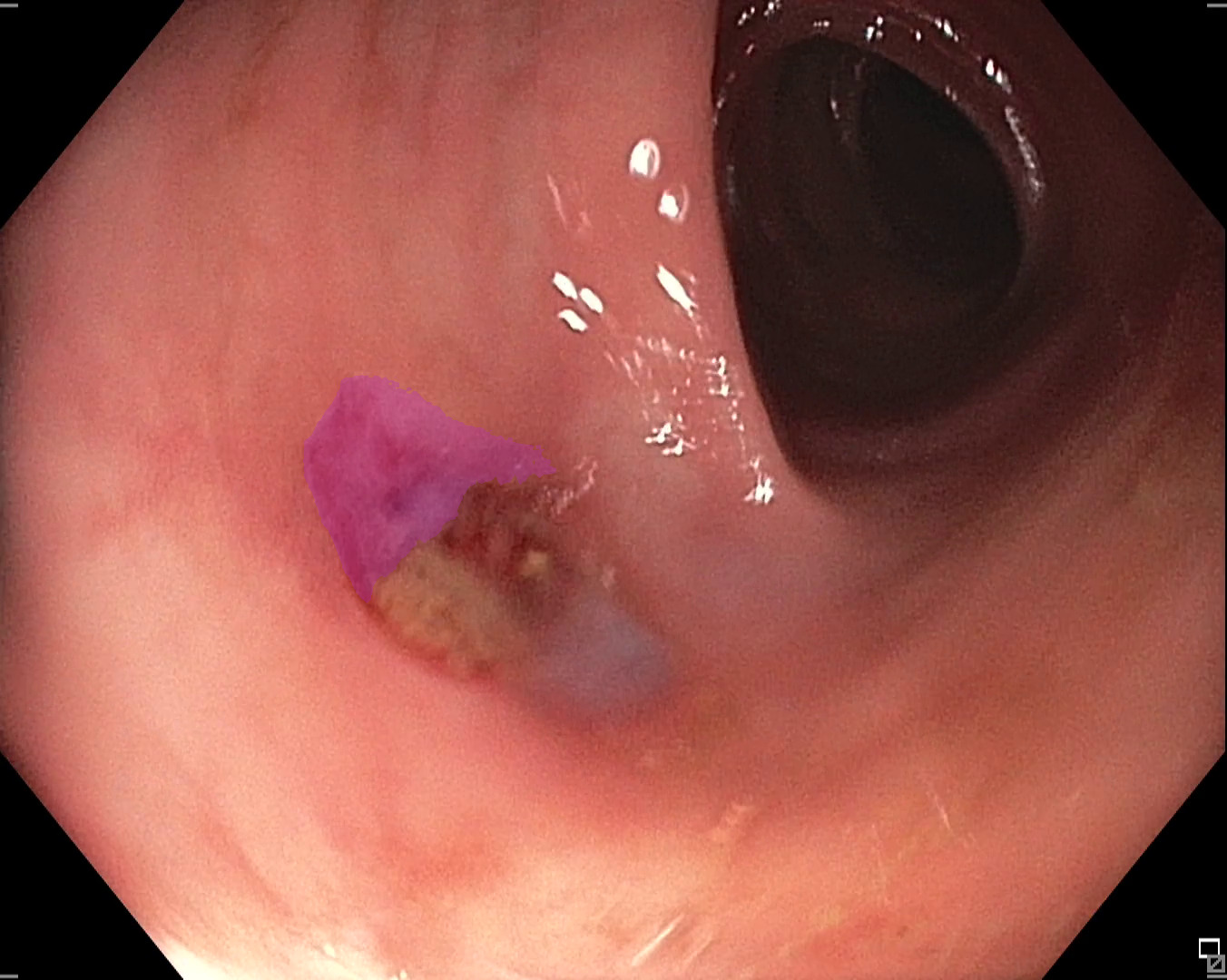} &
    \snap{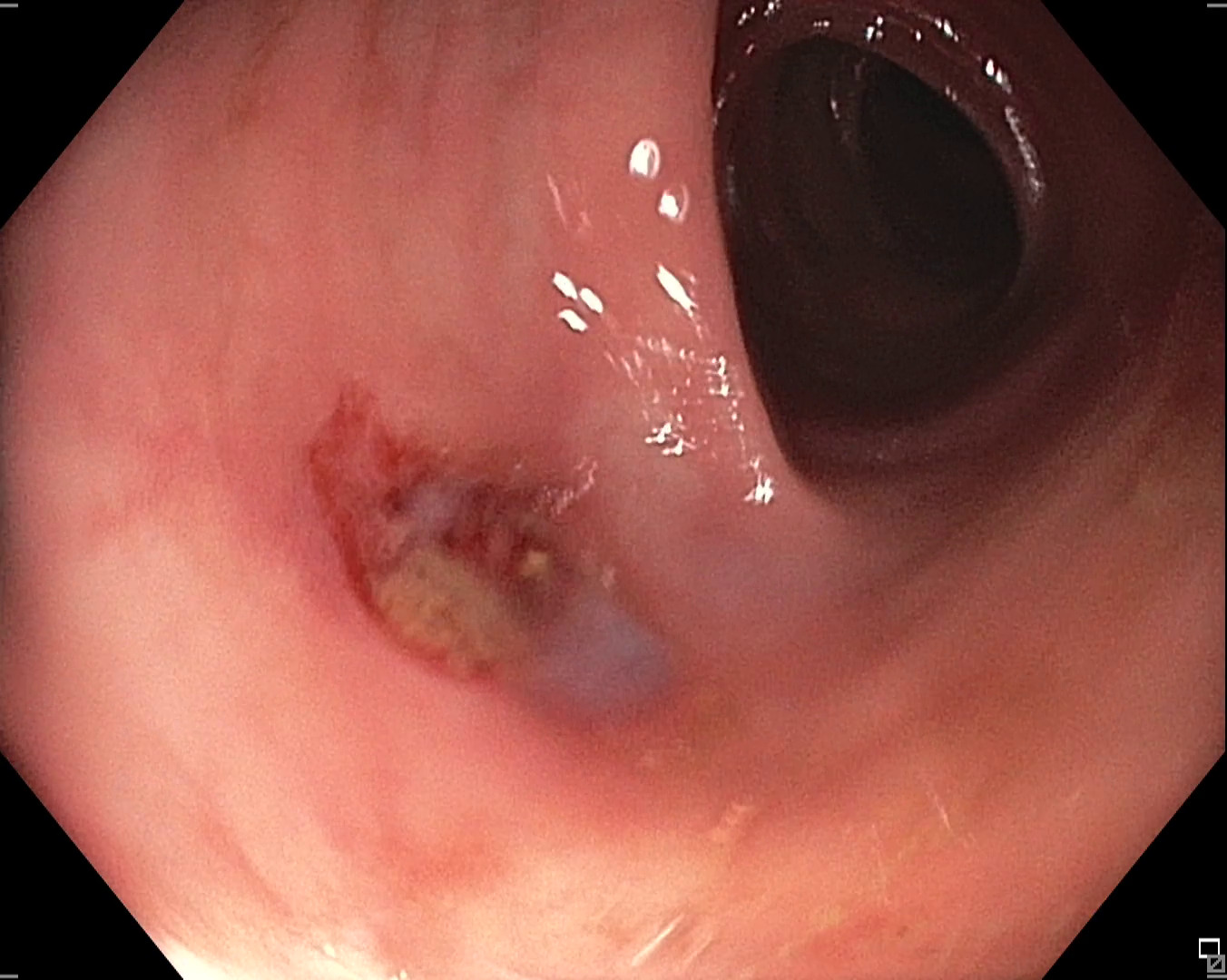} &
    \snap{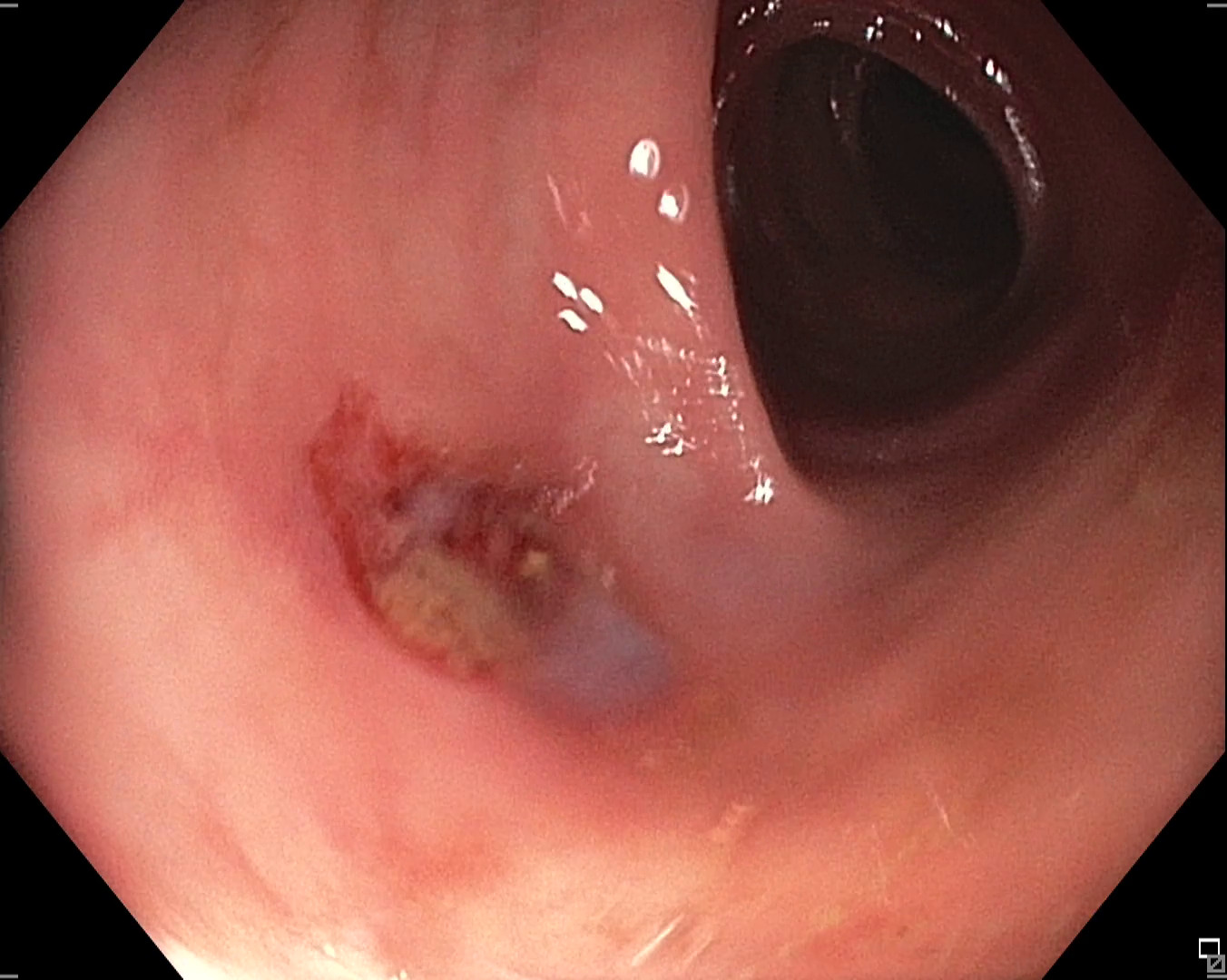} &
    \snap{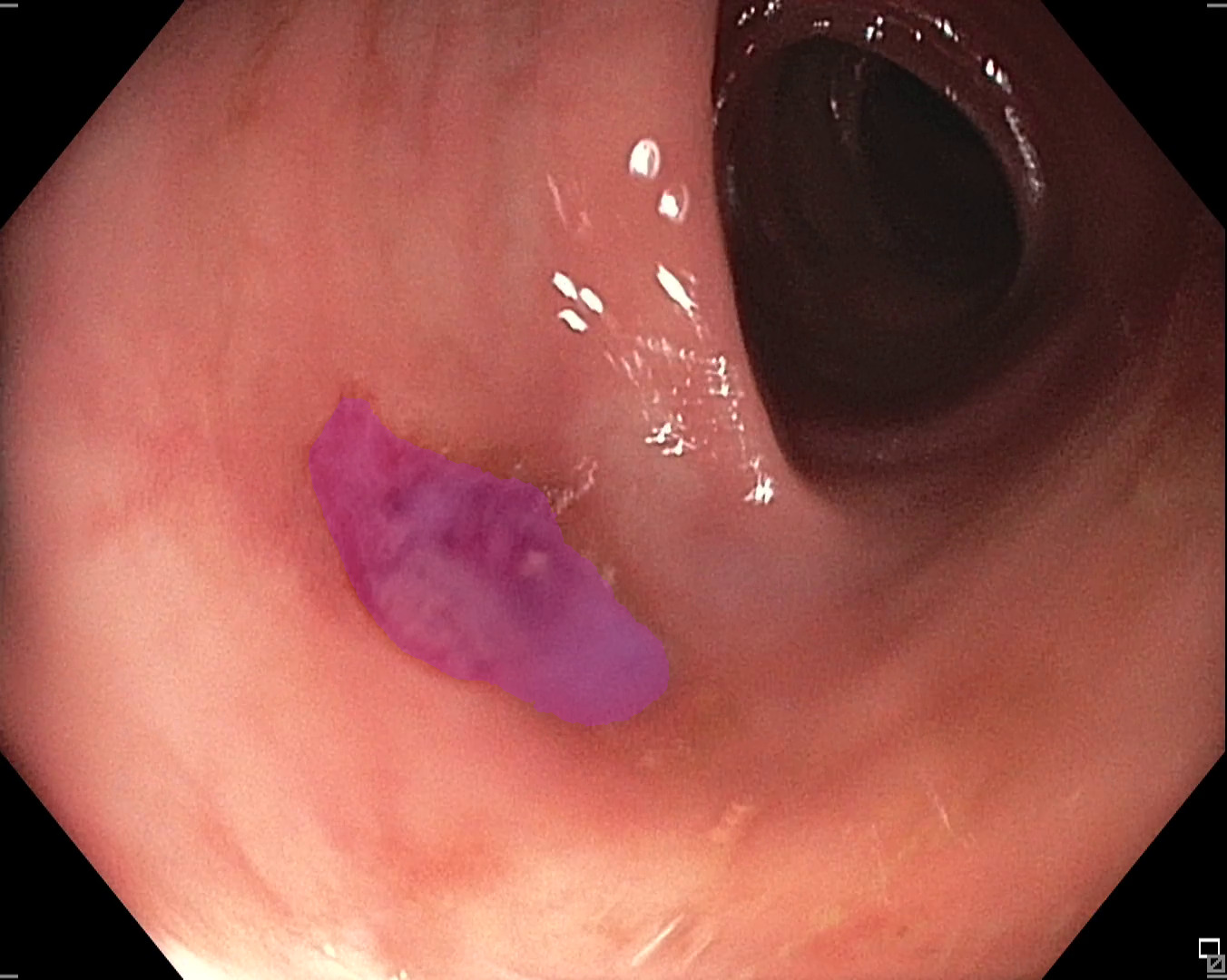} \\[1pt]
    \snap{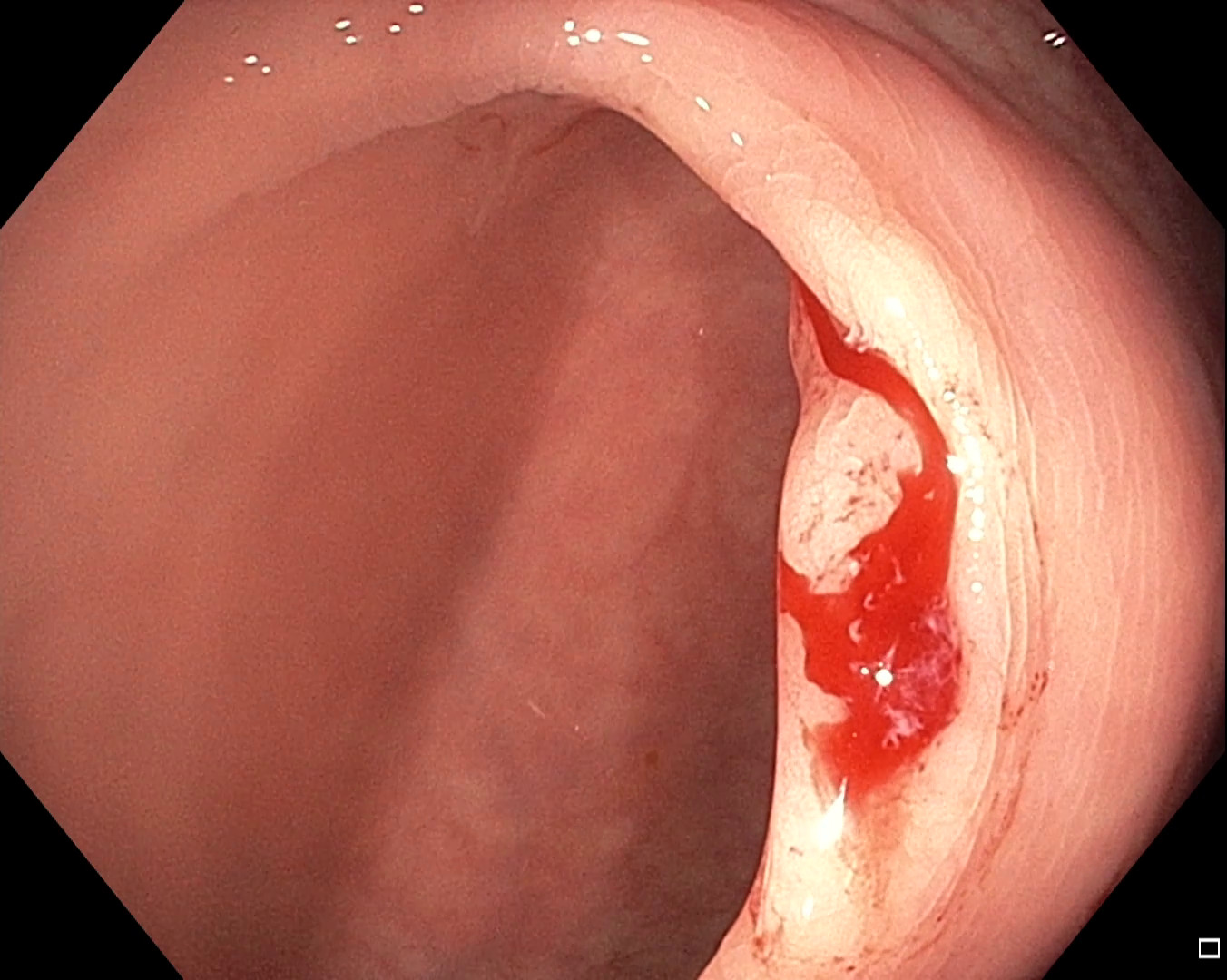} &
    \snap{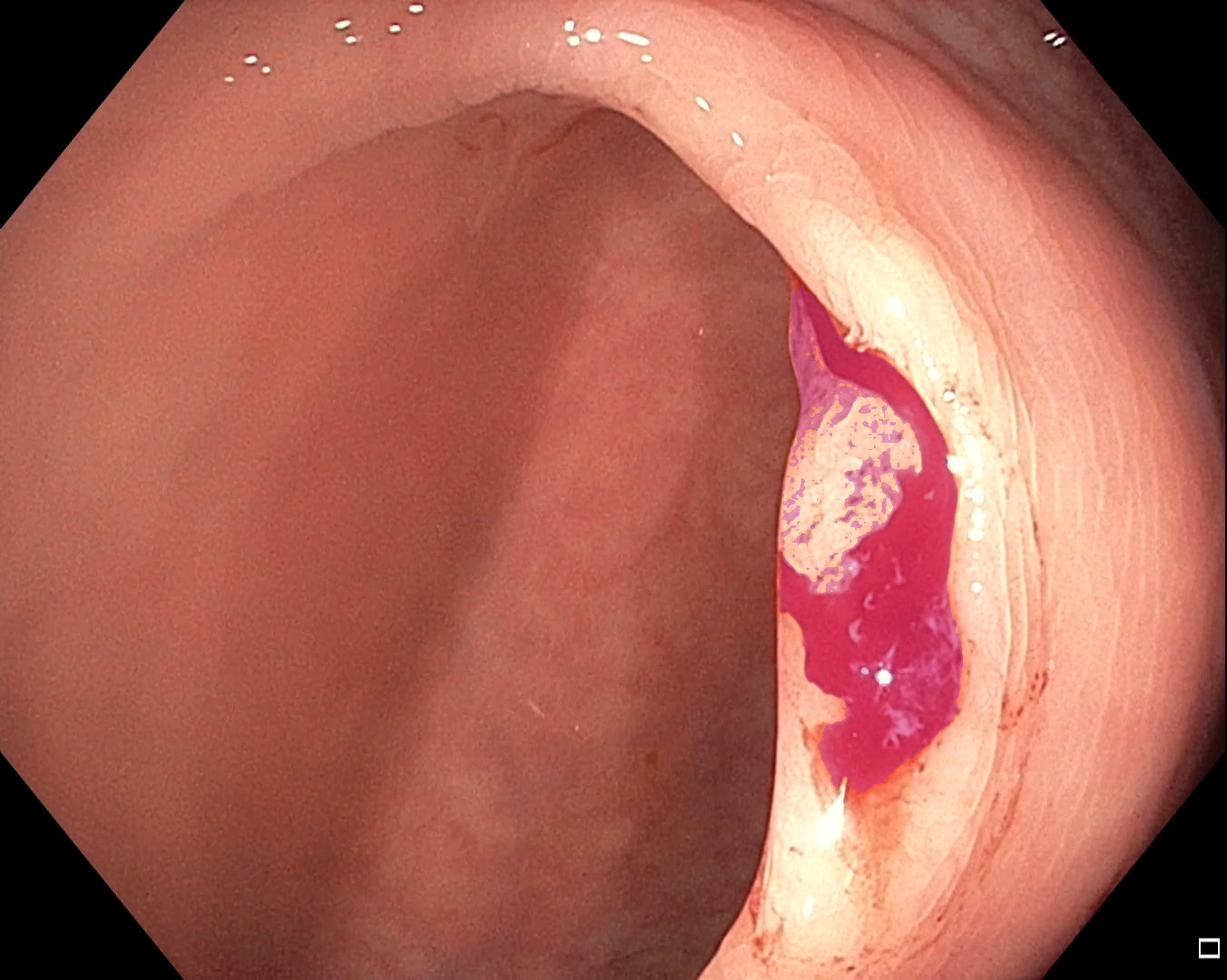} &
    \snap{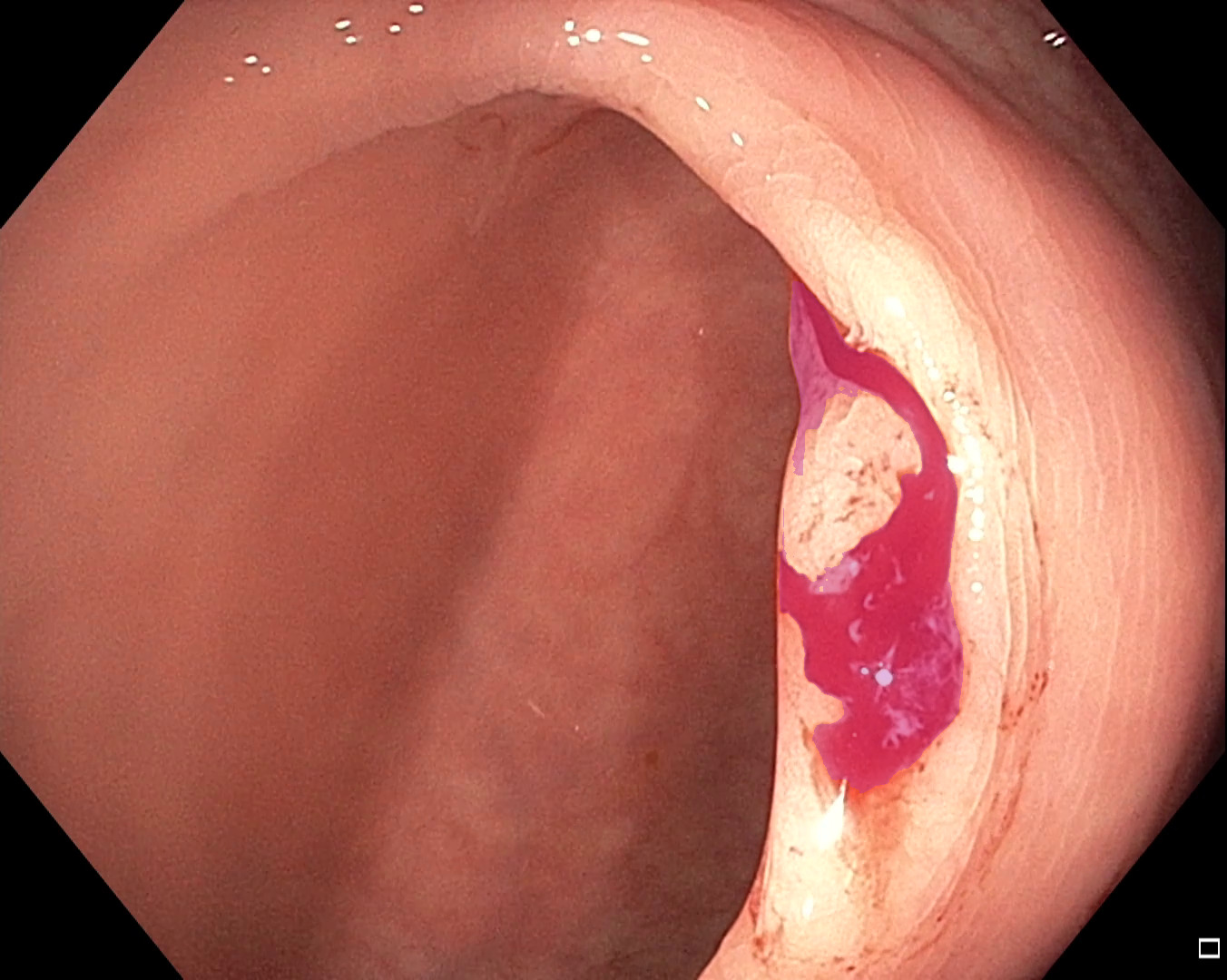} &
    \snap{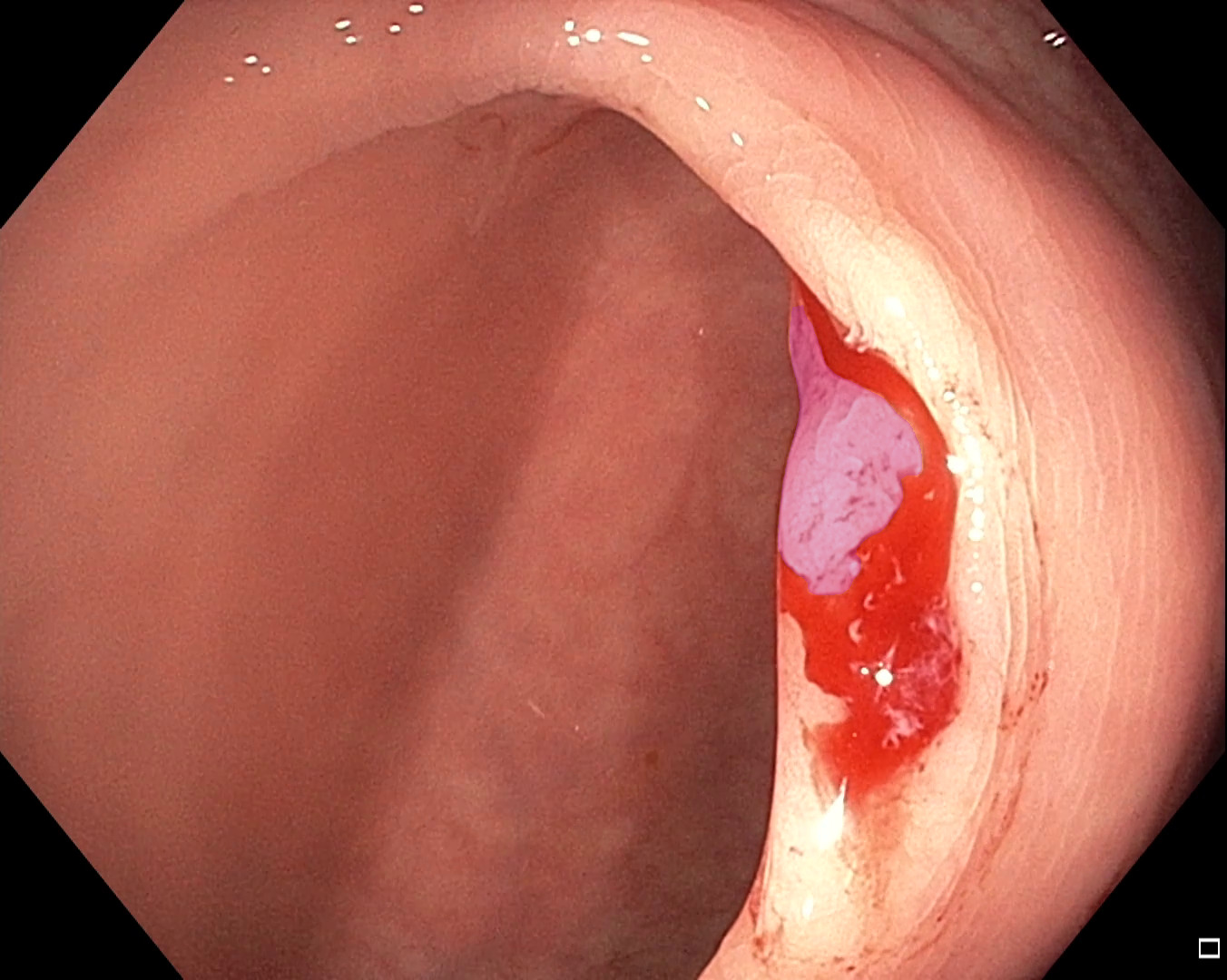} &
    \snap{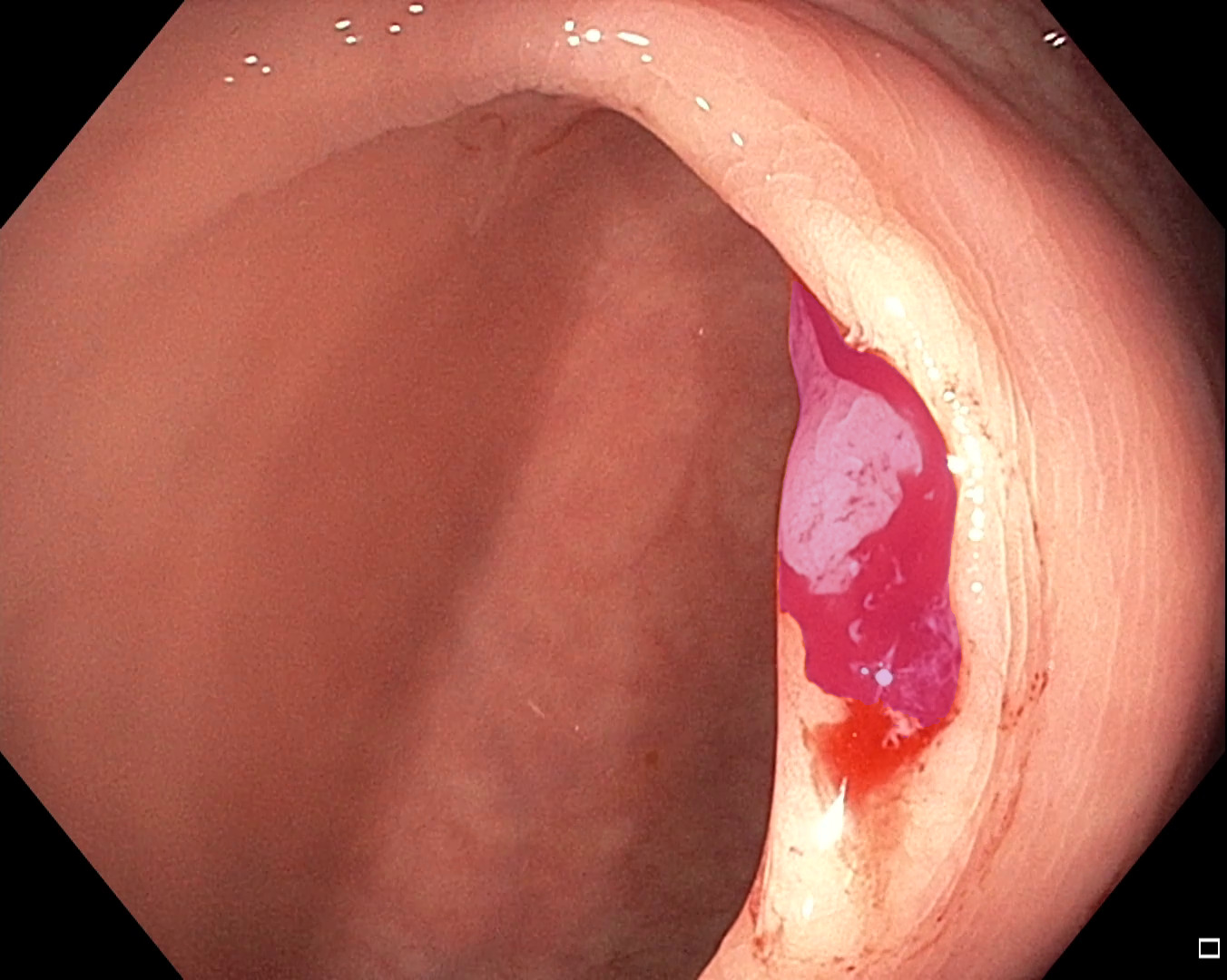} &
    \snap{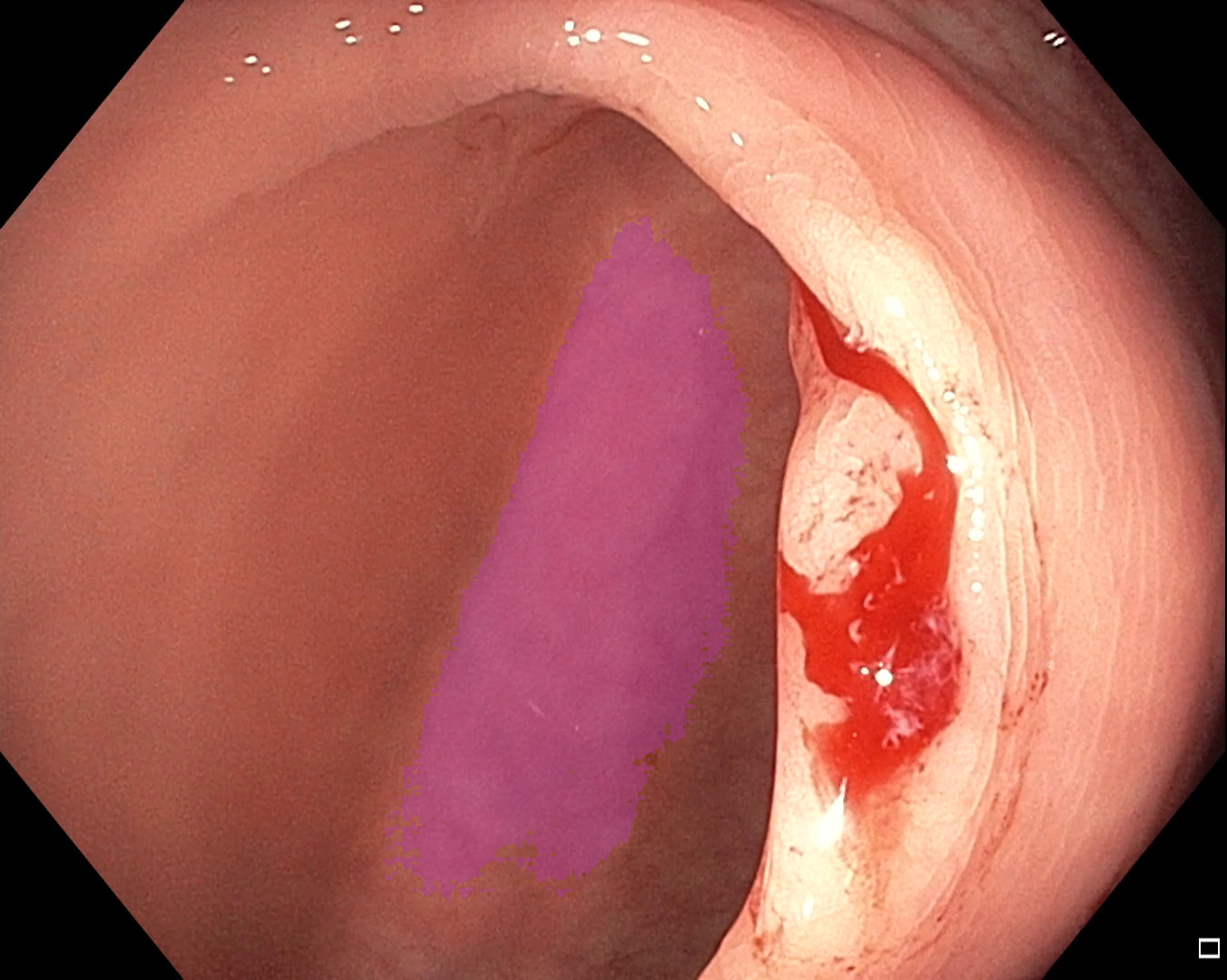} &
    \snap{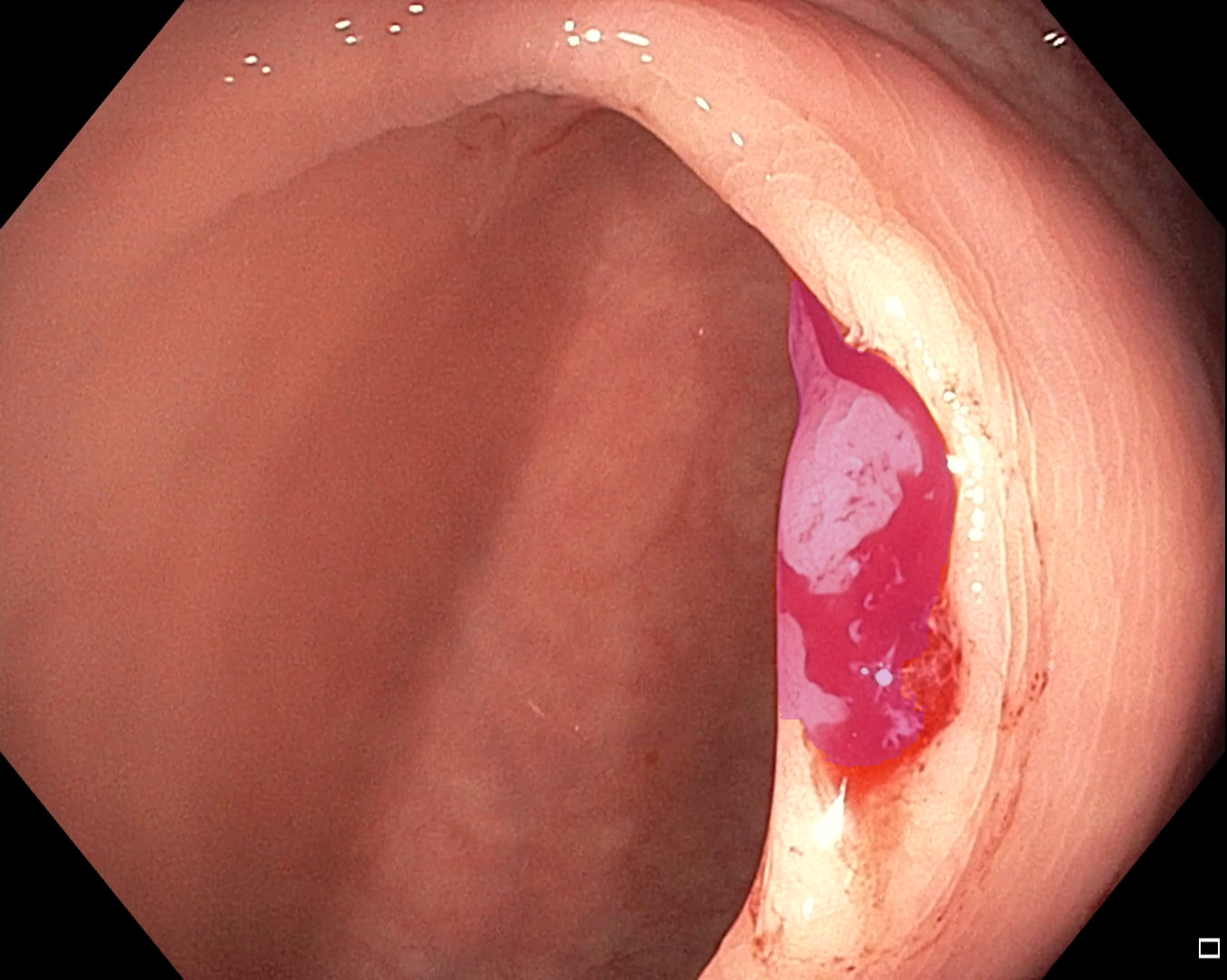} &
    \snap{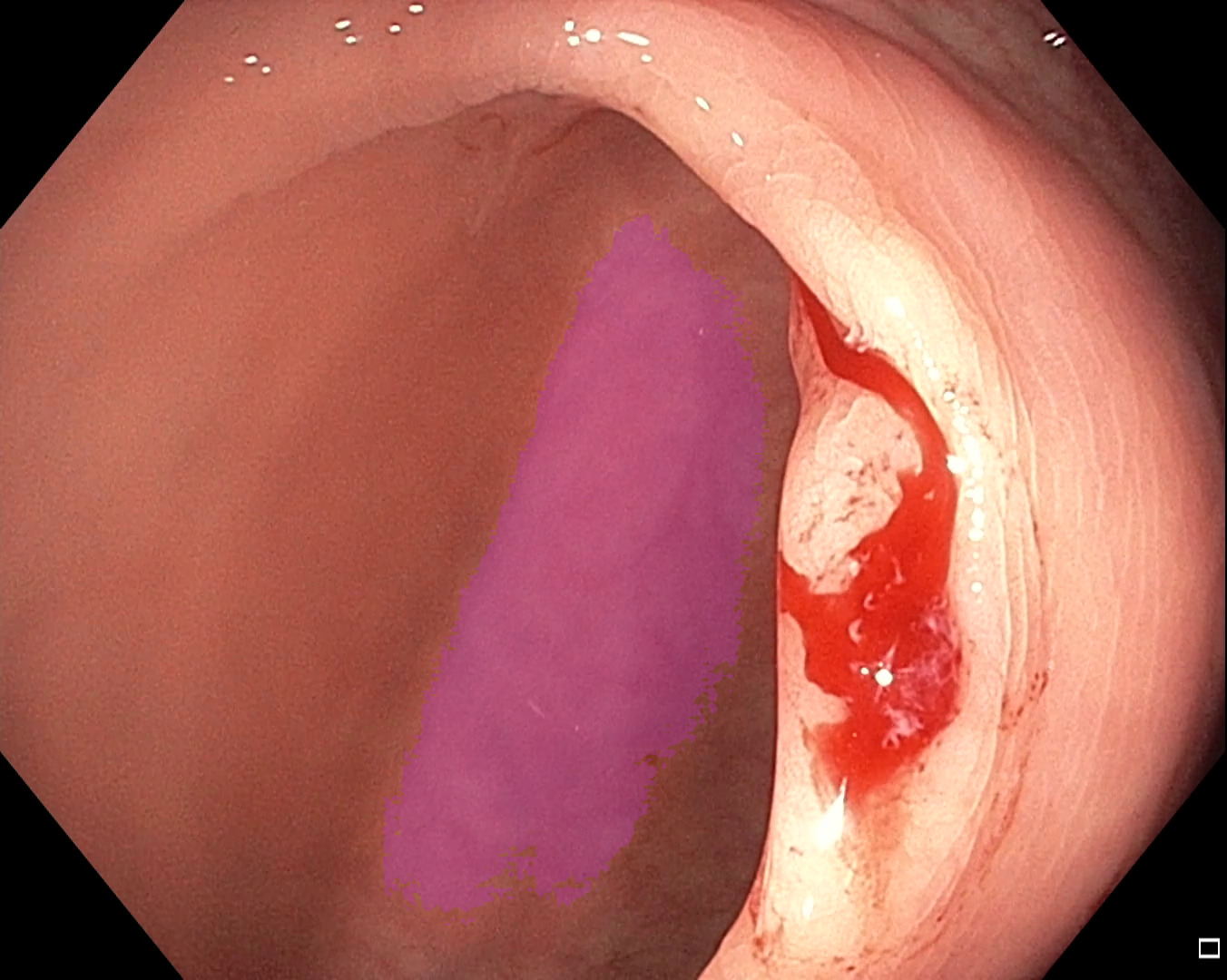} &
    \snap{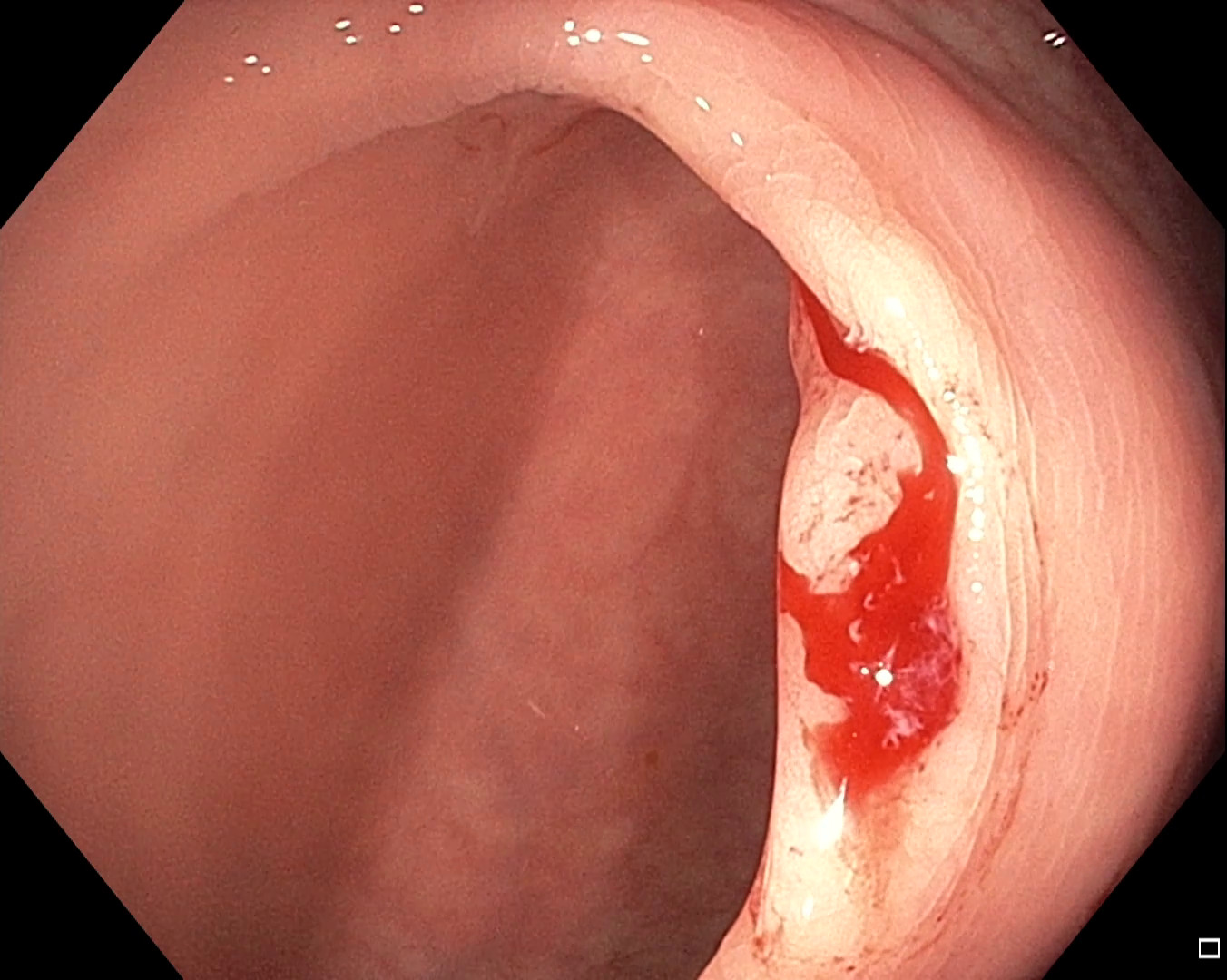} \\
  \end{tabular}}%
  \caption{\textbf{\methodname Qualitative Segmentation Comparison.} Each row shows a different endoscopy video frame:
    (1)~\textit{erythematous region},
    (2)~\textit{lipoma},
    (3)~\textit{sessile polyp},
    (4)~\textit{erosion},
    (5)~\textit{ulcer},
    (6)~\textit{erythematous region}.
    Columns show the raw input, ground-truth mask overlay, and predicted mask overlays from each model (including SAM\,3).
    Masks are rendered as semi-transparent purple overlays.%
  }
  \label{fig:qualitative-segmentation}
  \end{minipage}
\end{figure*}

\clearpage \clearpage
\section{Additional Analysis} \label{supsec:analysis}

\subsection{Ablation Study} \label{supsec:ablation}
\begin{figure}[h]
  \centering
  \includegraphics[width=0.55\linewidth]{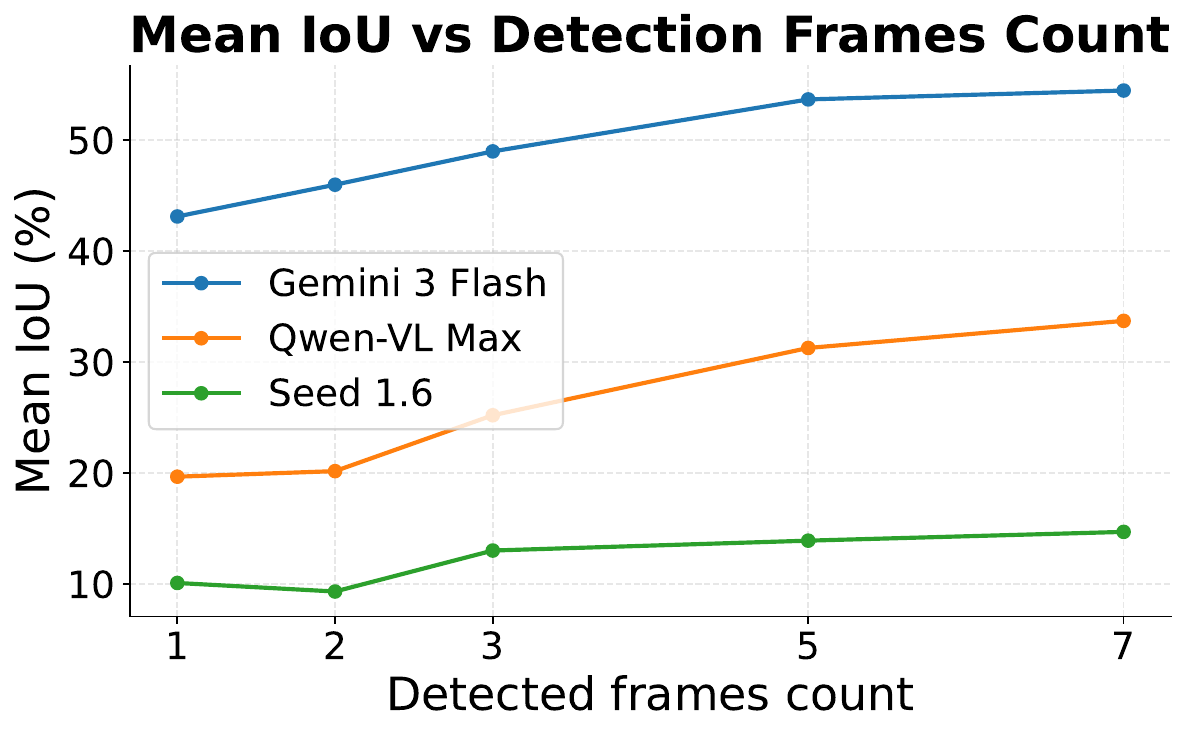}
  \caption{\textbf{Ablation Study on Lesion Segmentation.} We ablate the number of detection frames per windwo used in the segmentation benchmark, showing diminishing improvement in downstream segmentation quality.
  }
  \label{fig:ablation_frame_count_segmentation}
\end{figure}

We study four ablations with tabulated results in Tabs.~\ref{tab:exp1_frames_seg}-\ref{tab:exp4_temporal_vqa}. The detection frame count ablation shows steady gains in downstream segmentation quality, with Gemini~3 Flash improving mIoU/mDice as frames rise (Tab.~\ref{tab:exp1_frames_seg} and Fig.~\ref{fig:ablation_frame_count_segmentation}). The temporal context ablation for binary lesion classification compares video vs.\ single-frame inputs and finds temporal context is generally helpful, especially for Seed~1.6, while Gemini~3 Flash is slightly stronger on frames (Tab.~\ref{tab:exp2_temporal_cls}). The bounding-box annotation ablation for VQA shows only marginal changes, indicating limited dependence on explicit boxes (Tab.~\ref{tab:exp3_bbox_vqa}). The temporal context ablation for prompted VQA shows small, model-dependent shifts, with most models benefiting from video but a few showing minor gains on frames (Tab.~\ref{tab:exp4_temporal_vqa}).

\begin{table*}[h]
\centering
\caption{\textbf{\methodname Ablation: Detection Frames for Segmentation.} Each column pair reports Mean IoU and Mean Dice for the given number of detection frames. Best results per frame count are shown in \textbf{bold}.}
\label{tab:exp1_frames_seg}
\begin{tabular}{l c c c c c c c c c c}
\toprule
Model & \multicolumn{2}{c}{1 Frame} & \multicolumn{2}{c}{2 Frames} & \multicolumn{2}{c}{3 Frames} & \multicolumn{2}{c}{5 Frames} & \multicolumn{2}{c}{7 Frames} \\
 & mIoU & mDice & mIoU & mDice & mIoU & mDice & mIoU & mDice & mIoU & mDice \\
\midrule
Seed 1.6 & 10.1 & 12.4 & 9.3 & 11.8 & 13.0 & 16.6 & 13.9 & 17.9 & 14.7 & 19.2 \\
Qwen-VL Max & 19.7 & 22.6 & 20.2 & 23.4 & 25.2 & 29.3 & 31.3 & 36.3 & 33.7 & 39.4 \\
Gemini 3 Flash & 43.1 & 48.8 & 46.0 & 52.2 & 49.0 & 55.2 & 53.6 & 60.3 & 54.4 & 61.5 \\
\bottomrule
\end{tabular}
\end{table*}

\begin{table}[h]
\centering
\caption{\textbf{\methodname Ablation: Bounding-Box Effect on VQA.} ``With bbox'' uses annotated videos containing polyp bounding boxes, while ``Without bbox'' uses frame-generated videos of the same lesion windows without annotations (499 matched clip windows). $\Delta$ = Without $-$ With. Best results per condition are shown in \textbf{bold}.}
\label{tab:exp3_bbox_vqa}
\begin{tabular}{l c c c}
\toprule
Model & With bbox & W/o bbox & $\Delta$ \\
\midrule
Qwen-VL Max & 39.1 & 37.2 & -1.8 \\
Gemini 2.5 Flash Lite & 48.5 & 47.8 & -0.7 \\
Qwen3.5 397B & 49.0 & 48.1 & -0.9 \\
Gemini 3 Flash & 76.6 & 77.2 & +0.6 \\
\bottomrule
\end{tabular}
\end{table}

\begin{table}[h]
\centering
\caption{\textbf{\methodname Ablation: Temporal Context for Prompted VQA.} ``Video'' uses the full video clip and ``Frame'' uses a single representative frame over 1485 questions. $\Delta$ = Frame $-$ Video. Best results per condition are shown in \textbf{bold}.}
\label{tab:exp4_temporal_vqa}
\begin{tabular}{l c c c}
\toprule
Model & Video & Frame & $\Delta$ \\
\midrule
Qwen-VL Max & 37.5 & 34.1 & -3.4 \\
Gemini 2.5 Flash Lite & 46.7 & 48.0 & +1.3 \\
Qwen 3.5 397b & 49.6 & 42.6 & -7.0 \\
Gemini 3 Flash & 76.8 & 74.3 & -2.5 \\
\bottomrule
\end{tabular}
\end{table}

\begin{table}[h]
\centering
\caption{\textbf{\methodname Ablation: Temporal Context for Classification.} ``Video'' uses the full video clip and ``Frame'' uses a single representative frame. $\Delta$ = Frame $-$ Video. Best results per condition are shown in \textbf{bold}.}
\label{tab:exp2_temporal_cls}
\begin{tabular}{l c c c c c c}
\toprule
 & \multicolumn{3}{c}{Accuracy (\%)} & \multicolumn{3}{c}{F1 Score (\%)} \\
\cmidrule(lr){2-4} \cmidrule(lr){5-7}
Model & Video & Frame & $\Delta$ & Video & Frame & $\Delta$ \\
\midrule
Qwen VL Max & 65.6 & 49.6 & -15.9 & 0.0 & 0.0 & 0.0 \\
Gemini 3 Flash & 69.1 & 70.8 & +1.6 & 68.6 & 69.5 & +0.9 \\
Seed 1.6 & 82.4 & 60.0 & -22.4 & 69.9 & 60.8 & -9.1 \\
\bottomrule
\end{tabular}
\end{table}

\subsection{VQA Question Aspect Distribution}
\label{sec:vqa_aspect_distribution}

To characterize \emph{what} our VQA questions ask about, we categorize each question by the clinical/visual aspect it probes: lesion type, morphology, size,color/surface appearance, anatomical location, temporal reasoning, and procedural action/instrument use. Because a single question frequently targets multiple aspects (e.g., asking about both the morphology and color of a lesion), we adopt a \emph{multi-label} scheme based on keyword matching over the question text; questions matching no category are assigned to \textit{Others}. On average
this yields $2.11$ and $1.65$ labels per question for the prompted ($1{,}485$ questions) and unprompted ($2{,}740$ questions) splits, respectively.

\begin{figure}[t]
  \centering
  \includegraphics[width=0.8\linewidth]{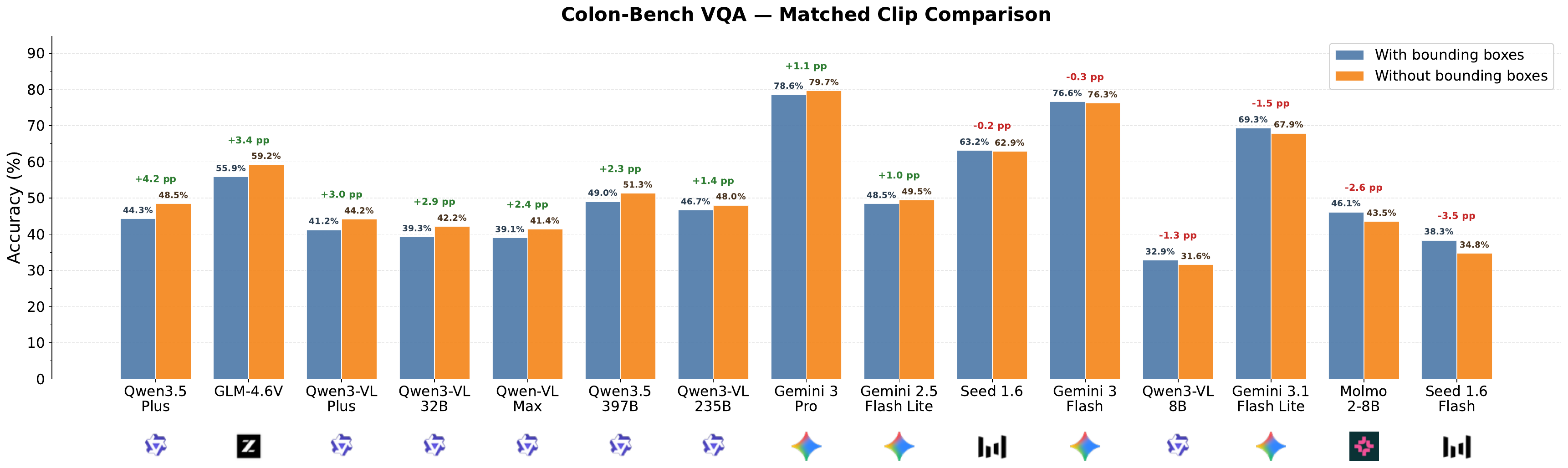}
  \caption{%
    \textbf{VQA Question Aspect Distribution.} Distribution of clinical/visual aspects probed by the VQA questions in \methodname, shown for the promptedand unprompted splits. Questions are multi-label, so a single question may contribute to several aspects.%
  }
  \label{fig:vqa_aspect_distribution}
\end{figure}

Figure~\ref{fig:vqa_aspect_distribution} shows that both splits are dominated by the same core aspects, lesion type ($39\%$/$43\%$, prompted/unprompted), color/surface appearance ($38\%$/$44\%$), and anatomical location ($30\%$/$35\%$), followed by morphology ($20\%$/$20\%$), size ($11\%$/$9\%$), and procedural action ($12\%$/$6\%$). The most pronounced difference is in temporal reasoning: explicit temporal references occur in $59\%$ of prompted questions but are almost entirely absent ($2\%$) from the unprompted split by design, since unprompted questions deliberately omit time cues to force the model to ground its answer in the video itself. The low \textit{Others} rate ($1.5\%$/$6.7\%$) confirms that the taxonomy covers the large majority of questions. Overall, this confirms that \methodname spans a broad and clinically meaningful range of question types rather than concentrating on a single skill.

\subsection{Colon-Skill for MLLMs}
\begin{figure}[htbp]
    \centering
    \includegraphics[width=0.99\linewidth]{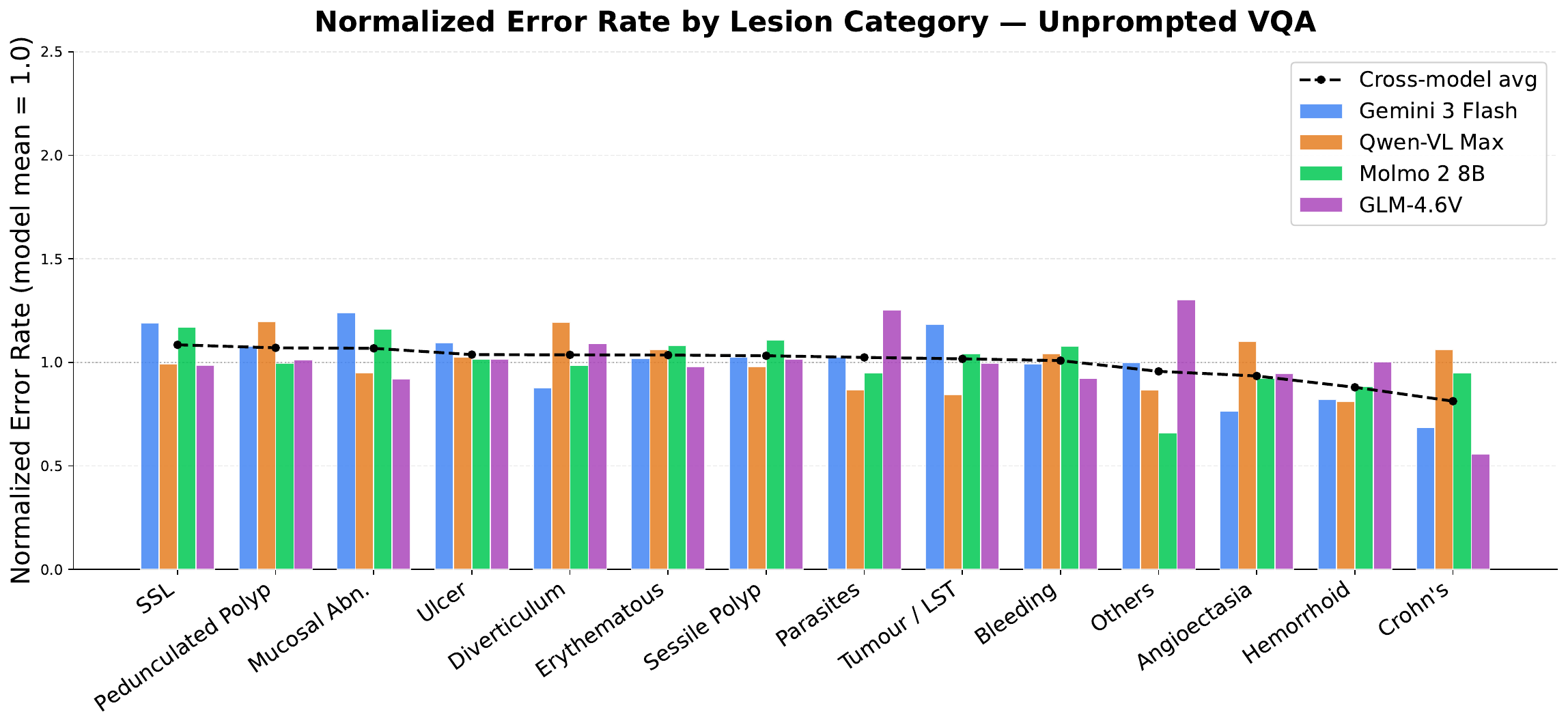}
    \caption{\textbf{Error Patterns: Unprompted Split.} Normalized error rate by lesion category (1.0 = model average).}
    \label{fig:exp5_error_hard}
\end{figure}

\begin{figure}[]
    \centering
    \includegraphics[width=0.99\linewidth]{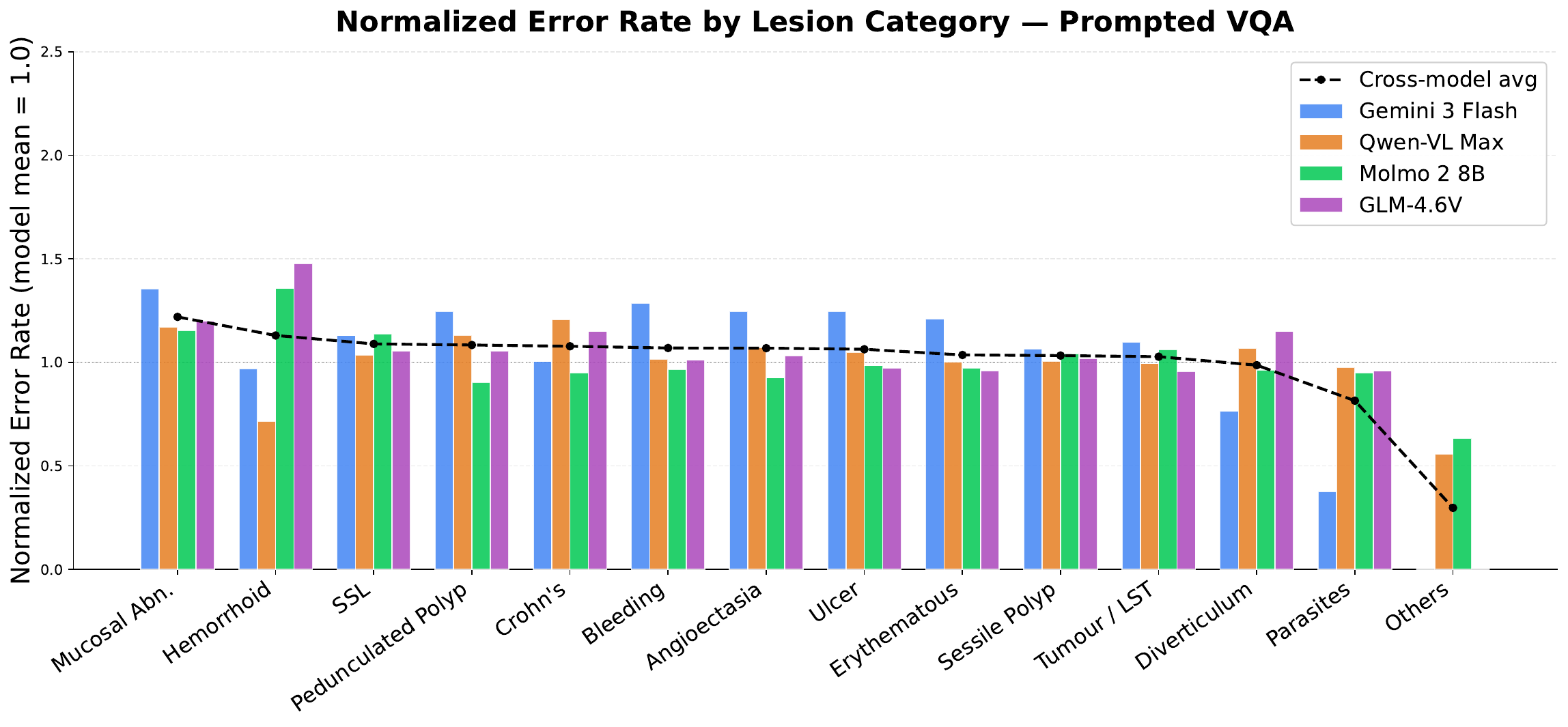}
    \caption{\textbf{Error Patterns: Prompted Split.} Normalized error rate by lesion category (1.0 = model average).}
    \label{fig:exp5_error_easy}
\end{figure}

We investigate whether MLLMs benefit from structured domain knowledge at inference time. We first collect per-model VQA predictions and stratify errors by lesion category (Figs.~\ref{fig:exp5_error_hard} and~\ref{fig:exp5_error_easy}), retaining questions that a majority of models answer incorrectly. A frontier LLM then synthesises these shared failure cases into a concise natural-language \emph{Colon-Skill} comprising morphological cues, common confusion traps, and a decision checklist. This skill context is prepended to every VQA prompt at inference time.

\begin{figure}[]
    \centering
    \includegraphics[width=0.99\linewidth]{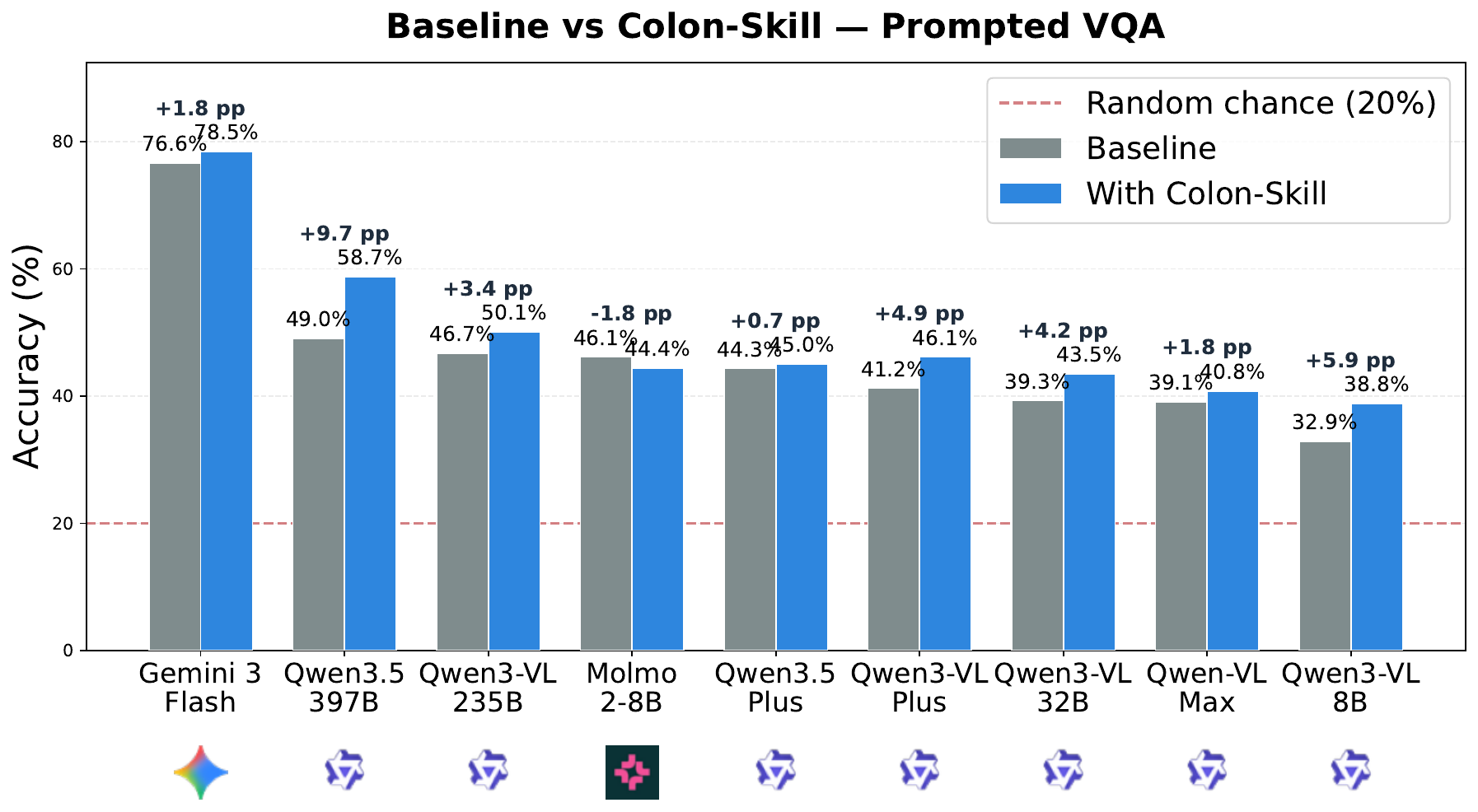}
    \caption{\textbf{Colon-Skill Effect: Prompted VQA.} Effect of skill context on VQA performance (\textit{prompted} split).}
    \label{supfig:exp6_skill_context_easy}
\end{figure}

\begin{figure}[]
    \centering
    \includegraphics[width=0.99\linewidth]{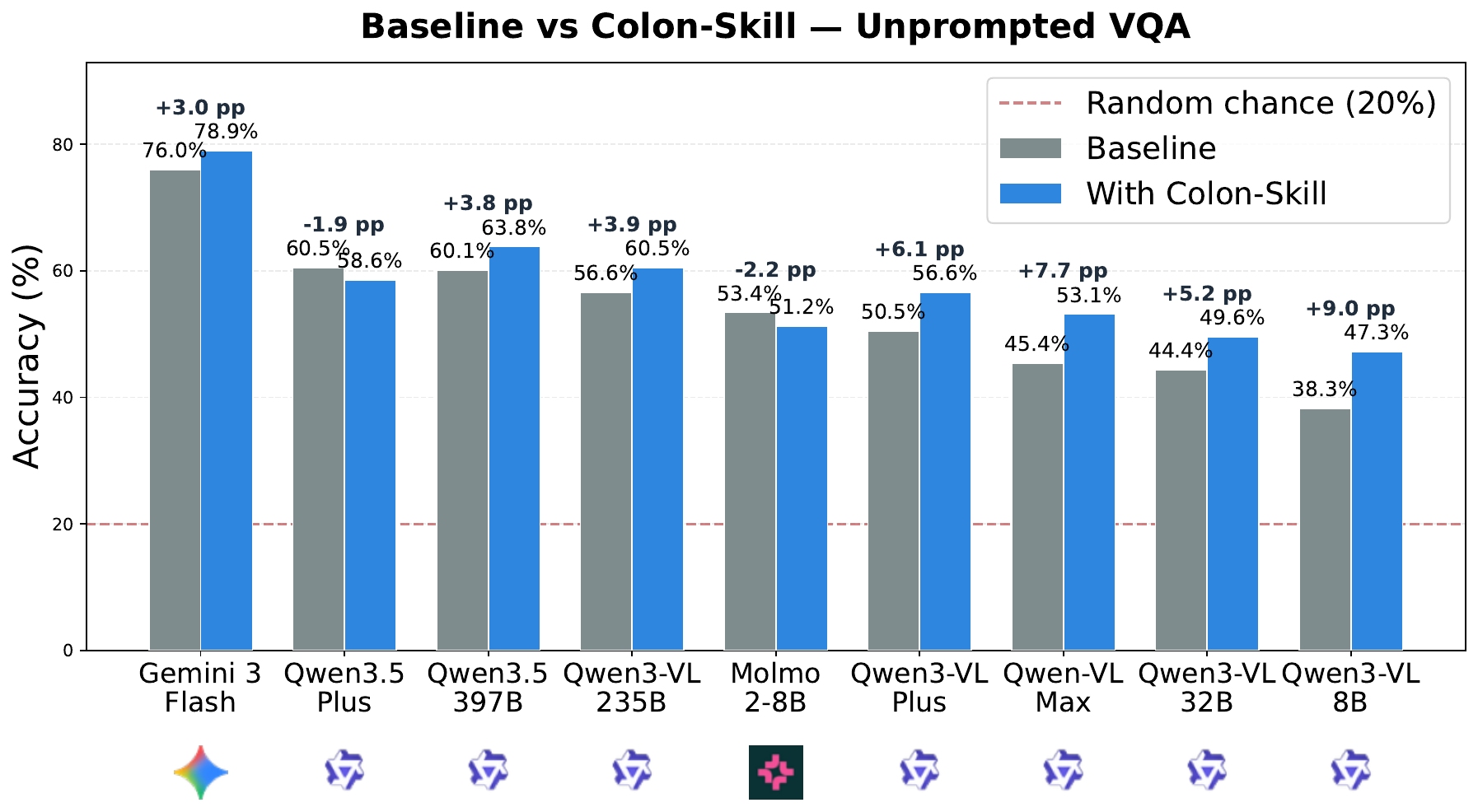}
    \caption{\textbf{Colon-Skill Effect: Unprompted VQA.} Effect of skill context on VQA performance (\textit{unprompted} split).}
    \label{supfig:exp6_skill_context_hard}
\end{figure}

Across nine MLLMs, skill-augmented prompting yields gains of up to +9.7\% (prompted) and +9.0\% (unprompted), with the effect most pronounced for higher-capacity models (Table~\ref{tab:exp6_skill_context_vqa_combined}, Figs.~\ref{supfig:exp6_skill_context_easy} and~\ref{supfig:exp6_skill_context_hard}). Smaller models such as Molmo-2-8B show marginal or slightly negative deltas, suggesting sufficient model capacity is needed to integrate the additional context.

\begin{figure}[t]
\begin{tcolorbox}[
  colback=gray!5,
  colframe=gray!60,
  fonttitle=\bfseries\small,
  title=Colon-Skill,
  boxrule=0.4pt,
  arc=2pt,
  left=4pt, right=4pt, top=2pt, bottom=2pt,
  fontupper=\ttfamily\tiny
]
\begin{lstlisting}[
  basicstyle=\ttfamily\tiny,
  breaklines=true,
  columns=fullflexible,
  keepspaces=true
]
## Universal Anti-Error Rules

*   **Do not hallucinate stalks:** The most common morphological error is calling a sessile (broad-based) polyp "pedunculated." If the lesion sits flatly on a haustral fold without a distinct, narrowed fibrous tether, it is a sessile or flat-elevated lesion.
*   **Correctly interpret NBI (Narrow-Band Imaging):** Under NBI, healthy mucosa appears greenish/cyan. Adenomas typically appear brownish with regular pit patterns, while hyperplastic or sessile serrated lesions (SSLs) appear pale or whitish. Do not confuse white-light colors with NBI colors.
*   **Differentiate holes from masses:** Models consistently mistake diverticula (dark, hollow outpouchings) for depressed flat lesions or dark polyps. If a finding is a perfectly circular dark shadow with smooth margins, it is a hole/pocket, not a lesion.
*   **Be conservative with size estimates:** Models routinely overestimate diminutive lesions. Without tool reference, subtle, dome-shaped nodules on folds are usually 3-5mm (diminutive). Lesions occupying a large portion of the lumen are >15mm.
*   **Accurately identify interventions:**
    *   *Water jet:* Used for irrigation/clearing mucus or blood (not for coagulation).
    *   *Needle catheter:* Injects fluid to create a blue submucosal cushion (lifts flat lesions).
    *   *Cold/Hot Snare:* Wire loop used for Endoscopic Mucosal Resection (EMR) or polypectomy.

## Lesion Morphology Cues by Category

*   **Sessile Polyps (Paris Is / IIa):** Broad base, typically situated on the crest of a haustral fold. Often pale, isochromatic (matches mucosa color), or slightly yellowish. Surface is smooth or subtly granular.
*   **Sessile Serrated Lesions (SSL):** Flat, highly subtle, pale/isochromatic lesions. Key visual signatures include a "cloud-like" surface texture, indistinct/blurred borders, and a distinct "mucus cap" (adherent yellow/white debris).
*   **Angioectasia / Angiodysplasia:** Sits completely flush with the mucosa. Appears as a bright, cherry-red, "fern-like" or stellate pattern of tortuous, dilated submucosal blood vessels.
*   **Pedunculated Polyps (Paris Ip):** Features a distinct, thick or thin stalk attached to the colonic wall. The head is usually bulbous, multi-lobulated, and noticeably redder (erythematous) than the stalk.
*   **Ulcers:** Deep or shallow punched-out depressions. Look for white/yellow necrotic slough or fibrin exudate in the center, surrounded by raised, rolled, and erythematous margins.
*   **Diverticula:** Distinct, dark, circular pockets with smooth, well-defined rims. Often seen in clusters.

## Common Confusion Traps and How to Resolve Them

*   **Trap: Angioectasia vs. Suction Artifact vs. Dieulafoy.** Models overwhelmingly fail here. *Resolution:* Suction artifacts are tiny, non-specific red petechial spots. Dieulafoy/ulcers involve tissue defects or active arterial spurts. Angioectasia is a flat network of *fern-like*, branching dilated vessels without a mucosal defect.
*   **Trap: Diverticulum vs. Depressed Lesion (Paris IIc).** *Resolution:* A diverticulum is a true anatomical hole (deep black center, hollow). A Paris IIc lesion is a mucosal depression with a visible base, often showing irregular margins and altered pit patterns.
*   **Trap: Mischaracterizing pale flat lesions.** Models frequently call pale, mucus-covered flat lesions "ulcers" or "tumors." *Resolution:* If a flat lesion is pale with a "cloud-like" surface and lacks rolled margins or central necrosis, it is an SSL, not an ulcer or malignant tumor.
*   **Trap: "Lobulated" vs. "Smooth" surface confusion.** *Resolution:* "Smooth" dome-shaped pale nodules are typically diminutive hyperplastic or small sessile polyps. "Lobulated" or "cerebriform" (brain-like) surfaces with redder hues indicate adenomatous polyps.
*   **Trap: Mistaking anatomical folds for masses.** *Resolution:* Prominent haustral folds curve predictably around the lumen. Do not label a normal fold as a "sessile mass" unless there is a localized change in color, vascularity, or elevation (nodularity).

## Fast VQA Decision Checklist

1.  **Assess the geometry:** Is it a mass (protruding), a defect (depressed/ulcerated), a hole (diverticulum), or a flat vascular anomaly?
2.  **Evaluate the attachment:** If it's a polyp, is it draped over/broadly attached to a fold (sessile) or hanging by a tether (pedunculated)?
3.  **Check the lighting mode:** Is the image under white light (pink/red hues) or NBI/BLI (green/brown/cyan hues)? Adjust color descriptions accordingly.
4.  **Examine the surface & margins:** Is the surface smooth, cloud-like, granular, or lobulated? Are the margins sharp, rolled, or indistinct?
5.  **Identify active tools or residue:** Are there white/yellow strings (mucus/stool), active red oozing (bleeding), a wire loop (snare), or blue fluid (submucosal injection)?
\end{lstlisting}
\end{tcolorbox}
\caption{\textbf{Colon-Skill Prompt Context.} The colon-skill \textit{SKILL.md} file is used as context augmentation for MLLMs in the VQA benchmarks. The skill is extracted by analysing error patterns across lesion categories and examples of failure modes.}
\end{figure}

\begin{figure*}[t]
\begin{tcolorbox}[
  colback=gray!5,
  colframe=gray!60,
  fonttitle=\bfseries\small,
  title=Automated Annotation Pipeline Prompts,
  boxrule=0.4pt,
  arc=2pt,
  left=4pt, right=4pt, top=2pt, bottom=2pt,
  fontupper=\ttfamily\tiny
]
\begin{lstlisting}[
  basicstyle=\ttfamily\tiny,
  breaklines=true,
  columns=fullflexible,
  keepspaces=true
]
## Stage 1 - Temporal Lesion-Window Detection (video clip)

You are an expert colorectal surgeon analyzing a colonoscopy video. This is clip [k] of [N] from a ~1-hour procedure. The clip was encoded at [FPS] FPS and is approximately [DURATION] seconds long.

TASK: Analyze the ENTIRE clip and divide it into 1-[MAX] time windows based on what you observe. Each window should cover several seconds where similar content/findings occur. Windows must be consecutive and span the full clip duration.

OUTPUT FORMAT: Return a JSON array with one object per time window, each with:
- start time (float): start in SECONDS; first window starts at 0.0
- end time (float): end in SECONDS; last window ends near the clip end
- lesion (int): 0 = normal/healthy tissue, 1 = lesion or abnormality detected
- description (string): brief description of observations in this window

IMPORTANT: Timestamps are in SECONDS, not minutes (e.g. 30 seconds = 30.0, not 0.30). Output ONLY valid JSON (no comments or extra text).

## Stage 2 - Lesion Verification and First-Appearance Timestamp (video clip)

You are an expert colorectal surgeon reviewing a colonoscopy video segment. This clip is [DURATION] seconds long.

TASK: Watch the ENTIRE video carefully. If there is a lesion, polyp, or abnormality:
1. Identify the EXACT timestamp (in seconds from start) when the lesion FIRST becomes clearly visible.
2. Provide a detailed description of the lesion including size, shape, color, location, and any notable features.

IMPORTANT: The timestamp should mark when the lesion appears and becomes clearly visible. Do not describe other events (e.g. instruments, movements) except for the lesion.

ANSWER FORMAT:
- If lesion found:  YES <timestamp_in_seconds> | <detailed lesion description>
- If no lesion found:  NO | <what you see instead>

## Stage 3 - Per-Frame Lesion Localization (single frame)

You are an expert colorectal surgeon. This is frame [i] out of [N] frames from a colonoscopy video.
Previous analysis identified: <lesion description from verification>

TASK: Locate and highlight all visible lesions, polyps, or abnormalities with bounding boxes.

ANSWER FORMAT:
- If lesion(s) found: return bounding box coordinates as [x_min, y_min, x_max, y_max] in pixels from the top-left corner.
- Example (lesion):     FOUND | 120,80,250,200 | Sessile polyp
- Example (no lesion):  NONE | No clear lesion visible in this frame

## Stage 4 - Lesion Confirmation on Tracked Clips (box-overlaid video)

You are an expert colorectal surgeon reviewing a colonoscopy video segment for lesion confirmation. This video shows RED BOUNDING BOXES overlaid on the frames, indicating where the automated system has detected and tracked a potential lesion.

IMPORTANT: Determine if a TRUE LESION exists anywhere in this segment. The box location may not be perfectly accurate; do NOT base your decision on whether the box precisely covers the lesion. If you see a lesion (even if the box is slightly off or has drifted), CONFIRM it and note any box inaccuracy.

CONFIDENCE THRESHOLD: Confirm only if you are at least 80%

DECISION CRITERIA:
- CONFIRMED: a true lesion (polyp, diverticulum, inflammation, abnormal growth, etc.) is visible with >=80%
- REJECTED: no lesion is visible or confidence is below 80%

ANSWER FORMAT:
- If lesion visible:    CONFIRMED | <brief clinical description; note if box location is inaccurate>
- If no lesion visible: REJECTED | <explanation of what is actually shown>
\end{lstlisting}
\end{tcolorbox}
\caption{\textbf{Automated Annotation Pipeline Prompts.} Core prompts driving the multi-stage lesion annotation pipeline. A vision--language model first (1) segments each colonoscopy clip into coarse temporal windows and flags candidate lesions, then (2) verifies each window and timestamps the lesion's first clear appearance, (3) localizes the lesion with frame-level bounding boxes that seed the tracker, and (4) confirms the tracked, box-overlaid clips with a conservative confidence threshold. Internal variable names and formatting tokens are abstracted as bracketed placeholders for readability.}
\end{figure*}

\begin{table*}[h]
\centering
\caption{\textbf{Colon-Skill VQA Gains.} Effect of skill-context prompting on VQA accuracy (\%) across Unprompted (2740 questions) and Prompted (1485 questions) splits. Baseline uses the default prompt; With Skill appends a distilled skill context. $\Delta$ is the absolute change in percentage points. Best results per split are shown in \textbf{bold}.}
\label{tab:exp6_skill_context_vqa_combined}
\begin{tabular}{l c c c c c c}
\toprule
& \multicolumn{3}{c}{\textbf{Unprompted}} & \multicolumn{3}{c}{\textbf{Prompted}} \\
\cmidrule(lr){2-4} \cmidrule(lr){5-7}
\textbf{Model} & \textbf{Baseline} & \textbf{w/ Skill } & \textbf{$\Delta$} & \textbf{Baseline } & \textbf{w/ Skill } & \textbf{$\Delta$} \\
\midrule
Gemini 3 Flash & 76.0 & 78.9 & +3.0 & 76.6 & 78.5 & +1.8 \\
Qwen 3.5 Plus & 60.5 & 58.6 & -1.9 & 44.3 & 45.0 & +0.7 \\
Qwen 3.5 397B & 60.1 & 63.8 & +3.8 & 49.0 & 58.7 & \textbf{+9.7} \\
Qwen 3 VL 235B & 56.6 & 60.5 & +3.9 & 46.7 & 50.1 & +3.4 \\
Molmo 2 8B & 53.4 & 51.2 & -2.2 & 46.1 & 44.4 & -1.8 \\
Qwen 3 VL Plus & 50.5 & 56.6 & +6.1 & 41.2 & 46.1 & +4.9 \\
Qwen VL Max & 45.4 & 53.1 & +7.7 & 39.1 & 40.8 & +1.8 \\
Qwen 3 VL 32B & 44.4 & 49.6 & +5.2 & 39.3 & 43.5 & +4.2 \\
Qwen 3 VL 8B & 38.3 & 47.3 & \textbf{+9.0} & 32.9 & 38.8 & +5.9 \\
\bottomrule
\end{tabular}
\end{table*}

\end{document}